\documentclass[a4paper]{article}
\usepackage[latin1]{inputenc}
\usepackage[T1]{fontenc}
\usepackage[italian,french,english]{babel}
\usepackage{amsmath}
\usepackage{amssymb,amsfonts,textcomp}
\usepackage{array}
\usepackage{supertabular}
\usepackage{hhline}
\usepackage{hyperref}
\usepackage[nottoc,numbib]{tocbibind}
\usepackage{subscript} 
\usepackage{comment}
\usepackage{graphicx}
\usepackage{txfonts}
\usepackage{pdfpages}
\usepackage{authblk}
\usepackage{gensymb}
\usepackage{fancyhdr}
\newcommand{\version}[1]{Version {#1}}

\usepackage[all]{background}
\SetBgContents{Living document}
\SetBgScale{3}
\SetBgOpacity{1}
\SetBgVshift{4mm}

\SetBgAngle{90}
\SetBgPosition{current page.east}

\usepackage[disable]{todonotes}

\hypersetup{
    colorlinks,
    linkcolor={blue},
    citecolor={blue},
    urlcolor={green}
}
\usepackage{tocloft}

\newenvironment{notes}{\sffamily\color{blue}}


\newcommand\textstyleInternetlink[1]{\textcolor{blue}{#1}}

\makeatletter
\newcommand\arraybslash{\let\\\@arraycr}
\makeatother
\makeatletter
\newcommand\ps@Conversioni{
  \renewcommand\@oddhead{}
  \renewcommand\@evenhead{\@oddhead}
  \renewcommand\@oddfoot{}
  \renewcommand\@evenfoot{\@oddfoot}
  \renewcommand\thepage{\arabic{page}}
}
\newcommand\ps@Standard{
  \renewcommand\@oddhead{}
  \renewcommand\@evenhead{\@oddhead}
  \renewcommand\@oddfoot{}
  \renewcommand\@evenfoot{}
  \renewcommand\thepage{\arabic{page}}
}
\makeatother
\newcounter{Table}

\newcounter{Figure}

\title{QUBIC Technical Design Report \\ \Large The QUBIC collaboration}
\date{\today}
\graphicspath{{.}{./QUBICTDRcompilation-img/}}


\renewcommand{\contentsname}{Whatever}
\newenvironment{incomplete}{\sffamily\color{orange}\large \bf }

\author[7]{J. Aumont}
\author[13]{S. Banfi}
\author[14]{P. Battaglia}
\author[17]{E.S. Battistelli}
\author[13]{A. Ba\`u}
\author[8]{B. B\'elier}
\author[15]{D.Bennett}
\author[5]{L. Berg\'e}
\author[9]{J.Ph. Bernard}
\author[14]{M. Bersanelli}
\author[1]{M.A. Bigot-Sazy}
\author[1]{N. Bleurvacq}
\author[1]{G. Bordier}
\author[1]{J. Brossard}
\author[16]{E.F. Bunn}
\author[17]{D. Buzi}
\author[18]{A. Buzzelli}
\author[1]{D. Cammilleri}
\author[14]{F. Cavaliere}
\author[1]{P. Chanial}
\author[1]{C. Chapron}
\author[12]{G. Coppi}
\author[17]{A. Coppolecchia}
\author[11]{F. Couchot}
\author[18]{R. D'Agostino}
\author[17]{G. D'Alessandro}
\author[17]{P. de Bernardis}
\author[18]{G. De Gasperis}
\author[17]{M. De Petris}
\author[1]{T. Decourcelle}
\author[14]{F. Del Torto}
\author[5]{L. Dumoulin}
\author[10]{A. Etchegoyen}
\author[13]{C. Franceschet}
\author[10]{B. Garcia}
\author[18]{A. Gault}
\author[15]{D. Gayer}
\author[13]{M. Gervasi}         
\author[1]{A. Ghribi}
\author[9]{M. Giard}
\author[1]{Y. Giraud-H\'eraud}
\author[15]{M. Gradziel}
\author[1]{L. Grandsire}
\author[1]{J.Ch. Hamilton}
\author[3]{D. Harari}
\author[12]{V. Haynes}
\author[11]{S. Henrot-Versill\'e}
\author[5]{N. Holtzer}
\author[1]{J. Kaplan}
\author[2]{A. Korotkov}
\author[17]{L. Lamagna}
\author[5]{J. Lande}
\author[1]{S. Loucatos}
\author[19]{A. Lowitz}
\author[18]{V. Lukovic}
\author[7]{B. Maffei}
\author[5]{S. Marnieros}
\author[7]{J. Martino}
\author[17]{S. Masi}
\author[12]{A. May}
\author[12]{M. McCulloch}
\author[6]{M.C. Medina}
\author[12]{S. Melhuish}
\author[14]{A. Mennella}
\author[9]{L. Montier}
\author[15]{A. Murphy}
\author[5]{D. N\'eel}
\author[12]{M.W. Ng}
\author[15]{C. O'Sullivan}
\author[17]{A. Paiella}
\author[9]{F. Pajot}
\author[13]{A. Passerini}
\author[17]{A.Pelosi}
\author[1]{C. Perbost}
\author[11]{O. Perdereau}
\author[17]{F. Piacentini}
\author[1]{M. Piat}
\author[12]{L. Piccirillo}
\author[4]{G. Pisano}
\author[1]{D. Pr\^ele}
\author[17]{R. Puddu}
\author[9]{D. Rambaud}
\author[5]{O. Rigaut}
\author[6]{G.E. Romero}
\author[17]{M. Salatino}
\author[17]{A. Schillaci}
\author[15]{S. Scully}
\author[1]{M. Stolpovskiy}
\author[10]{F. Suarez}
\author[1]{A. Tartari}
\author[19]{P. Timbie}
\author[11]{M. Tristram}
\author[2]{G. Tucker}
\author[14]{D. Vigan\`o}
\author[18]{N. Vittorio}
\author[1]{F. Voisin}
\author[12]{B. Watson}
\author[13]{M. Zannoni}
\author[17]{A. Zullo}
\affil[1]{APC, Paris, France}
\affil[2]{Brown University, Providence, RI, USA}
\affil[3]{Centro Atomico Bariloche, CNEA, Argentina}
\affil[4]{Cardiff University, Cardiff, UK}
\affil[5]{CSNSM, Orsay, France}
\affil[6]{IAR-CONICET, CCT-La Plata, UNLP, Argentina}
\affil[7]{IAS, Orsay, France}
\affil[8]{IEF, Orsay, France}
\affil[9]{IRAP, Toulouse, France}
\affil[10]{ITeDA, CNEA, CONICET, UNSAM, Argentina}
\affil[11]{LAL, Orsay, France}
\affil[12]{University of Manchester, Manchester, UK}
\affil[13]{Universit\'a degli Studi di Milano-Bicocca, Milano, Italy}
\affil[14]{Universit\'a Degli Studi di Milano, Milano, Italy}
\affil[15]{NUIM, Maynooth, Ireland}
\affil[16]{Richmond University, Richmond, VA, USA}
\affil[17]{Universit\'a di Roma La Sapienza, Roma, Italy}
\affil[18]{Universit\'a di Roma Tor Vergata, Roma, Italy}
\affil[19]{University of Wisconsin, Madison, WI, USA}

\begin{document}
%auto-ignore
\def\aj{AJ}%
          % Astronomical Journal
\def\actaa{Acta Astron.}%
          % Acta Astronomica
\def\araa{ARA\&A}%
          % Annual Review of Astron and Astrophys
\def\apj{ApJ}%
          % Astrophysical Journal
\def\apjl{ApJ}%
          % Astrophysical Journal, Letters
\def\apjs{ApJS}%
          % Astrophysical Journal, Supplement
\def\ao{Appl.~Opt.}%
          % Applied Optics
\def\apss{Ap\&SS}%
          % Astrophysics and Space Science
\def\aap{A\&A}%
          % Astronomy and Astrophysics
\def\aapr{A\&A~Rev.}%
          % Astronomy and Astrophysics Reviews
\def\aaps{A\&AS}%
          % Astronomy and Astrophysics, Supplement
\def\azh{AZh}%
          % Astronomicheskii Zhurnal
\def\baas{BAAS}%
          % Bulletin of the AAS
\def\bac{Bull. astr. Inst. Czechosl.}%
          % Bulletin of the Astronomical Institutes of Czechoslovakia
\def\caa{Chinese Astron. Astrophys.}%
          % Chinese Astronomy and Astrophysics
\def\cjaa{Chinese J. Astron. Astrophys.}%
          % Chinese Journal of Astronomy and Astrophysics
\def\icarus{Icarus}%
\def\jcap{J. Cosmology Astropart. Phys.}%
          % Journal of Cosmology and Astroparticle Physics
\def\jrasc{JRASC}%
          % Journal of the RAS of Canada
\def\mnras{MNRAS}%
          % Monthly Notices of the RAS
\def\memras{MmRAS}%
          % Memoirs of the RAS
\def\na{New A}%
          % New Astronomy
\def\nar{New A Rev.}%
          % New Astronomy Review
\def\pasa{PASA}%
          % Publications of the Astron. Soc. of Australia
\def\pra{Phys.~Rev.~A}%
          % Physical Review A: General Physics
\def\prb{Phys.~Rev.~B}%
          % Physical Review B: Solid State
\def\prc{Phys.~Rev.~C}%
          % Physical Review C
\def\prd{Phys.~Rev.~D}%
          % Physical Review D
\def\pre{Phys.~Rev.~E}%
          % Physical Review E
\def\prl{Phys.~Rev.~Lett.}%
          % Physical Review Letters
\def\pasp{PASP}%
          % Publications of the ASP
\def\pasj{PASJ}%
          % Publications of the ASJ
\def\qjras{QJRAS}%
          % Quarterly Journal of the RAS
\def\rmxaa{Rev. Mexicana Astron. Astrofis.}%
          % Revista Mexicana de Astronomia y Astrofisica
\def\skytel{S\&T}%
\def\sovast{Soviet~Ast.}%
          % Soviet Astronomy
\def\ssr{Space~Sci.~Rev.}%
          % Space Science Reviews
\def\zap{ZAp}%
          % Zeitschrift fuer Astrophysik
\def\nat{Nature}%
          % Nature
\def\iaucirc{IAU~Circ.}%
          % IAU Cirulars
\def\aplett{Astrophys.~Lett.}%
          % Astrophysics Letters
\def\apspr{Astrophys.~Space~Phys.~Res.}%
          % Astrophysics Space Physics Research
\def\bain{Bull.~Astron.~Inst.~Netherlands}%
          % Bulletin Astronomical Institute of the Netherlands
\def\fcp{Fund.~Cosmic~Phys.}%
          % Fundamental Cosmic Physics
\def\gca{Geochim.~Cosmochim.~Acta}%
          % Geochimica Cosmochimica Acta
\def\grl{Geophys.~Res.~Lett.}%
          % Geophysics Research Letters
\def\jcp{J.~Chem.~Phys.}%
          % Journal of Chemical Physics
\def\jgr{J.~Geophys.~Res.}%
          % Journal of Geophysics Research
\def\jqsrt{J.~Quant.~Spec.~Radiat.~Transf.}%
          % Journal of Quantitiative Spectroscopy and Radiative Trasfer
\def\memsai{Mem.~Soc.~Astron.~Italiana}%
          % Mem. Societa Astronomica Italiana
\def\nphysa{Nucl.~Phys.~A}%
          % Nuclear Physics A
\def\physrep{Phys.~Rep.}%
          % Physics Reports
\def\physscr{Phys.~Scr}%
          % Physica Scripta
\def\planss{Planet.~Space~Sci.}%
          % Planetary Space Science
\def\procspie{Proc.~SPIE}%
          % Proceedings of the SPIE
\let\astap=\aap
\let\apjlett=\apjl
\let\apjsupp=\apjs
\let\applopt=\ao

\pagestyle{empty} % Removes page numbers

\begin{center}
\begin{minipage}[t]{.2\linewidth}

{\includegraphics[height=18cm]{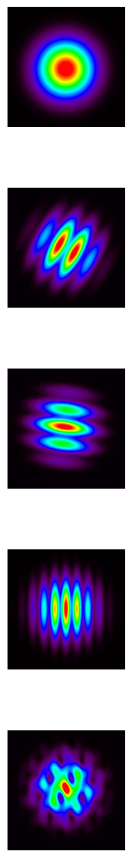}}% 
\end{minipage} % ne pas sauter de ligne
\hfill
\begin{minipage}[t]{.7\linewidth}

\hbox{ 
\hspace*{0.05\textwidth} % Whitespace between the vertical line and title page text
\parbox[b]{0.95\textwidth}{ % Paragraph box which restricts text to less than the width of the page
\centering\includegraphics[width=.4\textwidth]{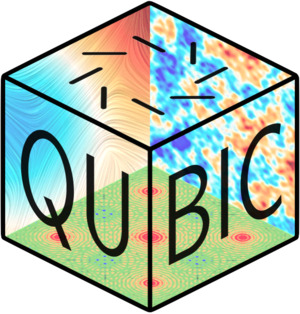}\\
%{\noindent\Huge\bfseries QUBIC \include}\\[2\baselineskip] 
{\large \textit{ Q\&U Bolometric Interferometer for Cosmology }}\\[4\baselineskip] % Tagline or further description
{\center \includegraphics[angle=90,width=0.6\textwidth]{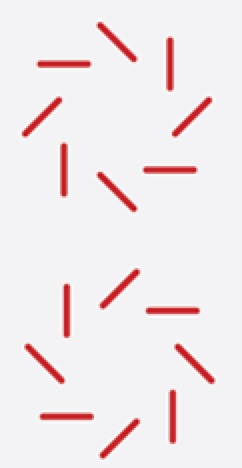}}\\[2\baselineskip]
{\noindent\Huge\bfseries Technical Design Report}\\[2\baselineskip]
{\Large \textsc{The QUBIC Collaboration}\\ \ \\
 \version{1.0}\\
\today}
\vspace{0.05\textheight} 
 }}

\vspace{0.2\textheight} % Whitespace between the title block and the publisher
\end{minipage}
\end{center}

\newpage

%%%%%%%%%%%%
\maketitle
%\newpage
%\pagestyle{plain}
\pagestyle{fancy}

\fancyhf{}
\lhead{QUBIC TDR}
\rhead{\version{1.0}\ (\today)}
\rfoot{Page \thepage}
%%%%%\pagestyle{scrheadings} 
%auto-ignore

\clearpage

\setcounter{tocdepth}{3}
\renewcommand\contentsname{Table Of Contents}
\tableofcontents

%% ADDED - OP 

\newpage
%auto-ignore
\section*{{Preface}}
\addcontentsline{toc}{section}{\protect\numberline{}Preface}%
\hypertarget{RefHeadingToc314322895}{}{
QUBIC, now in its construction phase, is dedicated to the exploration of the inflation age of the Universe. By detecting and
characterizing the Cosmic Microwave Background B-mode polarization, QUBIC will contribute to find the ``smoking gun''
of inflation and to discriminate among the numerous models consistent with current data. The primordial B-modes (as
opposed to E-modes) is the unique direct observational signature of the inflationary phase that is thought to have
taken place in the early Universe, generating primeval perturbations, producing  Standard Model elementary particles and
giving its generic features to our Universe (flatness, homogeneity{\dots}).}

Recent results from the BICEP2 and the Planck collaborations have brought the importance of the quest for B-modes to the
attention of a wide audience well beyond the cosmology community. The original claim from BICEP2, contradicted by Planck
later on has also shown how challenging the search for primordial B-mode polarization is, because of many difficulties:
smallness of the expected signal, instrumental systematics that could possibly induce polarization leakage from the
large E signal into B, brighter than anticipated polarized foregrounds (dust)  reducing to zero the initial hope of
finding sky regions clean enough to have a direct primordial B-modes observation.

{QUBIC is designed to address all aspects of this challenge with a novel kind of instrument, a
}{\textit{Bolometric Interferometer}}{, combining the
background-limited sensitivity of Transition-Edge-Sensors and the control of systematics allowed by the observation of
interference fringe patterns, while operating at two frequencies to disentangle polarized foregrounds from primordial B mode 
polarization. }

QUBIC is the only European ground based B-mode project with the scientific potential of discovering and measuring
B-modes. It is the natural project for the European CMB community to continue at the edge-cutting level it has reached
with Planck. 

With the measurement of the Cosmic Microwave B-mode Polarization in two bands at 150 and 220
GHz, with two years of continuous observations from Alto Chorillos  near San Antonio de los Cobres, Argentina, the first QUBIC module would be able to
constrain the ratio of the primordial tensor to scalar perturbations power spectra amplitudes  with a   
conservative projected uncertainty of $\sigma(r)=0.02$, while having a good control of foregrounds contamination thanks to its dual band
nature.
%\todocomment{Aniello: I have a problem with this sentence for two reasons: first because talking about QUBIC discussing alternative sites before PNRA is brought officially in the discussion and accepts even the possibility of an alternative site can be very dangerous for the continuation of the italian support. Second because the sentence is written in a way that already suggests a preference from the point of view of the collaboration, and I don't think this can be stated, even implicitly.
%When talking about the site I would like to see only the current baseline mentioned, i. e. DOME-C , until there is a formal and public decision that we can go elsewhere.} 

%Two sites are currently considered for QUBIC: Dome C, Antarctica, offering exquisite atmospheric conditions but complex logistics and Alto Chorillos near San Antonio de los Cobres, Argentina, of%fering easier logistics and very high quality atmospheric conditions.

Depending on the scientific and technological results of the first module we could envidage to construct more QUBIC modules operating at three frequencies (90, 150 and 220 GHz) that could feature design upgrades in order to achieve a higher sensitivity, and could preferentially be deployed 
in Antartica to take benefit of its exquisite atmospheric conditions.  These could include different detectors (eg MKIDs), larger horn arrays or number of detectors, different optical combiner design, ... QUBIC is therefore a project dedicated to grow and could be a Europen Stage-IV CMB Polarization experiment.

QUBIC has been and will be implemented through successive steps:
\begin{enumerate}
\item R$\&$D to design the instrument (now finalized)
\item Validation of the detections chain (now finalized)
\item Validation of the technological demonstrator (less detectors and horns than the final instrument, but in the nominal cryostat). This will occur in the course of 2017.
\item Construction and operations of the of the first module which will happen in the second half of 2017.
\item Optionaly, construction and operations of a number of additional modules to complete the QUBIC observatory.
\end{enumerate}

More details can be found on the QUBIC website : \url{http://qubic.in2p3.fr/QUBIC/Home.html}

\newpage
\section{Science Case}
%auto-ignore

\subsection{Context: the Quest for primordial B-modes}
\label{bkm:Ref311965747}
\subsubsection{Primordial Universe, Inflation and the CMB Polarization}
Our understanding of the origin and evolution of the Universe has made remarkable progress
during the last two decades, thanks in particular to the observations of the Cosmic Microwave Background (CMB). 
The diverse and more and  more numerous probes, such as CMB anisotropies, SNIa, BAO (...)
give complementary informations, enabling consistency tests of the 
standard cosmological model (aka $\Lambda \hbox{CDM} $ model). This concordance model is based
on General Relativity and is parameterized, in its simplest form, with six parameters.
From the determination of those cosmological parameters using the observations, we have learned that the Universe is spatially
flat, contains a large fraction of dark matter, and experiences accelerated expansion. The latter can be accommodated
within the Friedman-Lemaître framework through the presence of a mysterious dark energy $\Lambda $ (or cosmological
constant). 

\begin{figure}[h]
% \centering
\resizebox{\hsize}{!}
{ \includegraphics[width=8.5cm]{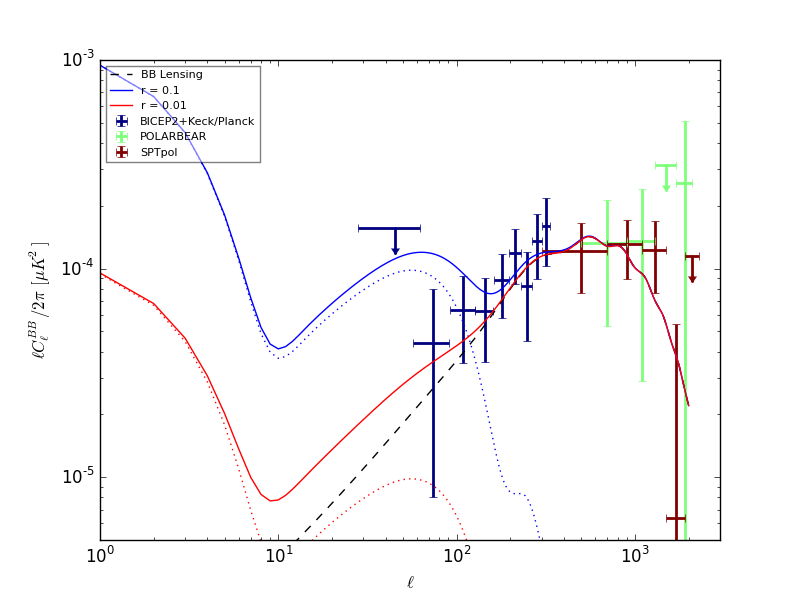} 
 \includegraphics[width=10cm]{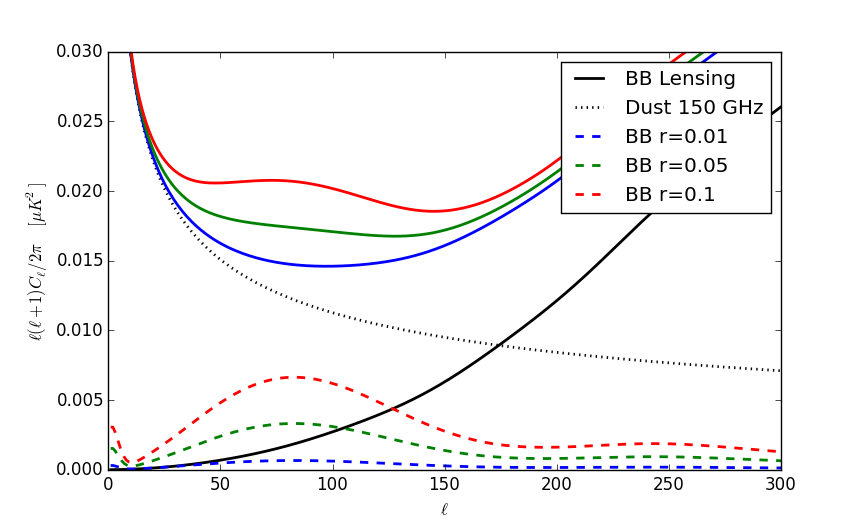} }
%{Figure \stepcounter{Figure}{\theFigure}}
\caption[Recent
B-modes measurements ; Total
B-modes  expected signal]{(left) Recent
B-modes measurements compiled in ~\cite{sptpol}, with results from BICEP2/Keck~\cite{bicep2} (dominated by dust), PolarBear~\cite{PolarBear}, ACTpol~\cite{actpol} and SPTpol~\cite{sptpol} due to lensing. The reionisation and recombinations bumps 
can be seen at respectively $\ell\simeq 7$ and $\ell\simeq 100$. (right) Total
B-modes (primordial + lensing + dust) expected signal for different values of the tensor-to-scalar ratio with dust level set
by Planck and BICEP2 measurements~\cite{Ade:2015tva} (black dotted line). ~\label{fig1}}
\end{figure}

Regarding the most primordial Universe history (i.e. shortly --
10\textsuperscript{{}-38} sec -- after the Big Bang), all the
observational data are up to now perfectly consistent with the inflation paradigm in which the young Universe undergoes
a period of accelerated expansion that results in a flat, almost uniform space-time when inflation ends. Besides
explaining flatness and homogeneity (which has originally motivated its introduction), inflation appears as the best theory able to
produce the observed almost scale invariant spectrum for the Gaussian primordial density fluctuations without
fine-tuning, only relying on the evolution of the quantum fluctuations of the scalar field(s) driving inflation. One of the most important
predictions of inflation is that, on top of the density anisotropies (corresponding to scalar perturbations of the
metric), it is expected to produce primordial gravitational waves (corresponding to tensor perturbations of the
metric). This specific prediction of inflation remains to be tested and is the core
motivation for the QUBIC instrument, together with the measurement of their amplitude to further constrain
the inflation models.

CMB polarization can be decomposed into two modes of opposite parities: E-modes (with
even parity) and B-modes (with odd parity). In total, four different power spectra describe
the correlations of CMB temperature (T) and polarization (E and B) anisotropies. Density perturbations only give rise
to E modes, while gravity waves are also source of B-modes. In other words, an
observation of primordial B-modes would be the ``smoking gun'' 
betraying the presence of primordial gravity waves generated by inflation. 
B-mode detection is today one of the major challenges to be
addressed in observational cosmology. This signal is parameterized using the tensor-to-scalar amplitude ratio $r$ 
which value would allow us to distinguish between
various inflation scenarios and is directly related to the energy scale of inflation~\cite{baumann}. Theoretical predictions on the
tensor to scalar amplitudes ratio are rather weak however but the simplest inflationary models predict r to be higher than 0.01 as the
corresponding energy level would be too low with respect to Grand Unified Theories  for smaller values or $r$. We plan to explore
this very range (between 0.01 and 0.1) with QUBIC. 

\subsubsection[A major observational challenge]{ A major observational challenge}
Unfortunately, B-modes appear to be very difficult to detect because of their small amplitude: a tensor-to-scalar ratio
of 0.01 corresponds to polarization fluctuations of the CMB of a few nK while the well observed temperature
fluctuations are around 100 microK. Even if such a sensitivity can be achieved using background limited detectors such
as bolometers from low-atmospheric emission suborbital locations or from a satellite, the challenge to face for this
detection remains huge because of two main reasons: instrumental systematics and foregrounds.

Instrumental systematic effects of usual telescopes (sidelobes, cross-polarization) may become too large to be
disentangled from a small primordial B-mode signal. Indeed any instrument, even designed with care, exhibits
cross-polarization, beam mismatch, inter calibration uncertainties, cross-talk, {\dots} All of these instrumental
systematics mix the electric fields in the two orthogonal directions inducing a mixing between the Stokes parameters Q
and U and possibly a leakage from intensity into polarization. This induces leakage from I and E into B-modes that,
given the smallness of the primordial B-modes, may completely overcome those B-modes. A new generation of instruments
achieving an unprecedented level of control of instrumental
systematics
is therefore needed for the B-mode quest. QUBIC was precisely designed with this objective.

B-modes anisotropies are also produced by foregrounds (summarized in the right panel of Figure~\ref{fig1}): 

\begin{enumerate}
\item {
\ The lensing of the B-modes by intervening large scale structure in the Universe converts part of the E-modes into
B-modes, mostly at small scales($\ell \gtrsim 300$). The spectrum of those lensing B-modes however has a well defined shape and has been
detected recently by PolarBear~\cite{PolarBear}, ActPol~\cite{actpol} and SPTpol~\cite{sptpol} 
(see Figure~\ref{fig1}, left panel). This contribution is not expected to affect the primordial B-modes detectability on the large
scales observed by QUBIC (around the so-called recombination peak at l=100) if the tensor-to-scalar ratio is
sufficiently high, but would become a strong limitation if r is below $\sim 0.01$.}
\item {
\ Thermal emission from dust grains in the Galaxy is expected (and measured) to be linearly polarized due to the
elongated shape of the grains which align along the magnetic field. 
The dust e.m. spectrum is
different from the CMB one so that multiple frequencies (above 150 GHz) can be used to remove it and obtain cleaned
maps of the CMB B-modes. This was the motivation for adding a 220 GHz channel in QUBIC besides the initial 150 GHz one.}
\item {
{\ Synchrotron emission from electrons swirling around magnetic fields in the Galaxy is also 
expected to produce B-modes. The synchrotron EM spectrum is falling with frequency so that it can be
}{monitored with channels at lower frequencies that the CMB ones the same way as dust.
Synchrotron polarization is not expected to be highly significant at 150 nor
at 220 GHz, the QUBIC operation frequencies~\cite{Krachmalnicoff:2015xkg}~\cite{Fuskeland}.}}
\item {
{\ For ground based observations, atmosphere is also a possible source of contamination. However,
the main effect of atmosphere is to increase the loading on the detectors in a time-variable manner that increases the
variance of the data. A recent study with PolarBear data~\cite{JosquinAtmo} has shown that the polarization
induced by atmosphere remains at a small level when observing from the Atacama plateau, which is known to be worse than 
in South Pole.}}
\end{enumerate}

Searching for B-modes in the Cosmic Microwave Background polarization is therefore a major challenge that requires
instruments observing at multiple frequencies with high sensitivity and unprecedented control of instrumental
systematics. The current best upper-limit on r is r {\textless} 0.07 at 95\% C.L. ~\cite{Ade:2015tva} and is obtained by combining BICEP2, Keck Array and Planck data.

\subsubsection[Ongoing and planned projects]{ Ongoing and planned projects}
\label{bmodeprojects}
Two kinds of instruments have been used so far in the Cosmic Microwave Background polarization observations:

\begin{itemize}
\item {\textbf{\textit{Imagers}} where an optical system (reflective as in
Planck or refractive as in BICEP2) allows us to form the image of the sky on a focal plane equipped with high sensitivity
total power detectors. Bolometers have been successfully used because their intrinsic noise is lower than the photon
noise of the observed radiation (so-called «background limited»). This is achieved by cooling
the bolometers down to sub-Kelvin temperatures. The detection principle is that incoming radiation heats the bolometers
whose temperature is being monitored through the variation of a resistance (resistively or using the normal-superconducting
transition). Recently, Kinetic Inductance Detectors (KIDs) have been developped, they present the advantage of an easier fabrication
process and natural ability for multiplexed readout (a major issue at cryogenic temperatures). Imagers directly measure
the temperature on the sky in a given direction (with a resolution given by that of the telescope and horns) and
therefore allow building maps of the CMB Stokes parameters I, Q, and U that further enables us to reconstruct T, E and B power
spectra.}
\end{itemize}
\begin{itemize}
\item { \textbf{\textit{Interferometers}} where the correlation between two
receivers allows us to directly access the Fourier modes (known as visibilities) of the Stokes parameters I, Q and U
without producing maps. The observation of interference fringes with an interferometer allows for an extra control of
systematic effects in comparison with an imager. That explains why interferometers 
were used for the first measurements of sub-degree temperature
anisotropies (with VSA~\cite{vsa}) and E-mode polarization (with CBI~\cite{cbi} and DASI~\cite{dasi}). However, they suffered from a degraded sensitivity due to their heterodyne nature: signals at the frequency
of the CMB (from a few GHz to a few hundreds of GHz) need to be amplified and down-converted to lower frequencies
before being detected. This amplification process adds an irreducible amount of noise that prevents such interferometers from
being background limited. Furthermore, the complexity of traditional CMB interferometers (based on multiplicative
interferometry, making the correlation by pairs of detectors) prevent them from growing to the large number of
receivers that is now required to achieve the sensitivity needed for the B-mode quest (if N is the number of channels,
their complexity increases as N\textsuperscript{2} while that of
an imager grows as N). This is the reason why, despite their better ability to handle instrumental systematics,
interferometers have no longer been considered, until QUBIC, for CMB polarization observations.
}
\end{itemize}

\begin{table}
\begin{centering}
\begin{tabular}{|l|l|l|l|l|l|l|l|c|}
\hline 
Project & Country & Location & Status & Frequencies & \multicolumn{2}{l|}{$\ell$ range }&  \multicolumn{2}{c|}{ $\sigma{(r)}$ goal }  \\ 
 &  &  & & (GHz)  & value & Ref. &   no fg. & with fg.  \\ \hline
\hline
QUBIC          & France    & Argentina &  & 150,220          & 30-200 & & 0.006 & 0.01 \\
Bicep3/Keck    & U.S.A.    & Antartica & Running  & 95, 150, 220$^1$          & 50-250 & ~\cite{BicepMoriond} & $2.5\ 10^{-3}$ & 0.013 \\
CLASS          & U.S.A.    & Atacama   & $\geq 2016$ & 38, 93, 148, 217 & 2-100 & ~\cite{Class}  & $1.4\ 10^{-3}$ & 0.003 \\
SPT3G          & U.S.A.    & Antartica & 2017     & 95, 148, 223     & 50-3000 & ~\cite{SPTMoriond} & $1.7\ 10^{-3}$ & 0.005 \\
AdvACT         & U.S.A.    & Atacama   & Starting & 90, 150, 230     & 60-3000 & ~\cite{ACTMoriond} & $1.3\ 10^{-3}$ & 0.004 \\
Simons Array   & U.S.A.    & Atacama   & $\geq 2017$         & 90, 150, 220     & 30-3000& ~\cite{SimonsArrayPres} & $1.6\ 10^{-3}$ & 0.005 \\
LSPE           & Italy     & Artic     & 2017     & 43, 90,  140, 220, 245 & 3-150&~\cite{LSPE} & 0.03${*}$  &  \\
EBEX10K        & U.S.A.    & Antartica & $\geq 2017$ & 150, 220, 280, 350 & 20-2000 & ~\cite{EBex10K} & $2.7\ 10^{-3}$ & 0.007 \\
SPIDER         & U.S.A.    & Antartica & Running  & 90, 150          & 20-500 & ~\cite{SpiderMoriond}& $3.1\ 10^{-3}$ & 0.012 \\
PIPER          & U.S.A.    & Multiple  &$\geq 2016$ & 200, 270, 350, 600 & 2-300& ~\cite{Piper}& $3.8\ 10^{-3}$ & 0.008\\
\hline
\end{tabular}
\caption[Summary of the main ground and balloon  projects aiming at measuring B-modes]{Summary of the main ground
and balloon projects aiming at measuring B-modes. The label ``fg'' or ``no fg'' corresponds to the assumption
on the foregrounds, numbers have been extracted from~\cite{Josquin}. [$*$] The LSPE value is an upper limit at 
99.7$\%$CL. [1] Ref. ~\cite{Josquin} did not include this frequency.

\label{table1}.} 
\end{centering}
\end{table}

Most of the on-going or planned projects are lead by U.S. teams. They are all based on the
concept of a traditional imager with a broad variety of technical choices regarding the modulation of the polarization,
the optical setup, the detector technology, the frequency coverage or the instrument location. They also use different instrumental apertures, that sets the angular accuracy hence the multipole coverage and therefore are
optimized for different science goals: high angular resolution instruments are better suited
for the lensing B-modes study (allowing one to constrain neutrino masses for instance), and have published results on
this (PolarBear, SPTpol, ACTpol) while low resolution suborbital instruments aim at detecting the recombination peak of
the primordial B-modes at l=100. Satellite missions are considered by the community and aim at covering both science
goals with the additional advantage of a full sky coverage allowing one to search for the reionization peak at l=7.
However, no such mission has been selected up to now by Space agencies, neither in the U.S.A. nor in Europe. LiteBird is a
possible mission to be flown in the early 2020 by the Japanese Space Agency (JAXA) and would be an extremely sensitive
project (targeting r=0.001) with low angular resolution, therefore only focused on primordial B-modes.

Table~\ref{table1} summarizes the situation in terms of competitors for QUBIC. We
know since the BICEP2/Planck controversy that foregrounds cannot be neglected. This is why, when the foreground-free forecasted sensitivity of the QUBIC first module, from Argentina, is $\sigma(r)=0.01$, we can only achieve $\sigma(r)=0.02$ when accounting for realistic foregrounds.
The observation efficiency is taken to be $30\%$ in those QUBIC sensitivity forecasts. 
Besides BICEP/Keck~\cite{Ade:2015tva} on the ground and the ballon-borne SPIDER experiment~\cite{fraisse_spider}
which has already taken data in the same multipole range as QUBIC (namely
targeting the recombination peak at $\ell\sim 100$), it is clear from this table that QUBIC is competitive and timely with
respect to other competitors with the same target. High resolution experiments are more suited to the measurement of
the lensing B-modes which should provide very exciting neutrino constraints. Although these projects claim they will 
measure primordial B-modes,  this is not their primary goal and that they focus on the smaller
angular scales because large angular scales are harder to reconstruct due to 1/f noise (from electronics and/or
atmosphere). As a matter of fact, these experiments have never published data, even with temperature only, below a
multipole of $\sim 300$. While having comparable sensitivity with the other experiments, QUBIC will offer this improved
control of instrumental systematics that may be a decisive factor when reaching very low tensor-to-scalar ratio
sensitivity. 

%auto-ignore
\subsection{Bolometric Interferometry and QUBIC}
\label{bolinterfero}
{Most of the current projects aiming at detecting the B-mode radiation are based on the architecture of an imager because
of its simplicity and the high sensitivity allowed by bolometers. However, imagers do not allow for the same level of
control of instrumental systematics and could potentially reach a sensitivity floor because of E-modes leaking into
B-modes. \ Bolometric Interferometry is a novel concept combining the advantages of bolometric detectors in terms of
sensitivity with those of interferometers in terms of control for systematics. It was initially proposed in 2001 by
Peter Timbie (University of Wisconsin) and Lucio Piccirillo (University of Manchester). Two collaborations on both
sides of the Atlantic (BRAIN in Europe and MBI in the U.S.A.) started to develop the concept and decided to merge their
efforts in the QUBIC project in 2008. The QUBIC collaboration now includes six laboratories in France, all  members of the
CNRS (APC in Paris, LAL, IAS and CSNSM in Orsay and IRAP in Toulouse), three \ Universities in
Italy (Universitá di Roma -- La Sapienza, Universitá Milano Bicocca and Statale in Milano), Manchester and Cardiff
Universities in the UK and NUI/Maynooth in Ireland, three universities in the USA (University of Wisconsin at Madison, WI ;
Brown University at Providence, RH ; Richmond University, VI). NIKHEF (Netherlands) have joined QUBIC in 2014.}
%\todocomment{Check the instutes e.g. Institut Néel in Grenoble}

\subsubsection[The QUBIC design]{ The QUBIC design}
{
{QUBIC will observe interference fringes formed altogether by a large number of
receiving horns with two arrays of bolometric detectors (operating at 150 and 220 GHz) at the focal planes of an
optical combiner. The image on each focal plane is a synthesized image in the sense that only specific Fourier modes
are selected by the array of receiving horns. A bolometric interferometer is therefore a synthetic imager whose beam is
the synthesized beam formed by the array of receiving horns. The interferometric nature of this synthesized beam allows
us to use a specific self-calibration technique that permits to determine the parameters of the systematic effects channel by
channel with an unprecedented accuracy~\cite{Charlassier}~\cite{Bigot}. 
As a comparison, an imager can only measure the effective beam of each channel.
%while with bolometric interferometry, one can both measure it and reconstruct it from individual measurements of the
%gains, cross-polarization and locations of entry horns. 
We therefore have an extra-level of systematics control. The use of bolometric detectors
allows us to reach a sensitivity comparable to that of an imager with the same number of receivers~\cite{Hamilton}. }}

\begin{figure}
  \centering
 \includegraphics[width=13.194cm,height=9.823cm]{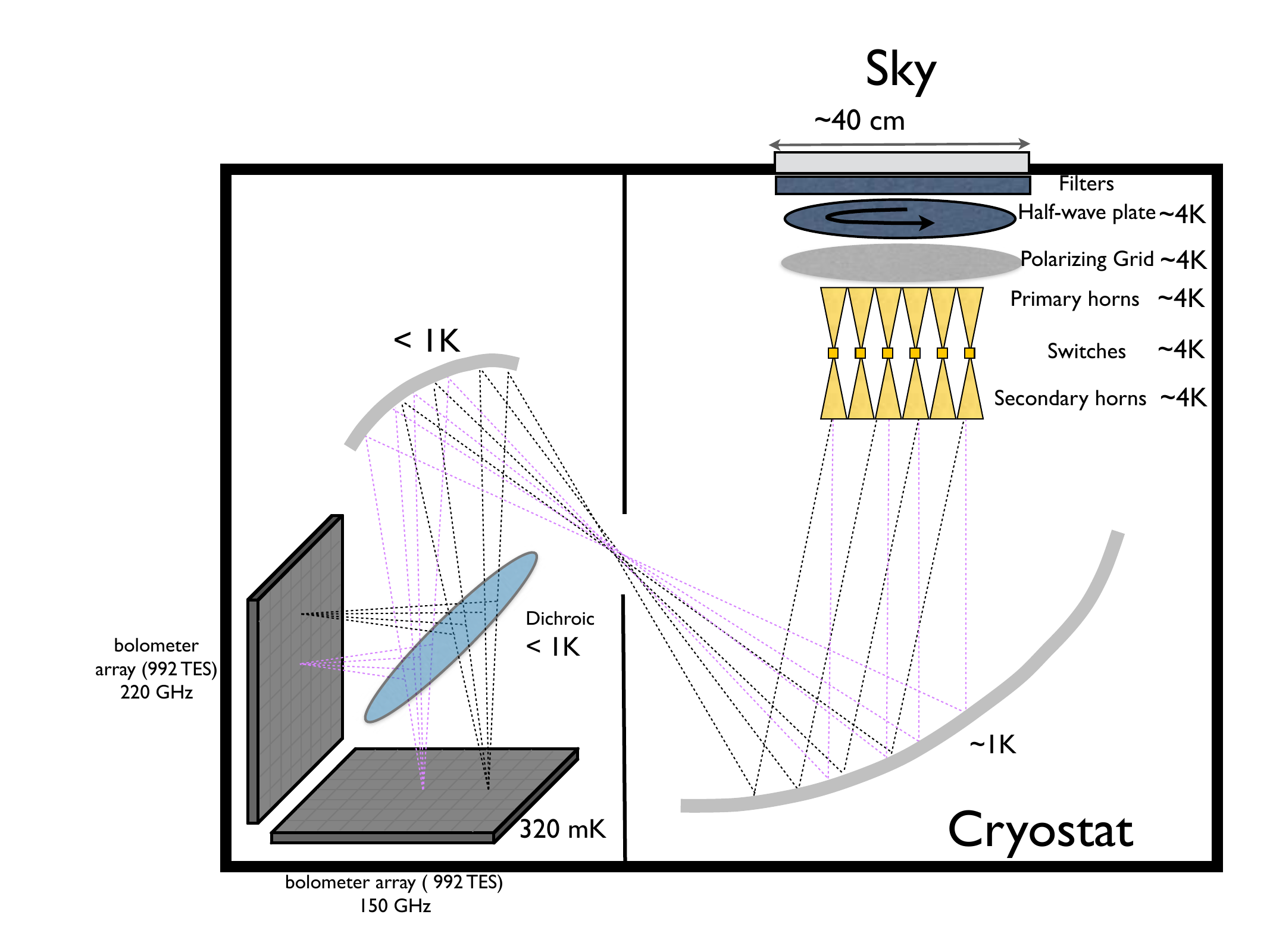}
\hypertarget{RefHeadingToc314323034}{}
\caption{ Sketch of QUBIC (see text for explanation)}\label{qubicsketch}
\end{figure}

{
{The QUBIC instrument is made (see Figure~\ref{qubicsketch}) of a cryostat cooled down to 4K using pulse-tubes.
The cryostat is open to the sky with a 45 cm diameter window made of 
high-density polyethylene (HDPE) providing an excellent transmission
and mechanical stiffness. Right after the window, filters ensure a low thermal load inside the cryostat and a rotating
Half-Wave-Plate (HWP) similar to that of the Pilot instrument~\cite{salatino} modulates
the polarization. Then, a polarizing grid selects one of the two polarization angles w.r.t the instrument. An array of
400 corrugated horns (called «~primary horns~» designed to be efficient throughout the 150 and 220 GHz bands with a 
$\approx 13$ degrees FWHM at 150 GHz) selects the baselines }{observed by QUBIC. These primary horns are
immediately followed by back-horns re-emitting the signal inside the cryostat towards an «~optical combiner~» which is
simply a telescope that combines on the focal plane the images of each of the secondary horns in order to form
interference fringes. Before the focal plane, a dichroic plate splits the signal into its 150 and 220 GHz components
that are each imaged on a focal plane equipped with 1024 Transition-Edge-Sensors  (TES) 
from which 992 are exposed to the sky radiation (blind ones are used for systematics studies)
cooled down to 320 mK 
 and read using a
multiplexed cryogenic readout system based on SQUIDs and SiGe ASIC operating at 4K. Finally, the signal measured by
each detector p at in the focal plane with frequency $\nu $ at time }{\textit{t}}{
is:}}

\begin{equation}
R(p,\nu,t)\ = \ S_I(p,\nu) + \cos\left[4\varphi_{\mathrm{HPW}}(t)\right]S_Q(p,\nu)+\sin\left[4\varphi_{\mathrm{HPW}}(t)\right]S_U(p,\nu)
\end{equation}

{
{where $\varphi_{\mathrm{HWP}}(t)${ is the angle of the HWP at time }{\textit{t}}, $S_{I/Q/U}$ the sky signal at frequency $\nu$ convolved with the synthesized beam (see Figure~\ref{primaryhorns})}{. With a scanning
strategy offering a wide range of polarization angles on the sky and thanks to the HWP rotation, one can
recover}\footnote{{It is worth noting that given the approximate cost of 5 k}{\texteuro}
{for a traditional correlator, a 400 elements traditional interferometer would require
\~{}80000 of them (one per baseline) and would therefore cost the amazing price of \~{}400
M}{\texteuro}{. Using an optical combiner as in QUBIC therefore appears as a very cheap way
(by a factor \~{}100) of performing interferometry with a large number of channels, leading to a better sensitivity
thanks to the use of bolometers.}}{ the
synthesized images of each of the three Stokes parameters I, Q and U. In contrast with traditional interferometry, the
observables of QUBIC are not the visibilities (Fourier Transform of the observed sky for modes corresponding to the
baselines), but the synthesized image, which is nothing else but the observed sky filtered to the modes
corresponding to the baselines allowed by our instrument. This particular feature is a crucial one in QUBIC as each of
these modes can be calibrated separately using the «~self calibration~» procedure (see section~\ref{selfcalib} 
and~\cite{Bigot}) allowing QUBIC to reach an unprecedented level of instrumental systematics control.}}

{
One important aspect of the QUBIC design is the presence of the polarizing grid right after the half-wave plate, {\em ie} very close to the sky. It may appear undesirable from the sensitivity point of view to reject half of the photons at the entrance of the instrument. However, this a very nice feature from the point of view of polarization systematics because this is associated with bolometers that are {\bf not} polarization sensitive: the rejection of the undesidered polarization with the polarizing grid is very efficient and whatever the cross-polarization of the rest of the instrument, the detectors will measure the polarized sky signal
modulated by the HWP. This means that we expect a very low level of instrumental cross-polarization for QUBIC.
}

\subsubsection{The QUBIC synthesized beam and map-making}
\label{beamsynth}
In QUBIC, each primary horns pair defines a baseline (a Fourier mode on the sky) that is
transmitted through the instrument and forms an interference fringe on the focal planes. In the standard «~sky
observing~» mode, the fringes formed by all the baselines are coherently combined on the focal and form a synthesized image of
the sky, which is  the sky image convolved by the QUBIC synthesized beam than can be calculated from the combination of all
baselines.

\begin{figure}
\centering
 \includegraphics[width=11cm]{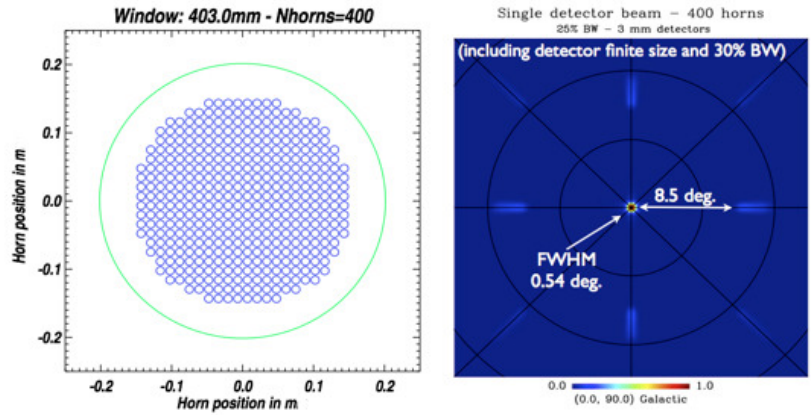} 

\caption{QUBIC primary horn array and corresponding beam on the sky for the central detector of any of the two focal planes\label{primaryhorns}.}
\end{figure}

%{\sffamily\itshape
%\. QUBIC primary
%horn array and corresponding beam on the sky for the central detector of any of the two focal planes.}

{
The QUBIC horn array and synthesized beams are shown in Figure~\ref{primaryhorns}. 
As can be seen on this Figure (left panel), the horn array, although enclosed in a circle to optimize the window occupation, respects a regular
square grid pattern that has been shown to ensure a coherent summation of redundant baselines which is the key aspect
offering to a bolometric interferometer a comparable sensitivity to an imager~\cite{Hamilton}~\cite{Charlassier}.

The synthesized beams shape is significantly different from the beam offered by a classical
imager and typical of that of an interferometer: it has a central peak, with 0.54\degree\ FWHM and has replications
around, damped by the primary 14\degree\ FWHM that are due to the fact that the primary horn array has finite extension.
These replications are not sidelobes as they are a
desired feature of an interferometer that only observes well defined and well «~calibrable~» baselines (see
Sect.~\ref{selfcalsect}). It however makes the map-making procedure much more complicated than with an 
imager as it involves partial deconvolution to disentangle the small contamination by secondary peaks with respect to
the main one. 

We have shown that using super-calculators and  a specific
map-making algorithm based on «~inverse problem solving~»~\cite{chanial}, one can recover the input
I, Q and U maps provided the fact that the scanning strategy offers a wide enough variety of polarization angles on the
sky (which is ensured by the combination of sweeps in azimuth with constant elevation and the rotation of the
Half-Wave-Plate, cf.~Figure~\ref{fig4} and~Figure~\ref{fig5}).

\begin{figure}
\centering
 \includegraphics[width=7.cm]{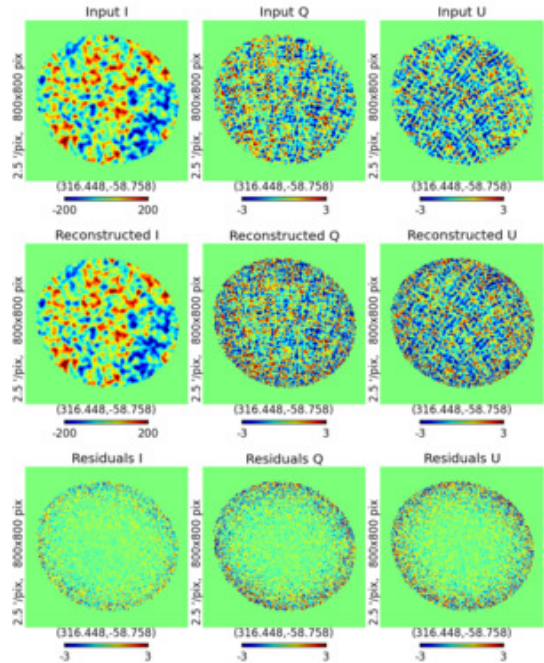} 

\caption[Current
results with the QUBIC mapmaking under the Gaussian peaks assumption]{\label{fig4}(left) Current
results with the QUBIC mapmaking under the Gaussian peaks assumption. First row shows the input I,Q and U maps in the region observed by QUBIC, second row shows the recovered maps using the full simulation pipeline, last row shows the residuals w.r.t. the input maps.}
\end{figure}

\begin{figure}
\centering
 \includegraphics[width=15cm]{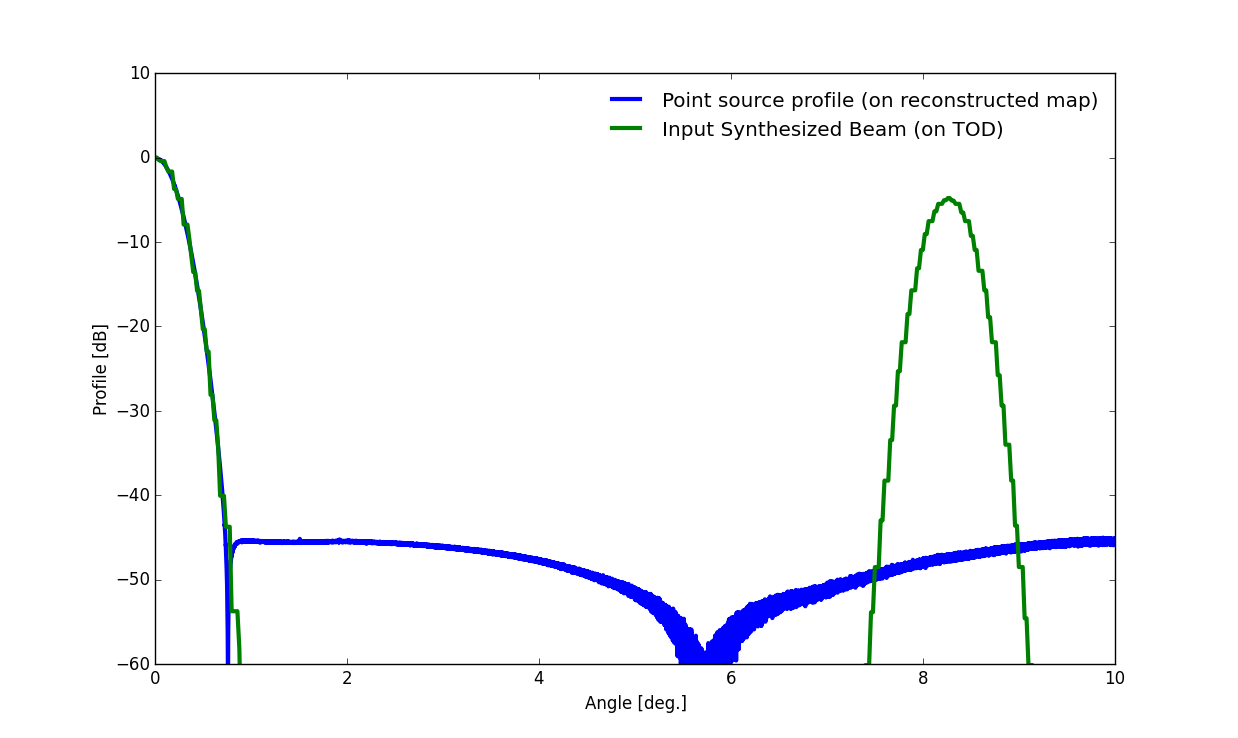} 
\hypertarget{RefHeadingToc314323037}{}
\caption[Results of the
current version of the map-making]{\label{fig5}Results of the
current version of the map-making: the multiple --peaked feature of the synthesized beam (green) present in the TOD is
deconvolved efficiently in the maps where we show the recovered profile for a point source (blue line). The achieved resolution (FWHM of the blue central peak) is 23.5 arcmin.}
\end{figure}

\subsubsection{Self-calibration and the systematic effects mitigation with QUBIC}
\label{selfcalib}
Interferometry is known~\cite{Timbie} to offer an improved control of instrumental systematics with respect
to direct imaging thanks to the observation of individual interference fringes that can be calibrated individually.
This feature is conserved with bolometric interferometry, in QUBIC, thanks to the presence of  electromagnetic
switches  between the primary and secondary horns (cf. sections ~\ref{horns} and ~\ref{switch}). This apparatus consists  
in a waveguide that is closed or open
using a cold (4K) shutter operated by solenoid magnets. In the  self-calibration  mode, pairs of horns are                                                  
successively shut when observing an artificial partially polarized source (we do not need to know its polarization). As                                       
a result, we can reconstruct the signal measured by each individual pairs of horns in the array and compare them. As                                          
 redundant baselines  correspond to the same mode of the observed field, a different signal between them can only be
due to photon noise or instrumental systematics. Using a detailed model of the instrument incorporating all possible
systematics (through the use of Jones matrices for each optical component), we have shown that we can fully recover all
of these parameters through a non-linear inversion involving hundreds of parameters (horn locations and beams,
components cross-polarization, detector inter calibration, {\dots}). The updated model of the instrument can then be
used to reconstruct the synthesized beam and improve the map-making, reducing the leakage between Stokes parameters. We
have shown in~\cite{Bigot} that with 2.5\% of the observing time, we can reduce the impact of the
instrument systematics on the E to B leakage to a level allowing us to measure the B modes down to r=0.05 (see Figure~\ref{selfcalres}).
No such feature exists with a usual imager justifying the fact that QUBIC will have
extra-control on instrumental systematics with respect to all the other running or planned instruments listed in Table
~\ref{table1}.

\begin{figure}
\centering\includegraphics[width=10cm]{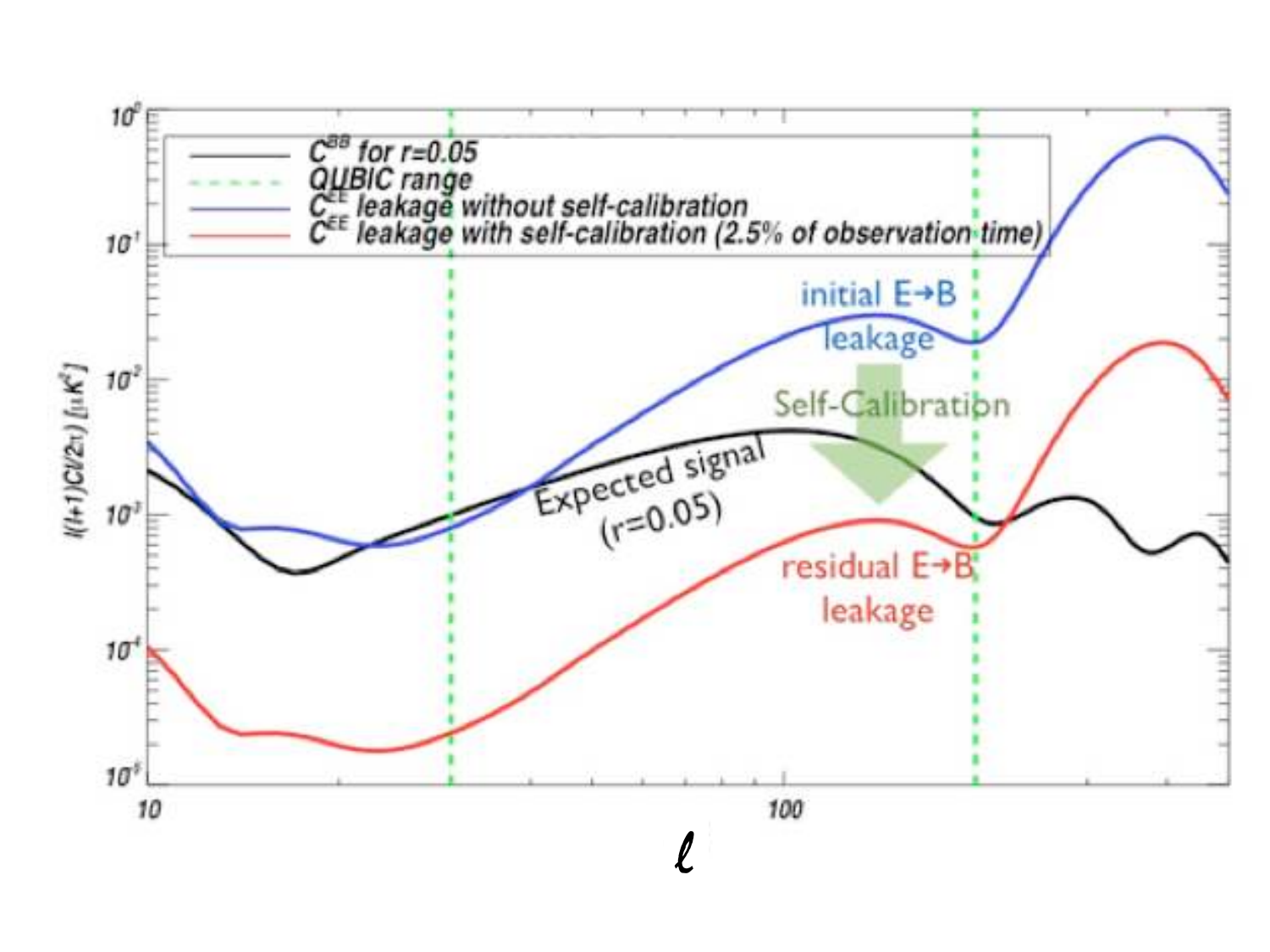}
\caption[``Self-calibration'' simulated result]{``Self-calibration'' simulated result~\cite{Bigot}. The leakage from E to B due to
instrumental systematics is reduced by more than an order of magnitude spending only 2.5\% of the observation time on
self-calibration.\label{selfcalres}}
\end{figure}

%auto-ignorex
\subsection{QUBIC sensitivity to B modes}
The first module of QUBIC will be installed in Argentina, on the Puna plateau
in the Salta Province, near the city of San Antonio de Los Cobres, on the site of the LLAMA experiment 
(cf. Sect.~\ref{argentina} for more details). Still, in the initial phase of the project,
we considered installing QUBIC in Dome C, Antartica. For this reason, some results presented below
refer to this site.

\subsubsection{E/B power spectra from realistic simulations}
\hypertarget{RefHeadingToc314322907}{}{
The outputs of the mapmaking are I, Q and U Stokes parameters maps. However, the
polarized fields of interest for cosmology are the scalar E and B fields instead of the spin-2 Q and U. They are
related through a non-local transformation in harmonic space that is trivial when full sky Q and U maps are available.
However, on a cut sky (a few percent of the celestial sphere for QUBIC), this transformation cannot be applied anymore since the
Spherical Harmonics are no longer a complete basis. As a result, some of the modes are ambiguous
(neither E nor B) and even in the absence of instrumental systematics, the cut sky induces massive leakage of E into B
when just expanding the cut sky Q and U maps onto E and B power spectra. This mixing is however easy to revert as we
know the exact geometry of the cut-sky. Unfortunately, although this inversion is unbiased and allows to recover
unaltered E and B fields in average, the variance of the recovered fields contains contribution from both the sample
variance of E and B so that the uncertainty on the small B field is largely dominated by the E sample variance~\cite{Tristram}.
It is nonetheless possible to reduce the non-optimality of the B measurement by
applying apodization functions~\cite{smith}~\cite{smith-zalda}~\cite{grain}. Finally, near-optimality can be reached 
(within a factor  $\sim 2$) but
requires a large amount of work with simulations in order to find the optimal apodization scheme for the Q and U maps.}

%{
%As of today, we have successfully implemented XPol which, in the case of the almost Gaussian apodization imposed by the
%QUBIC scanning strategy, is not far from being as optimal as XPure. We however still need to implement Pure (which is
%being done now) in order to make detailed comparisons.}

{
Figure~\ref{fig7} shows in red the anticipated error bars on the 150 GHz channel assuming a perfect cleaning of the dust
by the 220 GHz. They have been calculated using a full end-to-end Monte-Carlo Simulation (from time-ordered data to
maps) for Dome C. }

\begin{figure}
\centering
 \includegraphics[width=7cm]{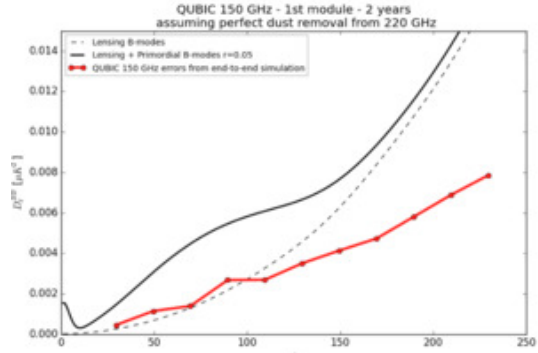} 
%\hypertarget{RefHeadingToc314323039}{}
\caption[Full end-to-end Monte-Carlo simulation  error-bars (dispersion) at 150 GHz]{Full end-to-end Monte-Carlo simulation
(from time-ordered data to maps and then to power spectra) error-bars (dispersion), at 150 GHz, assuming a perfect dust
cleaning from the 220 GHz channel, are shown in red along with the expected signal from B-modes for r=0.05 for observations
from Dome C.\label{fig7}}
\end{figure}

\subsubsection{QUBIC Sensitivity to B-Modes}
%\hypertarget{RefHeadingToc314322908}{}{
%{
Thanks to the extreme dryness of the Dome C site in Antarctica, the atmospheric emission in
the millimeter wavelengths is extremely small~\cite{Tomasi}~\cite{Battistelli}. The Precipitable Water
Vapor average in Dome C has been measured to be 0.6mm in January and well below 0.5mm the rest of the time. By
comparison, it is below 0.5mm only 50\% of the time in Chajnantor, Chile, where a number of B-modes experiments are
installed (see Table~\ref{table1}). The QUBIC detectors, cooled down to 320mK,
will be background
limited, where the background is dominated by the atmosphere. We will therefore  fully benefit from the former extreme
location which would ensure QUBIC to have an exquisite sensitivity.  
Another advantage of being close to the South Pole is that the interesting
fields in the sky (with minimal dust contamination~\cite{Adam:2014bub}) are not far from the
Southern Equatorial Pole and therefore visible 100\% of the time above 30 degrees elevation from Concordia, while this
is not the case from Argentina, nor Chile where several other CMB observatories are based, forcing these experiments to define multiple 
observation fields which is not optimal.  Expected errors on the B mode spectra obtained from Dome C are shown on the left 
panel of  Figure~\ref{fig8}.  %\todocomment{Definitely need to be coherent..}%}}
%perform a factor three faster from the noise point of
%view than the same instrument located in Chile..}}

We have performed full likelihood forecasts for QUBIC including lensing B-modes and dust
foregrounds at the level measured by Planck in the BICEP2 field [Planck Intermediate Results XXX, 2014]. We use our two
bands to form three cross power-spectra (150x150, 150x220 an 220x220) and multipole range from 25 to 300 to constrain
the dust spectral index and primordial tensor-to-scalar ratio (dust amplitude is fixed by Planck 353 GHz). Those results
(see Figure~\ref{fig8} right) show that using the two bands of QUBIC alone (blue) or with Planck 353 GHz added (red) allows to 
reach $\sigma(r)=0.01$ while this value goes down to
$\sigma(r)=0.006$ in the absence of foregrounds (green). These forecasts assume two years of continuous observations from Dome C, Antarctica and an overall 30\% efficiency. It
is worth noting that Planck 353 GHz does not bring much gain with respect to QUBIC dual band (difference between red
and blue). 

Finally,
when accounting for all the aspects, QUBIC, when deployed in Argentina,  will 
reach $\sigma(r)=0.01$ in two years of observation with an overall  30\% efficiency  as quoted in Table~\ref{table1}. (see also  Figure~\ref{fig107}, more
details on the site comparison can be found in Section~\{siteComparison}). 

%More detailed comparison of the foreseen QUBIC sensitivity in Antartica and Argentina are presented in Section~\ref{siteComparison}.

\begin{figure}
  \centering
 \includegraphics[width=8.cm]{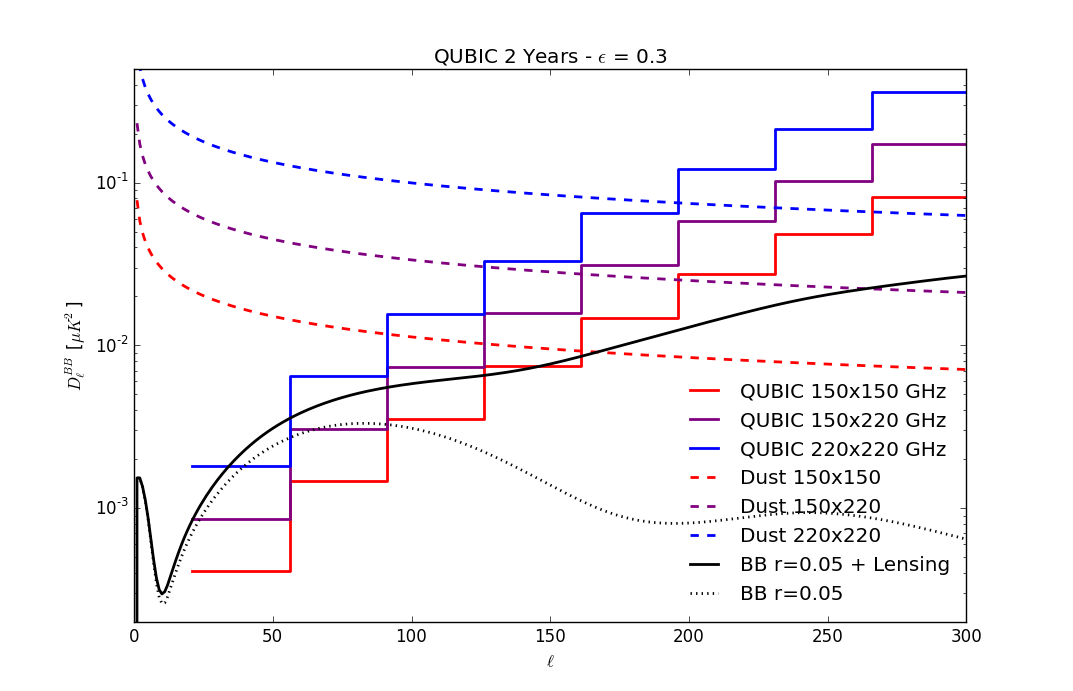} 
\includegraphics[width=8.5cm]{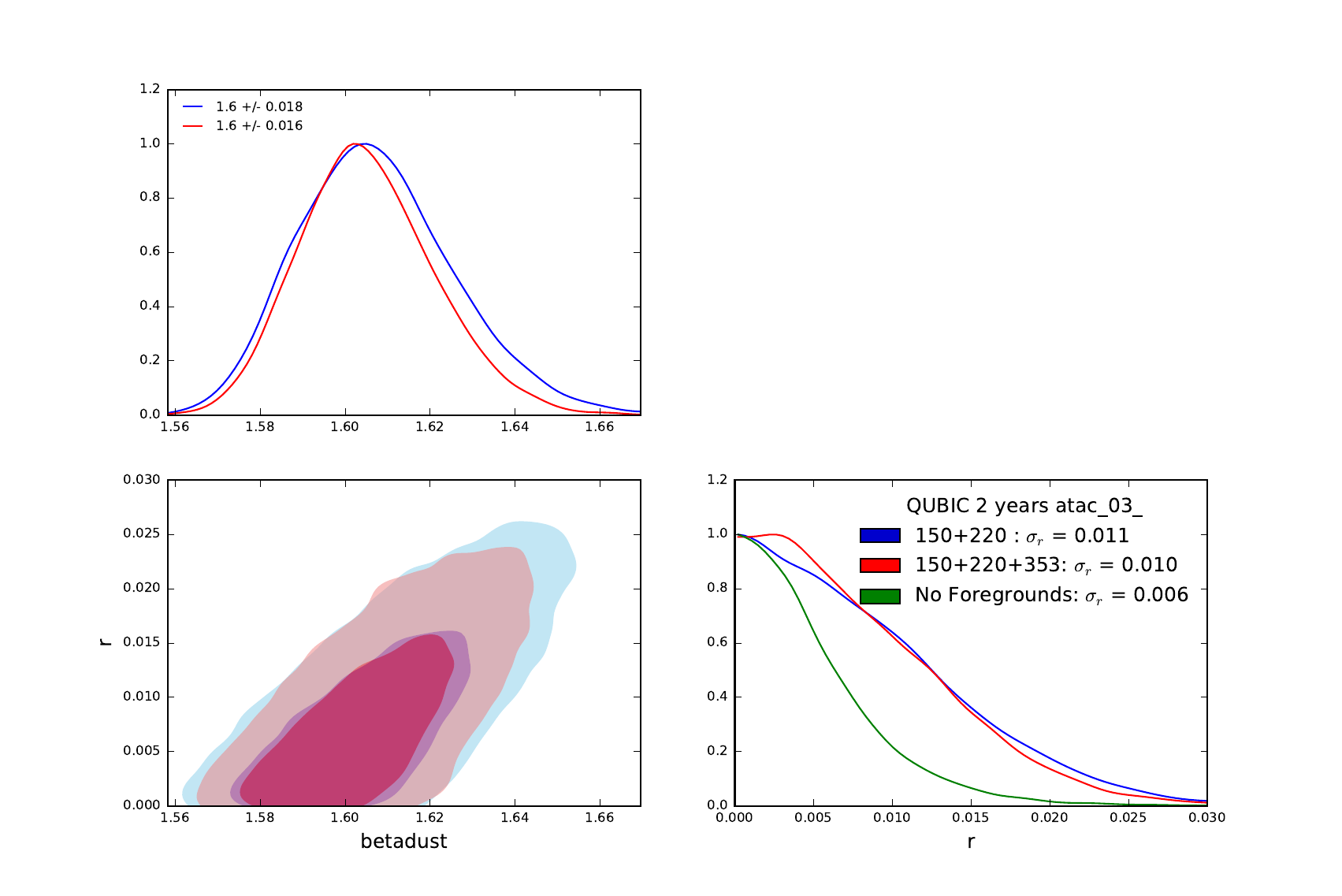} 
%\hypertarget{RefHeadingToc314323040}{}
\caption[Expected noise-only error bars for
QUBIC ; Expected forecasts on dust spectral index and tensor-to-scalar
ratio with QUBIC]{(left) Expected noise-only error bars for
QUBIC for the three BB cross-spectra that can be formed along with r=0.2 and dust levels extrapolated following~\cite{Adam:2014bub}  for the BICEP2 field. (right) Expected forecasts on dust spectral index and tensor-to-scalar
ratio with QUBIC using a full likelihood with three configurations: QUBIC alone (blue), QUBIC with Planck 353 GHz
information added (red), QUBIC if no foregrounds were present (green). The sensitivities on r are estimated
from a marginalization over the
dust spectral index and assume 2 years of continuous observations from Dome C and an overall efficiency of 0.3.\label{fig8}}

\end{figure}

\section{Overall Description of QUBIC}
%auto-ignore

\subsection{Main characteristics of the Instrument}

%{\notes \it There should probably be somewhere a description of which part of the instrument
%in cooled down to which temperature. Hmm what about cryo part ?}
%\todocomment{Aniello: The obvious comment is that several requirements are not
%specified yet in the document. Is it just a matter of filling in
%existing numbers?}

The final sensitivity and deep control of systematics quoted in the previous section assumes that a series of requirements are fulfilled. They are listed in
this section. The technical details on how these requirements are fulfilled in the instrument design are detailed in
the rest of the Technical Design Review. 
%The general requirements on the bolometric interferometer design 
Basic characteristics of the instrument are summarized in table~\ref{table2}.  
%\todocomment{ESB: put a table with temperatures of various elements?}. 
The QUBIC detectors (TESs) are cooled down to 320 mK thanks to a $^3$He/ $^4$He adsorption refrigerator. They are illuminated by an 
optical system  (optical combiner, horns,...) cooled down to 1K. This experimental system is  encased in a liquid-free cryostat housing a Pulse Tube cryocooler with base temperatures of 40K and 4K respectively for the 1st and 2nd cryogenic stage. Summaries of the characteristics of 
these various parts are listed below, and detailed further in this document.

\begin{table}
\centering
\begin{tabular}{|m{7.0480003cm}|m{5.321cm}|}
\hline
{ Frequency channels} &
{ 150 and 220 GHz}\\\hline
{ Bandwidth} &
{ 25\%}\\\hline
{ Number of horns (interferometric elements)} &
{ 400}\\\hline
{ Primary beam FWHM at 150 GHz} &
{ 12.9 degrees}\\\hline
{ Primary beam FWHM at 220 GHz} &
{15 degrees (not gaussian) }\\\hline
{ Number of detectors} &
{ 2x1024}\\\hline
\end{tabular}
\caption{General requirements on the bolometric
interferometer design.\label{table2}}
\end{table}

%\subsection{{Bolometric interferometry architecture [F0E0] Michel }}
%\hypertarget{RefHeadingToc314322921}{}\label{bkm:Ref314298797}
The QUBIC instrument is composed  of the following elements (see Figure~\ref{qubicsketch}) : 
\begin{description}

\item{Optical Chain :\\}
\hypertarget{RefHeadingToc314322911}
The optical chain of the QUBIC instrument starts from the window, opportunely coated with antireflection coating, directly observing the sky
and extends to the detectors. It also  includes the external baffling of the instrument that prevents ground pickup on the
detectors.

%\item{Cold Optics and Horns :\\}

\hypertarget{RefHeadingToc314322912} 
QUBIC horns are quasi diffraction-limited apertures at 150GHz. This implies a relationship  
%Given the fact that horns are quasi diffraction-limited apertures, there is a relationship
between their operating frequency, beam FWHM and aperture size: S$\Omega \sim \lambda^2$ which conditions their size to be  13.3mm (an internal diameter of 12.3mm to which 1 mm of metal wall thickness is added)
for single mode operation at 150 GHz (HE11). The same horn structure support three modes at 220 GHz (HE11, TM02, EH21), with a consequently larger FOV (15 degrees, as shown in Table~\ref{table2}) and increased throughput.  This
gives its dimensions to the whole instrument. 
%\begin{figure}
%\centering
%\includegraphics{QUBICTDRcompilation-img/hornsJC}
%\caption{horns positions\label{hornsposition}}
%\end{figure}

The size of the horn array is thus 33.078 cm diameter 
%(see document \href{https://atrium.in2p3.fr/nuxeo/nxdoc/default/8bb70808-5349-41a5-a945-92b5c6d50c66/view_documents}{\textstyleInternetlink{ATRIUM-77657}}
%annexed to this document)
as shown on Figure~\ref{primaryhorns} (left hand side), driving the 
requirements summarized in table~\ref{table3}.
%aperture window diameter to  {\bf XX} cm and the filters, polarizer and Half-Wave Plate
%diameters to  {\bf XX}cm. The cross-polarization of the Half-Wave plate is required to be less than  {\bf XX} dB across the band
%while its transmission is required to be better than  {\bf XX} dB. Similar requirements hold for the filters and polarized.
%\todocomment{AT: The cross-polarization is undefined at 220, since here we have three modes. 
%Therefore, I wouldn't put this number. The cross-pol will be in any case defined by the polarizing grid. I would remove this number from the following table as well.}
The horns need to have a low level of cross polarization ({\textless} -25dB) and secondary lobes ({\textless} -20dB),
and to transmit a large fraction of the incoming power (Return Loss {\textless} -25dB) across both 150 and 220 GHz
bands.% (see document
%\todocomment{SHV: there is almost nothing in the document in attachment...is there something missing otherwise I would suggest to remove it ...}
%\href{https://atrium.in2p3.fr/nuxeo/nxdoc/default/8bb70808-5349-41a5-a945-92b5c6d50c66/view_documents}{\textstyleInternetlink{{ATRIUM-77658}}}
%annexed to the present document). 
%These requirements are summarized on Table~\ref{table3}.

\begin{table}
\centering
\begin{tabular}{|m{7.0480003cm}|m{5.321cm}|}
\hline
{ Horn diameter (internal)} &
{ 12.33 +/- 0.1 mm}\\\hline
{ Back-to-Back Horn array diameter} &
{ 33.078 cm}\\\hline
{ Horn Return loss across the bands} &
{ {\textless} -25 dB}\\\hline
{ Horn secondary lobe level} &
{ {\textless} -20 dB}\\\hline
{ Horn cross-polarization level} &
{ {\textless} -25 dB }\\\hline
{Horn interaxis} & 
{ {\textless} 14 mm}\\ \hline
\end{tabular}
\caption{Requirements on horns\label{table3}}
\end{table}

\begin{table}
\centering
\begin{tabular}{|m{7.0480003cm}|m{5.321cm}|}
\hline
{ Window diameter} &
{  { 39.9} cm}\\\hline
{ Filters diameters} &
{  { 39.2} cm}\\\hline
{ Polarizer diameter} &
{  {32.6} cm}\\\hline
{ Half-Wave plate diameter} &
{  {32.7} cm}\\\hline
{ Half-Wave plate, filters and polarizer transmission} &
{  {-0.2} dB}\\\hline
{ Half-Wave plate, filters and polarizer cross-polarization} &
{  -20 dB}\\\hline
\end{tabular}
\caption{Requirements on cold optics chain
\label{table4}}
\end{table}

On the other hand, the detector size needs to be approximately the observed wavelength (2mm at
150 GHz) so that the overall $\sim$ 1k detectors array has a diameter of about 11 cm if it is maximally filled (which is of
course highly desirable). This implies a focal length for the optical combiner of $f\sim$ 330mm~\cite{OSulli2015}. Such a focal length was
found to be achievable with a minimal level of optical aberrations with an off-axis Gregorian system with the following
characteristics: 
(1) it is nearly telecentric, (2) it fulfills the Rusch and Mizuguchi-Dragone condition, (3) it features a field of view largely diffraction limited with with Strehl ratio >0.8 within +/- 4.9 degrees~\cite{Gayer}. The requirement for the amount of optical aberrations was  that the sensitivity loss is
less than 10\% when calculated by the ratio of the synthesized beam with and without optical aberrations. Requirements on cold optics and mirrors are summarized on Tables~\ref{table4} and \ref{table5}.

The different diameters have been calculated assuming that 95$\%$ of the power goes through the aperture, but similar 
values have been calculated to get 99$\%$ of the power.

\begin{table}
\centering
\begin{tabular}{|m{7.0480003cm}|m{5.321cm}|}
\hline
{ Optical combiner focal length} &
{30 cm}\\\hline
{ Number of mirrors} &
{ 2}\\\hline
{ M1 shape and diameter} &
{ 480mm x 600mm - }\\\hline
{ M2 shape and diameter} &
{ 600mm x 500mm - }\\\hline
{ Optical combiner sensitivity loss from aberrations} &
{ {\textless} 10\%}\\\hline
\end{tabular}
\caption{Requirements on mirrors and optical
properties\label{table5}}
\end{table}

{
{The possibility to monitor departure from idealities is provided by the self-calibration procedure. This procedure  (see Sect.~\ref{selfcalsect}) is indeed  one of the main advantages of QUBIC with
respect to other more traditional designs (see Sect.~\ref{bmodeprojects}). In order to perform it efficiently, one needs to be able to switch on and off some of the horns
while observing a calibration source. This requires waveguide switches placed in between the back-to-back horns. Such
switches need to be closed enough when in off position (-80 dB) while open enough when set to the on position (-0.1 dB).
Both of these criteria need to be fulfilled simultaneously across the 150 and 220 GHz bands. The switches also need to have low cross talk
between neighbouring switches. The switching between on and off needs to
dissipate minimal power at the 4K stage (60 mW) in order not to heat this stage and perturb observations. Such requirements are summarized on Table~\ref{table6}}}

\begin{table}
%\tablefirsthead{}
%\tablehead{}
%\tabletail{}
%\tablelasttail{}
\centering
\begin{tabular}{|m{7.0480003cm}|m{5.321cm}|}
\hline
{ Switches OFF transmission} &
{ -80 dB}\\\hline
{ Switches ON transmission} &
{ -0.1 dB}\\\hline
{ Switches Cross-talk} &
{ -40 dB}\\\hline
\end{tabular}
\caption{Requirements on switches\label{table6}}
\end{table}
{
{External shields are required to prevent ground pickup in the detectors and make sure that
photons coming from a large angle with respect to the optical axis are absorbed or reflected before entering the
cryostat. This is achieved thanks to:}}

\begin{itemize}
\item {
a cylindrical forebaffle attached to the cryostat with a 1m length and a $14 \deg$  opening angle. This allows to reduce
by more than 20dB the radiation coming from 20deg {\textless} $\theta $ {\textless} 40deg from the optical axis, and
by more than 40dB beyond.}
\item {
{an external shield around the instrument mount or the experiment module's roof (therefore
fixed with respect to the ground) that allows a reduction of the radiation by another 40dB beyond 80 degrees from the
zenith and minimize scan synchronous pick-up.}}
\end{itemize}

\begin{table}
\centering
\begin{tabular}{|m{7.0480003cm}|m{5.321cm}|}
\hline
{ Baffling reduction 20deg {\textless} $\theta $ {\textless} 40deg} &
{ {}-20 dB}\\\hline
{ Baffling reduction 40deg {\textless} $\theta $ {\textless} 80deg} &
{ {}-40 dB}\\\hline
{ Baffling reduction $\theta $ {\textgreater} 80deg} &
{ {}-80 dB}\\\hline
\end{tabular}
\caption{Requirements on the external shields\label{table7}}
\end{table}

\item{ Detectors :\\}
\hypertarget{RefHeadingToc314322913}{}{
{Transition Edge Sensors (TES) are the state of the art of bolometric detectors already employed in several millimetric and sub-millimetric astronomical  experiments all over the world. They have been chosen as detectors for the QUBIC
first module, relying on the extensive developments made in France over the last few years. We may however consider other
types of detectors such as KIDs (Kinetic Inductance Detectors) for future QUBIC modules as they may offer an easier
fabrication and readout, and larger scalability although they  are not yet completely competitive in terms of noise with the TES.}}

A QUBIC TES focal plane is made of an array of 4$\times$256-pixels arrays disposed in an overall diameter of the order of 110 mm. The 
TES matrix for one focal plane of "QUBIC 1st module" is made of four identical pieces. The full focal plane TES matrix will have a quasi-circular shape as shown in Figure~\ref{fig:tes_foc_plane}.
\begin{figure}
\centering\includegraphics[width=.7\textwidth]{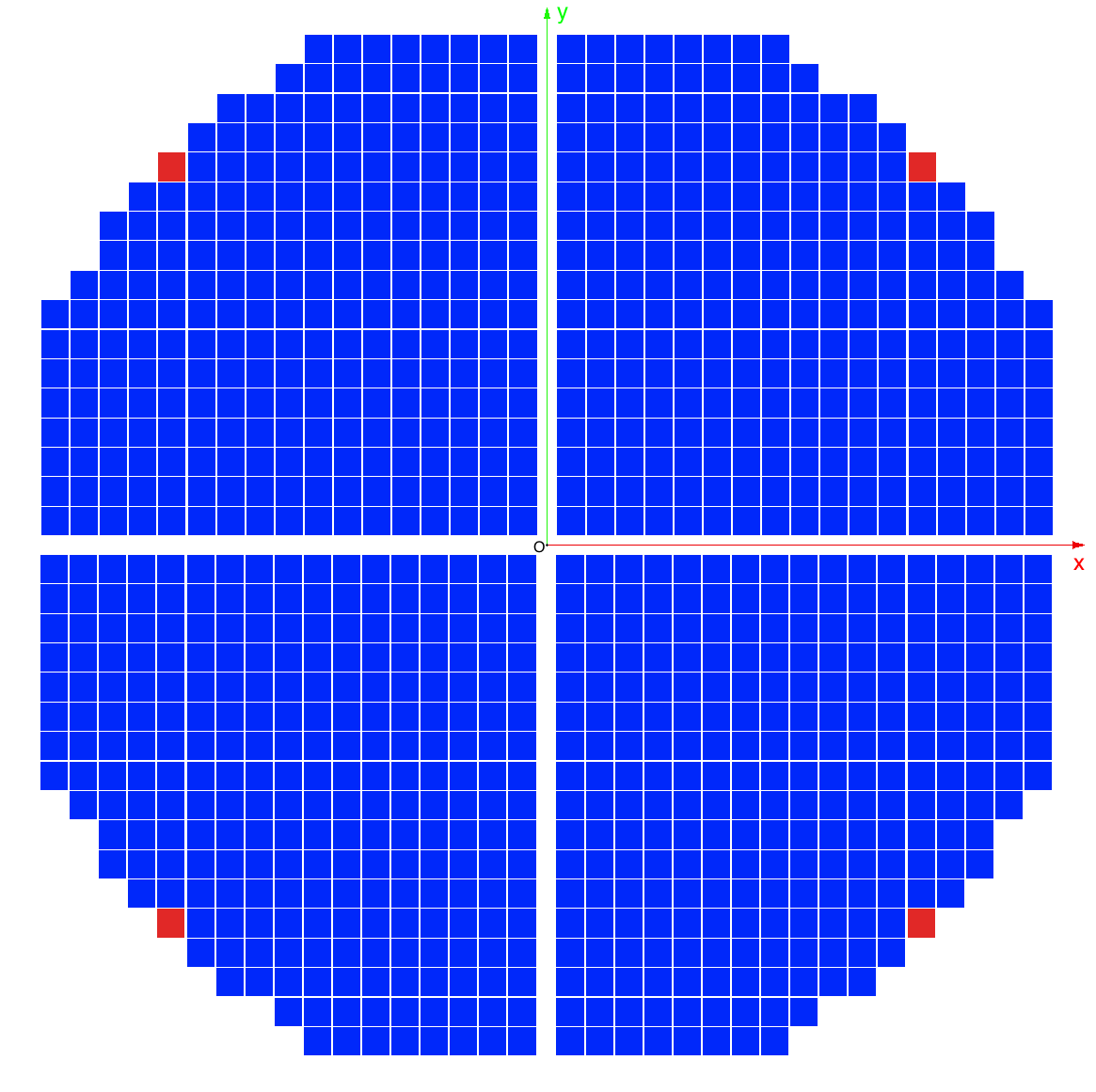}
\caption{schematic "top-view" of the TES bolometers for one focal plane of the QUBIC 1st module. Active detectors are shown in blue.\label{fig:tes_foc_plane}}
\end{figure}

%{The QUBIC TES dimensions are detailed in the document
%}\href{https://atrium.in2p3.fr/nuxeo/nxdoc/default/b0dd2bc5-5610-48ad-b84f-e81acec23b92/view_documents}{\textstyleInternetlink{{ATRIUM-77684}}}{
%annexed to this document.}}
A quarter of a focal plane is composed by 248 "usable" TES elements plus 8 blind sensors for $1/f$ noise monitoring. 
Thus a full focal plane include 992 "usable" TES bolometers, and the  QUBIC 1st module will have 1984 usable TES.
A quarter of a focal plane is presented in Figure~\ref{fig:tes_quarter_foc_plane}.
\begin{figure}
\centering\includegraphics[width=10 cm]{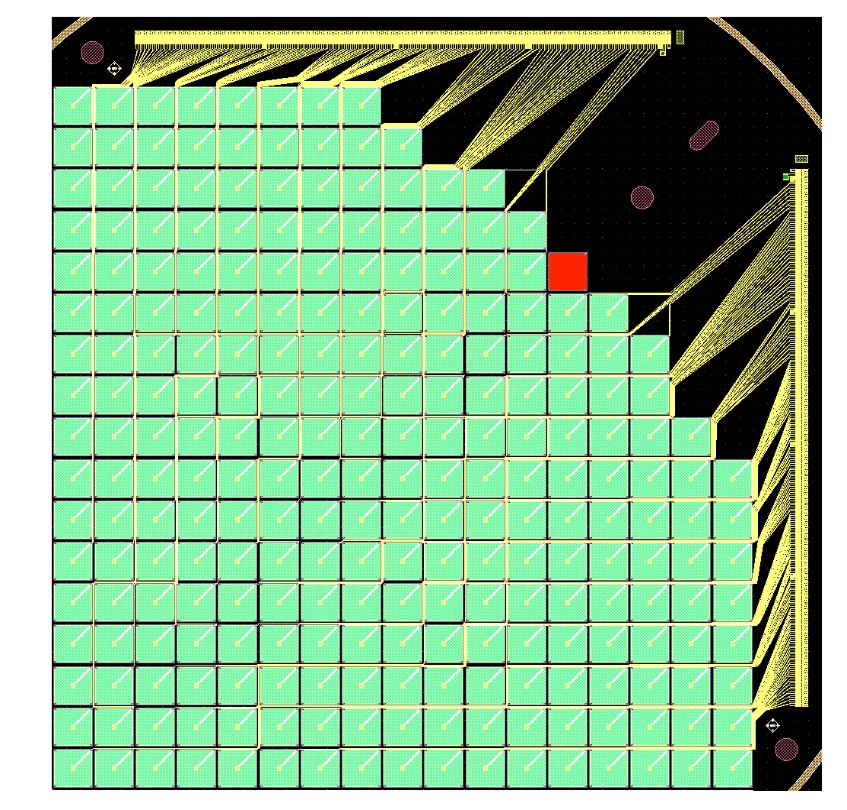}
\caption{Picture of a TES array covering a quarter of the focal plane. Yellow lines are wires used for the reading of the TESes signal. The TES in red is not used.\label{fig:tes_quarter_foc_plane}}
\end{figure}

 The shape of one single TES and its electromagnetic wave absorber part are shown on Figure~\ref{fig:tes_details}.

\begin{figure}
\centering
\begin{tabular}{cc}
\includegraphics[width=.5\textwidth]{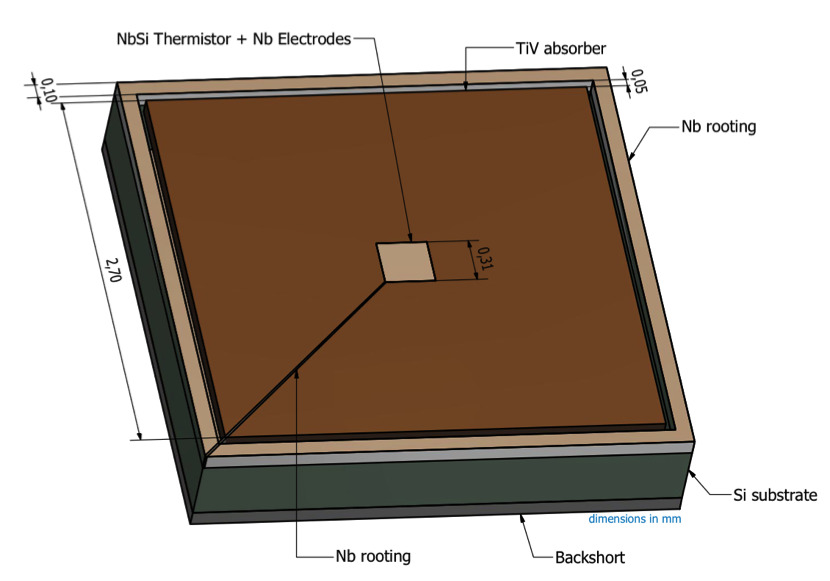}
&
\includegraphics[width=.4\textwidth]{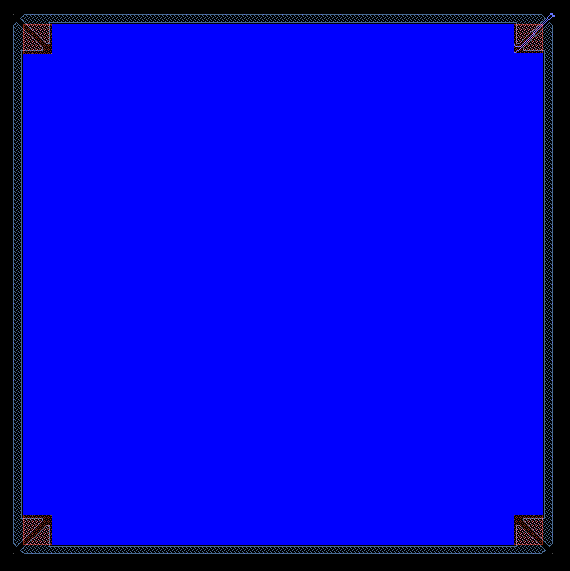}
\end{tabular}
\caption{(left) Picture on one of the QUBIC focal plane TESes ; (right) Absorbing part of one TES (in blue).\label{fig:tes_details}}
\end{figure}

\begin{figure}
\centering
\includegraphics[width=9cm]{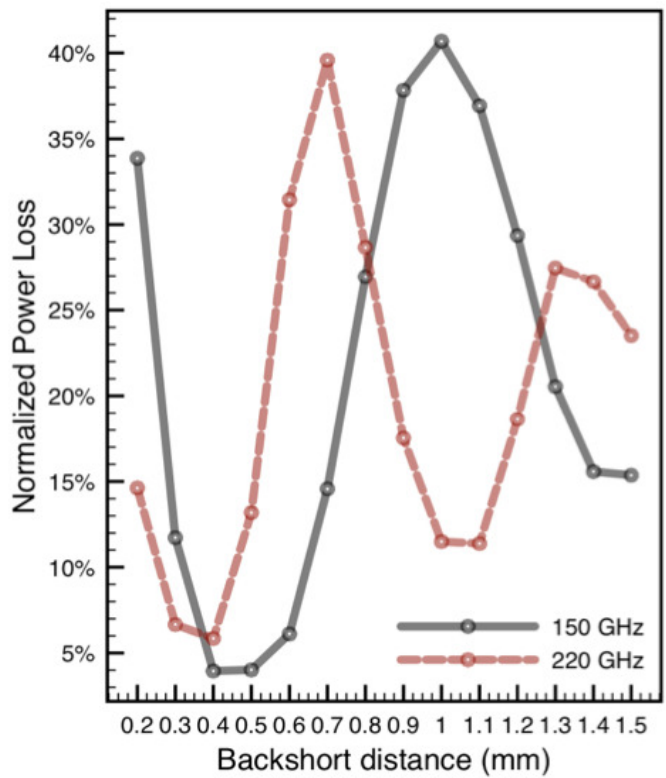}
\caption{Simulated power loss of a detector at 150GHz and 220GHz with respect to backshort distance. An optimal value for the backshort is 400$\mu$m.\label{simu_michel}}
\end{figure}

As mentioned above, the detector size is approximately defined by the central  wavelength of the 150 GHz band, namely 2.7mm. Detectors, 
however, need to exhibit the same efficiency at 220 GHz as at 150 GHz. This efficiency is driven by the thickness of the backshort
below the detector plane. We have set the power loss requirement at 10\% for each 
 and as it will be seen in Figure ~\ref{simu_michel}, we achieve
4\% at 150GHz and 6\% at 220 GHz. The number of detectors is determined by the required fraction of the secondary beam from
the horns to be integrated in the focal plane. The requirement of 80\% of the power integrated sets the number of
detectors to 992, namely 4 wafers of 256 TES assembled together (minus the 8 blind detectors per wafer). 
We require that the fabrication yield of the TES is larger than 90\%.

\begin{table}
\centering
\begin{tabular}{|m{7.0480003cm}|m{5.321cm}|}
\hline
{ TES size} &
{ 2.6 mm}\\\hline
{ Power loss on TES} &
{ {\textless} 10\%}\\\hline
{ Power integrated on focal plane} &
{ {\textgreater} 80\%}\\\hline
{ Number of bolometers / focal plane} &
{ 1024} \\\hline
{ Number of 256 TES wafers} &
{ 4}\\\hline
{ Fraction of operational detectors / wafer} &
{ {\textgreater} 90\%}\\\hline
\end{tabular}
\caption{Requirements on the TES detectors\label{table8}}
\end{table}
{
{In order to ensure a fruitfull exploitation of the QUBIC instrument data, the detectors sensitivities  need to be close to the 
background limit, despite the fact that the focal planes are cooled down to 320 mK. %for cost and design simplicity reasons (reaching 100 mK would have made them completely background limited). 
Such a situation is achieved with TES noise below 
$5.10^{-17}W/\sqrt{Hz}$. We also require the time constants
to be less than 10ms. Accordingly, the data rate for scientific data is required to be 100 Hz.}}

\begin{table}
\centering
\begin{tabular}{|m{7.0480003cm}|m{5.321cm}|}
\hline
{ Detector stage temperature spec.} &
{ 350 mK}\\\hline
{ Detector stage temperature goal} &
{ 320 mK}\\\hline
{ Bolometers NEP} &
{ {$5.10^{-17}\hbox{W.Hz}^{-1/2}$}}\\\hline
{ Bolometers time constant} &
{ {\textless} 10 ms}\\\hline
{ Number of bolometers / focal plane} &
{ 1024}\\\hline
{ Number of 256 TES wafers} &
{ 4}\\\hline
{ Scientific Data sampling rate} &
{ 100 Hz}\\\hline
\end{tabular}
\caption{Requirements on the sensitivity\label{table9}}
\end{table}

\item{ Cryogenics :\\ }
\hypertarget{RefHeadingToc314322914}{}{
{The whole instrument will be integrated in a cryostat that needs to be operated without the
use of cryogenic liquids in order to be usable in any remote observation site. The 4K stage is therefore
ensured thanks to a Pulse Tube Cooler achieving at least 1 W of cooling power at 4K. The electrical consumption of the
Pulse Tube Cooler was required to be less than 15kW\footnote{This was needed in the case of an installation of a QUBIC module in Dome C.}. A further requirement on the Pulse
Tube Cooler is that it remains with unchanged cooling efficiency when the instrument is tilted in elevation during
observations in the range required by the scanning strategy (30 to 70 degrees elevation, so $\pm$ 20 degrees). The 1K
stage (secondary and primary mirrors, dichroic and detector structure) will be achieved using a $^4$He sorption fridge.
% with minimal {\bf XX} W
%cooling power. 
The cryogenic stage for detectors will be ensured through a $^3$He/$^4$He sorption cooler achieving a cooling power of at least
20$\mu$W at this temperature.}}

{
{In order to be easily transported, the outer dimensions of
 the diameter of the cryostat
is required not to exceed 1.6m and the height 1.8m. This requirement sets the dimensions of  the whole internal
cryogenic architecture. The weight of the instrument should not exceed  800 kg in order to be still transportable by
helicopters if needed.}}

\begin{table}
\centering
\begin{tabular}{|m{7.0480003cm}|m{5.321cm}|}
\hline
{ 4K cooling} &
{ Pulse Tube Cooler}\\\hline
{ Pulse Tube Cooler 4K cooling power} &
{ {\textgreater}1 W}\\\hline
{ Pulse Tube Cooler Electrical consumption} &
{ {\textless} 15 kW}\\\hline
{ Pulse Tube Cooler angle range} &
{ +/- 20 degrees}\\\hline
{ 1K stage refrigerator} &
{ $^4$He sorption fridge }\\\hline
{ 1K cooling power} &
{ {\textgreater}2 mW}\\\hline
{ detector stage refrigerator} &
{ $^3$He/$^4$He Sorption Cooler}\\\hline
{ detector stage cooling power} &
{ {\textgreater} 20$\mu$W}\\\hline
{ Instrument Diameter} &
{ {\textless} 1.6m}\\\hline
{ Instrument Height} &
{ {\textless} 1.8m}\\\hline
{ Instrument Weight } &
{ {\textless} 800 kg}\\\hline
\end{tabular}
\caption{Requirement on cryostat and cryogenics\label{table10}}
\end{table}

The overall internal structure of the cryostat will hold the horns+switches assembly, the mirrors, the dichroic and the
detectors. It is cooled down to 1K. Such an assemble needs to weight less than 150 kg %\todocomment{EDP: seems to large - check this} 
in order to prevent a too long
cooling time for the cryostat. It also needs to bend by less than 400$\mu$m when the elevation of the instrument varies in the
observation range (30 to 70 degrees). The heat conduction of the attaches of this structure need to be less than 2$\mu$W.

\begin{table}
\centering
\begin{tabular}{|m{7.0480003cm}|m{5.321cm}|}
\hline
{ Internal Structure weight} &
{ {\textless} 150 kg}\\\hline
{ Internal Structure temperature spec.} &
{ {\textless}1.4K }\\\hline
{ Internal Structure temperature goal} &
{ 1 K}\\\hline
{ Internal Structure bending for +/- 20 deg.} &
{ {\textless} 400$\mu$m}\\\hline
{ Internal Structure attaches heat conduction } &
{ {\textless} 2$\mu$W}\\\hline
{ Internal Structure rotation } &
{ {\textless} 0.2\degree}\\\hline
\end{tabular}
\caption{Requirements on instrument internal
structure\label{table11}}
\end{table}

\item{ Self-calibration source :\\}
\hypertarget{RefHeadingToc314322915}{}
%{{}- tower, sources, ...}
\label{calibSource}
{
This external calibrator is an active source able to radiate a typical power of few mW through a feedhorn with a
well-known beam, and a low level of cross-polarisation (typically {\textless} -30 dB). Two similar systems, including a
microwave sweeper followed by a cascade of multipliers, will be used to generate quasi- monochromatic signals to span
both QUBIC bands. The external calibrator will be in the far-field of the interferometer, which means at about 40m. For
this reason, it will be installed on top of a tower nearby the instrument. Due to the extreme environment conditions,
the sources will be installed in an insulation box, suitable to maintain the devices in the desired temperature range.}

{
%The calibration sources will be procured after a call for tender and some companies have in their catalogues
%off-the-shelf systems with specifications very close to ours. 
We resume the basic specification of the
sources in Table~\ref{table12}. More details are given in Section~\ref{calibSource2}.}

\begin{table}
\centering
\begin{tabular}{|m{7.0480003cm}|m{5.321cm}|}
\hline
{ Frequency coverage} &
{ 110-170 GHz \& 170-260 GHz}\\\hline
{ power output spec.} &
{  5 mW}\\\hline
{ power output goal } &
{ 1 mW }\\\hline
{ Operation modes} &
{ CW + amplitude modulation}\\\hline
{ Polarisation } &
{ Linear}\\\hline
{ Cross-polarisation} &
{ ${\leq}$ -30 dB}\\\hline
{ Weight (estim., including insulation box)} &
{ 10 kg}\\\hline
\end{tabular}
\caption{Requirements on calibration sources\label{table12}}
\end{table}

{
{The tower must be around 40 m tall, and endowed with a lift to carry the source box and other
equipment on top. A platform must be accessible at least for one person to operate the source and/or perform basic
maintenance and/or to switch from 150 GHz to 220 GHz channel if required (we might consider the option of a source
having a single microwave sweeper, but two different multiplier chains). }}

{
In order to avoid uncontrollable power fluctuations during self-calibration, we require stability against the wind:
%. In
%particular, in normal conditions (wind at about 20 km/h in Dome C), 
the lateral displacement of the platform on top
shouldn't exceed {\textpm} 20 cm with respect to the nominal position.}

\item{ Mount :\\}
%\todocomment{ il mettre un summary des reqs ?}
The main requirements on the mount system 
are summarized in table~\ref{tablemount}.
\begin{table}
\centering
\begin{tabular}{|l|l|}
\hline
Maximal diameter &	2500 mm\\ \hline
Maximal height &	2500 mm \\ \hline
Mass (without the instrument) &	< 2300 kg\\ \hline
Mass to be supported by the mount	&  700 kg\\ \hline
Diameter of the instrument &	1600 mm\\ \hline
Height of the instrument with forebaffle &	1800 mm\\ \hline
Electrical consumption of the mount &	< 1 kW\\ \hline
Rotation in azimuth &	-220$^\circ$ / +220$^\circ$\\ 
\hline
Rotation in elevation	 & +30$^\circ$ / +70$^\circ$\\ 
\hline
Rotation around the optical axis &	-30$^\circ$/ +30$^\circ$\\ 
\hline
Pointing accuracy (all axis)& 	< 20 arcsec \\ 
\hline
Angular speed (all axis)	& Adjustable between 0 and 5$^\circ$/s with steps < 0.2$^\circ$/s \\ 
\hline
\end{tabular}
\caption{General requirements on the mount
system.\label{tablemount}}
\end{table}

%
%are specified in the document
%}\href{https://atrium.in2p3.fr/nuxeo/nxdoc/default/dc7d80fd-cb35-425b-8026-97761da77434/view_documents}{\textstyleInternetlink{{ATRIUM-77706}}}{
%annexed to this document.}}

%\item{environmental conditions we foresee for QUBIC (operational temperature, overall description of the experiment module...)}
%{\missing to be completed}

\item{ Slow control / data storage :\\}

Four operating modes have been identified:
\begin{itemize}
\item Passive mode (no signal is acquired),
\item Diagnostic mode (acquisition of diagnostic data such as temperatures),
\item Calibration mode (used during observation of calibration sources, acquisition of bolometric, matrix thermometer, mount, switches, diagnostic and calibration sources data),
\item Observation mode (acquisition of science data during sky observation, i.e. bolometric, matrix thermometer, mount and diagnostic data).
\end{itemize}
%\hypertarget{RefHeadingToc314322917}{}{
%{The data rate, storage and types requirements are specified in the document
%}\href{https://atrium.in2p3.fr/nuxeo/nxdoc/default/a7bec091-723a-4c2d-b1fb-cde35aa537df/view_documents}{\textstyleInternetlink{{ATRIUM-77671}}}{
%annexed to this document. }}
%\todocomment{il faut pas mettre un summary des reqs la aussi ?}

In the nominal observation mode (with an acquisition frequency of the scientific signal tuned at 2 kHz), the data rate (including raw and scientific signals, excluding house keeping signals) of the instrument will be 0.6 Mo/s. At that acquisition frequency, the needed data storage will be 20 To/year (see also section~\ref{datastore} and tables~\ref{table16} and~\ref{table17}).

The slow control of the instrument allows to operate properly the overall system and especially the cryogenic system. It will be implemented in the QUBIC studio data acquisition system which has all the needed interfaces already implemented (serie, USB, GPIB...). All subsystems will provide their slow control system which will be further interfaced with QUBIC studio.

\end{description}

%\subsection{\missing System concept [F0E0] Michel}
%\hypertarget{RefHeadingToc314322919}{}

%{\missing nothing written ?}

%auto-ignore

\subsection{Cryogenic systems }

\subsubsection{Cryostat design / Mechanic architecture and CAD }
\label{cryo}

The cryogenic system of QUBIC aims at cooling the detector arrays at 0.3K, the beam combiner optics at  1K, and the rotating HWP, the polarizing analyzer, the horn array, and the switches at 4K. 
It is based on:

\begin{itemize}
\item {
A self-contained 3He refrigerator cooling the detector arrays}
\item {
A self-contained 4He refrigerator pre-cooling the 3He fridge and cooling a large 1K shield surrounding the optical
system (the beam combiner optic)}
\item {
{Two 1W pulse-tube (PT) refrigerators working in parallel and cooling the experiment volume at 3K
and the surrounding radiation shield at 40K respectively }}
\item {
A large vacuum jacket surrounding the entire system, including a large (50 cm) optical window}
\item {
Heat switches, Heaters, Thermometers, Control Electronics to run the system.}
\end{itemize}

In the following we describe the basic design choices, and the dimensions and interfaces of the cryogenic system.

\subsubsection{Cryostat vacuum }

The purpose of the outer shell of the cryostat is to allow the setup ot operate under  high-vacuum conditions in the internal volume of the
cryostat, to support all the internal elements, and to permit  mm-wave radiation under study to reach the cryogenic part of
the instrument through the optical window.
The size of the outer shell of the cryostat is driven by the volume of the cryogenic
instrument, which includes the polarization modulator, the horns array, the beam combiner mirrors, and the focal plane
assembly, for a total volume of the order of $1\ m^3$. The cryostat has been designed around
the cryogenic instrument, and its dimensions are a trade-off between the total size limit imposed by the 
transportation and the need for sufficient thermal insulation between the cryogenic instrument and the room-temperature
shell. 

The resulting vacuum shell has a diameter of 1.4m and a height of 1.55m. Its shape and
structure has been optimized for withstanding the stress from atmospheric pressure outside and vacuum inside, with
sufficient safety factors. The structure is made out of Aluminium  alloy sheets, roll-bent and welded, reinforced by a
stiffening ribs structure. The vacuum jacket is obtained by closing a vertical cylinder with two flanges (using indium
seals) as shown in Figure~\ref{fig42}. The axes of the two PTs are tilted by
$40\deg$ { with respect to the vertical, to allow
optimal elevation coverage during the observations of the sky at the latitude of operation, while maintaining the Pulse Tube head close to the vertical position where its operational performance are maximized.

Figure~\ref{fig42} also shows the two pulse-tube (PT) heads, mounted on dedicated flanges on the cylinder. The top flange differs from the
bottom one because it includes the vacuum window.

\begin{figure}
\centering  
\includegraphics[height=6.5cm, angle=90]{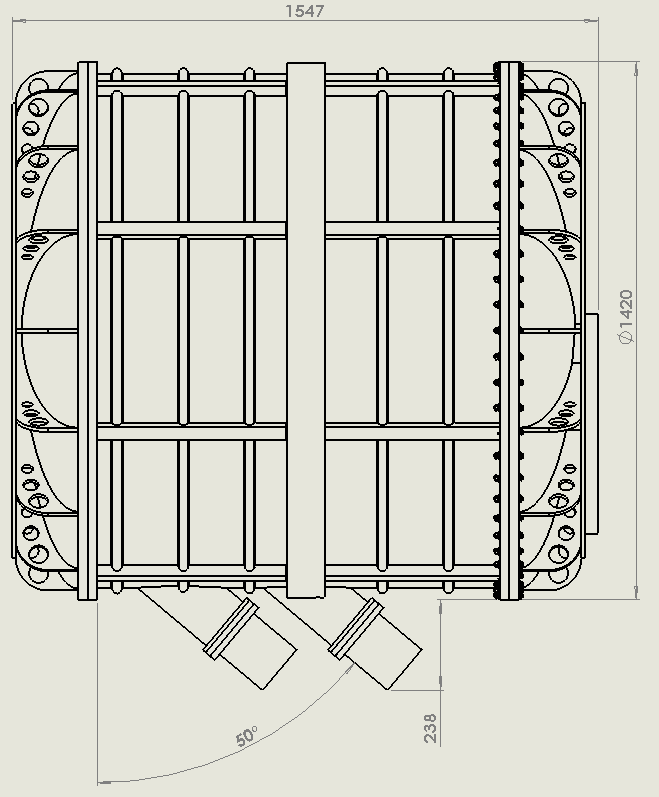}
\includegraphics[width=10.cm]{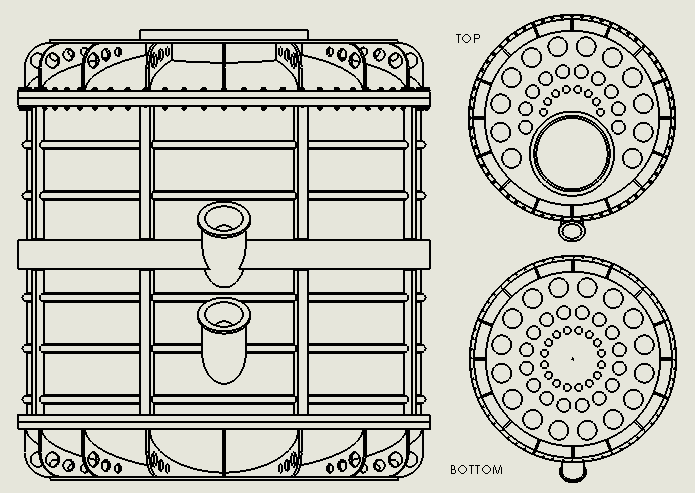} 
\includegraphics[width=7cm]{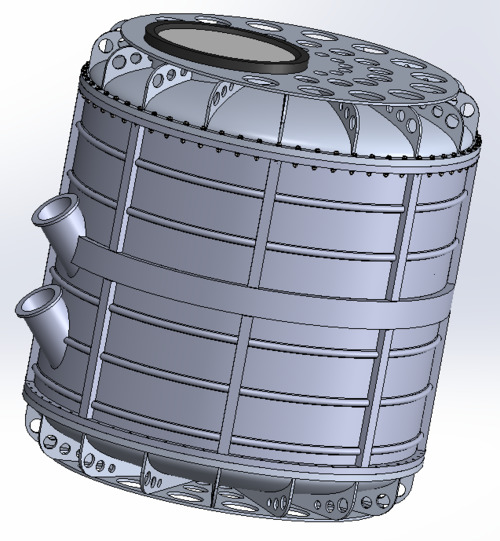} 

\caption{Schematics of the
cryostat shell, with outer dimensions indicated, including the two Pulse-Tube heads.\label{fig42}}
\end{figure}

\subsubsection{Main Cryostat Cooling System }

The cryogenic system is cooled down by two PTs, each providing cooling power of the
order of 1W at 4K and 30Wm at 40K.

The two-stages pulse tubes refrigerate two temperature stages: a 40K shield, surrounding the lower temperature stages and
intercepting warm radiation loads and supporting low-pass filters on the optical chain, and a 4K stage and shield,
surrounding the lower temperature stages, intercepting radiation loads, and supporting directly low-pass filters, the
horns array, the wave-plate rotator assembly, and the hexapod of the 1K stage.

A superinsulation blanket is placed between the outer shell and the 40K shield to reduce the radiative load. 
The two shields are shown in Figure~\ref{fig43} and Figure~\ref{fig44}.

\begin{figure}
\centering  
\includegraphics[width=10cm]{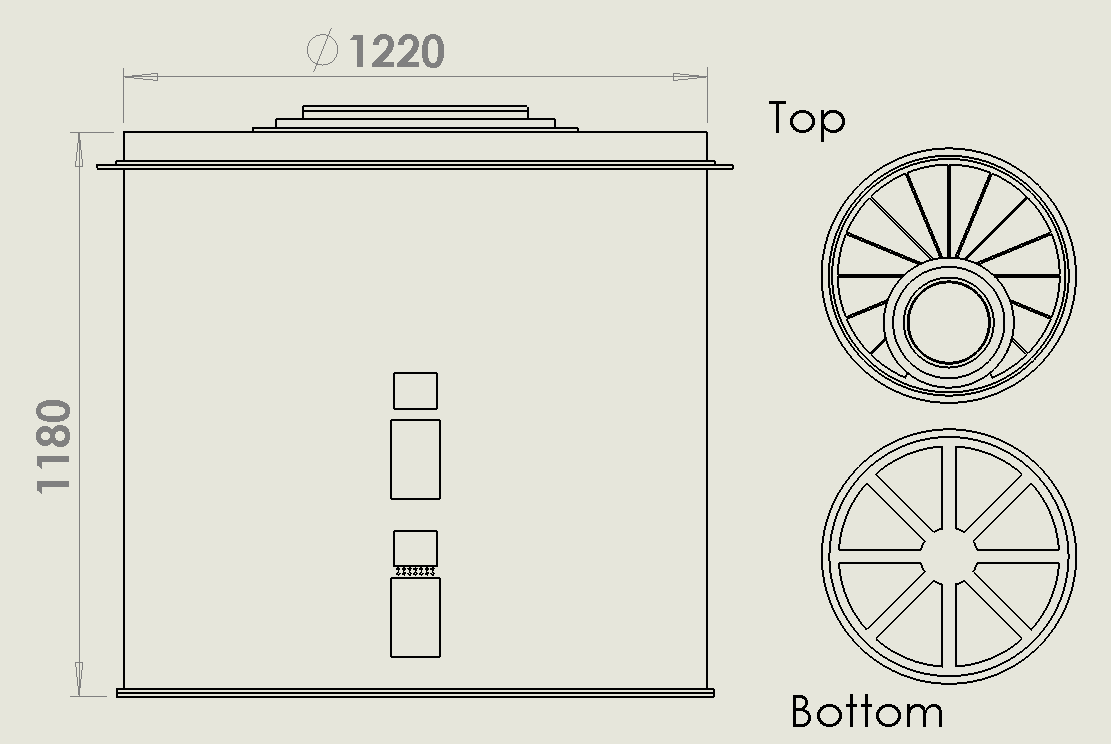} 
\includegraphics[width=6.cm]{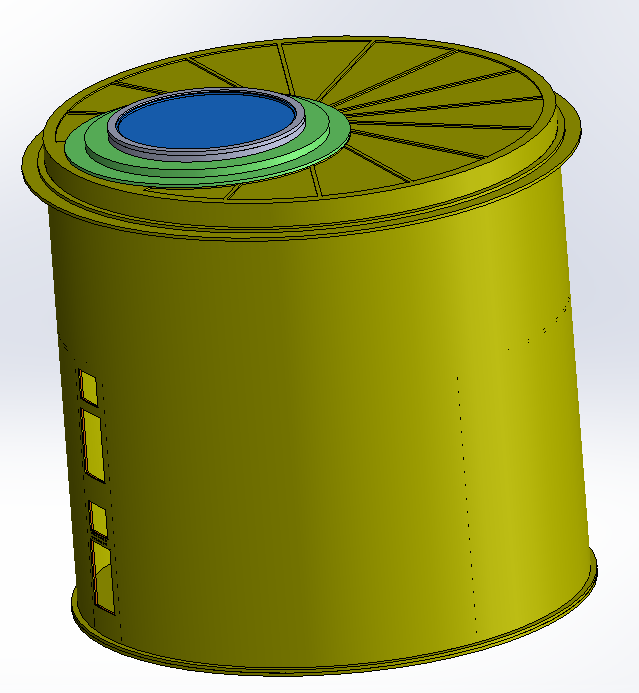} 
\caption{Views of the 40K
shield (first PT stage) with main dimensions. \label{fig43}}
\end{figure}

\begin{figure}
\centering  
\includegraphics[width=10cm]{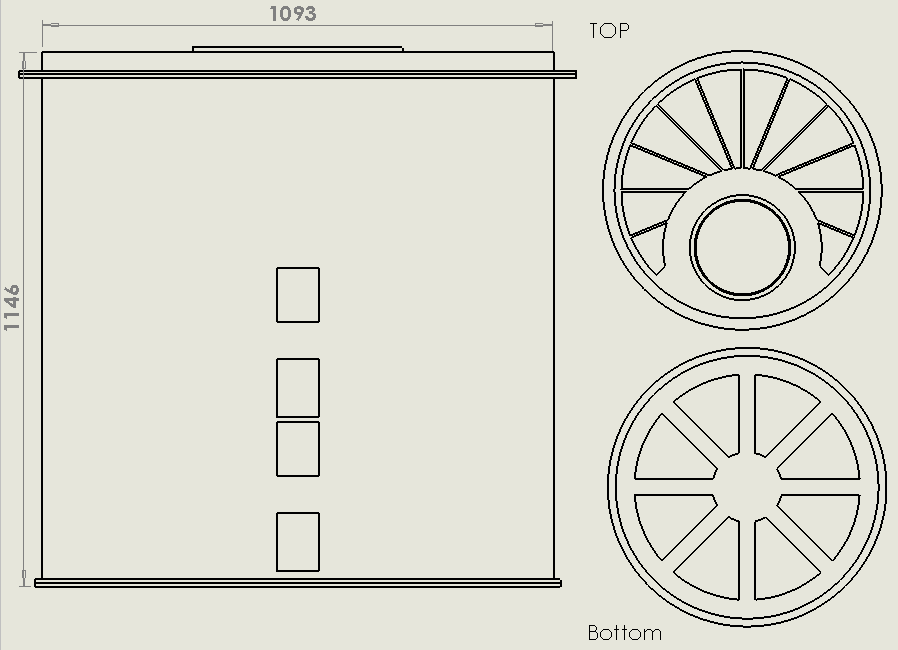} 
\includegraphics[width=7cm]{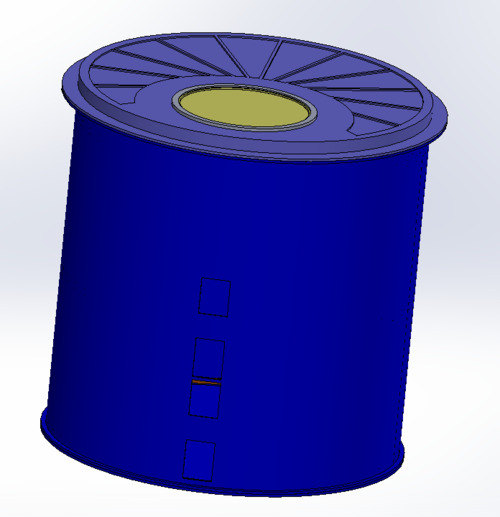}
\caption{Views of the 4K shield (second PT stage) with main dimensions.\label{fig44}}
\end{figure}

The interfaces between the PTs and the shields, flexible enough to accommodate for
differential thermal contraction of the cryostat parts are shown in Figure~\ref{fig45}. The key flexible conductive elements are
gold-plated copper flaps, optimized for flexibility and heat conduction. Further copper belts are used to thermalize
the large shields (especially the 4K one) as shown in the right panel of Figure~\ref{fig45}. 

\begin{figure}
\centering  
\includegraphics[width=6.cm]{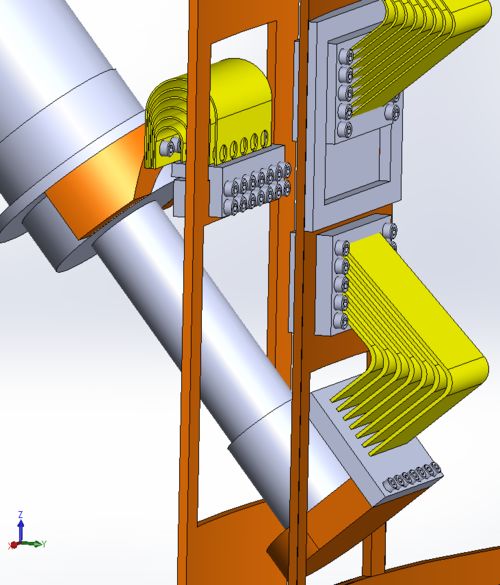} 
\includegraphics[width=7.cm]{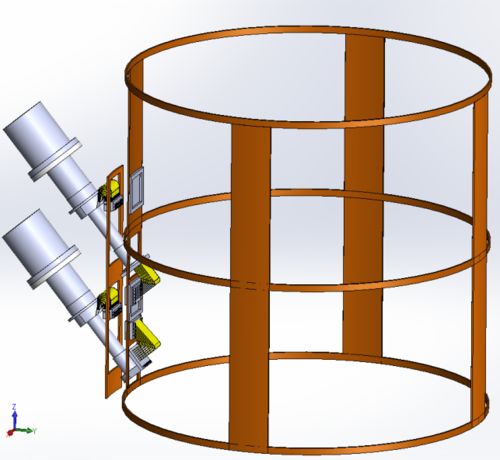} 
\caption{Left: flexible
thermal interfaces between the PTs and the shields. Right: system of copper belts used to thermalize the shields.\label{fig45}}
\end{figure}

The 40K stage is held firmly in place by a system of insulating fiberglass tubes assembled as
in a drum, as visible in Figure~\ref{fig46}. A similar drum is used to hold firmly in place the 4K stage. 
The support structure is completed by a system of radial fiberglass straps mounted on the bottom of the 40K and 4K
shields.

\begin{figure}
\centering  
\includegraphics[width=6cm]{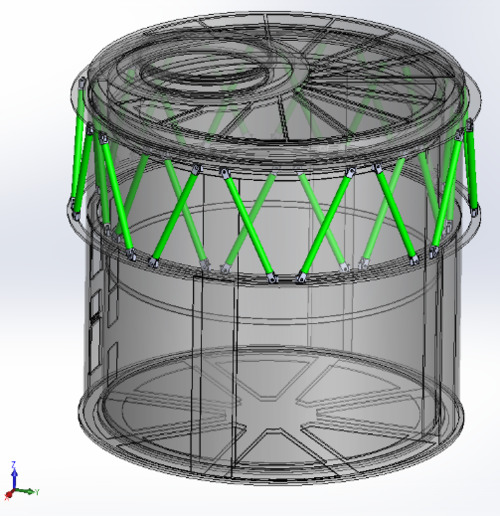} 
\includegraphics[width=10cm]{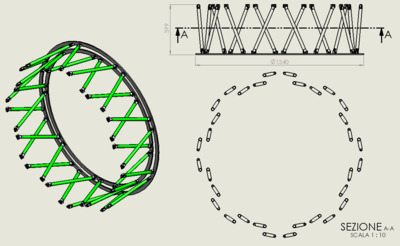} 

\caption{Views of the
fiberglass tubes ``drum'' holding firmly in place the 40K and 4K shields.\label{fig46}}
\end{figure}

Results from a  preliminary simulation of the heat loads on the two stages of the system are reported in Table~\ref{table18}.

\begin{table}
\centering
\begin{tabular}{|lcc|}
\hline
T1 & \multicolumn{2}{r|}{3.0 K} \\
T2 & \multicolumn{2}{r|}{40 K} \\
T3 & \multicolumn{2}{r|}{300 K} \\
\hline
top fiberglass tubes & \multicolumn{2}{r|}{16} \\
bottom fiberglass straps  & \multicolumn{2}{r|}{6} \\
area of 4K shield &  \multicolumn{2}{r|}{5.81 m$^2$}\\
area of 40K shield &  \multicolumn{2}{r|}{6.12 m$^2$}\\
window diameter &  \multicolumn{2}{r|}{0.50 m}\\
\hline
number of superinsulation shields 1-2 & \multicolumn{2}{r|}{10}\\
number of superinsulation shields 2-3 & \multicolumn{2}{r|}{30}\\
number of ASICs & \multicolumn{2}{r|}{16}\\
\hline
W cond (1,2) & 91.68 & 1615.68 mW \\
W wires (1,2) & 0.18 & 2180.00 mW \\
W rad window & 0.28& 901.77 mW \\
W rad (1,2) &  8.43 &  9369.72 mW \\
W ASIC (2) & &  1600.00 mW \\
Q dot (1,2)  &  100.58 & 15667.20 mW \\
\hline
\end{tabular}

\caption{Summary of simulated heat loads on 4K and 40K
stages.\label{table18} }
\end{table}

With a total load of about 0.1W on the 3K stage and of about \ 16W on the 40K stage, operation with a single pulse tube
is possible. We maintain the second pulse tube mainly to handle unexpected large thermal gradients in the system and
extra loads from the window and warm filters. Moreover, \ when cycling the sub-Kelvin \ fridges, operation with a
single PT would be marginal. 
Pre-cooling of the cryogenic sections of the systems is obtained through suitable gas switches.

\subsubsection{1K-box}
As shown in Figure~\ref{fig47}, the 1 K box contains the followings parts: 

\begin{itemize}
\item {
The primary and secondary mirrors}
\item {
The cold stop}
\item {
The dichroic}
\item {
The focal plane}
\end{itemize}

\begin{figure}
\centering 
 \includegraphics[width=10cm]{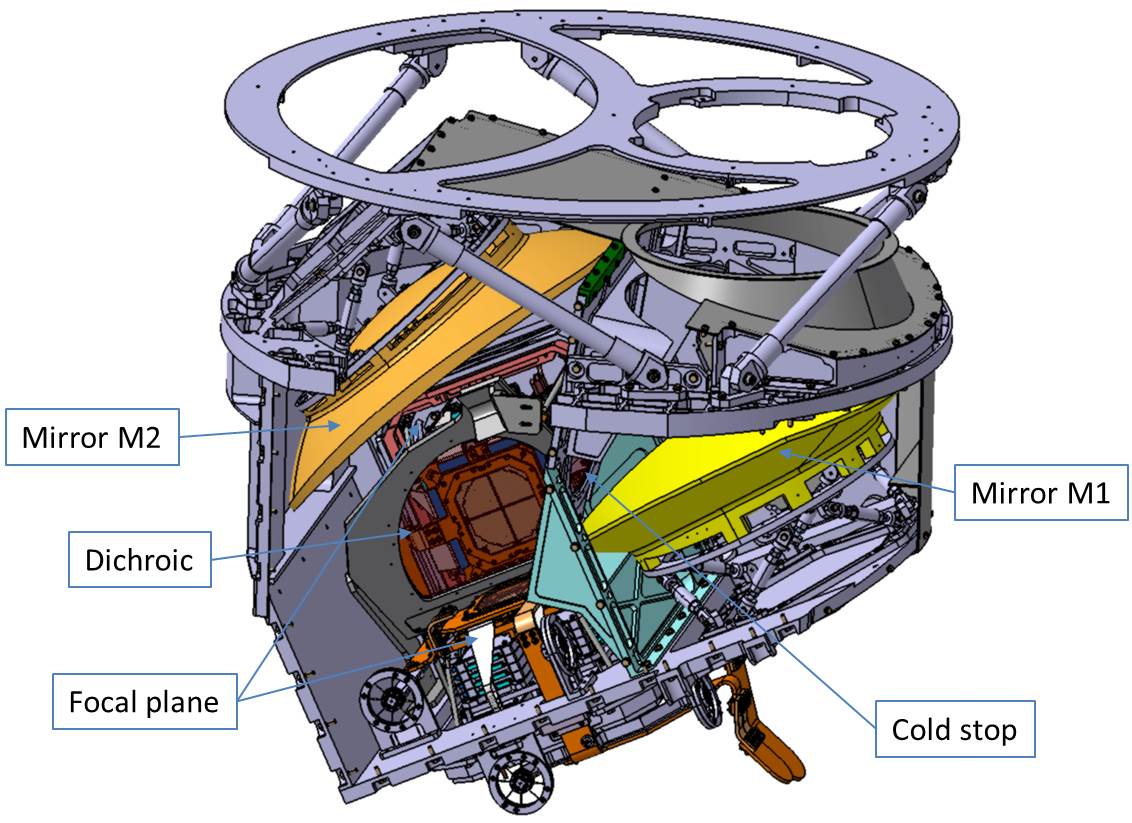}
\caption{1K box and its
inner parts\label{fig47}}
\end{figure}

{
The purpose of the 1K box is, on the one hand to assure the mechanical holding and the alignment of these different
parts, and on the other hand to ensure a thermal shielding at 1K. }

{
{The 1K box is fixed on the 4K stage of the cryostat through its upper support frame (see
Figure~\ref{fig48}) which will be made of Carbon fiber hexapods (which temperature will thus lie between 4K
and 1K). On this upper support frame will also be assembled the
}{horns and switches. The 1K box itself is assembled on its upper support frame by 6 carbon fiber  tubes of thin section for thermally insulating the 1K box of the 4K stage of the cryostat. The 1K box will be
connected to the 1K subsystem.}}

\begin{figure}
\centering  
\includegraphics[width=11.cm]{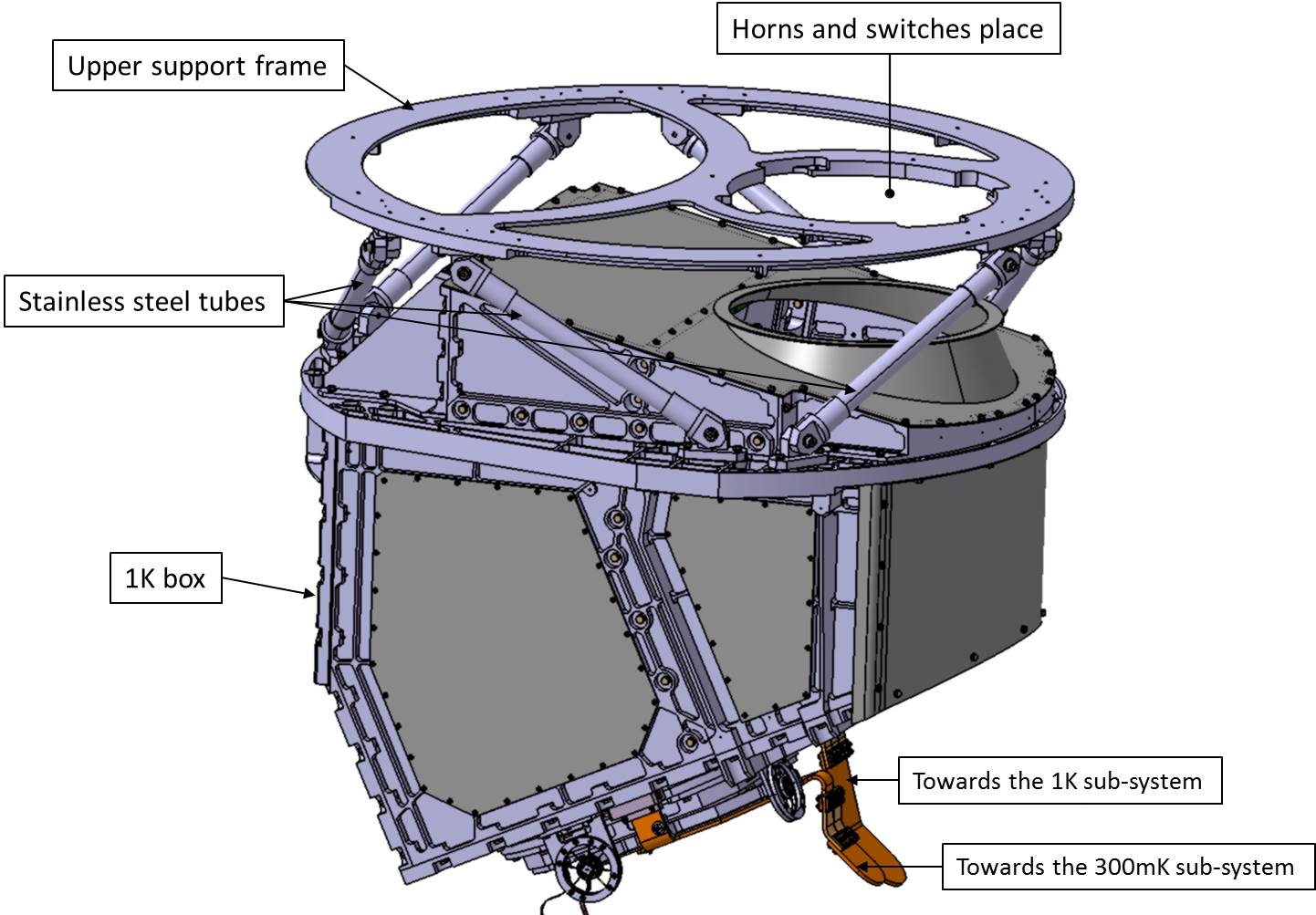}
\caption{Sketch of the 1K box\label{fig48}}
\end{figure}

The 1K box is made of aluminium alloy sheets and plates with stiffening ribs screwed between
them to allow their assembly and to mount and align inner parts (mirrors, cold stop, dichroic, focal plane {\dots}).
Its design is optimized to reduce its mass  (in particular for thermal reason), but also to
increase its stiffness. The requirements are that, under the effect of gravity during the displacement of the
instrument while scanning the sky, the 1K box must be stiff enough to guarantee the alignment of the optical components,
in particular the mirrors and the focal plane. Its dimensions are outlined by Figure~\ref{fig49} and summarized
in Table~\ref{table11}.

\begin{figure}
\centering  
\includegraphics[width=11.cm]{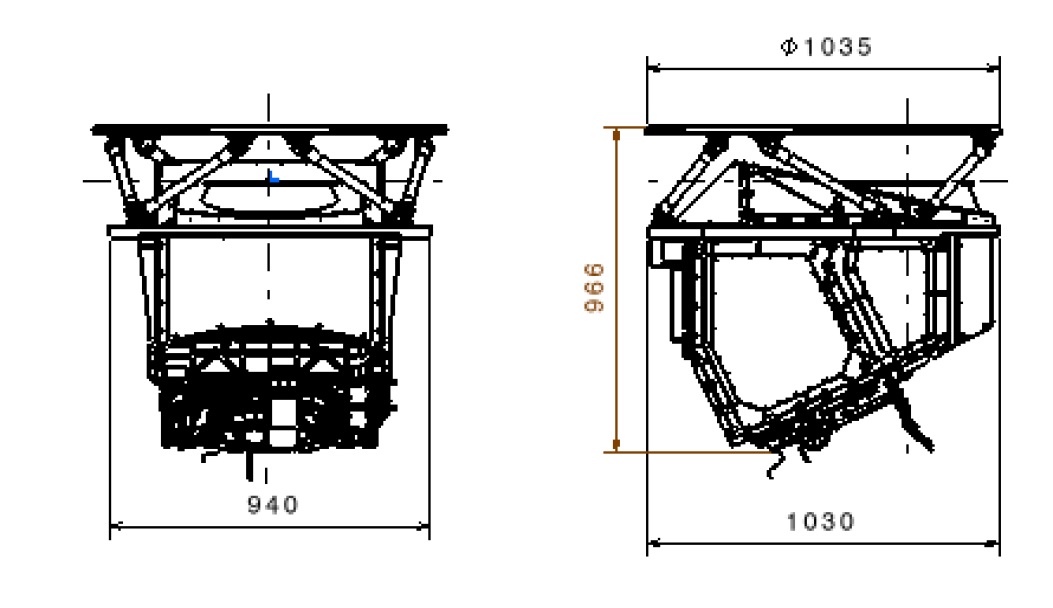}

\caption{Main dimensions of the 1K box.\label{fig49}}
\end{figure}

\subsubsection{1 K System}
\label{sec:1Ksystem}

\paragraph{Requirements}
\label{subsec:1Kreq}

This system is dedicated to cool down the optics box from $4.2$ K to $1$ K. 
Since the requested temperature is in the K-regime, the best option is use an 
$^4$He sorption cooler. The optics box is a system of about $165$ kg: $140$ Kg 
of Aluminium-6061 (Al6061) , $11$ Kg of Stainless Steel 304 (SS304), $10$ kg of 
Copper and $4$ kg of Brass. In order to support the optics box, there are two 
possibilities, the first one is the use of Stainless Steel 304 hexapod. Instead, 
the second one is the use of Carbon Fibre (CF) support. The difference between 
these two material is mainly due to the thermal load that will be on the optics 
box. Indeed, the SS304 will introduce a heat load of $168$ J/day at $1$ K, while 
the CF heat load will be of $43$ J/day at the same temperature. The minimum hold 
time of the fridge requested for this experiment is one day, plus the time of 
recycling. In order to cool down from $4.2$ K to $1$ K, the fridge should be 
able to provide $123$ J. Other contributions (such as radiative transfer from 
cold environment of from window) can be considered negligible. Indeed, heat load 
coming from these sources is less than $0.2$ J/day.

A typical $^4$He sorption cooler is able to provide a minimum cooling power of 2 
mW at $1$ K. Considering the latent heat of the $^4$He, an amount of $1.5$ moles 
of helium to keep the optics box at $1$ K for an entire day in case of the use 
of SS304 hexapod with a previous cooling power. While using CF, only $0.55$ 
moles of $^4$He are requested.
During the cooling phase, a certain amount of gas will evaporate to cool itself. 
In particular, it is possible to find that the number of moles evaporated is 
equal to $2.5$ mol for the SS304 support and less than $1$ mol for the CF (This 
value changes as function of temperature of the pulse tube cold head as it is 
possible to see in figure \ref{fig:TvsMol}).

\begin{figure}
    \centering
    \includegraphics[scale =0.5]{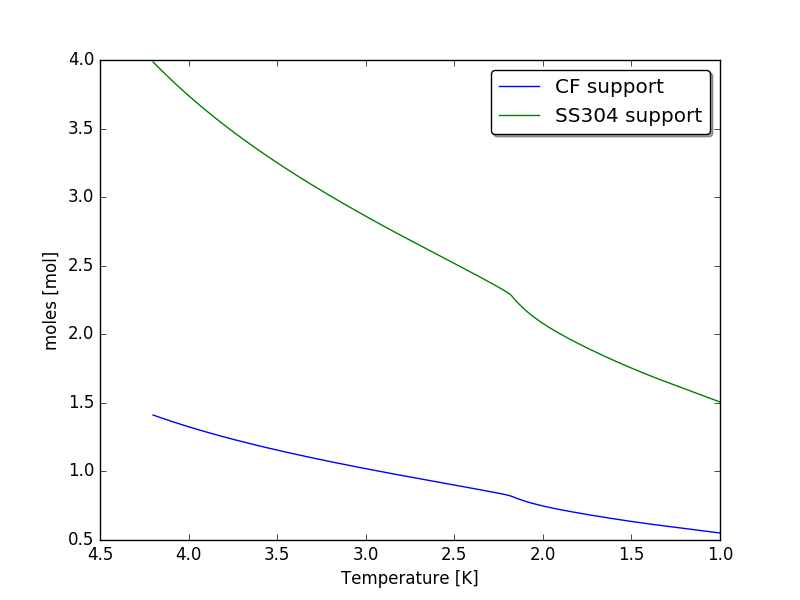}
    \caption{Moles of $^4$He as function of temperature during the cooling phase.}
    \label{fig:TvsMol}
\end{figure}

Therefore, the final requirements for the $1$ K fridge are:

\begin{itemize}
    \item cooling power of at least $2$ mW, \\
    \item total time of operation 24 hrs (hold time) plus cooling time, \\
    \item $4$ mol of $^4$He using SS304 or $1.5$ mol using CF.
\end{itemize}

\paragraph{Design}
\label{subsec:1Kdes}

To design a fridge that respects the previous requirements, there is the 
necessity to distinguish the two different solution for the support. In case of 
the SS304 hexapod, at the moment there is not a fridge able to contain almost 
$4$ mol, so the easiest way is to design two equal small fridges, each of $2$ 
mol. Instead in case of CF support, one fridge is enough. A CAD of a single 
fridge is presented in figure \ref{fig:1KFridge}. This fridge is designed to 
reach the requested temperature, and it has been already manufactured, as shown 
in Figure~\ref{fig:He4GM}. The condenser of the fridge will be attached to the $4.2$ K 
flange in order to condense the helium. In addiction to this connection, there 
will be an heat switch between the cryopump and the $4.2$ K. In order to allow 
the adsorption of the gas and reducing the temperature of the helium bath, the 
switch must be in the ON state to cool down the charcoal. When all the gas is 
adsorbed, a heater will be switched on (and the heat switch off) to increase the 
temperature of the charcoal up to $50$ K and allows the desorption of the gas. 
When all the gas is desorbed, the heater will be switched off, so the heat 
switch on. This phase is very delicate, indeed the charcoal pump will cool down 
from $50$ to $4$ K releasing a huge amount of energy, some thousands of Joule, 
on the $4.2$ K flange. This is due to heat capacity, and so the enthalpy 
difference, of the copper and of the stainless steel that are the main 
components of pump (in a first instance, it is possible to neglect the heat 
capacity of the charcoal which is significantly lower). The releasing of this 
energy will be in a short time corresponding to a power of $2-3$ W, which is 
greater than the cooling of the pulse tube ($1.4$ W). This means that the $4.2$ 
K flange will increase its temperature (with a steep spike) and all the other 
elements attached too. To avoid this problem (which is present in both the cases 
considered for the support), it is possible to use two different pulse tubes, 
one of them dedicated only to the $^4$He fridges (fridge). This implies that 
only the pulse tube attached to the fridges (fridge) will suffer the temperature 
drift, while the other components will remain at $4.2$ K thanks to the other 
pulse tube. 

\begin{figure}
    \centering
    %\begin{tabular}{cc}
    \includegraphics[scale =0.4]{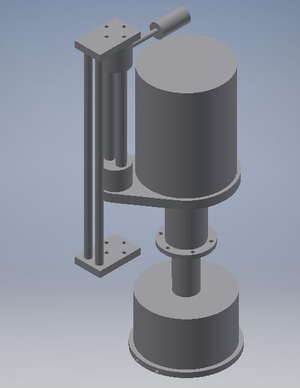}
    %&
    %and Photo of the $^4$He Sorption Cooler  (right) 
    %\end{tabular}
    \caption[CAD drawings  of the 1K Fridge]{CAD drawings of the 1K Fridge.}
    \label{fig:1KFridge}
\end{figure}

\paragraph{Testing}
\label{subsec:1Ktest}

The testing phase will start with the commissioning of the new 
cryostat. This cryostat will use a Gifford-McMahon (GM) mechanical cooler to 
precool the $^4$He fridge at suitable temperature to allow the condensation of 
the gas. The system is presented in figure \ref{fig:He4GM}. The sorption cooler will be 
attached to the cold stage of the GM cooler, which is the lowest copper flange 
in the picture on le left hand side of Figure~\ref{fig:He4GM}.

The $^4$He fridge is presented in figure \ref{fig:He4GM}. 
The indium tube coming from the top of the charcoal pump is visible. This will be connected 
to a gas line, in this way it is possible to charge, and consequentially test, 
the fridge with different quantities of the gas.

\begin{figure}
    \centering
    \begin{tabular}{cc}
    \includegraphics[width=4cm]{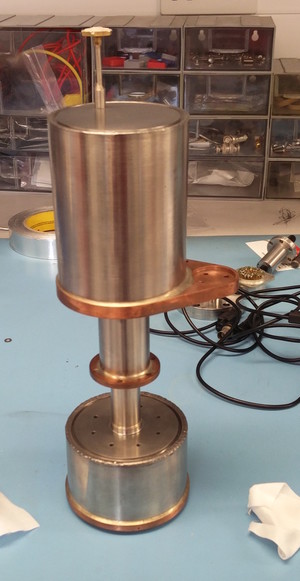}  
    &
    \includegraphics[width=5cm]{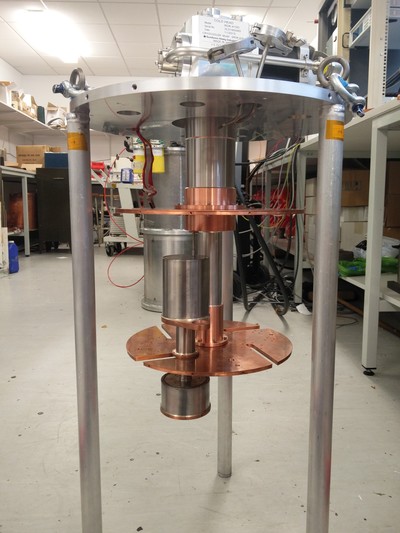}
    \end{tabular}
    \caption[Photos of the $^4$He Sorption Cooler alone and mountend on the 4K stage]{Photo of the $^4$He Sorption Cooler (left); 
Photo of the $^4$He Sorption Cooler mounted on the 4K stage (right).}
    \label{fig:He4GM}
\end{figure}

\subsubsection{sub-K systems}
\label{sec:subKsystem}

\paragraph{Sub-K System Description}
\label{subsect:subKdesc}

The $^4$He + $^{3}$He sorption fridge selected to cool the QUBIC focal plane was
manufactured by Chase Cryogenics and is shown in Figure~\ref{fig:ChaseLoadCurve} (left).%ChaseFridge}.
It is presently
installed within another experiment (cryogenic {\it electron paramagnetic resonance}).
We are negotiating a calendar for final operations of this system, but for now
we have only limited data from earlier tests.

\paragraph{Sub-K Performance Tests}
\label{subsect:subKperformance}

\begin{figure}
 \centering
 \begin{tabular}{cc}
\includegraphics[width=4cm]{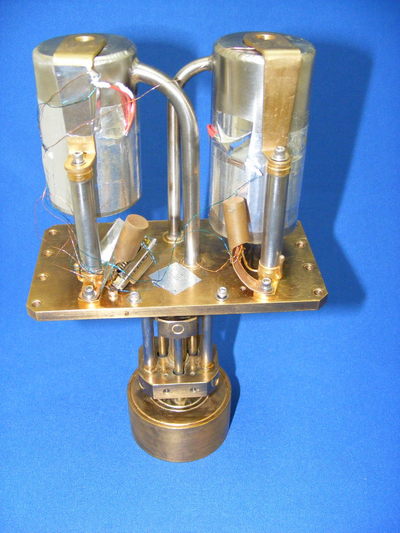}
 &
 \includegraphics[width=.5\textwidth]{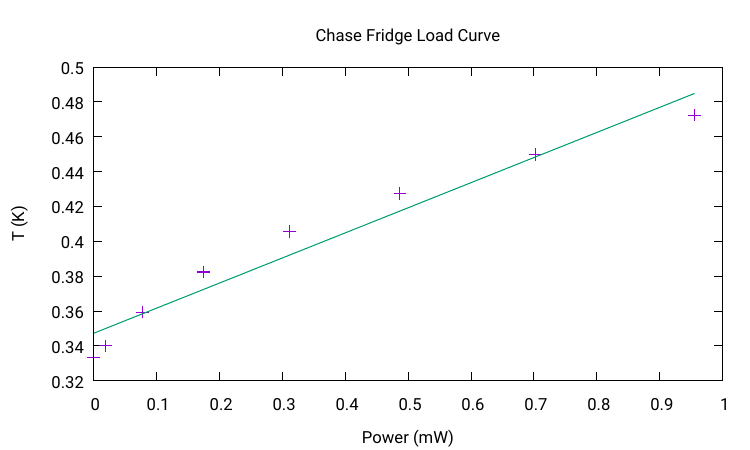}
 \end{tabular}

 \caption[Chase Cryogenics fridge picture and load curve]{Chase Cryogenics fridge for the focal plane (left), Cold stage load curve for Chase fridge (right)}
 \label{fig:ChaseLoadCurve}
\end{figure}

Figure \ref{fig:ChaseLoadCurve} (on the right) shows a load curve measured with the Chase fridge installed into the EPR system.
No loads were connected to the fridge (other than unintentional stray loads, e.g. radiative
loading, but we would expect these to be similar).

We do not at this stage have a predicted hold time for the power considered.
This fridge has a large charge, however.
We will run the fridge with the expected load applied to obtain an expected run-time figure at a later date.

\paragraph{Sub-K Interfaces}
\label{subsect:subKinterfaces}

The fridge requires a certain volume within the system and must be mechanically connected
to the 4-K PTC second-stage cold plate and to the focal plane attachment.
A crude CAD model of the fridge has been provided.
When we have full access to the fridge again we will verify the dimensions of these mechanical interfaces.

Heat will flow into the fridge from the focal plane attachment and from the fridge to the 4-K plate.
The heat lift from the FP will be characterized as described above.
The energy flow to the 4-K stage will be substantial.
This is illustrated by Figure~\ref{fig:GM2EPR}, which shows the response of the 2nd stage of a Sumitomo RDK415 GM cooler
from 3 cycles / part cycles of the Chase fridge.
Admittedly no particular effort has been made to be gentle with these cycles,
but the peak temperature of 7\,K corresponds approximately to a peak load of 7\,W.
For sure careful operation of the fridge can reduce this, perhaps by a factor of two.

\begin{figure}
 \centering
 \includegraphics[width=.5\textwidth]{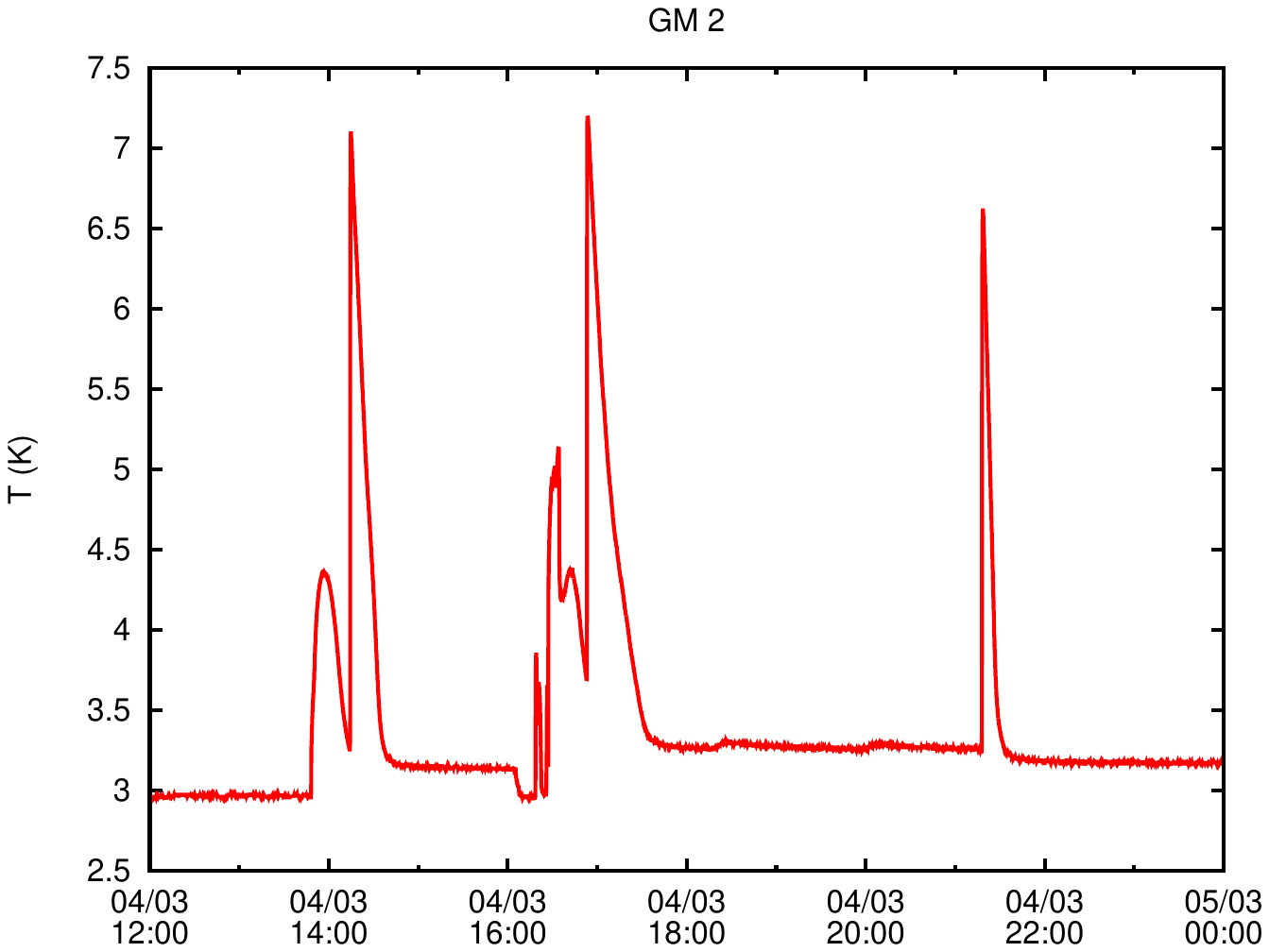}

 \caption{The effect on GM cooler 2nd-stage T from operation of the fridge}
 \label{fig:GM2EPR}
\end{figure}

\begin{table}
\centering
\begin{tabular}{c c c c}
\hline\hline
%Heading
Type & Location & No. Wire Pairs & Notes \\
\hline
Diode & $^4$He heat switch & 1 & \\
Heater & $^4$He heat switch & 1 &  \\
Diode & $^3$He heat switch & 1 & \\
Heater & $^3$He heat switch & 1 &  \\
Diode & $^4$He cryo pump & 1 & \\
Heater & $^4$He cryo pump & 1 &  \\
Diode & $^3$He cryo pump & 1 & \\
Heater & $^3$He cryo pump & 1 &  \\
RTD & Cold stage & 2 & Optional \\
RTD & Intermediate stage & 2 & Optional \\
Heater & Cold stage & 1 & Optional \\
\hline
\end{tabular} 
\caption{Sub-K fridge electrical interfaces}
\label{table:subKfridgeElec}
\end{table} 

The operating conditions of the heaters will be confirmed later, but are 25\,V max 100\,mA max.
Currently we are using 0.1-mm copper twisted pair to supply power to the heaters,
but in the past we have used 0.1-mm Manganin.

Operation of the fridge will require readout of the cold stage temperature.
This could be a thermometer mounted as close as possible to the cold stage for this purpose.
However, a thermometer elsewhere on the load should be adequate.
A thermometer on the intermediate stage can be useful.

A heater on the cold stage can be useful for verification of fridge operation (load curves) or warming up the system.
It could also be used for thermostatic control.
However, it is not vital.

Currently a micro-D connector is mounted to the fridge for these connections.
Gender and pin-out will be confirmed at a later date.

The readout of the thermometers can be by typical equipment (e.g. Lakeshore 370, 318).
We use computer-controlled heaters capable of driving up to 25\,V at 100\,mA.

Our in-house control system uses an xml script to describe a state machine for fridge cycling.
For example, one state might set heat switch drive Voltages, with a test condition that would
progress to a timed wait state once both heat switch thermometers are reading a high enough temperature.
This script should easily translate to whatever control system is employed.

\paragraph{Sub-K Verification}
\label{subsect:subKverification}

Tests as described in section \ref{subsect:subKperformance} have been conducted to check for adequate hold time at the expected power,
the results are summarized in tables~\ref{holdtime1} and ~\ref{holdtime2}.

\begin{table}
\centering
\begin{tabular}{|l|l|l|l|l|l|}
\hline
$P_{\hbox{load}}$ ($\mu$W) & Days & Hours & Seconds & Joules & T (mK)\\
\hline
19.5 &  3.75  & 89.92 & 323700 & 6.31 & 336 mK \\
43.9 & 2.45 & 58.83 & 211800 &  9.29 & 349 mK\\
\hline
\end{tabular} 
\caption{$^3$He hold times for two load values.
\label{holdtime1}}
\end{table} 

\begin{table}
\centering
\begin{tabular}{|l|l|l|l|l|l|}
\hline
$P_{\hbox{load}}$ (mW) & Days & Hours & Seconds & Joules & T (K)\\
\hline
4.39 &  0.03 &  0.70 &2520 & 11.05 &  1.2 K \\
\hline
\end{tabular} 
\caption{$^4$He hold times for one load value.
\label{holdtime2}}
\end{table}

\paragraph{Transportation Issues}
\label{subsect:subKtransport}
It should be noted that this fridge (and the 1-K fridge) relies upon thin-walled tubing to contain high-pressure gas.
This makes it necessary to take extra precautions whilst shipping.
The fridge will be shipped from Manchester with any shipping stays considered necessary, and comparable arrangements
must be put in place to stiffen the assembly sufficiently for onward shipping after integration.

\subsubsection{Heat Switches}
\label{sec:HS}
\paragraph{Heat Switches Description}
\label{subsect:HSdesc}

\begin{figure}
 \centering
 \includegraphics[width=4cm]{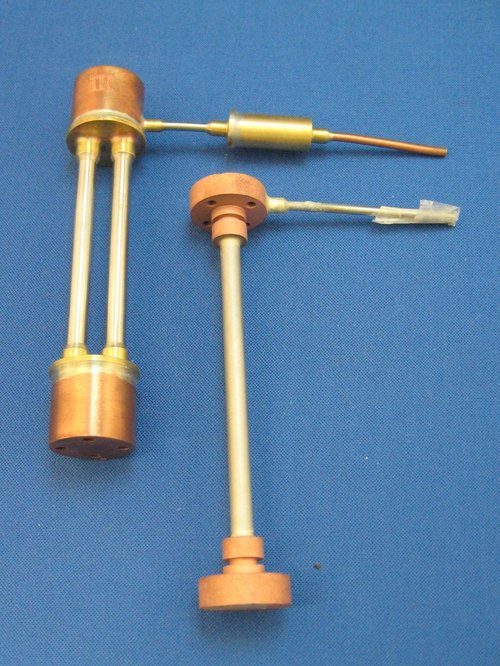}
 \caption{Both switch types}
 \label{fig:SwitchPhoto}
\end{figure}

For the base-line configuration two types of heat switch are considered (see Figure~\ref{fig:SwitchPhoto}):
\begin{description}
 \item[Convective Heat Switch] Two thermal stages (OFHC copper) are connected with a twin-pipe circulation system (thin-wall stainless steel).
 Helium is injected into the circuit using a small {\it cryo pump}.
 So long as the physically-higher stage is colder than the lower stage the gas will convect around the circuit.
 Gas is cooled by the upper stage and warmed by the lower stage, effecting a transfer of heat.
 If the switch is operated across a phase transition (i.e. the stages are above and below the boiling point)
 heat transfer is especially effective due to the latent heat taken / given up by condensing / boiling.
 \item[Minimal Gap Heat Switch] This new design uses a single stainless-steel tube, which is almost filled with
 a copper rod, with a small gap around the rod such that it is not in contact with the inside wall of the tube. 
 This is not connected to the bottom, but at room temperature it might just touch the bottom stage. At cryogenic temperatures 
 differential contraction opens a very small gap at the bottom of the rod.
 This means that the \textit{off} conductance will be determined solely by the conductance of the stainless steel tube.
 Helium from a {\it cryo pump} (not fitted in the photo) is released into the volume to turn the switch \textit{on}.
 Conduction across the small gap by helium gas is very effective.
 When the low end temperature is low enough to condense liquid to bridge the gap the conductance rises further,
 and further still with the formation of super-fluid helium.
\end{description}
\begin{figure}
 \centering
 \begin{tabular}{cc}
 \includegraphics[width=.5\textwidth]{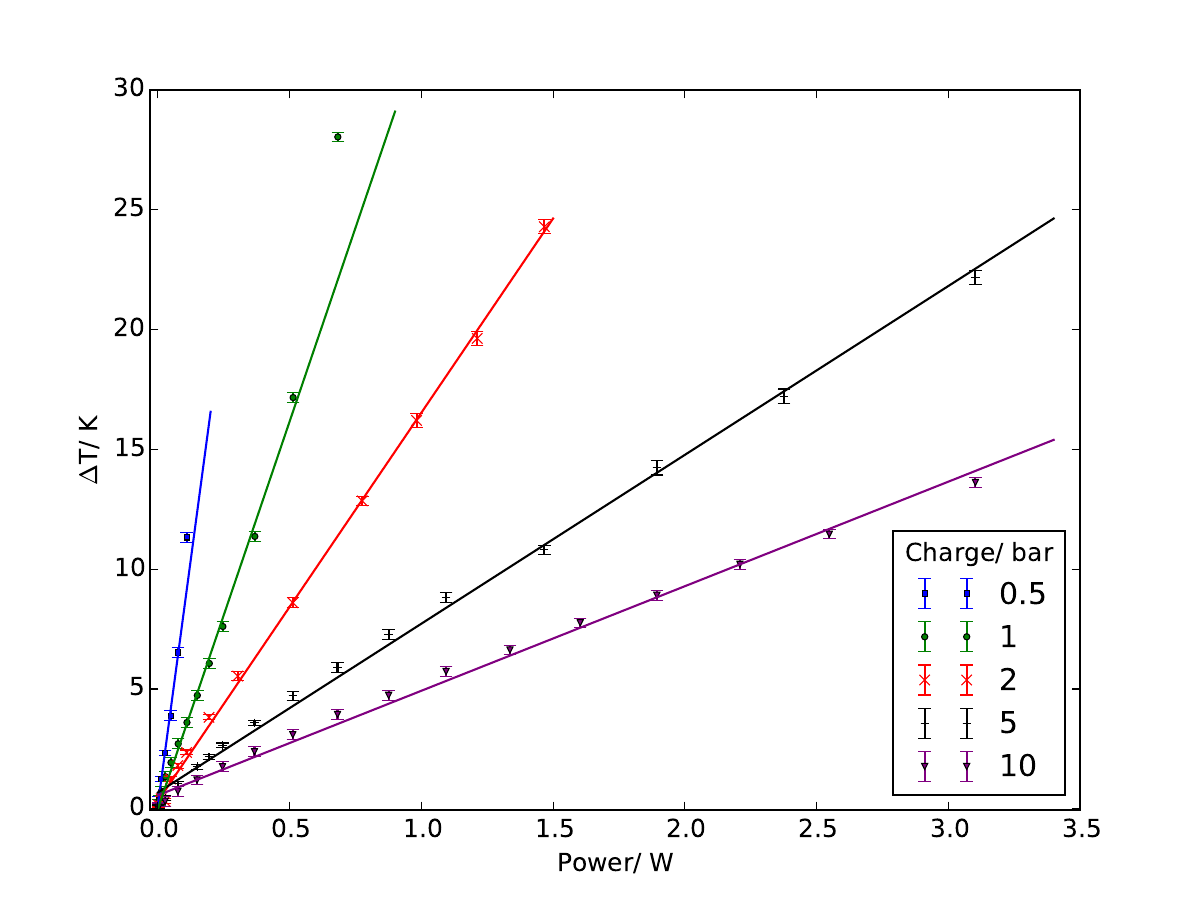}
 &
 \includegraphics[width=.5\textwidth]{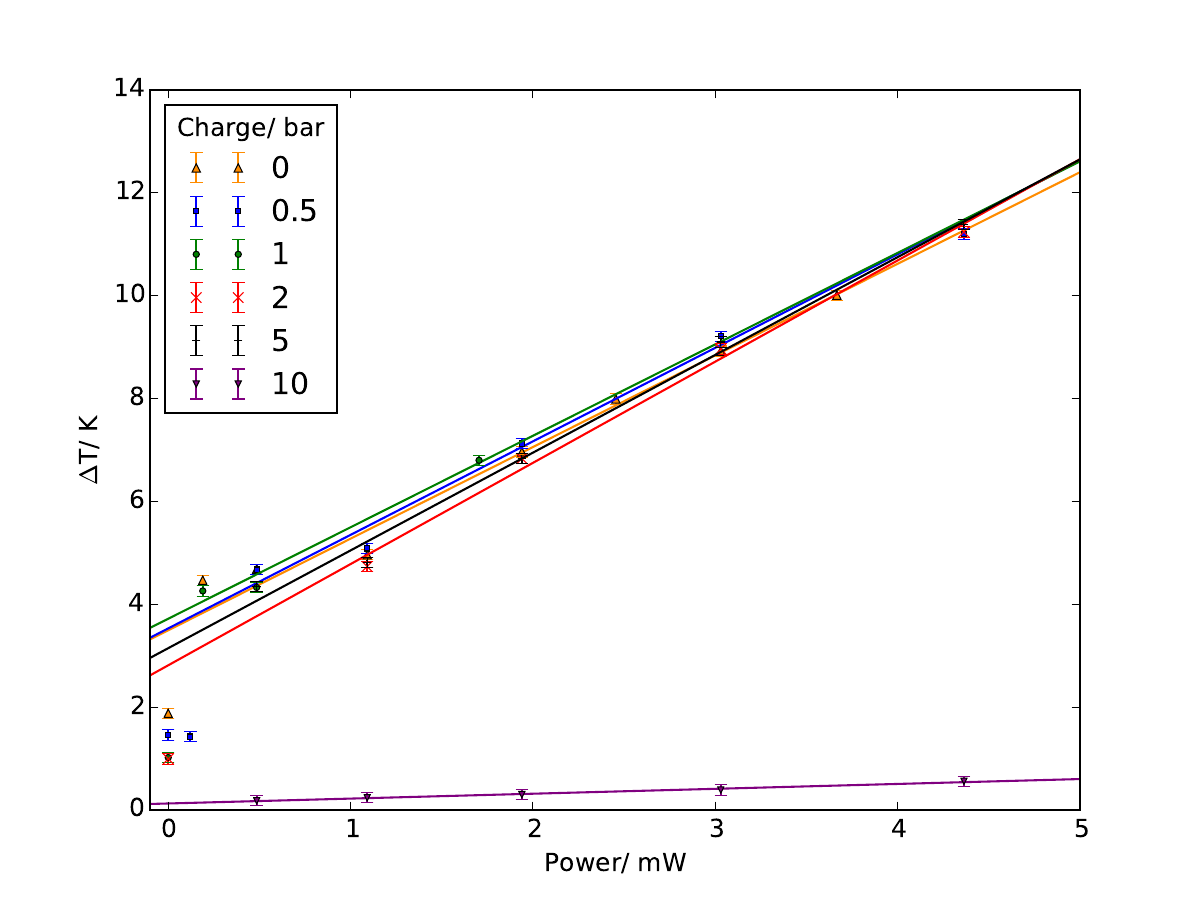}
\end{tabular}
 % allon.pdf: 0x0 pixel, 300dpi, 0.00x0.00 cm, bb=
 \caption[Heat switch on and off]{Heat switch on (left) and off(right)}
 \label{fig:ICEOxfrodHS_on}
\end{figure}

%\begin{figure}
% \centering
% \includegraphics[width=.5\textwidth]{HS_off}
% % off lin fit.pdf: 0x0 pixel, 300dpi, 0.00x0.00 cm, bb=
% \caption{Heat switch off}
% \label{fig:ICEOxfordHS_off}
%\end{figure}

We use convective switches routinely, for example to cool large sorption-fridge cryo pumps.
In fact the example presented here has been designed for use with the large $^4$He fridge we
propose for cooling the 1-K Box.
This design uses larger tubes than previously for higher heat transport.
For use at lower temperatures where the off resistance should be optimized we would probably choose finer tubes.

\paragraph{Heat Switch Performance Tests}
\label{subsect:HSTests}
Results from on and off conduction measurements of the convective switch are given in 
Figure~\ref{fig:ICEOxfrodHS_on} on the right and left hand side %and fig.\ref{fig:ICEOxfordHS_off} 
respectively (note the different power scales).
These were taken with a range of $^4$He charge pressures.
As may be anticipated increasing the charge results in higher heat transport.
However, the 10-bar charge clearly shows that the off conductance has been compromised.
We intend to repeat this test with a larger cryo pump.

Whilst we have not made a test with a negative temperature difference imposed on the switch
we would expect the residual conductance to be, at worst, no more than the y-intercept of
the \textit{off} measurements.
We would expect a reduction in practise, since whilst an residual vapour can contribute heat
transport by convection when the bottom stage is warmer than the top, with the bottom stage
now held cold than the top this should be suppressed.

Our tests of the minimal-gap arrangement have so far been unsatisfactory.
We will report further as this develops.
We expect to be able to provide adequate conductance with the convective switch if the MGHS is unsuccessful.

\paragraph{Heat Switch Interfaces}
\label{subsect:HSInterfaces}

The mechanical interfaces to the heat switches are the 4-hole mounting points at top and bottom.
Correct orientation is paramount, with the item to be cooled attached to the lower stage.
The height of the switches is \textit{to-be-decided}.
The volume taken by the switch may be inferred from the CAD model.
There is some flexibility in terms of reorienting the cryo pump,
but note that we might want to double the size over that shown in the model.
A weak link wire will be required to bring the cryo pump to 4\,K.

Thermally, the switch will accept thermal power at the bottom and couple it to the top.
The load on the fridge will be determined mostly by the power extracted from the cooled stage.
A small amount of power is added by the cryo pump.

Electrical interfaces are described in table \ref{table:HSElec}.
Provision of thermometers / heaters has not been discussed.
The type preferred elsewhere may be used for the switch, for operation from room temperature to 4\,K.
The maximum power to the heater is typically less than 500\,mW (actually more usually about 200\,mW)
but up to a few W can be useful when a rapid heating is desired.
We use 330R, max 32\,V 100\,mA with 0.1-mm copper wire (but we have used 0.1-mm Manganin in the past).

\begin{table}
\centering
\begin{tabular}{c c c c}
\hline\hline
%Heading
Type & Location & No. Wire Pairs & Notes \\
\hline
Diode & Heat switch cryo pump & 1 & \\
Heater & Heat switch cryo pump  & 1 &  \\
\hline
\end{tabular} 
\caption{Heat switch electrical interfaces (per switch)}
\label{table:HSElec}
\end{table} 

As for the fridge our usual computer control uses a state machine language.
Operation of a heat switch is trivial and this approach may readily be translated
to the language of choice.

%auto-ignore
\subsection{Optical chain }

As shown in Figure~\ref{qubicsketch}, the sky radiation experiences several steps as it propagates through the optics of the QUBIC
1}{\textsuperscript{st}}{ module; all of them  are described in this section.

The optical chain shown on Figure~\ref{fig9} is completed with a selection of spectral conditioning before the combiner : in intensity, by filters,
and in polarization, by a modulator (HWP) and a polariser. A dichroic, before the focal planes, splits the radiation
into the two bands at 150 and 220 GHz. Finally a couple of radiation shields in front of the cryostat and around the
whole instrument allow a reduction of local spillover.

\begin{figure}
\centering
\includegraphics[width=8cm]{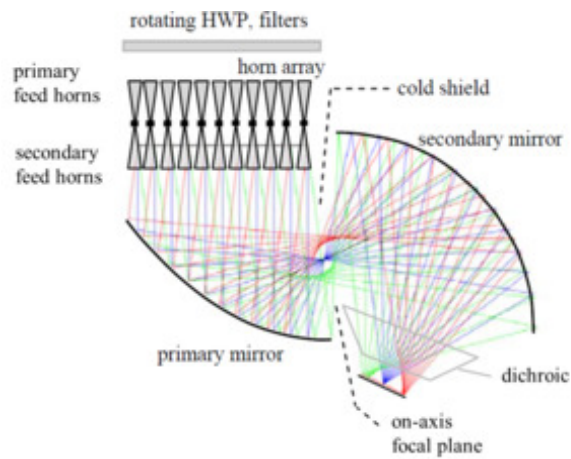} 

\caption[The off-axis dual
reflector chosen for the QUBIC beam combiner]{The off-axis dual
reflector chosen for the QUBIC beam combiner. The rays, at -7\degree (green), 0\degree (blue) and +7\degree (red) represent the beams
from the re-emitting horns.\label{fig9}}
\end{figure}

\subsubsection{Window}
The window is the first optical element encountered by the incoming radiation beam, and separates the high vacuum present in the cryostat jacket from the room-pressure environment, while allowing millimeter waves in the cryostat. A cylindrical slab (560 mm diameter, 20 mm thick) of high-density polyethylene (HDPE) has been used, as the best compromise between transparency at mm waves and stiffness (the window must withstand an inward force of about 2.4 tons due to atmospheric pressure). 

The HDPE slab is pressed against the top cover of the cryostat by an Al ring (see Figure~\ref{fig:window}) with a suitable number of screws. The vacuum seal is obtained using an elastomer o-ring for laboratory tests, and an indium seal for operation in Concordia, at very low ambient temperatures. The pressing ring is designed to mitigate the effects of differential thermal contractions, which is significant for HDPE vs aluminum.

\begin{figure}
\centering
\includegraphics[width=.4\textwidth]{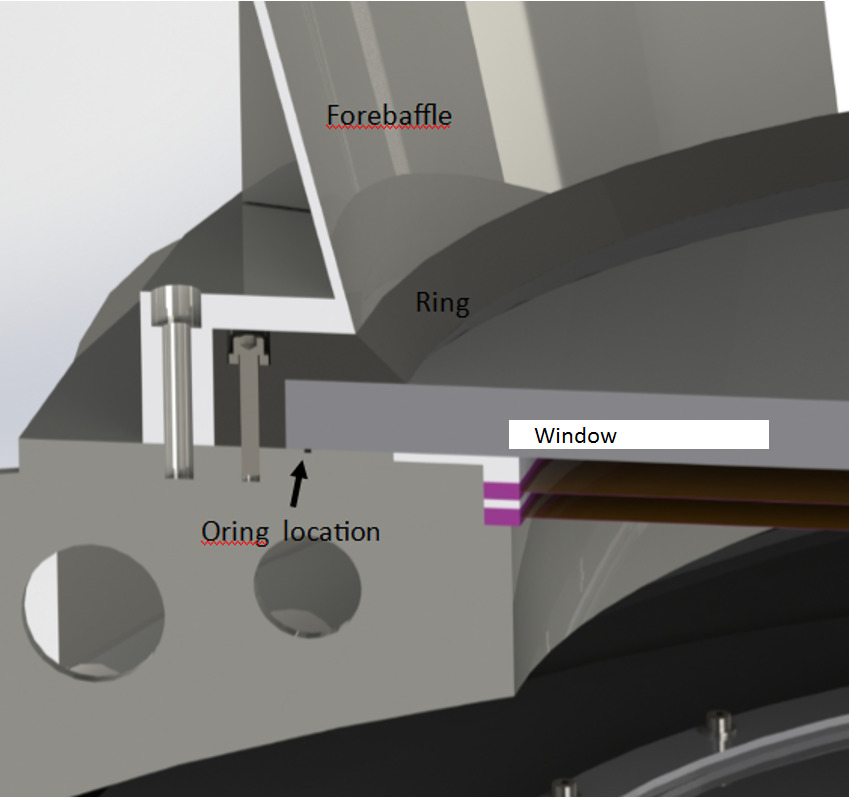}
\caption{\label{fig:window} Detail of the window mounting system.} 
\end{figure}

\subsubsection{Half Wave plate}
\paragraph{Mesh Half Wave plate}

The QUBIC mesh HWP is designed to work across the two bands of the QUBIC first module instrument (see Figure~\ref{table14}). This means
that good RF performance needs to be achieved across a large relative bandwidth, of the order of 73\%. The required
diameter is 500mm clear aperture.

\begin{table}
\centering  
\caption{Mesh-HWP bandwidth requirements. \label{table14}}
\begin{tabular}{|c|c|c|c|}
\hline
Channel &  $\nu_1$ (GHz) &  $\nu_2$ (GHz) & Bandwidth \\
\hline
150 GHz & 127 & 171 & 30\% \\
\hline
220 GHz & 192  & 272 & 34\%  \\  
\hline 
2 channels & 127 & 272 & 73 \% \\
\hline
\end{tabular}
\end{table}

The QUBIC HWP is based on metamaterials (Figure~\ref{fig18}). These devices are alternative solutions
to the more massive, expensive and limited-diameter birefringent Pancharatnam multi-plates. The metamaterials are
developed using the embedded mesh filters technology. Very large bandwidth mesh-HWPs ($\approx 90\%$) have been successfully
realised in the past. They have been used for millimetre wave astrophysical observations at the 30m IRAM telescope with
the NIKA and NIKA2 instruments\cite{NIKA2}. The measured performance of a typical prototype, in terms of transmissions and
cross-polarisation, are reported in Figure~\ref{fig19} which shows a very good agreement between model and data.

\begin{figure}
\centering  
\includegraphics[width=12cm]{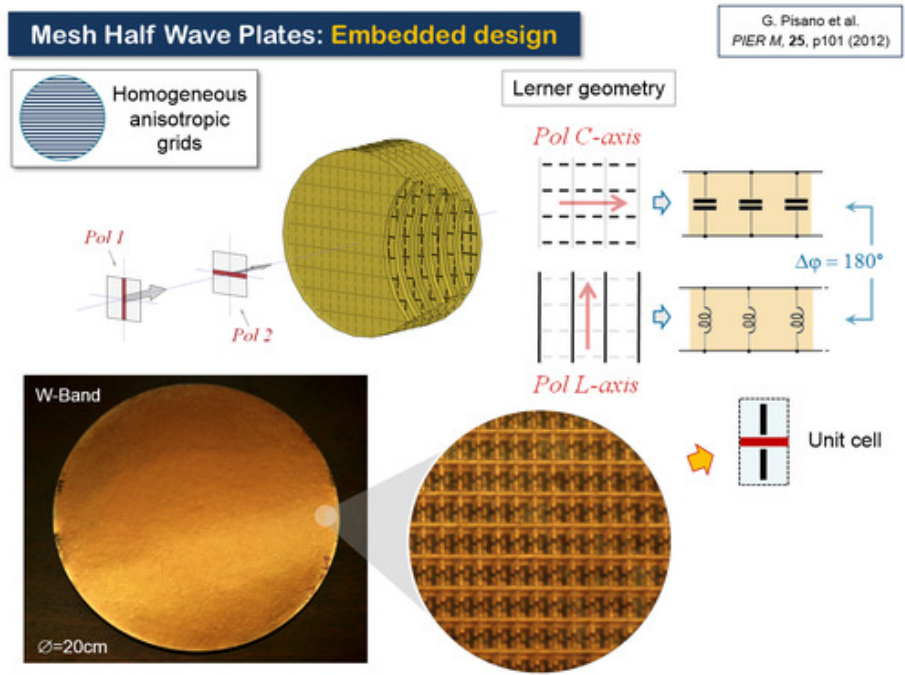}

\caption{Embedded mesh-HWP
based on metamaterials. \label{fig18}}
\end{figure}

\begin{figure}
\centering  
\includegraphics[width=10 cm]{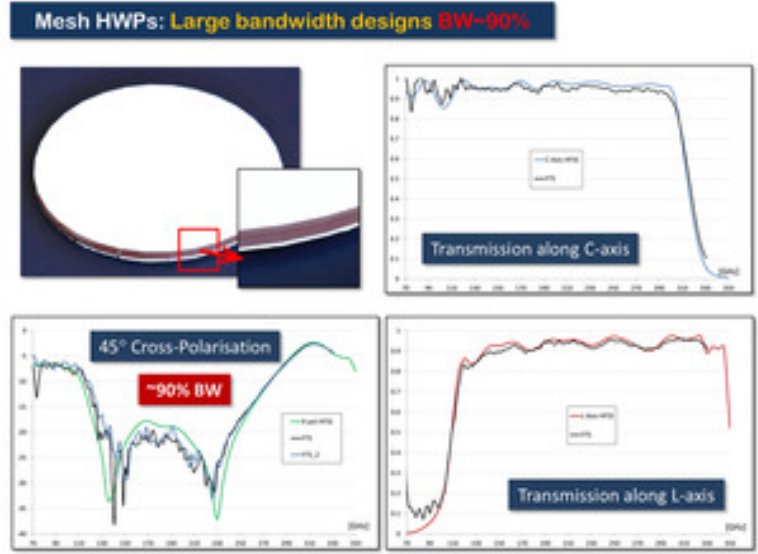}

\caption{Large bandwidth
(\~{}93\%) mesh-HWP prototype: modeled and measured performance.\label{fig19}}
\end{figure}

The QUBIC final design is very similar to the prototype discussed above. The bandwidth requirements
are less challenging and this allows to achieve better in-band RF performance. The design is based on 12 anisotropic
mesh grids and the overall thickness is of the order of 3.5mm. The expected performances of the QUBIC mesh HWP are
reported in Figure~\ref{fig20}. The averaged transmissions, absorptions, differential phase-shift and cross-polarization are
listed within the same figure.

\begin{figure}
\centering
\includegraphics[width=9cm]{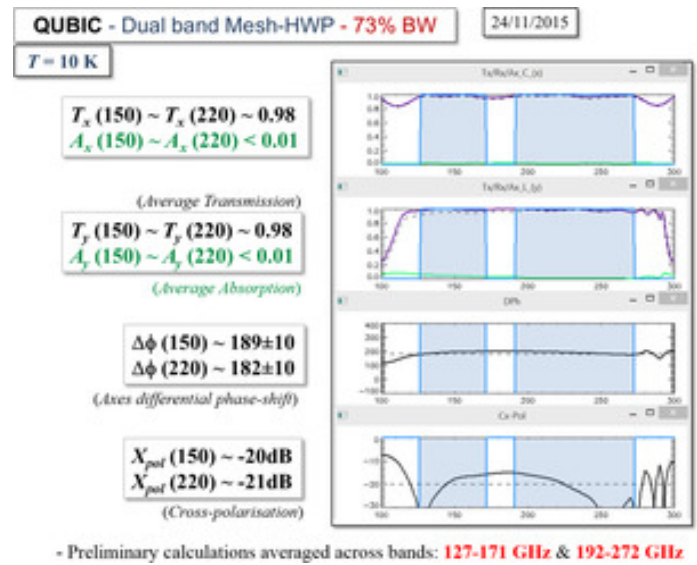} 

\caption{Expected
performance of the QUBIC broadband mesh-HWP design. \label{fig20}}
\end{figure}

\paragraph{ Rotational system for the HWP}
Polarization modulation is achieved by rotating a large diameter HWP (Half Wave Plate). 
Since the HWP is mounted on the 3K stage of the cryostat, a cryogenic rotation mechanism is needed. The one designed for QUBIC inherits several of the solutions developed for cryogenic rotator developed for the PILOT balloon-borne instrument successfully flown by CNES\cite{salatino2}
This is a stepping rotator, able to position the HWP in 8 different positions, in steps spaced by 11.25\degree, for redundant coverage of the needed position angles. The system is shown in Figure~\ref{fig:hpw_rotator}.  
The HWP is rotated by a stepper motor mounted outside the cryostat shell. Motion is transmitted through the shell by means of a magnetic joint. A fiberglass shaft transmits the rotation to the cryogenic part of the system, with negligible heat load, and rotates a pulley driving a Kevlar belt. The HWP support ring has a groove for the Kevlar belt, which is tensioned by a spring-loaded capstan pulley. 
The HWP support ring is kept in place by three spring loaded hourglass shaped pulleys at 120\degree, as shown in Figure~\ref{fig:hpw_rotator}. All the pulleys rotate on optimized-load thrust-bearings for minimum friction. 
The step positions of the HWP are set by holes sets precisely located on a section of the HWP support ring: this builds a 3-bits optical encoder read by optical fibers (see \cite{salatino2} for more details).

\begin{figure}
\centering
\includegraphics[width=.6\textwidth]{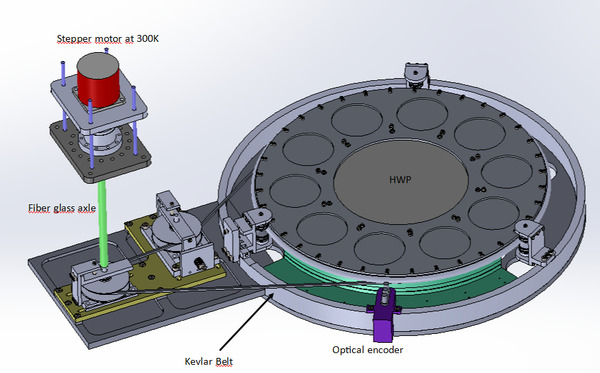}
\caption{General view of the cryogenic waveplate rotator.\label{fig:hpw_rotator}}
\end{figure}
The HWP is mounted on the ring by using a custom made block in order to reduce the differential thermal contraction between Al and Polyethylene (see Figure~\ref{fig:detail_rotator}).  
\begin{figure}
\centering
\includegraphics[width=.6\textwidth]{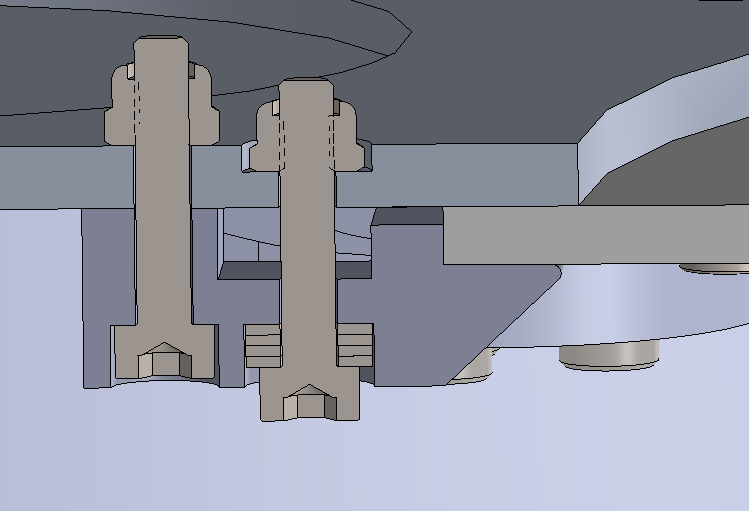}
\caption{Detail of the HWP clamp.\label{fig:detail_rotator}}
\end{figure}

\subsubsection{Filters / Polarizer /Dichro\"\i c}

The general philosophy in filter provision is two-fold: 
1. To minimise the thermal loading at the various temperature stages by sequentially rejecting short wavelength radiation.  This is achieved with thermal filters in combination with baffling and careful optical design, to ensure that the out of band and thermal load at the detector arrays is suitable for the scientific requirements.
2. To define the required spectral passband at the arrays and maximise the in-band optical transmission.  There will be an optimization procedure on the entire filter chain to maximise transmission (i.e. to manage the relative fringing between filters) and minimise out-of-band radiation.
In addition the filters must be able to withstand cryogenic cycling and maintain flatness within their mounts
These specifications have been proven in the past with the AIG's strong heritage in space mission filter production (e.g. ISO, Mars observer, Cassini, Herschel \& Planck Space Observatories) and with ground-based instruments, such as SCUBA2, BICEP and SPT. The QUBIC development puts in place the need for larger diameter components than the AIG have previously provided.

\paragraph{QUBIC optical configuration}

The QUBIC instrument optical layout/cryostat design is shown in Figure~\ref{fig9}. A series of band-defining, blocking and thermal (IR) filters will be mounted at different temperature stages, with a large photolithographic polarizer and a single rotating mesh HWP at 6K
We have allowed provision for a high number of filters at critical apertures and temperatures, although these may later prove unnecessary. 

\paragraph{Mesh Filter Specification }
The complete list of devices to be supplied by Cardiff University to QUBIC is given in Table~\ref{tab_metalmesh}. 
\begin{table}
\centering
\caption{Metal mesh devices to be supplied by Cardiff University to the QUBIC project. Z coordinates are given in the Global Reference Frame (GRF). (*) emissivity could be lower. (**) low frequency in reflexion, high frequency in transmission.\label{tab_metalmesh}}
\begin{tabular}{|c|c|c|c|c|c|c|}
\hline
Component  &   Temp  &   Transm   &   Emissivity   &   Coord. Z  GRF &   Optical diameter   &   Useful diameter \\
  &   K  &   $\%$  &   $\%$  &   mm  &   mm  &   mm \\	
\hline
\multicolumn{7}{|c|} { Common to 2 bands	} \\		
\hline				
Window        &   250  &   98  &   1  &   480.00 &   407  &   600  \\
IR blocker 1  &   250  &   98  &   1  &   460.00 &   401  &   600  \\
IR blocker 2  &   250  &   98  &   1  &   452.55 &   401  &   600  \\
IR blocker 3  &   100  &   95  &   1  &   342.10 &   385  &   430 \\
IR blocker 4  &   100  &   98  &   1  &   335.00 &   381  &   430 \\
IR blocker 5  &   100  &   98  &   1  &   327.10 &   381  &   430  \\
IR blocker 6  &   6    &   98  &   1  &   285.50 &   371  &   410 \\
12cm-1 LPE    &   6    &   95  &   2  &   276.30 &   371  &   410 \\
HWP           &   6    &   95  &   2.5&   237.80 &   361  &   380    \\
Polarizer     &   6    &   99  &   1  &   183.60 &   352  &   380  \\   
B2B horns + switches  &   6  &   99  &   5  &   0.0  &   330  &    (*) \\
Beam combiner  &   1  &   99  &   <1  &     &     &     \\
Cold stop 10cm-1 LPE  &   1  &   95  &   2  &     &     &   Ellipse 0.26 x 0.3  \\
Dichroic filter  &   1  &   95  &   2  &     &     &   Ellipse 0.253 x 0.482  (**)\\
\hline
\multicolumn{7}{|c|} { Band1: 150 GHz Singlemoded	} \\		
\hline				
6cm-1 edge  &   0.3  &   98  &   2  &     &     &        \\
~8cm-1 LPE  &   0.3  &   98  &   2  &     &     &        \\
%High frequency band definer  &   0.3  &   95  &   2  &     &     &   $\oslash$mini = 0.1036 \\
  &     &    &    &     &     &   $\oslash$maxi = 0.11486  \\
\hline
\multicolumn{7}{|c|} { Band2: 220 GHz Multimoded (bandpass 200-240 GHz)} \\		
\hline				
9cm-1 edge  &   0.3  &   98  &   2  &     &     &     \\
~11cm-1 LPE  &   0.3  &   98  &   2  &     &     &     \\
Band defining filters  &   0.3  &   80  &   2  &     &     &   $\oslash$mini = 0.1036 \\
 &    &     &     &     &     &   $\oslash$maxi = 0.11486  \\
\hline
\end{tabular}
\end{table}

\paragraph{Mesh Filter QO deliverables }

\begin{description}
\item{IR Blocking filters}\\
Up to 500mm active thermal filter devices are required of 2 or 3 basic pattern types.  The transmission performance of prototype (300mm) devices is given in Figure \ref{therfilttrans}. These thermal filters are single layer metal-mesh element devices that reflect a high proportion of the unwanted IR radiation.  They require a simple aluminium support ring and the filter element itself is only 4$\mu$m thick.
\begin{figure}
\centering
\includegraphics[width=.5\textwidth]{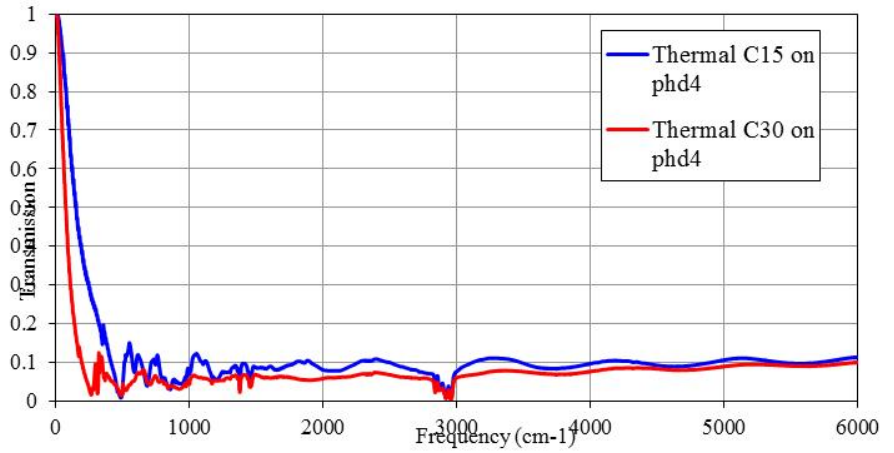}
\caption{Transmission performance of thermal filters for the QUBIC instrument.\label{therfilttrans}}
\end{figure}
\item{Blocking Filters (Low Pass Edges) - 6K, 1K, 0.3K stages }
Throughout each optical chain there will be a series of 4 low-pass edge filters per band/pixel. These will be located at the 6K and 1K stages and at the 320 mK array.  These are designed to block out-of-band FIR radiation, whilst maintaining high in-band throughput.  The transmission performance of the 3 possible large low-pass elements common to both bands is given in Figure~\ref{transproto}.  Although, for the purposes of all large scale CMB instruments, we have recently prototyped a new, multi-element 12cm$^{-1}$ LPE filter which will have improved FIR rejection (1 part in 104) - see Figure~\ref{trans12cm}.
\begin{figure}
\centering
\includegraphics[width=.5\textwidth]{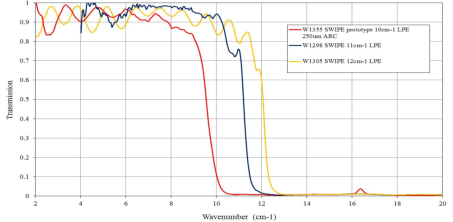}

\caption{Transmission performance of prototyped common LPE blocking filters\label{transproto}}
\end{figure}
\begin{figure}
\centering
\begin{tabular}{cc}
\includegraphics[width=.5\textwidth]{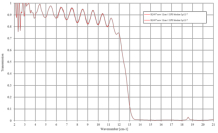}
& 
\includegraphics[width=.5\textwidth]{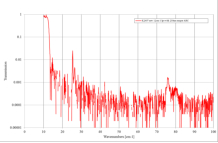}
\end{tabular}
\caption{Transmission performance of possible 12cm$^{-1}$ blocking filter.\label{trans12cm}}
\end{figure}
The thickness of these current filter elements is as follows:
\begin{itemize}
\item 10cm$^{-1}$ LPE	1.4mm without ARC, 1.8mm with ARC;
\item 11cm$^{-1}$ LPE	1.3mm;
\item 12cm$^{-1}$ LPE	2.2mm;
\end{itemize}

\item{Band Defining Filters - 150, 220 GHz}\\
Band-defining filters are required at 150 and 220GHz: \\
150 GHz is single-moded with a band-edge at 5.6 cm$^{-1}$ (25\% BW)\\ 
220 GHz  is multi-moded with 18\% bandwidth, requiring a band-pass filter or a high-pass low-pass combination (6.7 $cm^{-1}$ - 8.0 $cm^{-1}$).\\

Typical filter performance for a number of 150/220GHz options, with modelled atmosphere is given in Figure \ref{trans150220}.
\begin{figure}
\centering
\includegraphics[width=.5\textwidth]{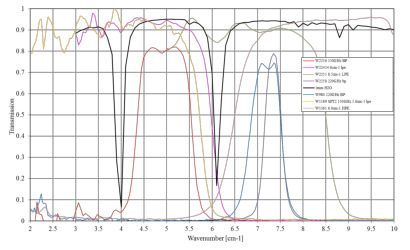}

\caption{Transmission performance of possible 12cm$^{-1}$ blocking filter.\label{trans150220}}
\end{figure}
\item{Dichroic}\\
This filter is designed to transmit (>90\%) the 220GHz band, whilst reflecting (>90\%) the 150GHz band.  Prototype hot-press and air-gap 6.5 cm$^{-1}$ HPE devices have been produced by Cardiff AIG and shown to be effective in both reflection and transmission at up to 
o\degree incidence. Figure \ref{transperf} show normal incidence transmitted performance.  
\begin{figure}
\centering
\includegraphics[width=.5\textwidth]{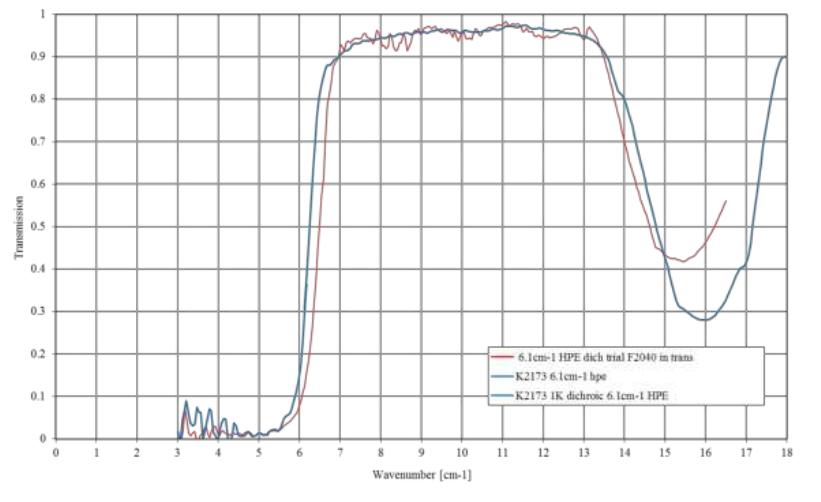}

\caption[Transmission performance of a prototype 6.5cm$^{-1}$ high-pass dichroic]{Transmission performance of a prototype 6.5cm$^{-1}$ high-pass dichroic in airgap (red) and hot-pressed (blue) options\label{transperf}}
\end{figure}

Further testing of prototype devices is underway, with emphasis on the mounting and flatness fo these large components.  A hot pressed device will be the preferred option, provided that the R and T performance are found to be comparable to that for an air-gap device and that flatness can be maintained through cryogenic cycling.  
 \item{Photolithographic polarizer}\\
A 10$\mu$m period wired polarizer is required for 6K operation.  A prototype has been made at 450mm diameter.  This is shown in Figure~\ref{examp45cm}.
\begin{figure}
\centering
\includegraphics[width=.5\textwidth]{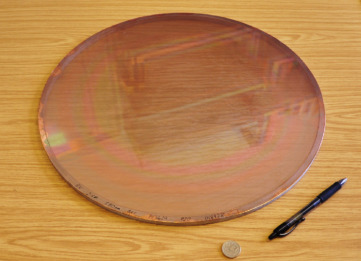}

\caption{Example previously manufactured large diameter polariser (45cm diameter).\label{examp45cm}}
\end{figure}

\end{description}

\paragraph{Quasi-Optical component fabrication, assembly and tests}

All the Quasi-Optical (QO) components (mesh filters, half-wave plate, polarizer and dichroic) are manufactured by the
Astronomy Instrumentation Group (AIG) in Cardiff. This group is the largest manufacturer and first world supplier of
metal-mesh components. Devices from this group have been successfully used in tens of astronomical experiments ranging
from ground-based to satellite missions.

The QO components are manufactured by using photolithographic techniques, specifically the mesh-technology. The TRL
level is 9 for mesh-filters, polarizers and anti-reflecting (ARC) coatings. Mesh-HWPs have been successfully used recently in
ground-based experiments.

The devices will be completely built within the cleanrooms of the Cardiff AIG. All the grids for all the device will be
visually checked. In addition, dimensional measurements will be required during the mesh-HWP development due to the
criticality associated to the phase response of each grid. The assembly and the bonding of the grids will be carried
out within our cleanrooms.

Depending on the type of device and on their thickness, different types of mechanical mounts will be used for the
different components. These metallic rings will guarantee rigidity, flatness and operation at cryogenic temperatures.

The filters, the HWP, the polarizer and the dichroic will all be tested and characterised in
our laboratories by means of different Fourier Transform Spectrometers. Different experimental setup will be adopted
for each type of device.% (one of these is schematically shown in Figure~\ref{fig94})\todocomment{do we need this figure ?}. 
For example, the HWP tests will require
transmission measurements along the two axes and cross-polarisation at 45 degrees rotation angle. The frequency range
of the tests will cover the QUBIC operational bandwidth. For the thermal filters, the measurements will be extended up
to the near-infrared region to check for unwanted leakage.

\subsubsection{Horns}
\label{horns}

Similarly to  the switches, we have been through a prototyping campaign of the QUBIC feed horn array.

This array is composed by two blocks of 400 horns each. The two blocks are placed back-to-back with a layer of switches
that can open or close the optical path to the radiation (cf. previous section and section~\ref{operation}). 

The feeds are corrugated horns optimized for a wide-band response (in the range 130-240 GHz).
They are based on a modification of a previous design, which was optimized for 150 GHz only, when QUBIC was still
designed for 150 GHz-only measurements. The left panel of Figure~\ref{fig52} shows the current profile of the QUBIC horns; we
call this the \textit{QUBIC2 } design. The right panel shows the
original design adopted for the 150 GHz -- only version of QUBIC; we call this the
\textit{QUBIC1 }design.

\begin{figure}
\centering
\includegraphics[width=7.6cm]{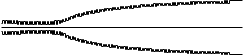} 
\includegraphics[width=8.0cm]{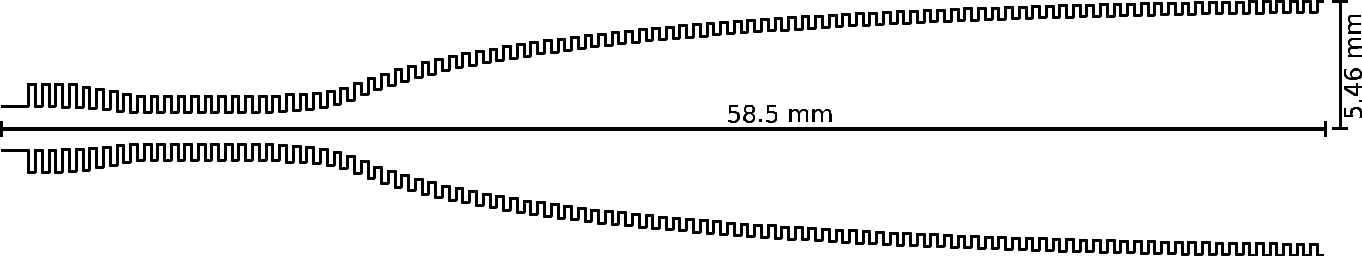} 

\caption[Designs of the QUBIC feed horns]{Left panel: the
current design of the QUBIC feed horns, optimized for wide band response. Right panel: the first design of the QUBIC
feed horns, optimized for 150 GHz measurements.\label{fig52}}
\end{figure}

\paragraph[2x2 prototypes array]{2x2 prototypes array}
The manufacturing of several horns with complex internal geometry with sub-mm tolerance is a
challenging task. To achieve the result we identified the
\textit{platelet} technique as the most suitable to build such a
large array with affordable cost. According to this technique the horn is built from suitably drilled metal platelets
that are subsequently stacked to form the horn structure.

{
For QUBIC, we chose Photochemical Etching and Milling to drill the platelets. This technology is applied to a wide range
of materials for the fabrication of highly complex objects, with an achievable precision of the order of {\textpm} 10\%
the material thickness.}

{
To verify the applicability of this technology to QUBIC horns we set up a prototyping campaign with the following
objectives:}

\begin{itemize}
\item {
{Verify the possibility to manufacture the horns and assess the achievable mechanical
tolerance,}}
\item {
{Assess the electromagnetic performance of the obtained horns in terms of return loss,
insertion loss and angular beam pattern and compare it with the simulations,}}
\item {
Verify the scalability of this technique to a large number of elements.}
\end{itemize}

{
{In this prototyping phase we built 6 prototypes, which are detailed in Table~\ref{table19}.}}

\begin{table}
\begin{tabular}{p{2.077cm}p{2.008cm}p{1.329cm}p{1.9129999cm}p{3.293cm}p{2.1759999cm}p{2.1829998cm}}
\hline
\centering{\bfseries Prototype n.} &
\centering{\bfseries \# of elements} &
\centering{\bfseries Design} &
\centering{\bfseries Material} &
\centering{\bfseries Scope} &
\centering{\bfseries Tests performed} &
\centering\arraybslash{\bfseries Status}\\\hline
\centering{ 1} &
\centering{ 4 (2x2)} &
\centering{ QUBIC1} &
\centering{ Brass} &
{ Check mechanical profile} &
{ Visual inspection } &
\centering\arraybslash{ Completed}\\\hline
\centering{ 2} &
\centering{ 4 (2x2)} &
\centering{ QUBIC1} &
\centering{ Anticorodal} &
{ Check mechanical tolerance and electromagnetic performance} &
{ Metrological measurements}

{ Return loss}

{ Insertion loss} &
\centering\arraybslash{ Completed}\\\hline
\centering{ 3} &
\centering{ 4 (2x2)} &
\centering{ QUBIC1} &
\centering{ Silver-plated anticorodal} &
{ Check effect of silver plating} &
{ Return loss}

{ Insertion loss} &
\centering\arraybslash{ Completed}\\\hline
\centering{ 4} &
\centering{ 4 (2x2)} &
\centering{ QUBIC1} &
\centering{ Aluminium} &
{ Check performance with pure aluminium} &
{ Return loss}

{ Insertion loss} &
\centering\arraybslash{ Completed}\\\hline
\centering{ 5} &
\centering{ 4 (2x2)} &
\centering{ QUBIC2} &
\centering{ Anticorodal} &
{ Double check with new design} &
{ Return loss} &
\centering\arraybslash{ In progress}\\\hline
\centering{ 6} &
{\centering 128 \par}

\centering{ (two 8x8 modules)} &
\centering{ QUBIC2} &
\centering{ Silver-plated anticorodal} &
{ Check of manufacturing scale-up}

{ Verification of interface with switches}

{ Verification of feed-switch functionality} &
~
 &
\centering\arraybslash{ In progress}\\\hline
\end{tabular}
\caption{List of QUBIC feed
horn array prototypes}
\label{table19}
\end{table}

{
{In Figure~\ref{fig53} and Figure~\ref{fig54} we show pictures of the QUBIC1 brass prototype, which has been cut
to perform a visual inspection of the corrugation details. }}

{
{Figure~\ref{fig54}, in particular, shows details of the inner structure of the feed corrugations,
revealing the presence of small }{\textit{cusps}}{ on the top of each
tooth and groove. These features likely result from a non-homogeneous action of the chemical agent on the metal.}}

{
{We have analyzed with simulations the impact of such defects on the feed-horn performance and
the result is that the impact is negligible. This analysis is presented in Section \ref{bkm:Ref435176639}}}

\begin{figure}
\centering
\includegraphics[width=6.0cm]{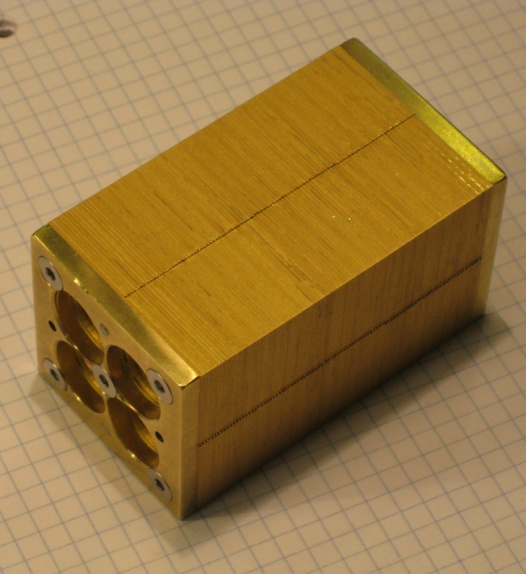}
\includegraphics[width=6.9cm]{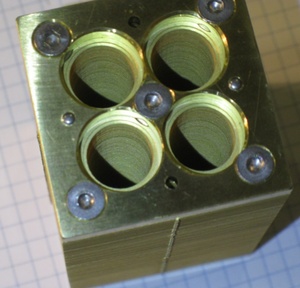} 

\caption{Outer view of the
brass QUBIC1 prototype.\label{fig53}}
\end{figure}

\begin{figure}
\begin{minipage}[c]{0.5\linewidth}
\includegraphics[width=7.308cm,height=5.847cm]{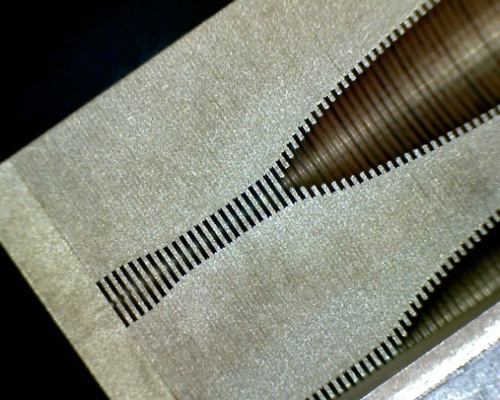} 
\end{minipage}
\begin{minipage}[c]{0.5\linewidth}
\includegraphics[width=7.315cm,height=5.851cm]{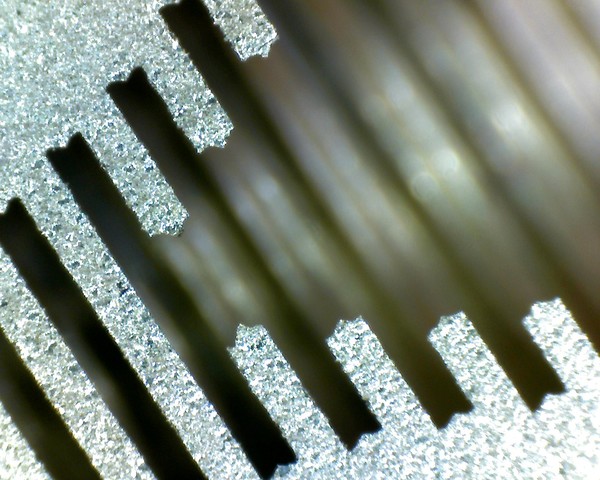}
\end{minipage}
\caption[Inside view of
the brass QUBIC1 prototype]{Inside view of
the brass QUBIC1 prototype. Left: Cusp in groove. Right: Cusp on tooth.\label{fig54}}
\end{figure}

{
{In Figure~\ref{fig55} and Figure~\ref{fig56} we show the complete set of 2x2 prototypes. }}

\begin{figure}
\centering  
\includegraphics[height=5.cm]{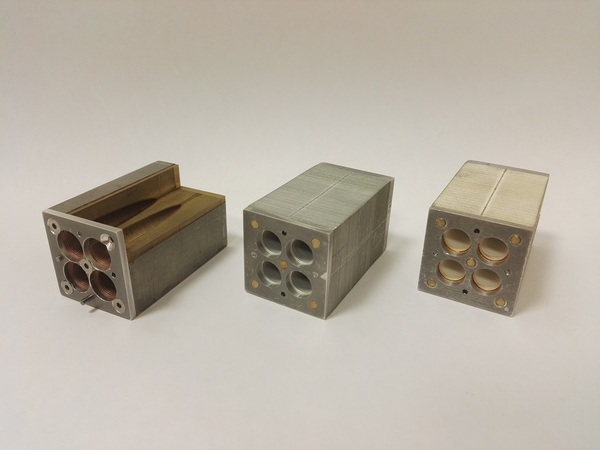}
\caption[Brass, anticorodal and silver-plated anticorodal prototypes]{Brass prototype
(left), anticorodal prototype (middle), silver-plated anticorodal prototype (right). \label{fig55}}
\end{figure}

\begin{figure}
\centering  
 \includegraphics[width=5.1cm]{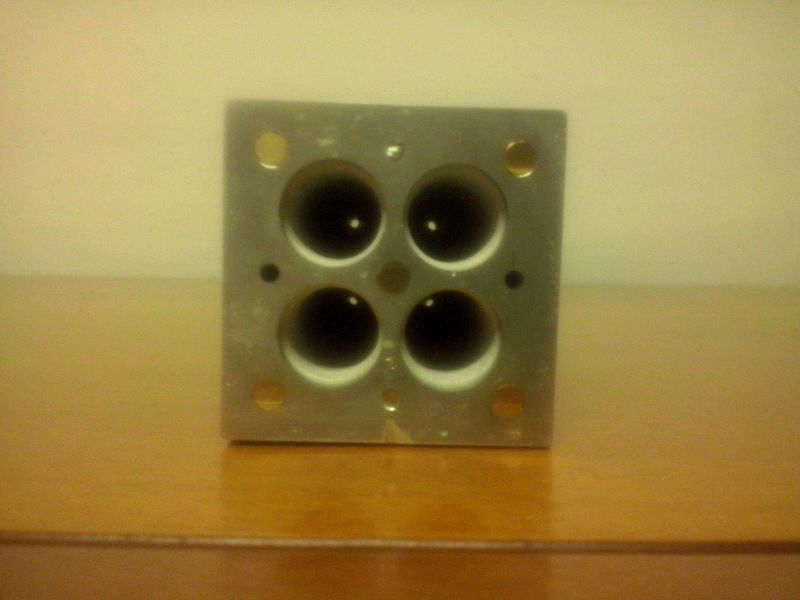}
\includegraphics[width=7.1cm]{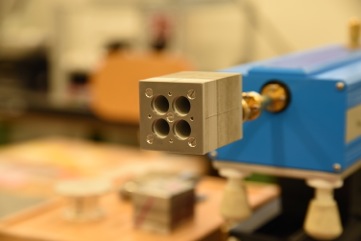} 

\caption[QUBIC1 aluminum and QUBIC2 anticorodal prototypes]{QUBIC1 aluminum
prototype (left), QUBIC2 anticorodal prototype (right). \label{fig56}}
\end{figure}

\paragraph[Metrological measurements]{ Metrological measurements}
\label{bkm:Ref435176639}
Here we report the results of metrological measurements performed on the platelets of the
brass module prior to its integration. We measured the diameters of the antenna holes with a Werth Scopecheck 200
metrological machine and compared measurements with the nominal values in the mechanical drawing. More
details are given in Figure~\ref{fig57} which shows the absolute and relative deviations of the measured diameters with respect to the nominal
one as a function of the diameter itself. From these results we can draw the following considerations:

\begin{enumerate}
\item {
The relative error is approximately constant, of the order of 0.5\% or less, apart from diameters smaller than $\approx$2
mm, for which the relative deviation is of the order of 1\%}
\item {
There is a systematic trend: smaller holes tend to be larger than the nominal, larger holes tend to be smaller. This
trend could be in principle corrected in the manufacturing if necessary by correcting the drawings. }
\end{enumerate}

\begin{figure}
\centering  
\includegraphics[width=9cm]{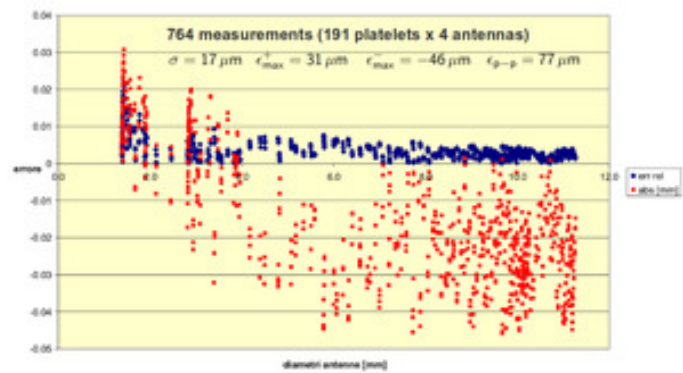}
\caption[Absolute and
relative deviations of measured diameters with respect to the nominal ones as a function of the diameter]{Absolute and
relative deviations of measured diameters with respect to the nominal ones as a function of the diameter. The plot shows a
systematic trend in these deviations that remain, however, at a level less than 1\%. \label{fig57}}
\end{figure}

Figure~\ref{fig58} shows the deviation of the measured position of the holes center from its nominal
value. We see that the distribution of these errors is not symmetrical around the origin but lies preferentially along
one direction. This asymmetry can be mitigated during integration, by turning every other plate by 90\degree.

We see that the r.m.s. values of these deviations are in the range 10 -- 20 {\textmu}m, so
about two order of magnitudes less than the wavelength. Therefore we expect that they will not impact the
electromagnetic performance of the feed significantly. This is confirmed by our return loss measurements reported in
Section \ref{bkm:Ref435184286}.

\begin{figure}
\centering  
\includegraphics[width=12cm]{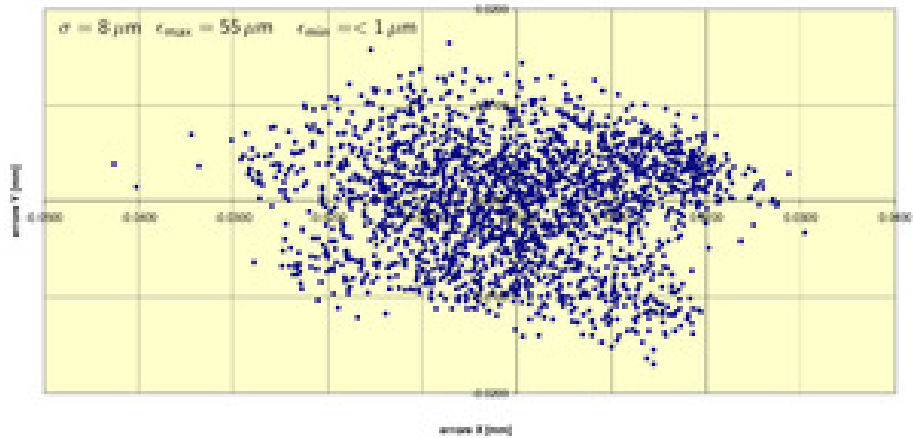}

\caption[Deviation of
the hole center position with respect to the nominal value]{Deviation of
the hole center position with respect to the nominal value. The distribution of these deviations is asymmetrical. This
asymmetry can be mitigated during integration by turning every other plate by 90°. \label{fig58}}
\end{figure}

\paragraph[Effect of mechanical non idealities on performance]{ Effect of non mechanical non
idealities on performance}
We have analyzed the effect of cusps on the feed-horn electromagnetic performance, namely: (i) E-plane and H-plane
radiation patterns, (ii) cross-polar maximum component in the 45 degree plane and (iii) return loss. This section discusses the
main results, the details of this work are reported in~\cite{Franceschet}. 
We have used the SRSR code, modifying the mechanical profile of the horn adding a triangular cusp with 60 {\textmu}m
height on the top of each tooth and groove. This value corresponds to the maximum height measured on the brass
prototype. We have considered three cases: 

\begin{itemize}
\item {
Nominal design with head plate of 0.3 mm thickness}
\item {
Baseline design with head plate of 3 mm thickness}
\item {
Baseline design with head plate of 3 mm thickness and 60 {\textmu}m cusps on both teeth and grooves of all
corrugations.}
\end{itemize}

We have run simulations at different frequencies: $f_0=150$GHz, $f_0\pm6.25\%$ and $f_0\pm12.5\%$. Our 
study shows that these defects produce only minor effects on the feed-optical performance. The main effect is an increased level of reflections (max $\approx$ 3dB) in the 135-160 GHz
frequency interval (see Figure~\ref{fig61}). The total return loss, however is always less than --20 dB, in line with the QUBIC
required value. The level of cross-polarization is practically unaffected (see Figure~\ref{fig60}) while the H-plane
beam pattern presents small differences from the ideal one only in the sidelobes below --40 dB (Figure~\ref{fig59}).

\begin{figure} \centering  \includegraphics[width=13.723cm,height=8.474cm]{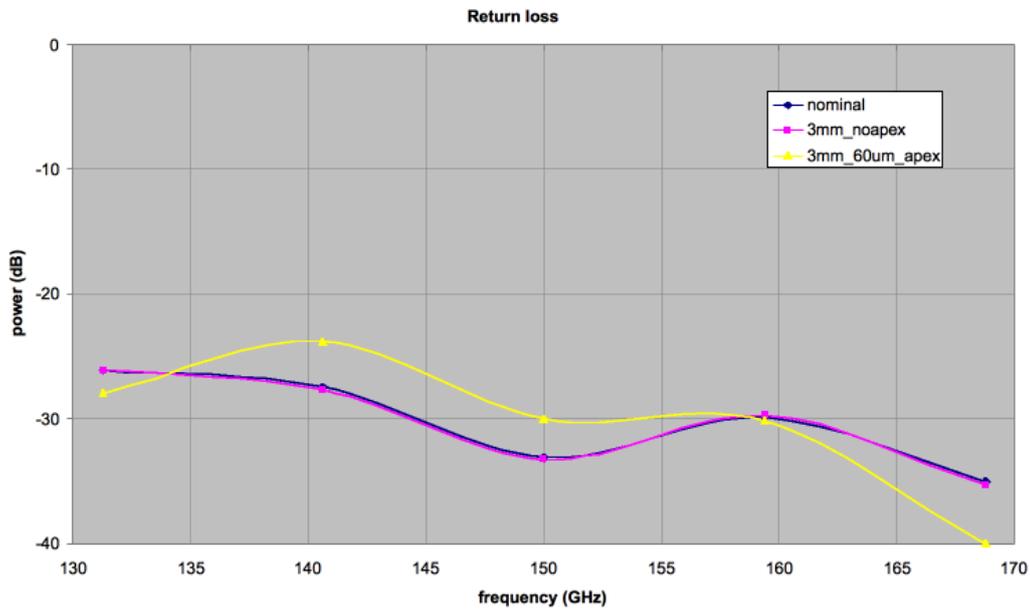}
\caption[Impact of cusps
on the feed-horn return loss]{Impact of cusps
on the feed-horn return loss. The effect is represented by the difference between the yellow and magenta curves. Here
we see that these defects slightly increase the reflections in the frequency range between 135 and 160 GHz.
\label{fig61}}\end{figure}

\begin{figure} \centering  \includegraphics[width=13.723cm,height=8.474cm]{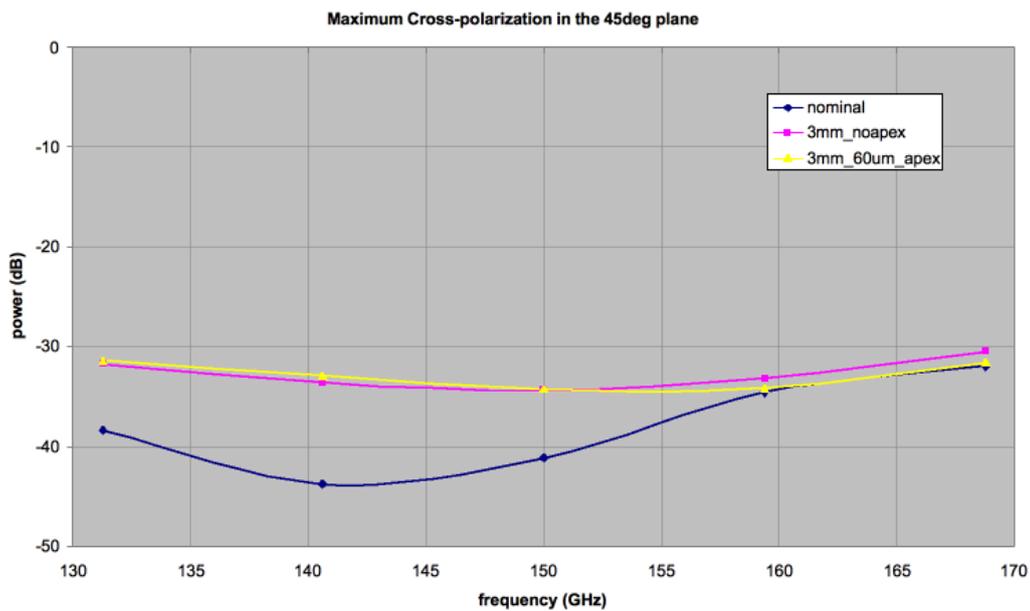}

\caption[Effect of cusps
on maximum cross-polarization]{Effect of cusps
on maximum cross-polarization. We see that cusps have practically no effect on the cross-polarization (compare yellow
and magenta curves). \label{fig60}}\end{figure}

\begin{figure}
\centering  \includegraphics[width=8.5cm]{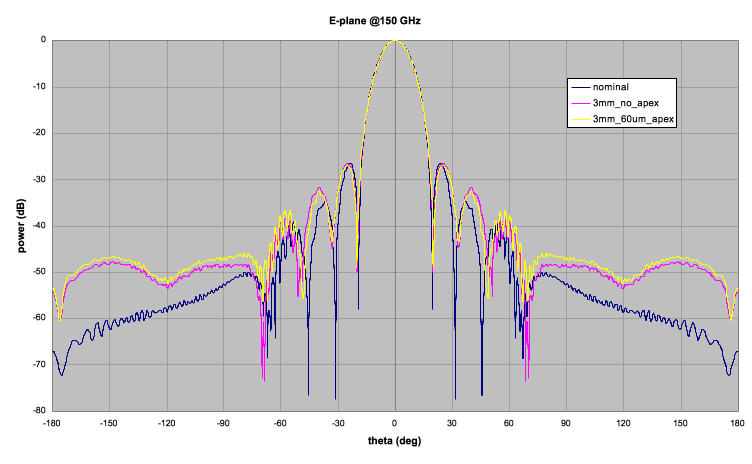} 
\includegraphics[width=8.3cm]{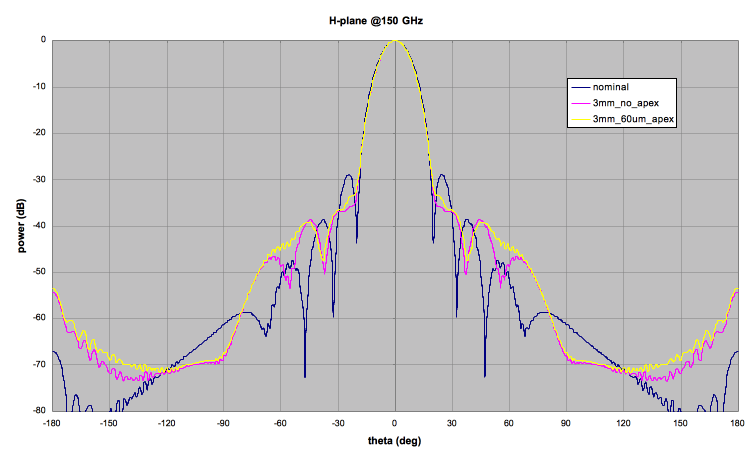} 

\caption[Simulated beam
patterns showing the effect of cusps on the E and H planes]{Simulated beam
patterns showing the effect of cusps on the E and H planes (compare magenta and yellow curves). The blue curve shows
the beam pattern of the nominal design with 0.3 mm head plate. The implemented design foresees a 3 mm head plate
thickness.\label{fig59}}
\end{figure}

\paragraph[Electro{}-magnetic measurements on QUBIC1 prototypes]{Electro-magnetic
measurements on \textit{QUBIC1}
prototypes}
\label{bkm:Ref435184286}
We have tested prototypes number 2, 3 and 4 to verify their electromagnetic performance in terms of return loss,
insertion loss and beam pattern. Because we realized the three prototypes with various materials (Anticorodal,
Silver-plated Anticorodal, Aluminium) we are also interested to check the impact of the material on the performance.
We summarize below the main results, while the reader can find more details about the experimental setup and the
analysis procedures in~\cite{Battaglia}.

\begin{description}
\item{Return loss :}

We have tested the return loss of the three prototypes with the Vector Network
Analyzer of the Milano Bicocca Radio Group. In~\cite{Battaglia} we discuss the details of the measurement
system calibration. In Figure~\ref{fig62} we show the measured return loss of the three prototypes. 
Our results shows essentially three things: the first is that the return loss does not depend
on the details of the three adopted materials, as was expected; the second is that the three prototypes have very
similar performance, demonstrating the repeatability of the manufacturing technique; the third is that the level of
return loss is of the order of -- 30 dB, in line with the simulations without the cusps (see the magenta line in Figure
\ref{fig61}) showing that the presence of such  defects produce only a minor impact on performance.

\begin{figure} \centering  \includegraphics[width=12cm]{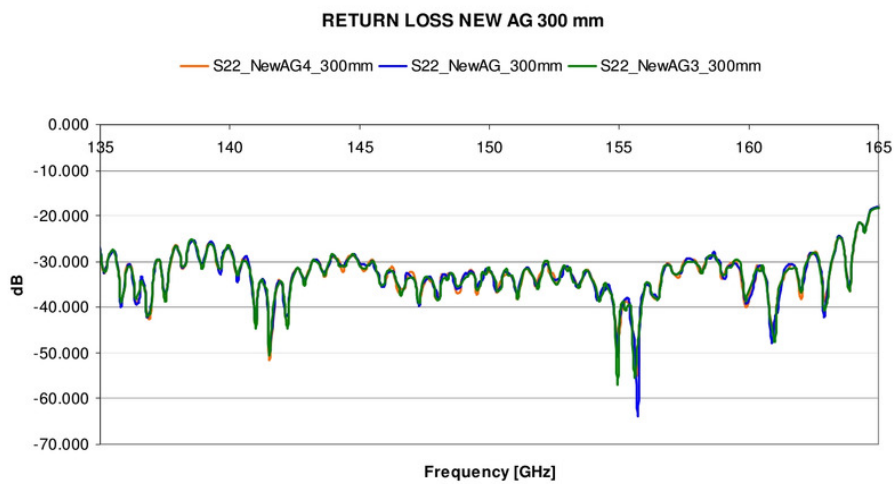}
\caption{ Return loss
measured on one horn of each of the three prototypes
\label{fig62} }\end{figure}

\item{Insertion loss :}

We have measured the insertion loss on the three prototypes. This is a key parameter for the scientific performance of
QUBIC and it is also difficult to measure with high accuracy. This is because the expected loss is less than 1 dB and
it is, therefore, difficult to disentangle from other experimental artefacts.
We used far field measurements, both in a laboratory optical bench and in an anechoic chamber, to measure the insertion
loss. The idea of the measurement is to take the difference of the power transmitted between to standard horns and the
power transmitted when one of the two horns is substituted with the QUBIC horn. Again, we refer to~\cite{Battaglia} 
for the details of the experimental setup and the data analysis.
In Table~\ref{table20} we summarize the insertion loss (IL) values measured on the three prototypes
according to the two methods. Notice that the IL of prototype n. 2 was not measured in the anechoic chamber, so we do
not report its value. The results show that, within the measurement uncertainty of {\textpm}0.3 dB, the insertion loss is less than 1~dB,
which is in line with the scientific requirements of QUBIC. Our measurements also suggest that silver-plating improve
electrical conductivity, although by a slight amount. This result led us to the decision to adopt silver-plating as our
manufacturing baseline.

\end{description}
\begin{table}
\begin{tabular}{|p{2.3cm}|p{2.4cm}|p{4.1cm}|p{4.1cm}|p{2.8cm}|}
\hline
{\bfseries Prototype n} &
{\bfseries Material} &
{\bfseries IL measured on lab bench [dB]} &
{\bfseries IL measured in anechoic chamber [dB]} &
{\bfseries Uncertainty [dB]}  \\
\hline
{ 2} &
{ Anticorodal} &
{ 0.15} &
{Not measured }&
\\\hhline{----~}
{ 3} &
{ Aluminium} &
{ 0.45} &
{ 0.5} &
{{\textpm} 0.3} 
\\\hhline{----~}
{ 4} &
{ Silver-plated Anticorodal} &
{ {\textless} 0.10} &
{ 0.3} &
\\\hline
\end{tabular}
\caption{Measured Insertion
Loss (IL) of the three QUBIC1 prototypes
\label{table20}}
\end{table}

\paragraph{Beam pattern: }

In the framework of the anechoic chamber measurements of the insertion loss we have also made
a preliminary measurement of the main beam pattern. In Figure~\ref{fig63} we show these measurements for the two tested prototypes. Our results show that,
at least at the level of the main beam, the pattern is as expected (compare with Figure~\ref{fig59}) and that the type of
material does not affect significantly the optical response.
We plan to perform more detailed measurements on the QUBIC2 prototype (which is the current baseline) to characterize
the beam sidelobes and the measurement errors. These measurements will be compared to the shape expected from
simulations.

\begin{figure}
\centering
 \includegraphics[width=7.cm]{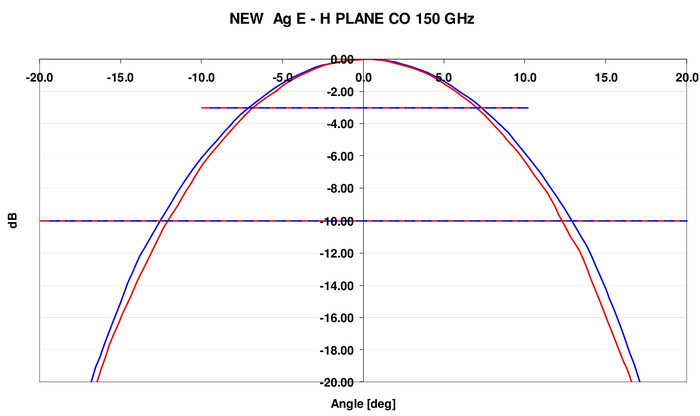} 
\includegraphics[width=7.cm]{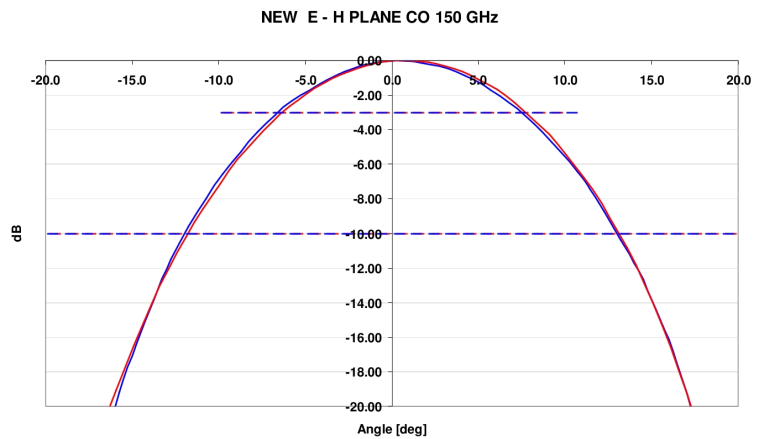}
\caption[Measured main
beam patterns]{Measured main
beam patterns. Left: Aluminum prototype. Right: Silver-plated anticorodal prototype. Blue curve: E-plane. Red curve:
H-plane. \label{fig63}}
\end{figure}

\paragraph{8x8 back-to-back horns :}

We have designed a 128-horn array arranged in two 8x8 blocks that will be interfaced with the
8x8 switch prototype. This is a complete and functional prototype of the QUBIC full array that will be integrated in the Technological Demonstrator (cf. section~\ref{techno}).
Figure~\ref{fig70} shows the CAD design of the 8x8 prototype interfaced with the switch block. On the left we show one of the two
64 horns blocks interfaced with the top flange of the switch array. Notice that we show the inside of the block,
constituted of drilled platelets. The platelets present two types of holes: one type reproducing the horn corrugated
structure, the second to lighten the structure. On the right we show the two horn arrays interfaced with the switches
block.

\begin{figure} \centering  \includegraphics[width=7.4cm]{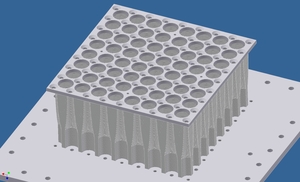} 
\includegraphics[width=7.4cm]{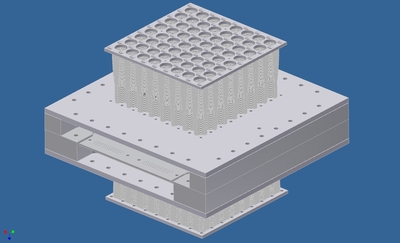} \caption[CAD design of
prototype \#6 (128 horn array)]{CAD design of
prototype \#6 (128 horn array). On the left we show one of the two 64 horns blocks interfaced with the top flange of
the switch array. On the right we show the two horn arrays interfaced with the switches block.
\label{fig70} }\end{figure}

Figure~\ref{fig71} shows the first realization of the back flange of the 8x8 prototype, which
interfaces with the switch block. 

\begin{figure} \centering
 \includegraphics[width=6.0cm]{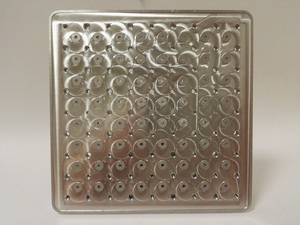}
\includegraphics[width=6.0cm]{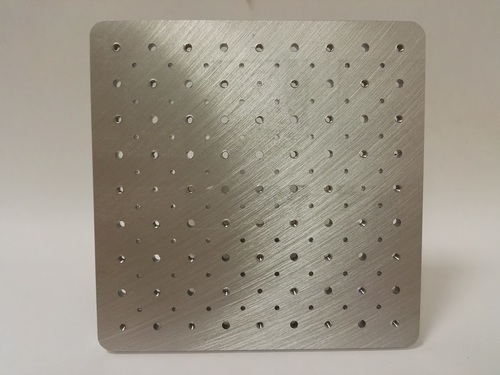} 
\caption{ First realization
of the back flange of the 8x8 prototype. \label{fig71}}
\end{figure}

\subsubsection{Switches}
\label{switch}

The QUBIC self-calibration technique is based on cross checking and comparing redundant baselines produced by equally spaced couples of horns. This requires the identification of the interferograms generated by equivalent baselines. 
The most obvious approach to do so is to enable only one couple of horns at a time (i.e. closing all but two horns), still, it has been shown 
that an equivalent option is to enable all the horns but that
particular couple\cite{BigotSazy:2012tr}. This second possibility provides the advantage to perform the self-calibration in very similar conditions with respect to the astronomical observations in terms of radiation loading on the detectors. In both cases a shutter for each back-to-back horn is needed. The QUBIC final module will have 400
horns with 400 switches working at 4K.

An important effort was made in these years to develop a low loss (mechanical), reliable, switch prototype compatible
with both the RF specification and cryogenic requirements.
A first single channel prototype was completed and succesfully tested at APC to test the working principle. It is a
single pole single through (SPST) realized by means of a blade (shutter) blocking the circular waveguide between the
back to back horns. The blade is activated by an electromagnet pushing and pulling a ferrite soldered to a hook
connected to the shutter (see Figure~\ref{fig64}). 

\begin{figure} \centering  \includegraphics[width=8.6cm]{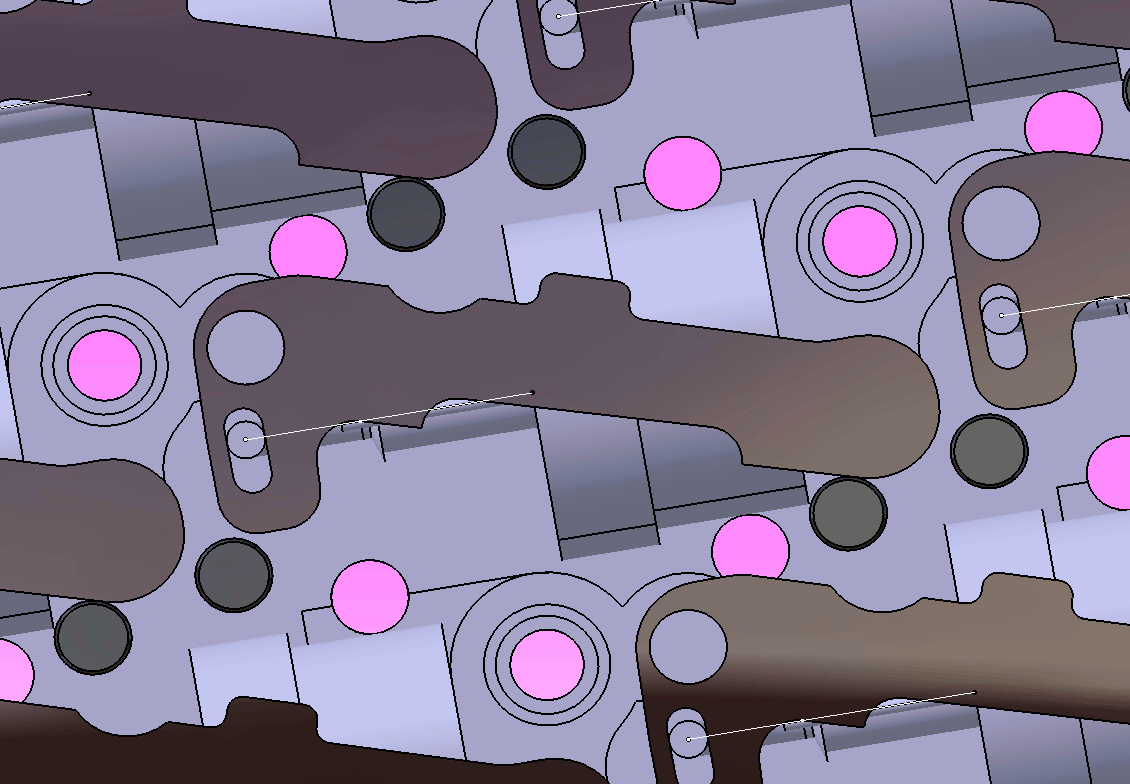}

\caption[Particular of the
waveguide shutter for the 8x8 prototype]{Particular of the
waveguide shutter for the 8x8 prototype built as a replica
of the single channel switch. In the center of this figure the blade (dark grey) and coil (light grey) can be seen.\label{fig64}}\end{figure}

The single channel prototype was designed to have a very good return loss %(S11) 
and low
insertion loss. 
Also the instrumental polarization (different phase delay of the propagating modes) must be kept
very low.  A first design was done in Manchester studying the effects of the waveguide gap with
respect to a shutter 100 {\textmu}m thick (Figure~\ref{fig65}). All the prototypes have been manufactured with 200{\textmu}m
gap.

\begin{figure} \centering  \includegraphics[width=10cm]{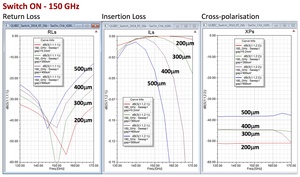}
\includegraphics[width=10cm]{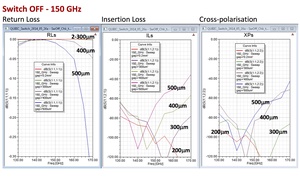}
\caption[Switch
Performance forecast when in ``ON'' (Top Panel) and ``OFF'' positions]{ Switch
Performance forecast when in ``ON'' (Top Panel) and ``OFF'' positions for various gap width between the two facing
waveguides. 200 {\textmu}m gap was chosen.
\label{fig65}}\end{figure}

The single channel prototype was successfully tested in liquid nitrogen to verify the
capability of the device to keep moving at cryogenic temperature. A RF test was also performed at room temperature to
verify the design performances (Figure~\ref{fig66}) that are encouraging.

\begin{figure}
\begin{minipage}[c]{0.45\linewidth}
\centering
\includegraphics[width=6.0cm]{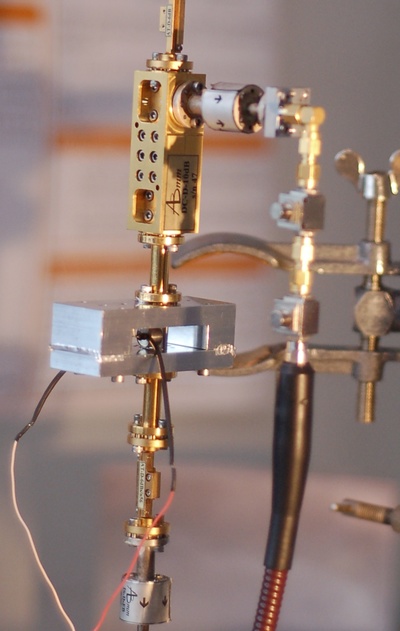} 
\end{minipage} \hfill
\begin{minipage}[c]{0.45\linewidth}
\includegraphics[width=7.0cm]{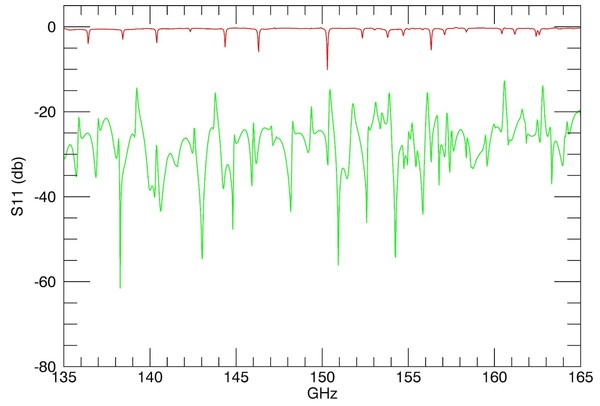}\\
\includegraphics[width=7.0cm]{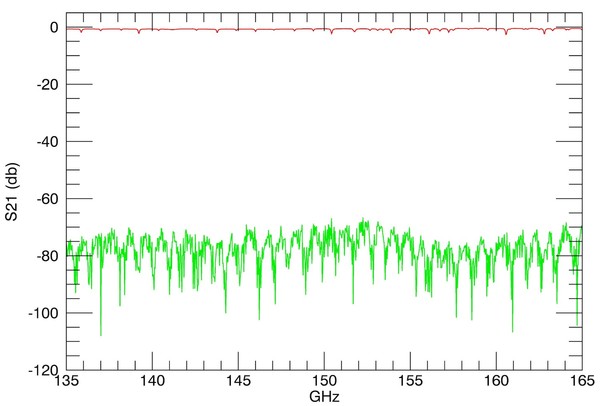}
\end{minipage} \hfill
\caption[instrumental set-up to measure the
Single switch S parameters,  return loss and insertion loss ]{On the left the instrumental set-up to measure the
Single Switch S parameters. On the right: Top Panel Return Loss (red for OFF position, green for ON position), Bottom
Panel Insertion Loss (red for OFF position, green for ON position).\label{fig66}}
\end{figure}

The University of Milano Bicocca (UNIMIB) is in charge of the realization of the electronics
to drive the switch coils and to acquire the shutters' positions. The idea is to excite the coil with a pulse and
acquire the response time that depends by the resistance R and inductance L. When the switch is open (ON position) the
ferrite is outside the coil and the inductance is lower than in the OFF position, causing a faster response time. When
in OFF position the current to drive the coil can cause self heating and the electronics can automatically compensate
the change of R with temperature. Another advantage of the possibility to switch in the OFF position only one couple of switches at the time, as opposed to switch the all but two, is the drastic reduction of the dissipation during the self-calibration. A board equipped with an FPGA (XILINX Spartan-6) and the driving circuits to operate
16 switches has been developed (Figure~\ref{fig67}). This is a scalar design which is ready for the 8x8 prototype simply using 4
boards  and can be adapted to the final 20x20 array. The FPGA is in charge of the calculation of the switch
positions and communicate with the rest of the slow control electronics by means of Ethernet transport. A PCB for the
8x8 switch prototype was also designed and realized. This PCB is used to distribute the bias to the 64 switches and
will be connected to four FPGA board. 

\begin{figure} \centering  \includegraphics[width=7cm]{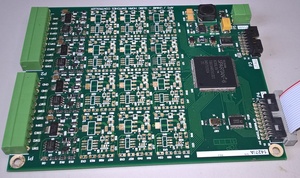} 
\includegraphics[width=7cm]{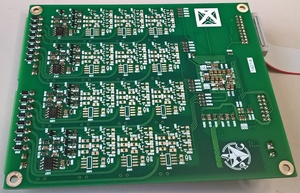} 
\caption[The two sides of
the board able to operate and acquire the position of 16 switches]{ The two sides of
the board able to operate and acquire the position of 16 switches. A SPARTAN-6 FPGA is used to calculate the switch
position and to communicate via Ethernet with the slow control electronics.\label{fig67}}
\end{figure}

The prototype of the 8x8 switch block  was also realized by a milling machine  at UNIMIB (Figure~\ref{fig69}) and it will be 
assembled and tested at room and cryogenic temperature inside the cryofacilities of the millimetric
lab. The main aim of the test is to verify the functionality of the 64 switches at 4K, make an estimate of the medium
time between failures of the coils and quantify the heat dissipation during a typical self-calibration cycle. 

\begin{figure} \centering  \includegraphics[width=6.985cm,height=4.623cm]{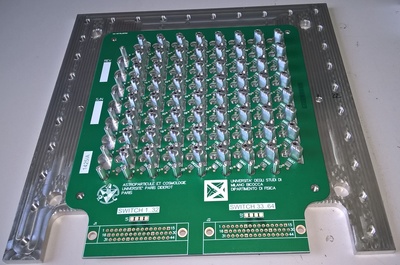} 
\includegraphics[width=8.174cm,height=4.607cm]{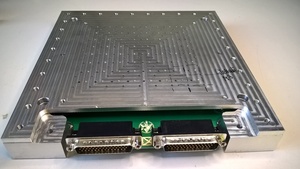} 
\caption[The 8x8 switch
block prototype]{ The 8x8 switch
block prototype with the PCB to bias the coils is visible in the left picture. No component is already mounted inside.
In the right picture the switch block is closed. 
\label{fig69}}\end{figure}

%%%%%%%%%%%%%%%%%%%%%%%%%%%%%%%%%%%%55555

\subsubsection{Optical combiner}
\paragraph{ Optical combiner configuration}

The interferometric requirements described in Section~\ref{bolinterfero}  have been
satisfied with a reflective configuration for the optical combiner. A comparative study among refractive and reflective
solutions preferred the latter options because they allow a more reliable and robust simulation and ensure a large
unobstructed aperture. The aberrations present in a fast off-axis system were the main cause of concern but adding
confocal subreflectors to a classical parabolic primary mirror gives the flexibility to cancel or reduce higher-order
aberrations; space limitations restricted us to dual-reflector designs. We studied several designs for QUBIC~\cite{Bennett}
 including
compensated classical Cassegrain (parabolic primary, confocal hyperbolic secondary), Gregorian (parabolic primary,
confocal elliptical secondary) and Dragonian (parabolic primary, confocal concave hyperbolic secondary) dual
reflectors. Both standard and crossed (front- and side-fed) geometries were considered.

A compensated off-axis Gregorian design (Table~\ref{table13}) was chosen that also obeyed the Rusch
condition for minimum spillover~\cite{Rusch}. A further
optimization of the mirror surfaces was carried out with the aid of commercial ray-tracing software (Zemax,~\cite{Zemax}) to improve the
diffraction-limited field-of-view (results of such simulations are shown on Figure~\ref{fig9}). The design is close to telecentric (distant exit pupil).

\begin{table}
\begin{tabular}{|l|l|l|l|l|}
\hline
mirror & Size (x dimension) & Size (y dimension) & conic constant & Focal length \\
\hline
{ Primary} &
{ 400 mm} &
{ 600 mm} &
{ {}-1} &
{ 164.48 mm (parent parabola)}\\
\hline
{ Secondary} &
{ 600 mm} &
{ 500 mm} &
{ {}-0.1294} &
{ 287.81 mm (focal point separation)}\\
\hline
\end{tabular}
\caption{\label{table13}Combiner
parameters (the primary mirror surface was further optimised using the Zemax ray-tracing package and the final design,
specified by a set of quadratic surface parameters differs slightly from the conic surface specified here).}
\end{table}

\paragraph{Mirrors}

The optical combiner is realized by two off-axis mirrors focusing the light remitted by the
back-horns onto the focal planes. The primary mirror (M1, cf. Figure~\ref{fig21}) and the secondary mirror, (M2, cf. Figure~\ref{fig22}) are machined in aluminium. They are attached by 9 points on their supports. The supports are two identical hexapods for the 2 mirrors.
Each hexapod has 6 degrees of freedom allowing the alignment of the mirrors and the correction some  possible minor errors in
the
manufacturing process.

\begin{figure}
\centering  
\includegraphics[width=11.225cm,height=7.576cm]{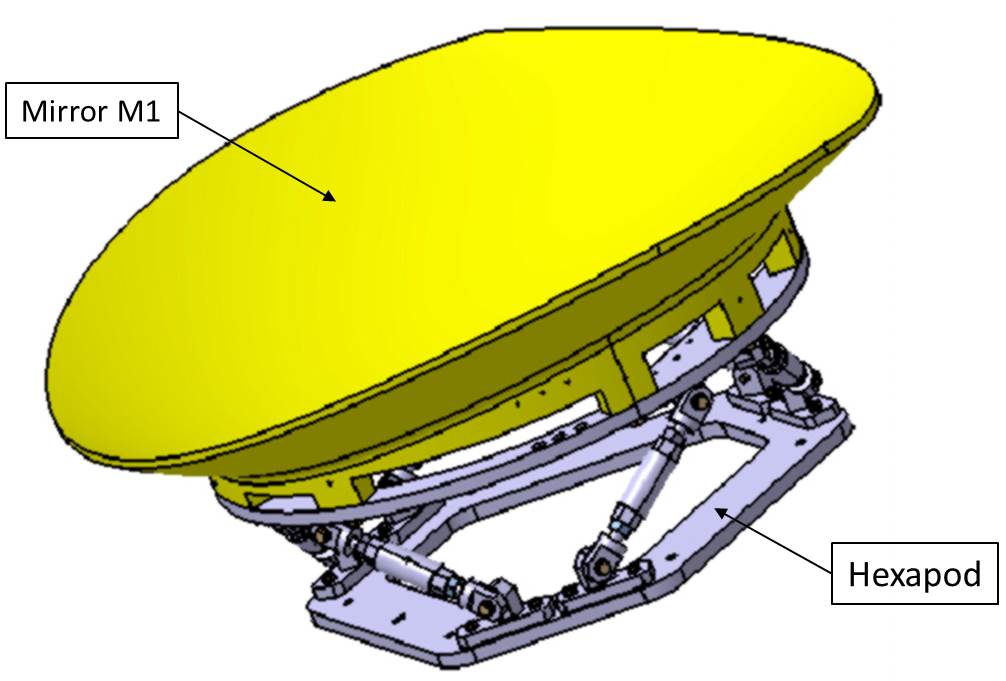}

\caption{M1, primary mirror. \label{fig21}}
\end{figure}

\begin{figure}
\centering  
\includegraphics[width=10.911cm,height=8.871cm]{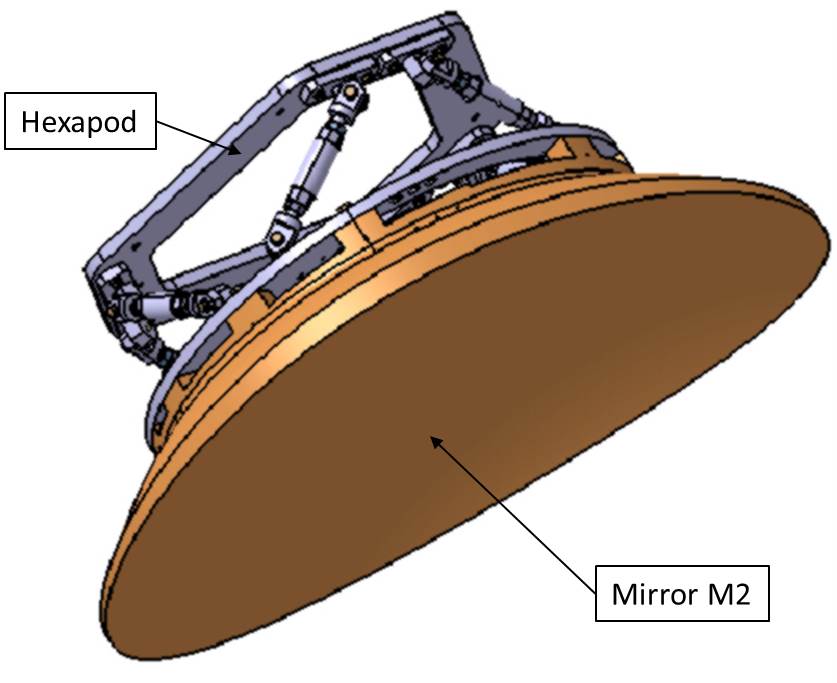}

\caption{M2, secondary mirror. \label{fig22}}
\end{figure}

The two mirrors, designed in Maynooth, will be machined at the University of Milano
Bicocca. The material will be Aluminium  processed with several thermal treatments from -200°C up to 340°C
during the different machining phases, in order to relax all the internal tensions and guarantee the desired final
profile. A reliable thermal sequence, conceived and optimized for VIRGO gravitational antenna [{\notes{Need a reference}}] and successfully
replicated by the Milano Bicocca team for the OLIMPO Fourier Transform Spectrometer\footnote{\url{http://planck.roma1.infn.it/olimpo/}}, will also be applied for the QUBIC
optical combiner. 
A set of reduced size mirrors will be used for the technological demonstrator.0

Once machined, the mirrors' shapes will be verified at room temperature with a home-made 3D gauge
realized by the Milano Bicocca team using a motorized 3 axis linear stage  with an accuracy of
\textcolor{black}\textpm}3$\mu $m over 1 meter moving a  digital
indicator with an accuracy of {\textpm}2.5$\mu $ over 60mm. The same team will evaluate the surface roughness of the
mirrors on some spots 1cm wide using interferometric techniques.

\subsubsection{Cold stop / internal screening}

In order to limit the beam propagation towards the detectors only due to the sky radiation passing through the feed horn array
and the optical combiner, an analysis focused on the straylight impact has been conducted and cold shields have been
considered in the final optical configuration. 
First an aperture between primary and secondary mirrors is inserted to minimize straylight and
all the potential diffracted light reaching the detectors. This shield is named ``Cold Stop'' and it is maintained
at 1K, such as the whole combiner. The size of this aperture has been inferred by a beam propagation analysis as
described in Section~\ref{bkm:Ref314304210}.

\begin{figure}
\centering  
\includegraphics[width=10.321cm,height=6.692cm]{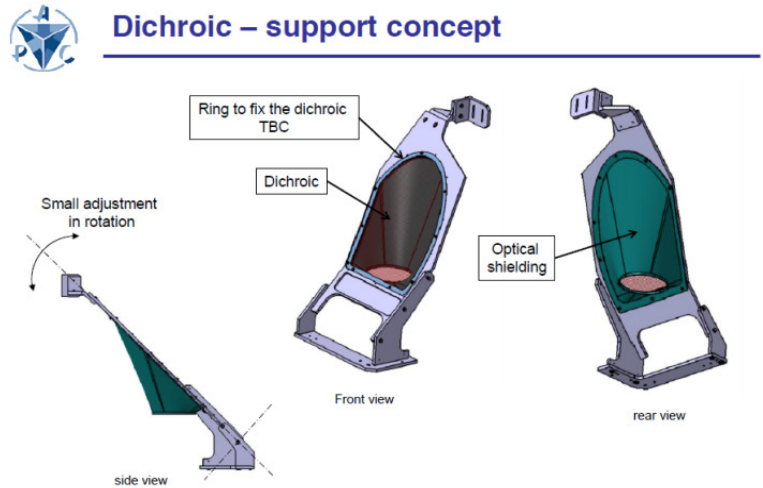}

\caption{ 3D
drawings of the 45 degrees dichroic mount and of the radiation shield for the on-axis focal plane.\label{fig23}}
\end{figure}

A second shield is also planned to avoid unwanted radiation on the on-axis focal plane as
shown in Figure~\ref{fig23}.

\begin{figure}
\centering  
\includegraphics[width=7.934cm,height=5.509cm]{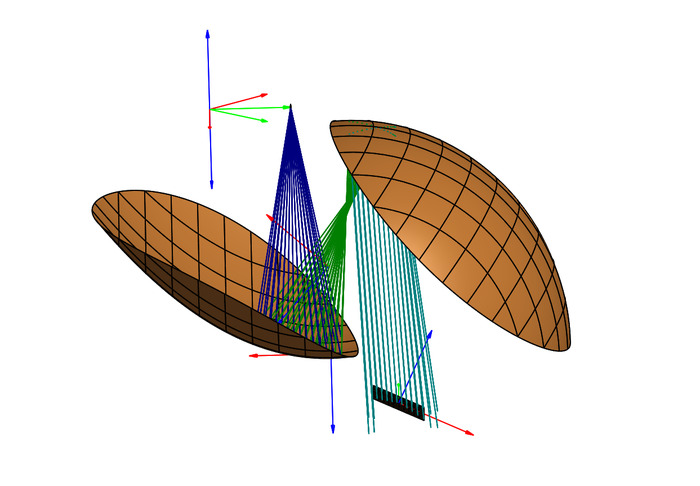} 
\includegraphics[width=7.934cm,height=5.509cm]{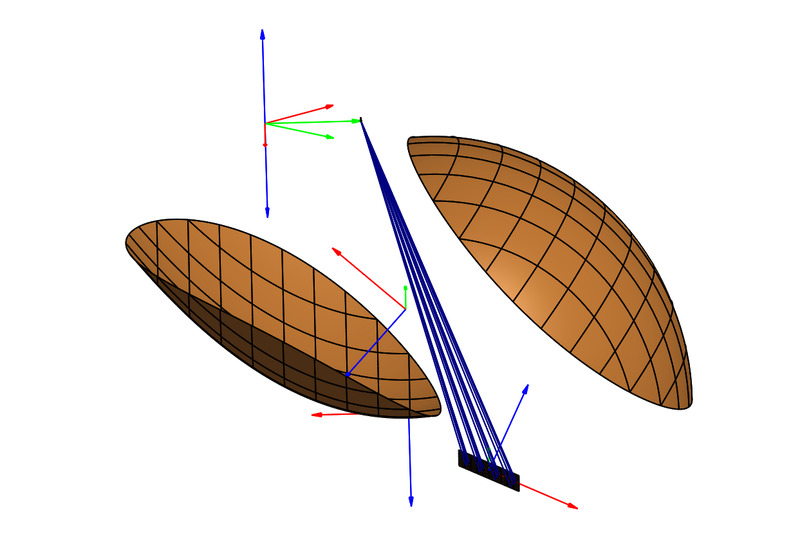} 

\caption[3D views of the
optical path through the combiner]{3D views of the
optical path through the combiner (left, only rays emerging with an angle lower than +/-14 deg) and for straylight
(right, angles between 15 and 21 deg) for the secondary feed horn ID-53.\label{fig24}}
\end{figure}

The decision to include these two screens has been derived from an analysis of the straylight
contribution from the feed horn array on the two focal plane arrays without passing through the optical combiner,
\textit{i.e.}  without reflection on both mirrors. At first we
employed a geometrical optic code written in Mathlab and after a Physical Optics approach with
GRASP~\cite{Grasp} see as an
example the rays in the 3D layout on the right of Figure~\ref{fig24} for a secondary feed horn on the edge of the array.

The additive noise in fringes generation and the unwanted radiative power extra-loading have been considered to estimate
the straylight impact in the final instrument optical performances. Neither the Cold Stop nor the dichroic were
included in this study then taking into account the worst situation.

\begin{figure}
\centering
\includegraphics[width=6.581cm,height=5.249cm]{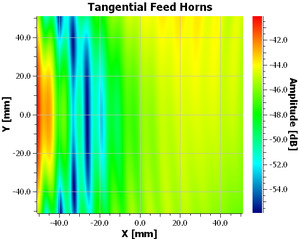} 
\includegraphics[width=6.581cm,height=5.249cm]{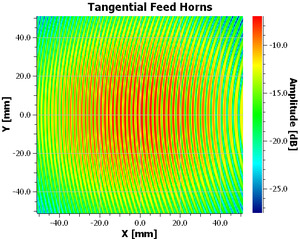}
\includegraphics[width=6.636cm,height=5.292cm]{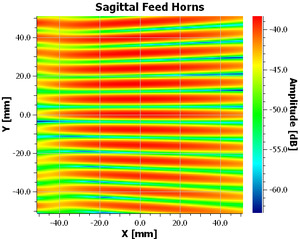} 
\includegraphics[width=6.636cm,height=5.292cm]{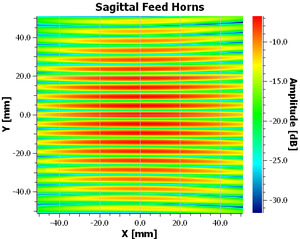} 

\caption[Maps of power on
the on-axis focal plane array considering a tangential couple and a sagittal couple of feed horns]{Maps of power on
the on-axis focal plane array considering a tangential couple of feed horns (Top panels) and a sagittal couple (Bottom
panels): \textbf{s\_fringes }on the left and \textbf{c\_fringes} on the right.\label{fig25}}
\end{figure}

The fringe patterns for two representative couples of feed horns (tangential and sagittal) are
analysed for straylight (\textbf{s\_fringes}) and for combiner
propagation (\textbf{c\_fringes}). The maps of the fringes are
plotted in Figure~\ref{fig25} for all cases.

The amplitude of \textbf{s\_fringes} is
always 4 orders of magnitude lower than \textbf{c\_fringes} 
assuming the same couples of feed horns and the spatial frequency of
\textbf{s\_fringes} is lower than
\textbf{c\_fringes} due to the different beam propagation. To
avoid that a source with strong intensity at low spatial frequency (such as CMB anisotropies) could contaminate target
observations (B mode), we decided to include shields to avoid straylight.

\begin{figure}
\centering
\includegraphics[width=7.cm]{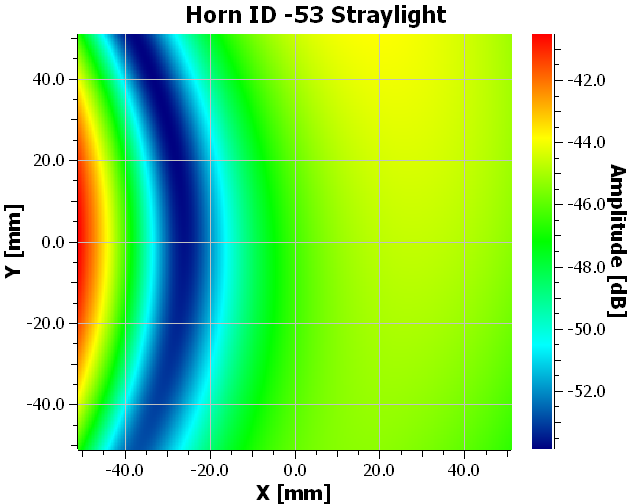} 
\includegraphics[width=7.cm]{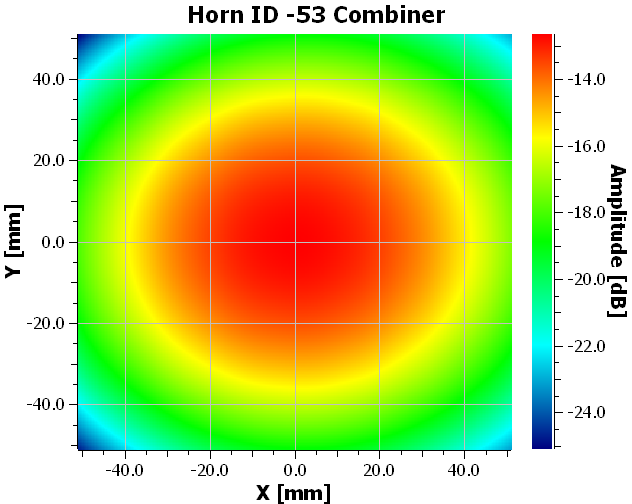} 

\caption[Power on the
on-axis focal plane collected from the feed horn ID-53 passing through the combiner and as straylight]{Power on the
on-axis focal plane collected from the feed horn ID-53 passing through the combiner (right) and as straylight (left).\label{fig26}}
\end{figure}

The increase in power due to straylight was estimated for a reference secondary feed horn on
the edge of the feed horn array, ID-53. The power collected on the on-axis focal plane is almost 3 orders of magnitude
lower respecting to the expected amount, see Figure~\ref{fig26}.

\subsubsection{Optical Simulations}
\label{bkm:Ref314304210}
The final optimisation of the dual-reflector combiner design was carried out using geometrical
optics (ray-tracing) in order to take advantage of the speed and optimisation routines available in commercial software
packages (e.g.~\cite{Zemax}).  However at
these operating frequencies component sizes are not very large compared with the wavelength of radiation and so, for
detailed analyses, techniques that do not ignore the effects of diffraction were used ~\cite{OSullivan}.

\begin{figure}
\centering 
 \includegraphics[width=9.354cm,height=3.748cm]{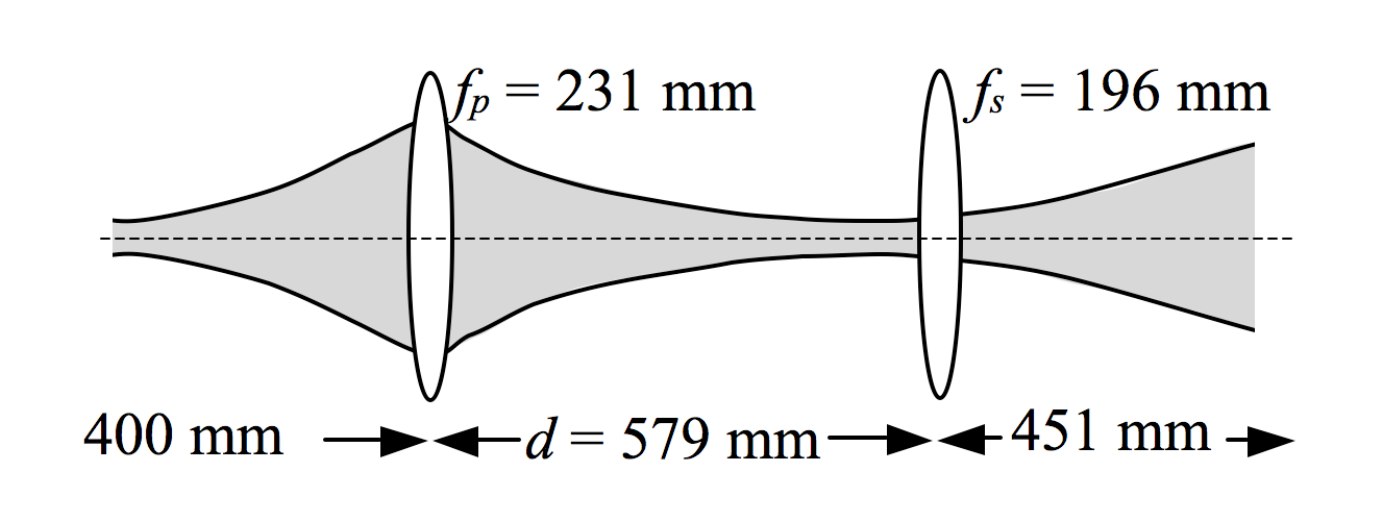}
\caption{Equivalent on-axis system for the QUBIC beam combiner. \label{fig10}}
\end{figure}

\begin{figure}
\centering 
\includegraphics[width=9cm]{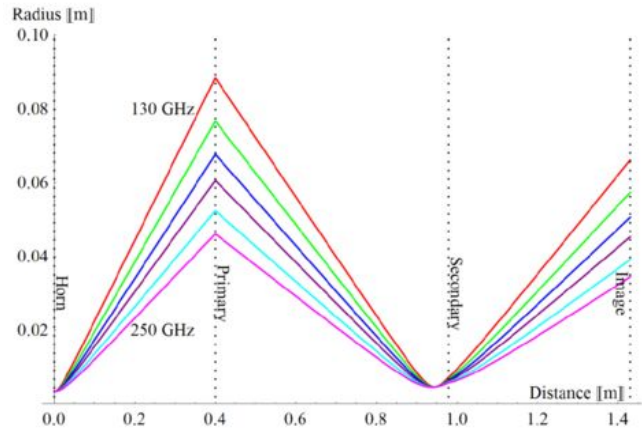} 
\caption[Gaussian beam
width ($w$) as a
function of propagation distance through the QUBIC beam combiner]{Gaussian beam
width ($w$) as a
function of propagation distance through the QUBIC beam combiner. \ The beam was chosen to have a far-field divergence
angle of 12.9\textrm{{\textdegree}} at 150 GHz. \ The range of frequencies shown covers that of the dual-band
instrument.\label{fig11}}
\end{figure}

We started with the beams emitted by the secondary feed horns and propagated them through the
optical system, primary then secondary mirror, and on to the focal plane.  Initially we used a best-fit Gaussian beam
for the horn beam and propagated it through an equivalent on-axis system using a Gaussian beam mode analysis and the
ABCD technique~\cite{Goldsmith}. Figure~\ref{fig10} shows
the QUBIC combiner's equivalent on-axis system and Figure~\ref{fig11} shows the Gaussian beam radius as a function of
propagation distance ($w(z))$) for a selection of frequencies and a beam with a far-field divergence angle of
12.9{\textdegree}. The initial waist radius $w(0)$ was calculated as 

$w(0)\sqrt{2\ln{2}}{\lambda}/{\pi\theta}$ 
 where $\theta$ is the far-field divergence angle of the beam (FWHM of intensity
in this case).  This  was useful for determining approximate beam sizes in the instrument and on the focal
plane. For example, at 150~GHz an $r=51$-mm focal plane can capture $\approx 80\%$ of the power in the 12.9{\textdegree} beam.

The
focal length of the system is given by :
\begin{eqnarray}
f_{eq}&=&(\frac{1}{f_p}+\frac{1}{f_s}-\frac{d}{f_p\times f_s})^{-1}\simeq 300\hbox{mm}
\end{eqnarray}
For a more accurate determination of system performance and for the optimisation of the
dichroic and cold stop size and location, a full vector physical optics (PO) analysis of all 400 beams was carried out
with Maynooth University's in-house software MODAL~\cite{Gradziel} and the
commercially available software package GRASP~\cite{Grasp} (Both use the same
modelling technique and MODAL has previously been tested against GRASP). The beam emitted by the horns was calculated
using a rigorous electromagnetic modematching technique~\cite{Murphy} that views the
corrugated structure as a sequence of smooth walled cylindrical waveguide sections each of which can support a set of
TE and TM modes. At each corrugation there is a sudden change in the radius of the cylindrical guide and this change
results in a scattering of power between the waveguide modes (the total power is conserved).

\begin{figure}
{\centering
 \includegraphics[width=5.976cm,height=6.02cm]{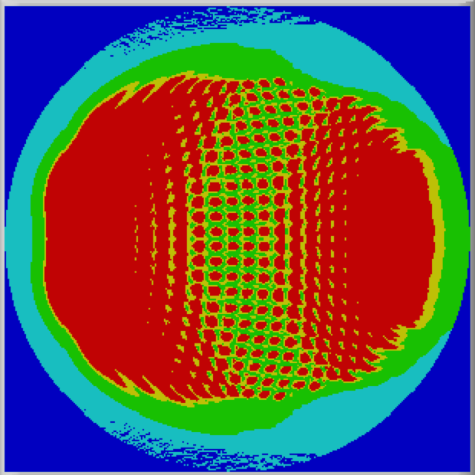}
\includegraphics[width=6.17cm,height=6.066cm]{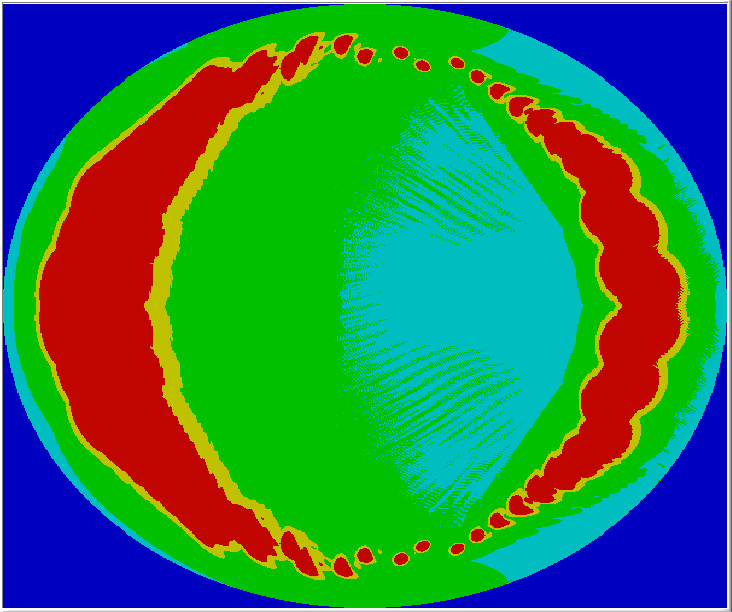} 

\caption[Footprint of 400 antenna beams on a 600-mm diameter secondary mirror, and the same simulation for the outer ring of 12.9{\textdegree} horn beams]{(left) Footprint of 400 antenna beams on a 600-mm diameter secondary mirror. \ The regions coloured red show where the
intensity of each beam is greater than $\exp(-2(0.8)^2)$ of its maximum. \ Yellow,
green and light blue correspond to $\exp(-2(1)^2),\exp(-2(2)^2)$ and $\exp(-2(3)^2)$, respectively~\cite{Scully}. This particular simulation was carried out for the
14{\textdegree} beams of the original QUBIC design. (right) The same simulation for the outer ring of 12.9{\textdegree}
horn beams (all 5 modes that could possibly propagate are included).\label{fig12}}}
\end{figure}

Calculating the footprint of these horn beams at various planes in the system allowed the
optimum size and location of components to be determined~\cite{Scully}. The example in
Figure~\ref{fig12} (left) shows the footprint of the beams on the secondary mirror. The figure is
coloured to show the region where the intensity of each beam is above $\exp(-2(r/w)^2)$ 
of its maximum (with $r/w=$ 0.8, 1, 2 and 3).  Green therefore shows the region where at least 99.9\% of the power
from each beam falls (this corresponds to much more than 99.9\% of the total power, of course, since most beams fall
entirely within this region).

\begin{figure}
\centering
\includegraphics[width=7.071cm,height=6.72cm]{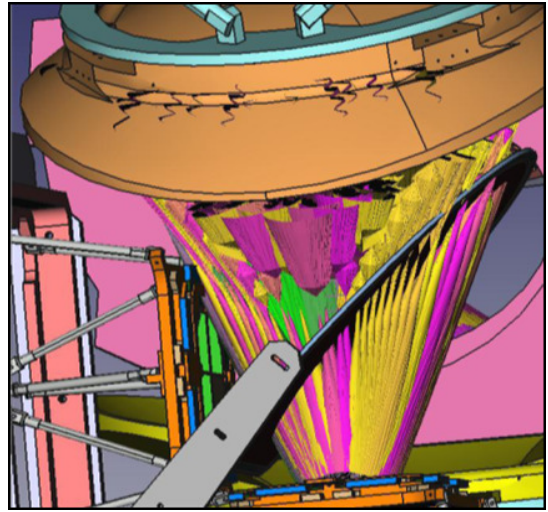}

\caption{3D
rendering of the PO beams in a CAD model of the QUBIC combiner~\cite{Gayer}.\label{fig13}}
\end{figure}

This work was further developed~\cite{GayerT}
to fit a surface to
the 'edge' of each of the beams in order to visualise their propagation through the optics (Figure~\ref{fig13}). \ The edge can
be defined as the points at which the intensity drops to a certain level or, for complex beam shapes, the radius
required to encircle a given percentage of power. The beams are first calculated at a series of planes in the system
using PO. The beam edges are joined using interpolation to check for beam truncation by supporting structures, electronics, ...

Once all the component sizes and locations were chosen, the 400 antenna beams were propagated
through the system to the focal plane. The beams on the focal plane for the final configuration are shown in Figure~\ref{fig14}
(150 GHz). The 400 plots of the focal plane beam are arranged so as to indicate the location of the horn antenna from
which the beam originated (the black line indicates the edge of the input array that will be used).  Here we can see
the effect of aberration and truncation on the beams, especially beyond the edge of the array that will be used. The
plot on the right in Figure~\ref{fig14} shows the percentage of power from each of the horns that is captured by the main focal
plane (limited to a maximum of approximated 80\% for an unaberrated beam by the physical size of the bolometer array).

\begin{figure}
\centering
\includegraphics[width=.5\textwidth]{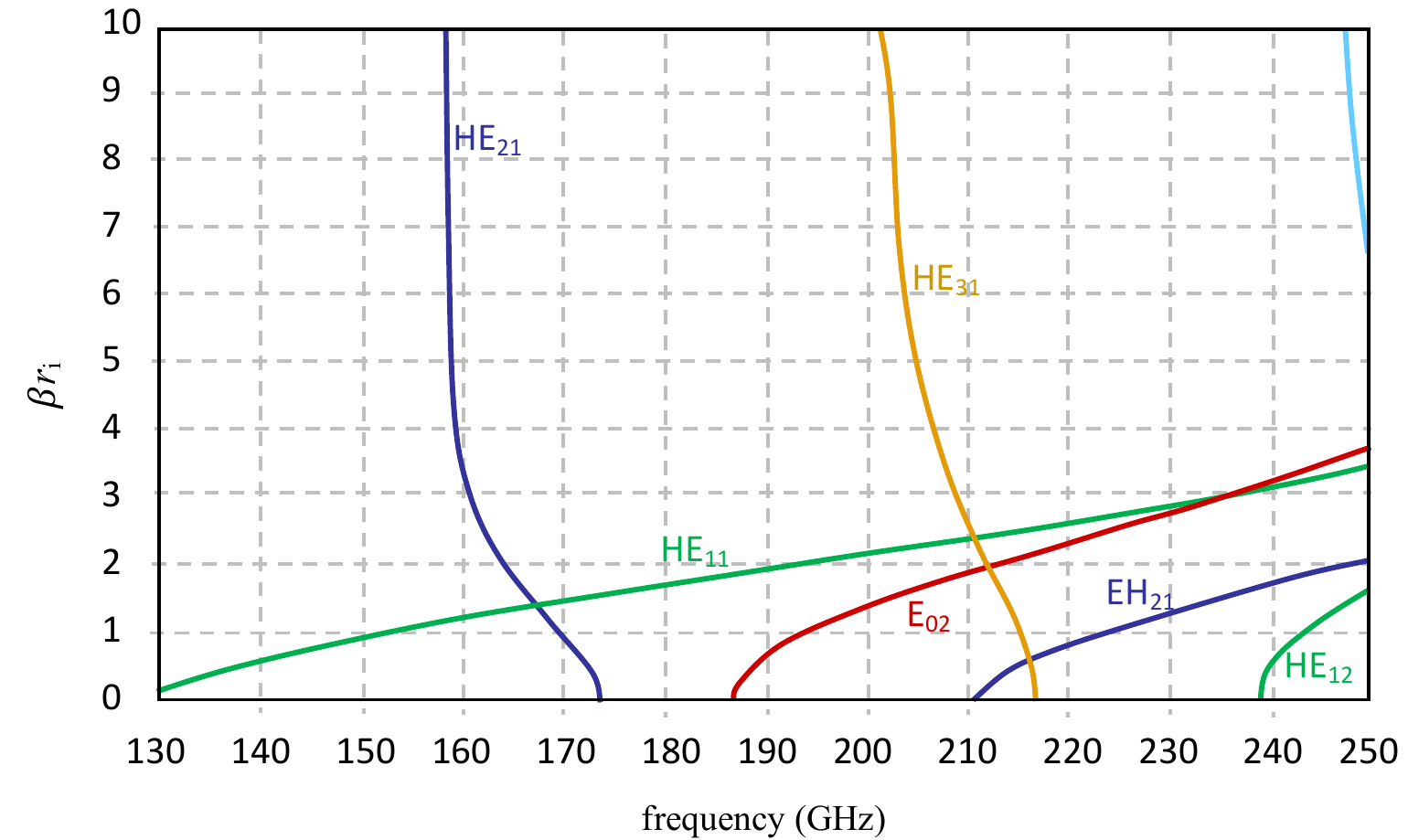}
\caption{(adapted from ref.~\cite{Scully})  Dispersion curves for the 12.9\degree\ feed horn showing the 
modes that can propagate ($\beta$ is the waveguide wave number for a given mode) at various frequencies in 
the narrowest point of the feed horn. 
In this case for for QUBIC's 12.9\degree\ feed horn the narrowest inner ($r_i$) and outer ($r_o$) 
radii are 0.684 mm and 1.394 mm respectively.\label{fig:creide1}}
\end{figure}

As the operating frequency of the instrument increases (up to 250 GHz) the back-to-back horns
allow more hybrid modes to propagate, making the instrument multi-moded. 
We have used the surface impedance model~\cite{claricoats} to calculate the dispersion curves of modes 
(up to azimuthal order  $n=4$) in the QUBIC band.  The surface impedance (hybrid mode) model is 
an approximate one that treats the corrugated wall of a horn as a surface with different average 
impedance in the longitudinal and azimuthal directions.  
It works well as long as there are several corrugations per wavelength but it cannot model detailed horn 
profiles.  
Here we have assumed that the waveguide section between the back-to-back horns is what allows modes through 
or not.  The resulting dispersion curves are shown in Figure~\ref{fig:creide1} which shows 
the HE11 mode propagating throughout the band. An $n=0$ mode cuts on at around 190 GHz, an  $n=2$ mode 
cuts on just above 210 GHz and an  $n=1$ mode cuts on at about 240 GHz.  There is 
a backward mode around 210 GHz.

\begin{figure}
\centering
\begin{tabular}{ccccc}

\multicolumn{2}{c}{$HE_{11}$} & $HE_{02}(=TM_{02})$ & \multicolumn{2}{c}{$EH_{21}$}
\\
\includegraphics[width=.2\textwidth]{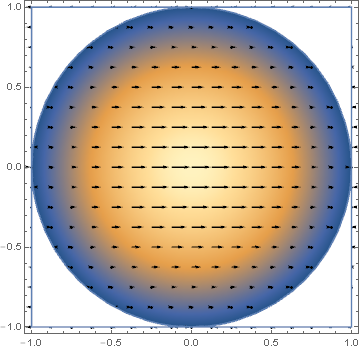}
&
\includegraphics[width=.2\textwidth]{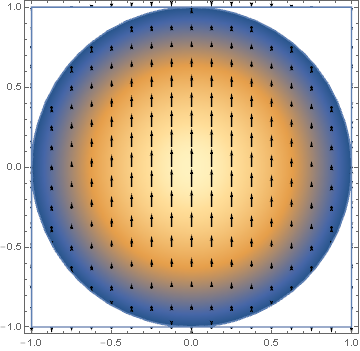}
&
\includegraphics[width=.2\textwidth]{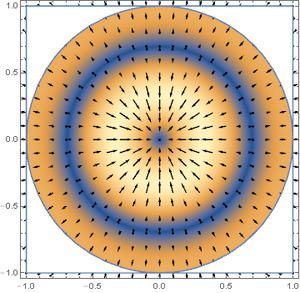}
&
\includegraphics[width=.2\textwidth]{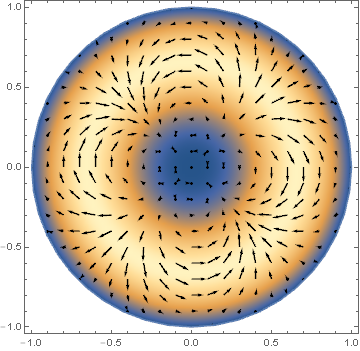}
&
\includegraphics[width=.2\textwidth]{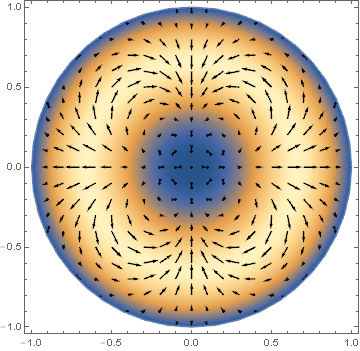}

\end{tabular}
\caption{Plots of hybrid mode (and) aperture fields for the 5 dominant modes of the QUBIC 220-GHz channel 
(orthogonal modes exist for  $n > 0$): the HE11 and its orthogonal mode, the E02 mode and the HE21 and its 
orthogonal mode.\label{fig:creide2}}
\end{figure}

We can use the more rigorous mode-matching technique to model the exact 12.9\degree\ QUBIC horn profile at a 
given frequency and to determine the relative power carried by each hybrid mode.  This was done at several 
frequencies across the 220-GHz band. The actual hybrid modes themselves have not 
been identified merely the weighting of modes at azimutal orders 0 to 4 (SCATTER uses many TE/TM modes to 
describe the aperture fields. The smaller number of hybrid modes can be reconstructed from these, if we 
want to identify them without using Figure~\ref{fig:creide1}, for example).   
Comparison of Table~\ref{tab:creide1} with Figure~\ref{fig:creide1} shows 
broad agreement between the two models.  It is clear that three modes dominate across most of the band 
(HE11, E02 and EH21) and carry approximately equal power.  The HE31 mode carries a little power and there is 
also evidence for modes cutting on at the upper end of the band.  For $n > 0$ modes, the orthogonal mode will also 
be supported.  The five important modes are plotted in Figure~\ref{fig:creide2}.

\begin{table}
\centering
\begin{tabular}{|c|c|c|c|c|c|c|c|c|}
\hline
azimuthal order $n$ & \multicolumn{2}{c|}{0} & \multicolumn{2}{c|}{1} &  \multicolumn{2}{c|}{2} &  \multicolumn{2}{c|}{3} \\
\hline
\hline 
190 GHz & $0.9928$ &\qquad \quad \quad & $0.9996$ & & $0.6\ 10^{-7}$ & & &\qquad  \quad \quad\\ 
\hline
200 GHz & $0.9866$ & & $0.9993$ & & $0.29\ 10^{-3}$ & & & \\
\hline
210 GHz & $0.9939$ & & $0.9998$ & &  &$0.9438$ &$0.300$ & \\
\hline 
212 GHz & $0.9825$&&$1.000$&&&$0.9835$&$0.2731$ & \\
\hline 
215 GHz&$0.9801$&&$0.9999$&&&$0.9911$& & \\
\hline  
217 GHz&$0.9958$&&$0.9999$&&&$1.000$&&\\
\hline
220 GHz&$0.9795$&&$0.9995$&&&$0.9868$&&\\
\hline
230 GHz&$0.9671$&&$0.9933$&&&$0.9989$&&\\
\hline
240 GHz&$0.8807$&&$0.9983$&&&$0.9856$&&\\
\hline
242 GHz&$0.3619$&&$0.9124$&&&$0.9709$&&\\
\hline
244 GHz&$0.5679$&&$0.7888$&$0.000$&&$0.9662$&&\\
\hline
245 GHz&$0.8807$&&$0.7884$&$0.001$&&$0.9852$&&\\
\hline 
247 GHz
&$0.0617$
&
&$0.9878$
&$0.0177$
&
&$0.9576$
&
&\\
\hline
248 GHz
&$0.4217$
&
&$0.9225$
&$0.3312$
&
&$0.9087$
&
&
\\
\hline
250 GHz
&$0.0017$
&
&$0.5540$
&$0.1972$
&
&$0.9393$
&
&\\
\hline
\end{tabular}
\caption{Weighting of hybrid modes in the 220 GHz band. (No  $n=4$ power).\label{tab:creide1}}
\end{table}

\begin{figure} 
\centering 
\begin{tabular}{cc}
\includegraphics[width=.4\textwidth]{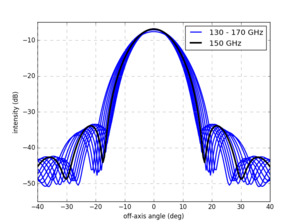} 
&
\includegraphics[width=.4\textwidth]{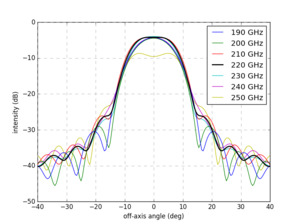} 

\end{tabular}
\caption{(left)Farfield beam patterns calculated across the 150-GHz band (4GHz intervals) where the horns are 
single-moded and (right) the beam patterns across the 220-GHz band. \label{fig:creide3}}
\end{figure}

\begin{figure}
\centering
\includegraphics[width=7.87cm,height=7.549cm]{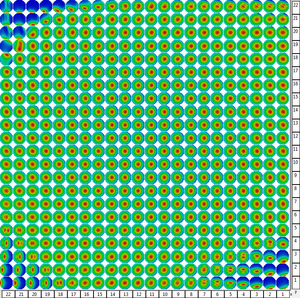}
\includegraphics[width=6.9cm,height=7.505cm]{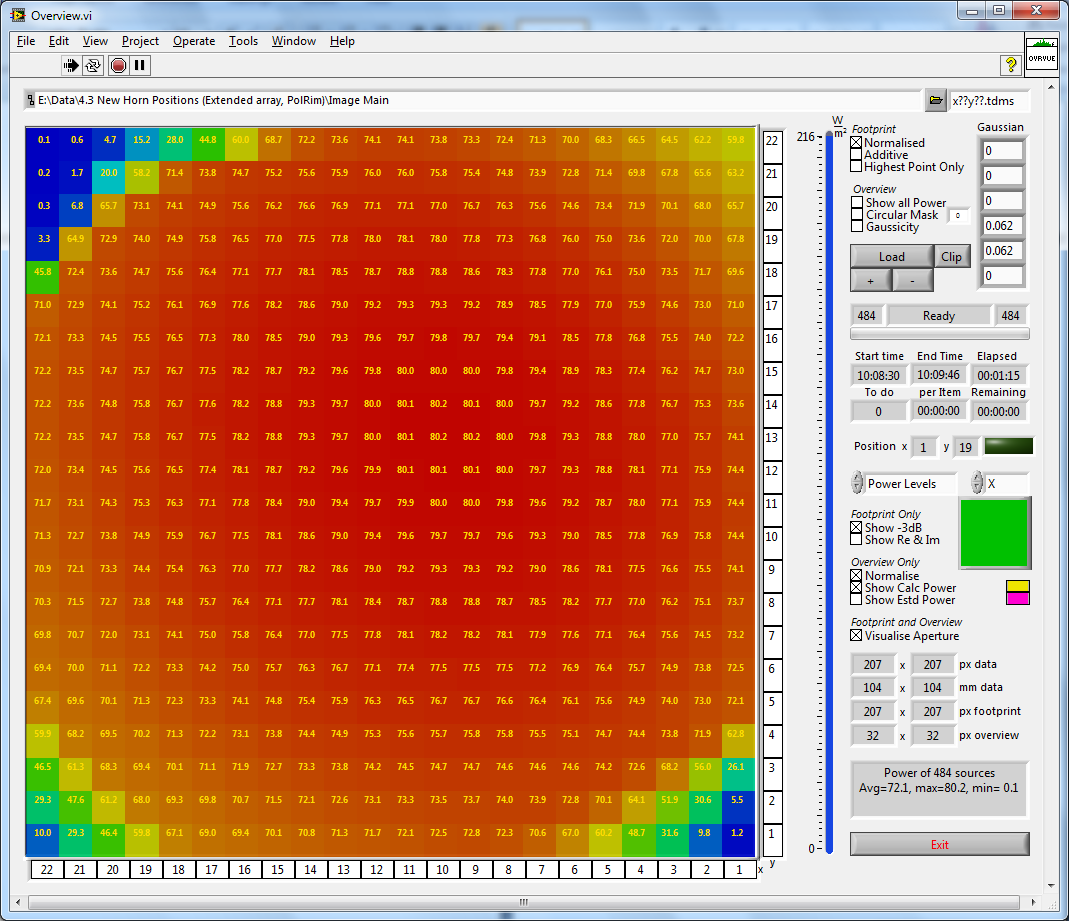} 

\caption[Plots of the beam pattern from each of 400 input horns on the circular focal plane ;  total percentage of power from each beam
that is integrated on the focal plane]{(left)
Plots of the beam pattern from each of 400 input horns on the circular focal plane. \ The location of the input horn in
the array is indicated by the placement of the focal plane plot. \ (right) The total percentage of power from each beam
that is integrated on the focal plane (limited to $\approx$80\% by the finite size of the focal plane). This calculation was
carried out at 150 GHz~\cite{Scully}.\label{fig14}}
\end{figure}

The upper-band farfield beam patterns were calculated (including both sets of orthogonal modes) and are shown 
in Figure~\ref{fig:creide3}.  The beam widens and flattens slightly towards the centre of the band and then 
narrows again.  The beam changes significantly with the cut-on of extra modes between 240 and 250 GHz.  
For this reason 240 GHz was chosen as the cut-off of the upper frequency band.

At high frequencies the slightly narrower beams mean that we expect more power to reach the focal plane when 
compared with the 150-GHz results in Figure~\ref{fig14}. The relative contribution of each of the five modes to the 
beam in the optical combiner will depend on how the input signal couples to the horns and so we have verified 
the combiner design (using their footprints) for each possible mode separately and for each of the 400 horns 
of the full instrument (including the the central 8x8 of the technical demonstrator).

\begin{figure}
\centering  
\includegraphics[width=7cm]{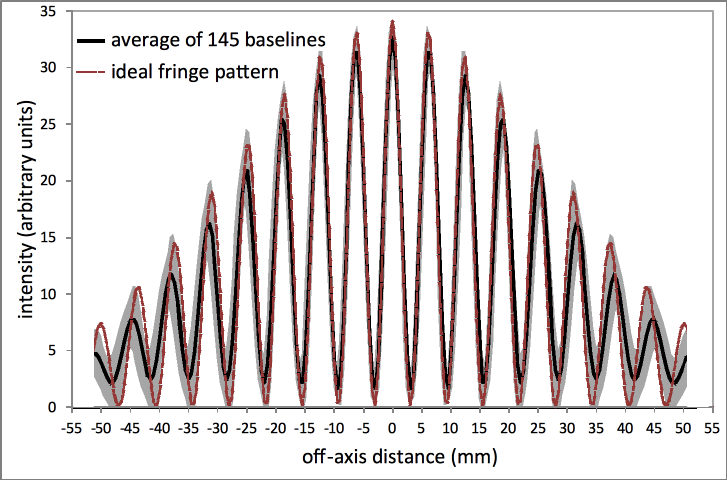}

\caption[Average
of the fringe patterns generated by 145 equivalent $\nu_\lambda= 50$
baselines]{ Average
of the fringe patterns generated by 145 equivalent $\nu_\lambda= 50$
baselines. The standard deviation of the patterns is indicated by the grey shading and the dashed line shows the ideal
fringe pattern.\label{fig15}}\end{figure}

It was not possible to design a diffraction-limited imager for an instrument with such a low
$F/D$ ($\approx$1), where $F$ is the effective focal length and $D$ the entrance pupil aperture,   and wide field-of-view (12.9{\textdegree} at 150 GHz) and so we know that the combiner will be affected by
aberrations at some level. To illustrate the effect of such aberrations on the operation of the beam combiner we
generated fringe patterns from a selection of 145 baselines ($\nu_\lambda = 50$ in this case).
Figure ~\ref{fig15} shows the
average pattern and also the standard deviation of intensity measurements on the focal plane. \ Because of aberrations,
the fringe patterns are not identical, particularly at the edges of the focal plane, and so there interference is neither fully constructive nor fully destructive  - the fringes become degraded.  This means that the sensitivity of the
interferometer to the corresponding angular scale will be reduced.

\begin{figure}
\centering  
\includegraphics[width=8cm]{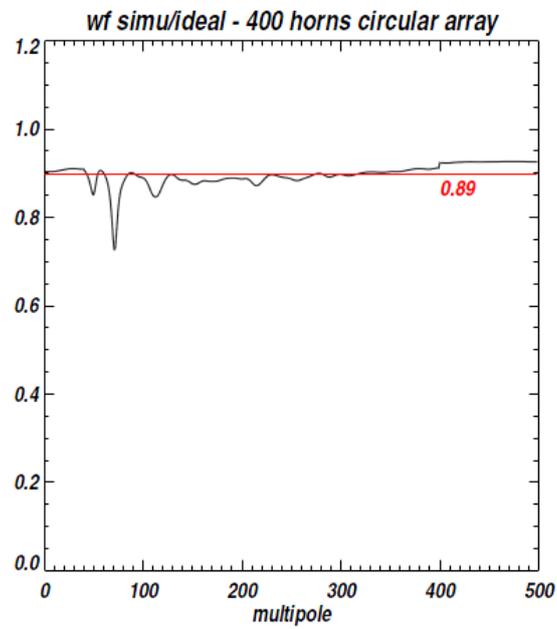}
\caption[Diagonal
window function of the real beam combiner divided by that of an ideal non-aberrating instrument]{Diagonal
window function of the real beam combiner divided by that of an ideal non-aberrating instrument~\cite{Bigot}.\label{fig16}}
\end{figure}

The resulting loss in sensitivity at all angular scales was calculated by considering the
synthesised beam and resulting window function of the instrument~\cite{Bigot}. The result for the
original 14{\textdegree} beams, plotted in Figure
~\ref{fig16} shows that the
effect of the aberrations is to reduce the sensitivity of the instrument by 10\%. \ The loss for the narrower beams is
expected to be less than this.

\begin{figure}
\centering  
\includegraphics[width=7.422cm,height=5.973cm]{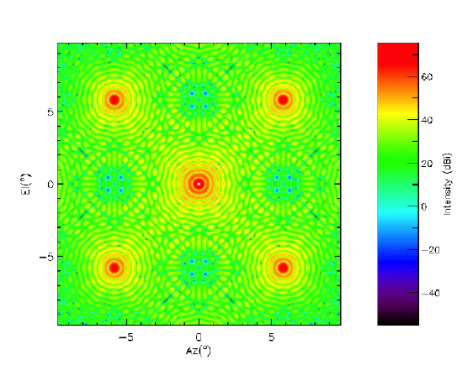} 
\includegraphics[width=7.583cm,height=5.874cm]{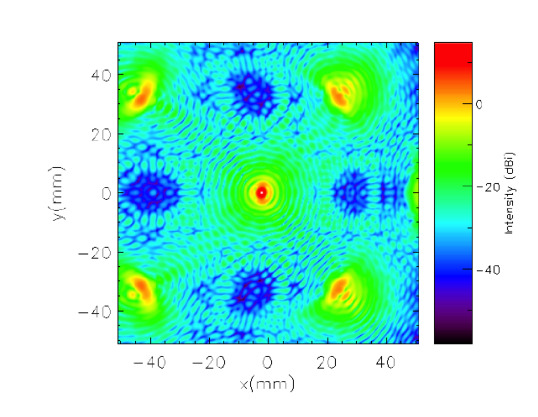} 

\caption[Simulated
QUBIC PSF for the dual band combiner, operated at 150 GHz, for an on-axis source or central pixel]{Simulated
QUBIC PSF for the dual band combiner, operated at 150 GHz, for an on-axis source or central pixel: ideal PSf (left) and
aberrated PSF (right). \label{fig17}}
\end{figure}

The point-spread-function (PSF) calculated by exciting the input horn array with an on-axis
plane wave and summing the 400 focal plane patterns is shown on the right
of Figure~\ref{fig17}. This
example is the PSF of the dual-band instrument operating at 150 GHz. The location of the subsidiary peaks depends on
the separation of horns in the aperture array and the width of the peaks depends on the number of horns. The amplitude
of the peaks is determined by the amplitude of the horn beam pattern on the focal plane.

%auto-ignore

\subsection{Detection chain}

\subsubsection{TES }

\begin{figure}
\centering  
\includegraphics[width=9cm]{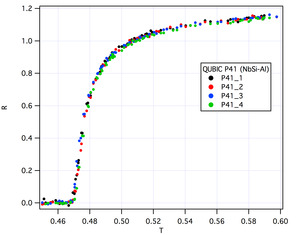}

\caption{Superconducting transition
characteristics of four Nb\textsubscript{0.15}Si\textsubscript{0.85} TES distributed far away from each other on a 256
pixel array.\label{testransi}}
\end{figure}

\paragraph{Description}
The QUBIC instrument detectors  for the 150 and 220 GHz frequencies are composed of four 256-pixel arrays
assembled together to obtain a 1024-pixel detector at the focal plane. The first QUBIC module is split into two focal
planes for a simultaneous scan of the sky at  both frequencies. The detectors are  Transition Edge Sensors
(TES) with a critical normal-to-superconducting temperature close to 500 mK, as illustrated by Figure~\ref{testransi}. Voltage biasing of the sensors allows
operation on the well known ``extreme electro-thermal feedback'' mode with increased bandwidth, direct power
calibration and self-regulation of the TES at the superconducting transition temperature. The TES are made with a 
Nb\textsubscript{x}Si\textsubscript{1-x} 
amorphous thin film (x${\approx}$0.15 in our case), a compound that has been extensively studied and whose production
is well mastered. Its transition temperature T\textsubscript{c}
and normal state resistivity R\textsubscript{n} can be easily
adjusted to meet the QUBIC requirements for optimum performances and multiplexed read-out.

Given the expected  background power of the QUBIC setup (5-50 pW in the 150-220 GHz range) an
extremely low thermal coupling between the TES and the cryostat is needed to optimize signal to noise ratio. This is
obtained using 500 nm thin SiN suspended membranes, which exhibit thermal conductivities between 50 and 500 pW/K depending
on the precise pixel geometry. The total Noise Equivalent Power (NEP) is of the order of 
$5.10^{-17}W/\sqrt{Hz}$ at 150 GHz, with a time
constant in the 10-100 ms range. The pixels have 3 mm spacing while the membranes structure is 2.7 mm wide.

Light absorption is achieved using a Palladium metallic grid placed in a quarter wave
cavity in order to optimize the  absorption efficiency. A distance of 400 ${\mu}$m between the grid and the rear reflector
is a good compromise for both 150 and 220 GHz photons. The array is not intrinsically sensitive to polarization.

The routing of the signal between the TES and the bonding pads at the edge of the array is
realised by superconducting Aluminium lines. These lines are patterned at the front of the array, on the Silicon frame
supporting the membranes.

\begin{figure}
\centering
 \includegraphics[width=.8\textwidth]{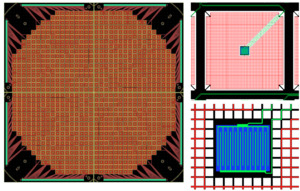} 
\caption[CAO of the 1024 array,  of
one pixel and of the TES with its interleaved electrodes ]{CAO of the 1024 array (left), of
one pixel (top right) and of the TES with its interleaved electrodes (bottom right). On the bottom right, TES is in blue,
the Aluminium interleaved electrodes in green and the Pd grid in red. SiN membrane is in black.\label{fig:tes}}
\end{figure}

\paragraph{Fabrication and AIT}
The fabrication of the TES arrays is based on commercially available silicon-on-insulator
(SOI) wafers. The available TES electron-beam deposition machine at the CSNSM laboratory limits the maximum size of the
wafers to 3 inches. Upgrade to 6 inches wafer technology in order to process a monolithic 1024-pixel array is possible
but needs modification of several fabrication and test devices, including some of the cryogenic test facilities. For
QUBIC it was decided to assemble four 256-pixel arrays for each of the focal planes of the instrument. 
The QUBIC TES design is shown on Figure~\ref{fig:tes}.

\begin{figure}
\centering
 \includegraphics[width=7.cm]{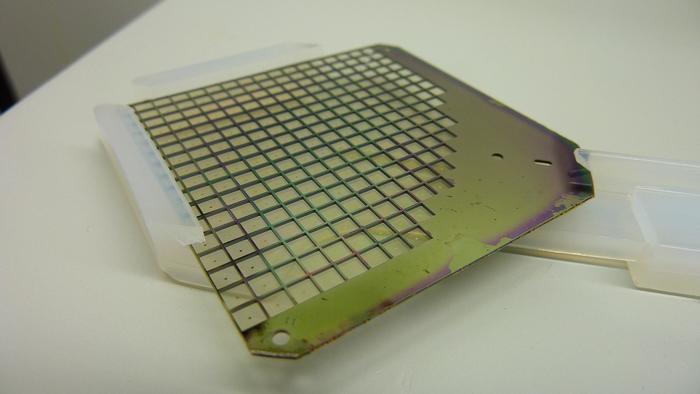}
\includegraphics[width=7cm]{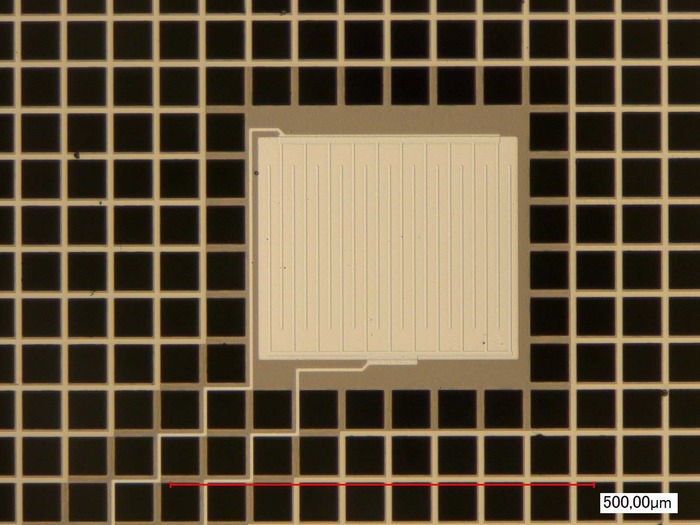} 
\caption[Picture of the Deep etching of the Si to
form the membrane of the pixels,  microscope image of the TES and the comb shaped electrodes]{Picture of the Deep etching of the Si to
form the membrane of the pixels (left). Microscope image of the TES and the comb shaped electrodes (right).\label{fig:tesetching}}
\end{figure}

The detector is realized using lithography process at the IEF-Renatech nanofabrication facility and electron-beam film
deposition technology at CSNSM. The fabrication process can be resumed in the following main steps:

\begin{enumerate}
\item {
{Commercial supply of 3 inches SOI wafers with a deposited layer of 500 nm ultra-low-stress SiN
on both sides, using LPCVD (low pressure chemical vapour deposition). The SOI wafers are composed of a 400 ${\mu}$m
thick Si substrate, followed by a 1 ${\mu}$m thick
SiO}{\textsubscript{2}}{ buried oxide (BOX), and a final 5 ${\mu}$m
thick Si device layer. In our case, the choice of SOI wafers is related to the need of a stop-etching layer (the BOX)
during the deep-etching of the 400 ${\mu}$m Si substrate.}}
\item {
{Electron-beam deposition of a bilayer composed of 30nm
Nb}{\textsubscript{x}}{Si}{\textsubscript{1-x}}{
(TES) followed by 200nm of Aluminium (comb TES electrodes + Routing + bonding pads) without braking the vacuum.}}
\item {
{Wet etching of the Al and reactive ion etching (RIE) of the NbSi. The routing is composed of
the
Nb}{\textsubscript{x}}{Si}{\textsubscript{1-x}}{{}-Al
bilayer, characterized by a superconducting transition temperature very close to that of pure Al (${\approx}$1.2 K). At
the end of this step the TES with its routing is patterned and can be tested. Using a bi-layer resolves the problem of
the contact resistance and the step between the superconducting electrodes and the TES.}}
\item {
{Realisation of the light absorption metallic grid by lift-of of a 10nm thick Pd layer. The
grid has a filling factor of 4\% in order to get a square electrical resistivity that matches the vacuum impedance (377
${\Omega}$ square).}}
\item {
{The next steps are related to the realisation of the micro-meshed membranes. We begin with the
back-side deep-etching of the Si substrate (DeepRIE-ICP) followed by the etching of the BOX. The back SiN layer is also
removed. This operation is illustrated by Figure~\ref{fig:tesetching}.}}
\item {
{Front RIE of the SiN to get the micro-meshed pattern (50 ${\mu}$m x 50 ${\mu}$m square
pattern).}}
\item {
{Dry etching of the 5 ${\mu}$m Si device layer using
XeF}{\textsubscript{2}}{. The Si device layer is completely removed
after this step and we obtain the SiN meshed membrane.}}
\item {
{Residual resist removal using Oxygen plasma treatment. Removal of the photolithography resist
using solvents is prohibited at this stage because it will damage the membranes.}}
\end{enumerate}

\begin{figure}
\centering
\includegraphics[width=7cm]{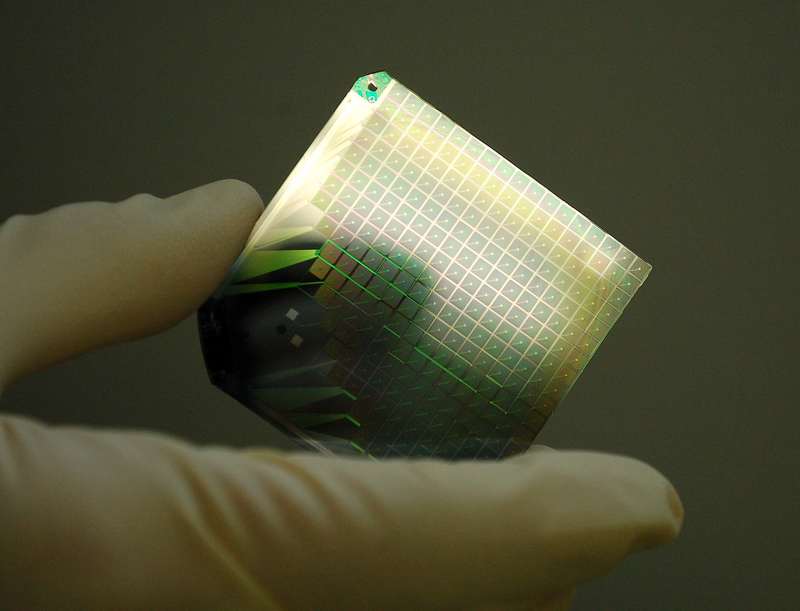}
\includegraphics[width=7cm]{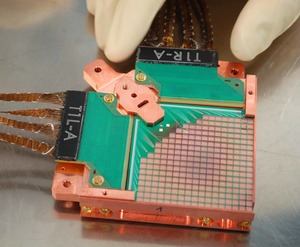} 
\caption{Pictures of the 256 TES array being
processed (left) and being integrated for the test (right)\label{fig:tesinteg}}
\end{figure}

The overall fabrication process takes typically two weeks if there is no testing of the TES characteristics at low
temperature. We are usually processing two wafers in parallel without considerably increasing the fabrication delays.
The 256-pixel array is finally integrated within the focal plane holder and electrically connected to a printed circuit
board (PCB) using ultrasonic bonding of Aluminium wires. The latest upgrade of the process allows very satisfying fabrication
quality with a dead pixels yield as low as 5\%. Pictures of TES in their final state are shown on Figure~\ref{fig:tesinteg}.

\subsubsection{Cold electronics}
\begin{figure}
\centering
\includegraphics[width=.8\textwidth]{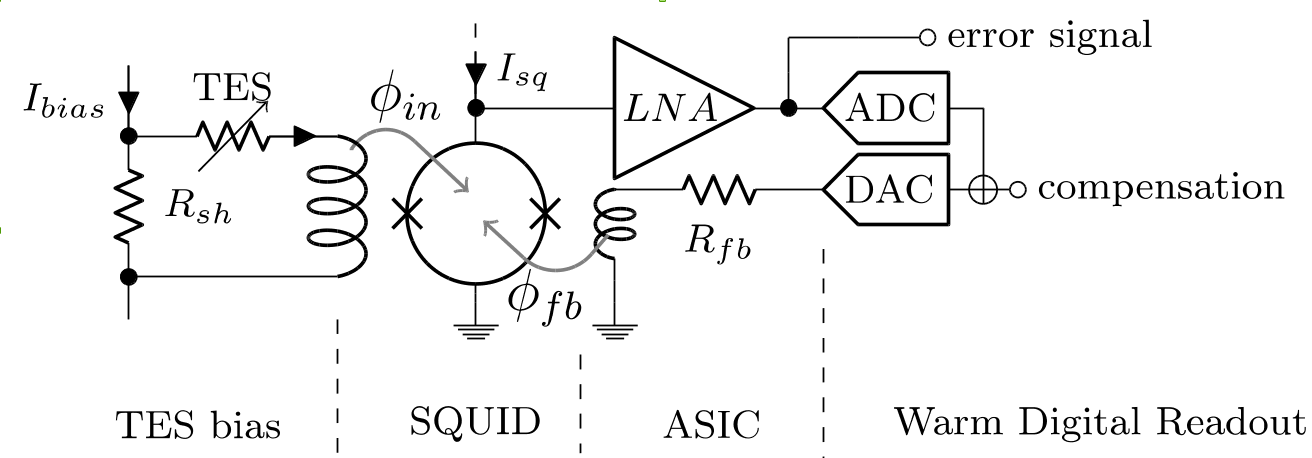}
\caption[ Detection Chain including TES, TES biasing ($R_{sh}$), SQUID, ASIC (LNA) and Warm Digital Readout (ADC+FPGA)]{\label{elec_scheme_cold_elec} Detection Chain including TES, TES biasing ($R_{sh}$), SQUID, ASIC (LNA) and Warm Digital Readout (ADC+FPGA). Feedback with DAC and $R_{fb}$ is also shown.}
\end{figure}

The detection chain of the QUBIC instrument, shown on Figure~\ref{elec_scheme_cold_elec},  can be decomposed in 5 parts:
\begin{enumerate}
\item TES			\hfill					320 mK
\item  TES voltage biasing and SQUID multiplexer		\hfill			1 K
\item  ASIC (LNA i.e. low noise amplification + Biasing + multiplexer clocking)	\hfill			77 K
\item Warm LNA				\hfill					300 K
\item Warm Digital ReadOut (ADC + FPGA) 		\hfill			300 K
\end{enumerate}
Therefore, the cold readout described in this section refers to the TES voltage biasing, the SQUID multiplexers and the ASIC.

\begin{description}
\item{TES voltage biasing :}
This  is the first part of the readout chain. Indeed, TES is wildly used when large number of detector is needed due to 
the strong Electro Thermal Feedback (ETF) which homogenize the detector responses even if unavoidable fabrications 
inhomogeneity's exists. However, strong ETF is obtained only if TES are voltage biased. This means that the voltage 
across the TES must be fixed independently to the TES resistance (which varies with the noise, the scientific signal 
and the background). To passively ensured such fixed voltage across a TES (operating resistance about 100 mK) the TES 
voltage biasing sources must have a Thevenin's resistance smaller than the TES operating resistance. To provide such 
extremely low output resistance voltage sources at deep cryogenic temperatures, a simple current biased $I_{bias}$ 
shunt (very low value) resistor is used. This shunt resistance is thus the Thevenin's resistance of the obtained 
TES biasing and is chosen with a value = 10 m$\Omega$ (which is lower than the TES operating resistance, that amounts to a few hundreds of $m\Omega$). The TES voltage is then roughly 
fixed to $V_{TES} = R_{sh} \times  I_{bias}$. The 10 m$\Omega$ shunt  resistors are placed in the SQUID PCB 
(Printed Circuit Board) at 1K. The $I_{bias}$  current is provided by the Warm Digital Readout adjusted by the FPGA 
trough a  specific slow differential DAC.
\item{SQUID stage :}
SQUIDs fabrication and testing is described in Sect.~\ref{sect_squid}.  We concentrate here on 
  the  current front-end readout of the TES that also provides the multiplexing.
\begin{description}
\item{Current front-end readout :}
 We previously discussed the needed TES voltage biasing, so this voltage biasing leads to a current readout: 
The voltage is kept constant across the TES and we measure the fluctuation of the current induced by the TES's 
resistance fluctuation as function of the scientific signal. Moreover, the input impedance of the SQUID must be smaller 
than the Rsh shunt resistor to avoid adding a significant resistance in the biasing circuit. 
We discuss the impedance because this requirement is 
needed in DC, but also over all the TES bandwidth to avoid instability (TES current biased is unstable). The input 
impedance of the SQUID corresponds to the impedance of the SQUID input loop Lin. That loop is made using 
superconducting material (Nb) and introduces 0 resistance in the TES biasing circuit. However, regarding to the 
frequency response, that loop introduces an impedance which increases with frequency as $Z_{in} = L_{in}\times 2\pi f$. 
Therefore, we have to chose $L_{in} < 10 m\Omega / 2\pi BW$ where $BW$ is the bandwidth response of the TES. 
Apart these considerations, $L_{in}$ is also used to convert the TES current in flux $\phi_{in}$ into the SQUID, then 
the SQUID provides an output voltage: trans-impedance amplification (gain)
of the order of 100 V/A. However, this gain is strongly non-linear as shown on Figure~\ref{squid_tf}.
\begin{figure}
\centering
\includegraphics[width=10cm]{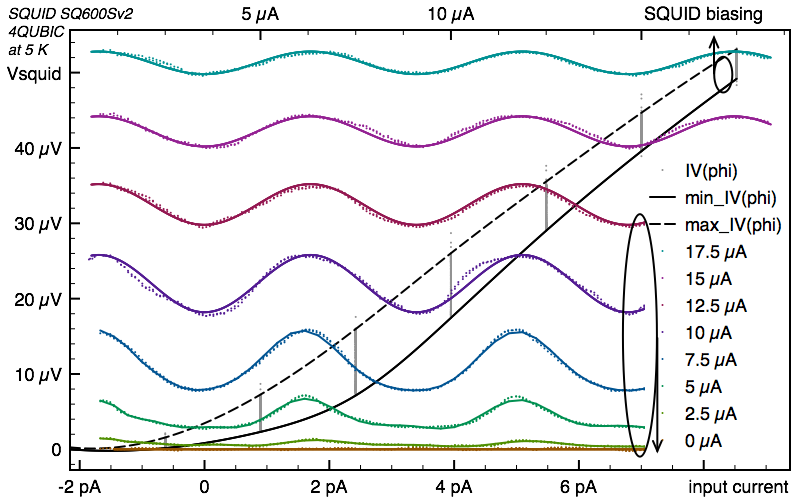}
\caption{Flux to voltage SQUID transfer function for current biasing from 0 to 17.5 $\mu A$. And I(V) SQUID output characteristic.\label{squid_tf}}
\end{figure}

To linearize the SQUID transfer function, the operating point is maintained in a steeper part of the flux-to-voltage 
transfer function. To counteract the input flux (coming from TES current fluctuation) a feedback flux is applied trough 
the feedback coil and the feedback resistance. Thanks' to this feedback techniques, a wide linear range is provided to 
readout TES as shown on Figure~\ref{squid_tf_linr}.
\begin{figure}
\centering
\begin{tabular}{cc}
\includegraphics[width=.4\textwidth]{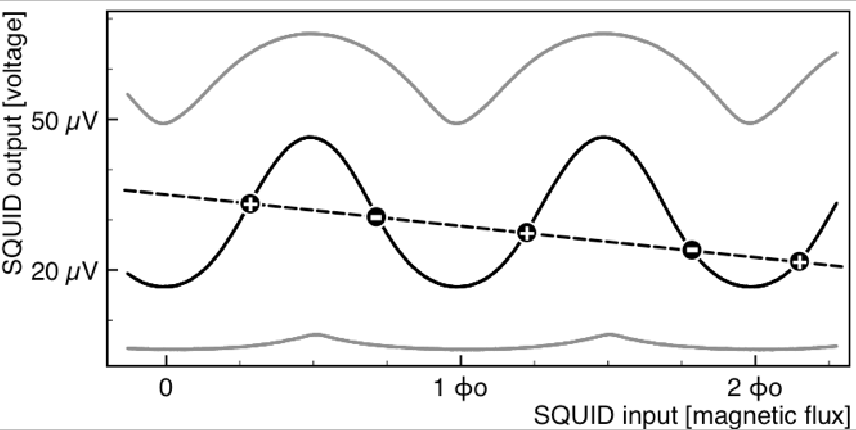}
&
\includegraphics[width=.4\textwidth]{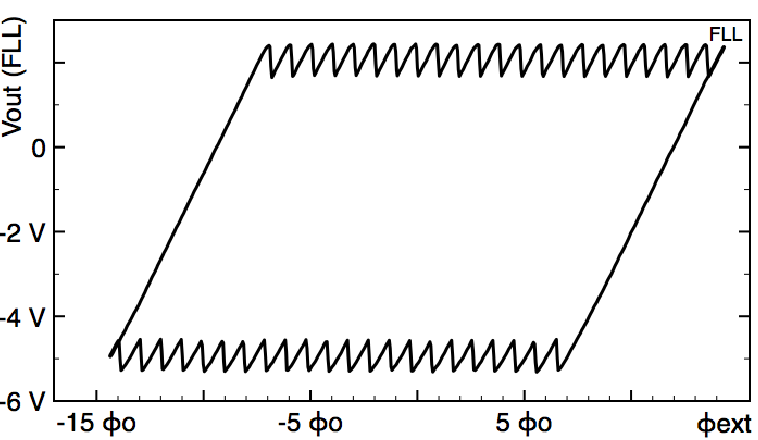}
\end{tabular}
\caption[flux-to-voltage open loop SQUID gain and feedback transfer function ;  FLL flux-to-voltage linearized transfer function]{Left : flux-to-voltage open loop SQUID gain and feedback transfer function (dashed line) highlighting operating points; right : FLL flux-to-voltage linearized transfer function.\label{squid_tf_linr}}
\end{figure}
\item{Multiplexing :} 
 More than a cryogenic amplifier, SQUIDs also enable the multiplexing thank to their large bandwidth. Indeed, the SQUID stages of 32 TES are connected 
together to readout successively each of this 32 TES. In addition, a 4 to 1 multiplexed LNA readout sequentially 4 
columns of 32 SQUID each. The multiplexing factor is at the end up to 128. This scheme is shown on Figure~\ref{fig84}. 
The low noise amplification (LNA), the sequentially biasing of the SQUID and the overall clocking of this 128:1 
sub multiplexer is obtained thanks to an ASIC (operated at cryogenic temperature).
%\begin{figure}
%\centering
%\includegraphics[width=\textwidth]{multiplex_scheme}
%\caption[Multiplexing scheme]{Multiplexing scheme: 32 SQUID in column provide a first 32 to 1 time domain multiplexing. A 4 to 1 multiplexed LNA provide the second multiplexing stage.\label{squid_multiplex}}
%\end{figure}
\end{description}
\item{ ASIC :}
The ASIC is described in details in Sect.~\ref{sect_asic}. 
\end{description}

\subsubsection{{SQUIDs}}
\label{sect_squid}
\paragraph[Providing the SQUIDs by StarCryo]{{Providing
the SQUIDs by StarCryo}}

\begin{figure}
\centering
 \includegraphics[width=7cm]{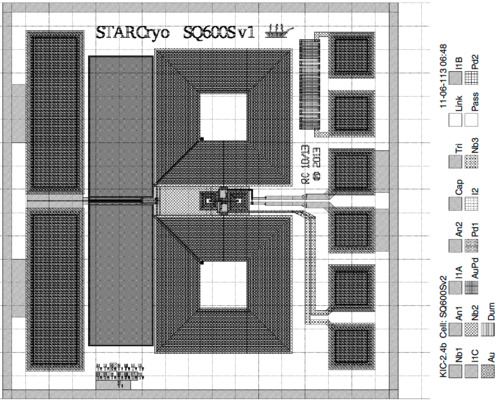}
\includegraphics[width=7cm]{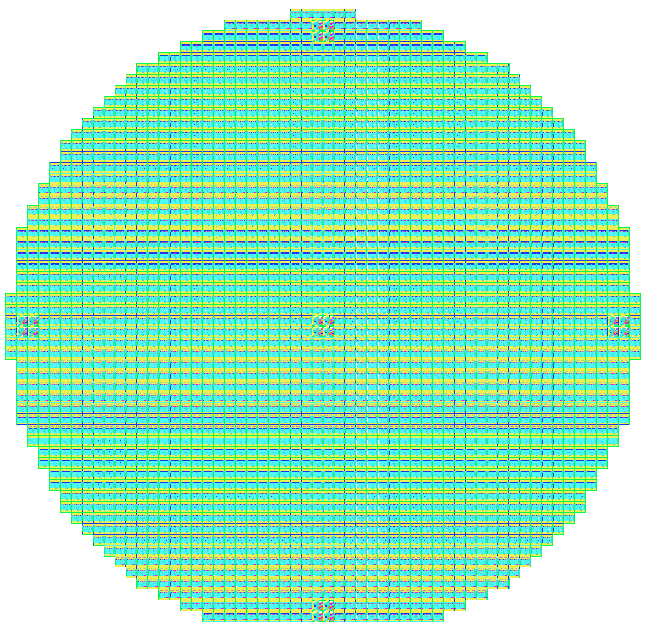} 
 \includegraphics[width=7cm]{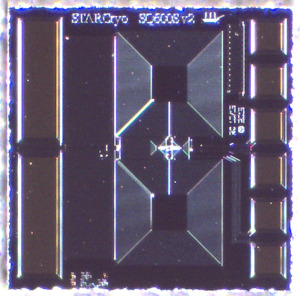}
\includegraphics[width=10cm]{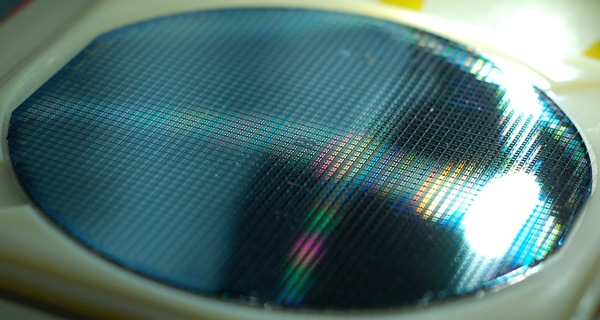} 
\caption[Layout  and photo  of the
SQUID and the full wafer]{Layout (top) and photo (bottom) of the
SQUID (left) and the full wafer (right). SQUID: \ input pads (left), Washer and flux loop (middle), heater (top right),
SQUID access (middle right) and feedback pads (bottom right). A square of the grid is 100$\mu$m. 
The SQUID chip has thus a size slightly smaller than 2mm. Wafer: about 54 SQUIDs on the diameter -{\textgreater} 2000 SQUIDs on the wafer. \label{fig:squids} }\end{figure}

The SQUIDs are based on the SQ600S commercial design provided by StarCryoelectronics (starcryo.com). However, this
design has been modified to remove an input transformer (for ``current-lock'' CL operation) not used in the QUBIC readout
chain (based on flux feedback). In addition, size of the pads has been reduced to $200\mu$ side and all the design has
finally be compacted to reduce the area of silicon need for each SQUID and put about 4000 SQUIDs on 2 custom wafers.
 Even taking into account a 80\% yield and the realization of spare cards, these two wafer
should be enough for one QUBIC module. Scheme and pictures of the SQUIDs aand their wafer are shown on Figure~\ref{fig:squids}.

\paragraph[Room temperature test and cleaning]{ Room temperature test and cleaning}

Before any use,  any SQUID must be removed from an adhesive layer used to maintain them while sawing. After
that a resin layer is removed chip by chip through first an acetone bath, and then a methanol bath, before drying the SQUID
chip with a nitrogen flux. 
Visual inspection allows removing part of the SQUIDs that clearly show defects, especially from the side of the wafer. 
A test probe-station equipped with a multimeter  
 allows to test the electrical
characteristics of the SQUIDs, with the criteria listed in Table~\ref{table25}.

\begin{table}
\begin{center}
\begin{tabular}{llc}

Heater & Rh  & 300-500 $\Omega $  \\

SQUID &  Rsquid  &  100-200 $\Omega $ \\

Rfb &  Rfb &  300-500 $\Omega $ \\

Rin & Rin & 10-15 \ \ k$\Omega $ \\

Insulation In &  Rsq/in &  {\textgreater} 10  M$\Omega $ \\

Insulation Fb &  Rsq/fb &  {\textgreater} 40  M$\Omega $
\end{tabular}
\end{center}

\caption{\label{table25}Requirements criteria for selection of
SQUIDs}
\end{table}

During these room temperature tests, many precautions should be followed to prevent ESD
damages on SQUID.

\paragraph[Cryogenic test:]{ Cryogenic test:}

\begin{figure}
\centering
\includegraphics[width=15.882cm,height=5.436cm]{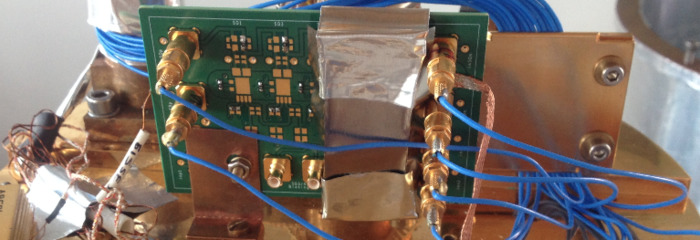} 
\includegraphics[width=8.259cm,height=4.935cm]{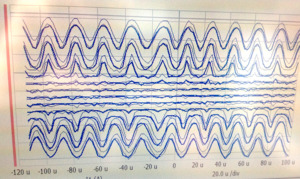} 
\includegraphics[width=7.627cm,height=5.267cm]{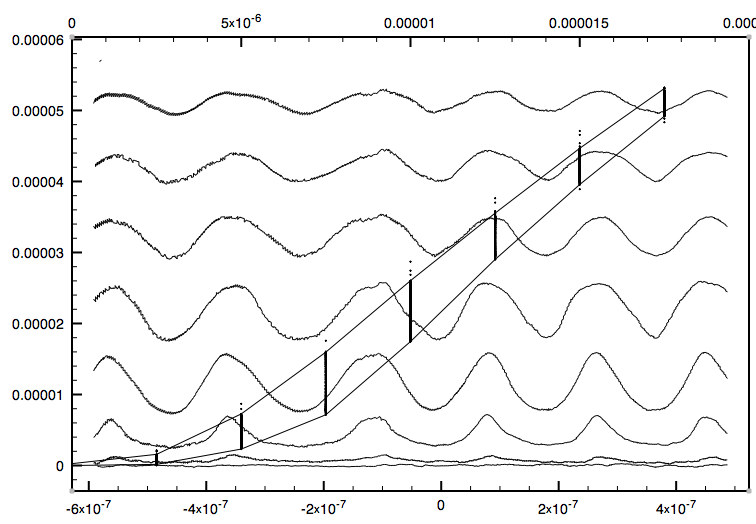} 
\caption[$I(\phi)$ measurement using a vector % B1500A Agilent
analyzer allows to obtain the SQUID $V(\phi)$ transfer function]{$I(\phi)$ measurement using a vector % B1500A Agilent
analyzer allows to obtain the SQUID $V(\phi)$ transfer function (left and bottom right). From this measurement given for
different SQUID biasing, the SQUID $I(V)$ curve can be reconstruct as shown in the right: Y axes is $V_{squid}[V]$, X bottom
is $I_{in}[A]$ equiv. to $\phi$ and X top is the SQUID biasing $I_{squid}[A]$.\label{fig:squidcoldtest}}
\end{figure}

One or two SQUIDs per wafer are tested at cryogenic temperature to determined the critical current and the swing
($\Delta V_{squid}$) of each wafer, as illustrated on Figure~\ref{fig:squidcoldtest}. To do that, a SQUID chip is glued and wire bonded in ``4 points''. A single $V(\phi)$
measurement allows to determine all the parameters.

These measurements without filtering and cryogenic ASIC are very noisy and give  significantly less precise measurements than
those obtained later in the full QUBIC readout chain. Nevertheless, this test gives an order of magnitude of the critical
current and of the swing of the output voltage which are important parameters for the use of the SQUID in the QUBIC readout
chain.

\paragraph[Integration]{ Integration}

\begin{figure}
\centering
\includegraphics[width=9.5cm]{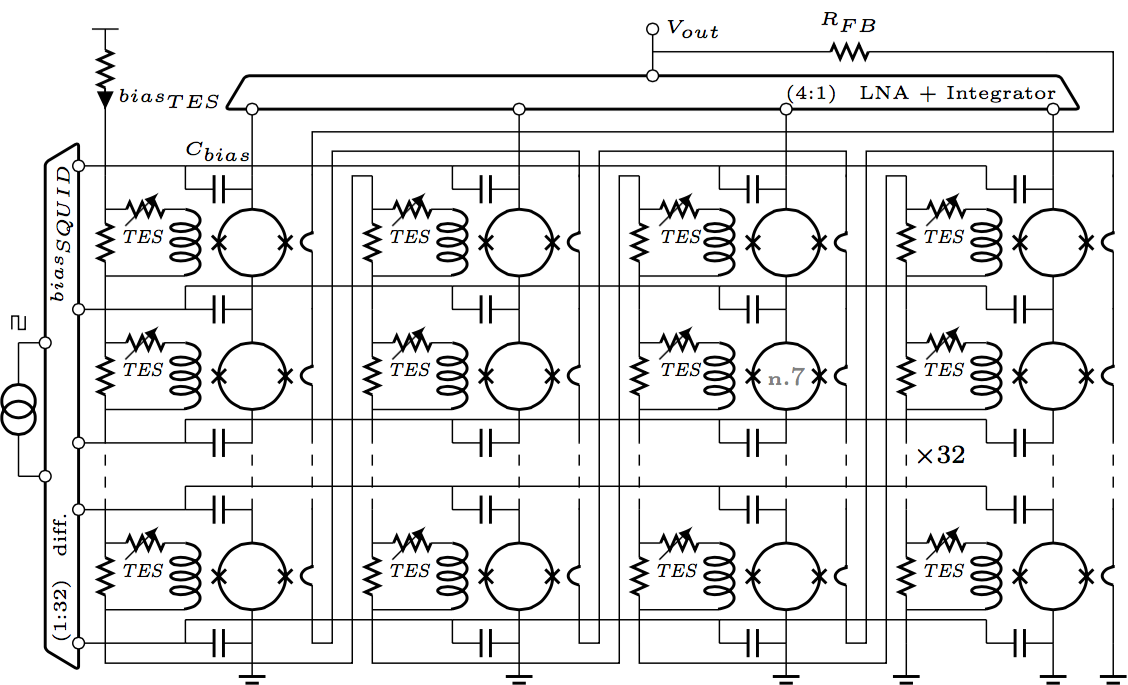} 
\includegraphics[width=6.4cm]{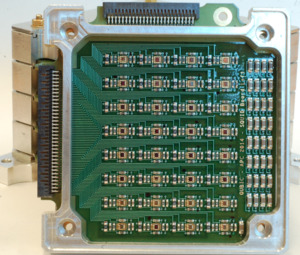} 
\caption{\label{fig84} Topology of the
128 to 1 multiplexer sub-system (4x32 SQUID + 1 ASIC). Integration of 32 SQUIDs (1 collumn) with bias capacitors and
filter devices.}
\end{figure}

Schematic of the SQUID multiplexer is given in Figure~\ref{fig84}. It is composed of 4 columns of 32
SQUIDs each. 32 SQUIDs are connected together using one PCB.

\begin{figure}
\centering
\includegraphics[width=5.983cm,height=6.158cm]{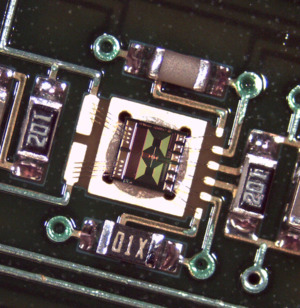} 
\includegraphics[width=9.541cm,height=6.075cm]{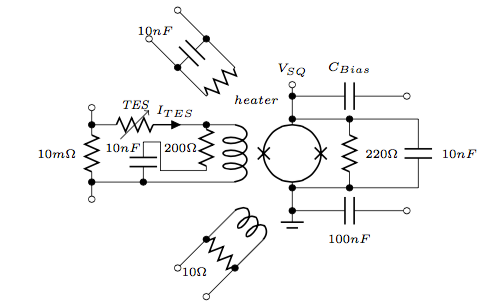} 
\caption{SQUID glued and wire bonded with filter
devices. Values of the filter devices put close to SQUIDs.\label{squidglue}}
\end{figure}

A SQUID is a very sensitive device and its input  must be filtered to keep nominal critical current. So,
capacitors are added in parallel to the SQUID, input loop and heater to avoid radio frequency parasitic signals close
to the SQUID washer. A resistor is added in series to the capacitors in the input inductor to damp LC resonances. 
Moreover, a SQUID is composed of two Josephson junctions (nm insulators) that are very sensitive to
electrostatic discharge. So a 220$\Omega $ resistor is put in parallel to the SQUID to deviate peak current. 220$\Omega
$ value is chosen as it is  much larger than  the typical 2$\Omega $ SQUID fragile shunt resistors, in order to neglect the
voltage division introduced by this resistor. The final circuit is oulined on Figure~\ref{squidglue}.

Finally, a 10$\Omega $ resistor is put in parallel with the feedback loop to ensure feedback signal even if one of the
feedback connection is open. Indeed, in the multiplexing scheme, all the feedback loops are connected in series.
Without this 10$\Omega $ resistor, only one open feedback loop would lead to the loss of 128 pixels.

\begin{figure}
\centering
 \includegraphics[width=6.022cm,height=4.754cm]{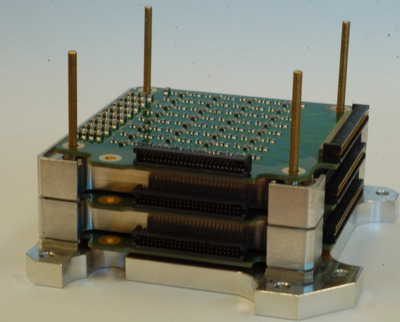}
\includegraphics[width=5.509cm,height=4.771cm]{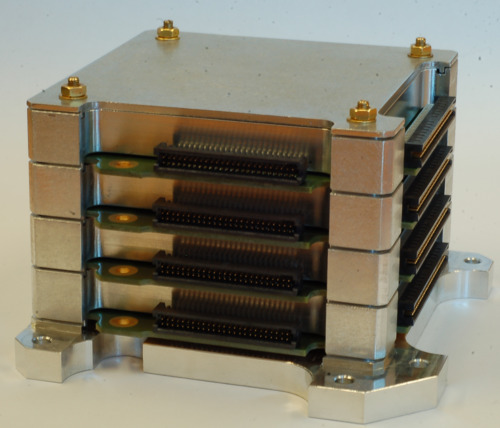}
\includegraphics[width=3.522cm,height=4.895cm]{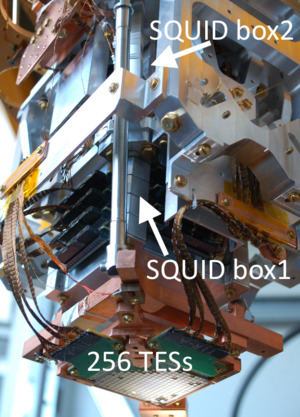} 
\caption[Two SQUID boards
stacked to finally obtain a SQUID box composed of 4 PCBs,  TES thermo-mechanical
structure showing the 2 SQUIDs boxes near the TES array]{Two SQUID boards
stacked (left) to finally obtain a SQUID box composed of 4 PCBs, and thus  128 SQUIDs (center). TES thermo-mechanical
structure showing the 2 SQUIDs boxes near the TES array.\label{fig86}}
\end{figure}

{
{As seen in Figure~\ref{fig84}, SQUID columns (32 SQUIDs) are connected together with a PCB board. 4 of
this PCB are needed to readout 128 pixels. So, 4 of this PCB are staked in a ``SQUID box'' as shown in Figure~\ref{fig86} (left
and center). Finally, 2 of this SQUID box are placed below the 256 TES array in the cryo-mechanical structure (Figure~\ref{fig86}, right panel).}}

SQUID are thermalized to 1K whereas TESs are at  320 mK.

\subsubsection{{ASIC}}

\label{sect_asic}
\paragraph[Technology and design approach]{ Technology and design approach}

The ASIC is designed in full-custom using CADENCE CAD tools. % (Version: 5.10.41\_USR5.90.69).}
The used technology is a standard $0.35\mu$  BiCMOS SiGe from Austria MicroSystem (AMS).  The access to this technology was made possible through the services of the
{\textquotedbl}Circuits Multi Projects{\textquotedbl} (CMP) of Grenoble. 
This technology consists of p-substrate, 4-metal and 3.3~V process. It includes standards complementary MOS transistors
and high speed vertical SiGe NPN Heterojunction Bipolar Transistors (HBT). Bipolar transistors are preferentially used
for the design of analog parts because of their good performances at cryogenic temperature. Due to kink effect in MOS
transistors resulting from carrier freeze-out phenomenon in semiconductors below 30~K, the use of these transistors is
preferentially reserved for the design of digital blocks and limited to PMOS current mirrors almost exclusively for
analog parts. The design of the ASIC is based on pre-experimental characterizations results, and its performance at cryogenic temperature
is extrapolated from simulation results obtained at room temperature, using CAD tools.

\begin{figure}
\centering
 \includegraphics[width=15.995cm,height=6.6cm]{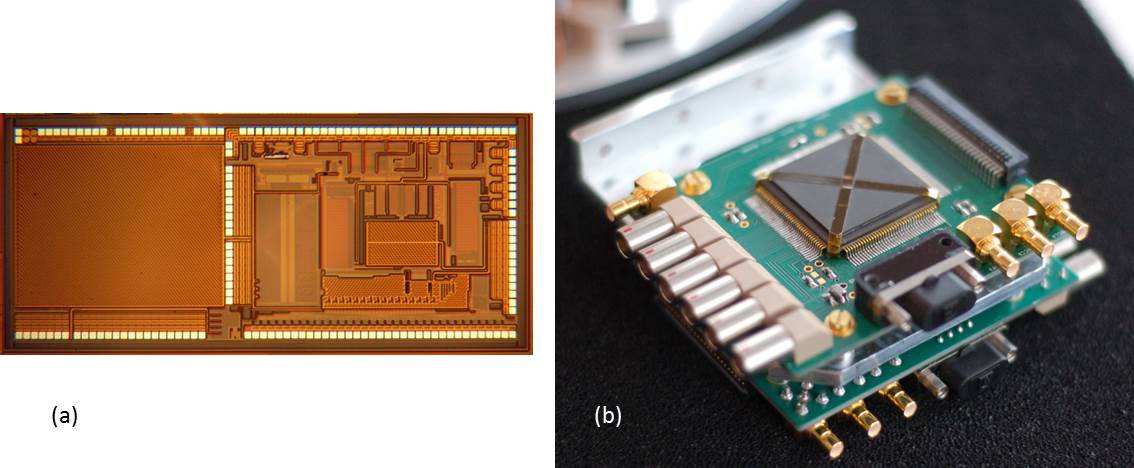} 

\caption[Microphotography of cryogenic ASIC
designed to readout 4x32 TES/SQUID pixels]{ (a) Microphotography of cryogenic ASIC
designed to readout 4x32 TES/SQUID pixels;(b) ASIC module assembly used for QUBIC experiment: chips are assembled in
CQFP144 packages, soldered on standard 6 layers FR-4 Printed Circuit Board (PCB) with the needed differential LEMO,
SMB, Micro Sub-D and high density SAMTEC connectors. Each module consists of 2 back to back of these PCB which is able
to readout 256 TES/SQUID pixels i.e. one quarter of a QUBIC telescope focal plan.\label{fig:cryoasic}}
\end{figure}

\paragraph[Implemented functions:]{ Implemented functions:}

Our Time Domain Multiplexer (TDM) readout is based on 4 columns of 32 SQUID in series associated to a cryogenic ASIC.
\begin{table}
\centering
\begin{tabular}{|l|l|}
\hline
Multiplexing factor	 & 128 :1 \\ 
Multiplexing frequency	& 100 kHz  \\
ASIC power dissipation @ 40K &	16 mW / ASIC \\
ASIC noise level& 	0.3 nV/$\sqrt{Hz}$ at 77 K\\
\hline
\end{tabular}

\caption{Main characteristics of the cold electronics \label{elecmain}}
\end{table}

\begin{figure}
\centering  \includegraphics[width=12.cm]{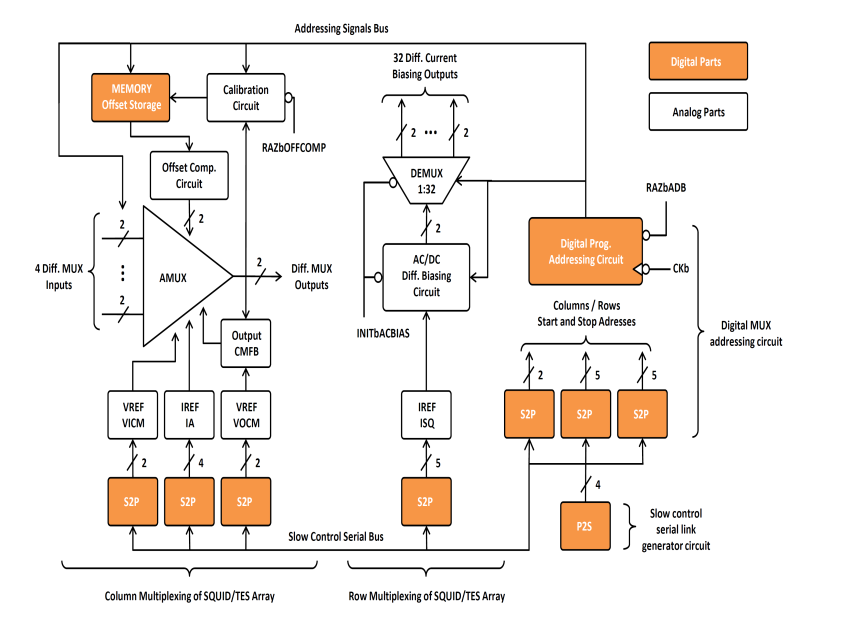}
\caption{Functions implemented in the ASIC.\label{fig:asicfunc}}
\end{figure}

SQUID boards are thermalized on the 1K stage whereas the ASIC are on the 40K stage.
This cryogenic ASIC, shown on Figure~\ref{fig:cryoasic}, integrates all parts needed to achieve the readout, the multiplexing and the control of an array up
to 128 TES/SQUID. Its functions are outlined on Figure~\ref{fig:asicfunc}. 
It
operates from room temperature down to 4.2K, thanks to a low power dissipation (16 mW per ASIC typically, whatever
the number of columns to readout). 
It includes a differential switching current source to address sequentially 32 lines of SQUID,
achieving a first level of multiplexing of 32:1. In this configuration, the SQUID are AC biased through capacitors which
allows satisfying both, good isolation (low crosstalk between SQUID columns) and no power dissipation. 
A cryogenic SiGe low noise amplifier (0.3 nV/$\sqrt{Hz}$), with 4 multiplexed inputs, performs a second multiplexing
stage between each column. 

This cryogenic ASIC includes also the digital synchronization circuit of the overall
multiplexing switching (AC current sources and multiplexed low noise amplifier). A serial protocol allows to focus on sub-array as well as to adjust the amplifiers and current sources with a reduced
number of control wires. 
This ASIC includes also an 8 bits analog-to-digital converter, register (memory) and digital-to-analog converter to
measure and store the offsets during the slow control, and dynamically compensate offset during observation time. 
As the digital side takes a large part, we have developed a full custom CMOS digital library dedicated to cryogenic
application and ionizing environments (rad-hard full custom digital library). 
%This digital library includes inverter,NAND, transmission gate, latch, counter and register.
 The main strategy consists to enclose each MOS transistors,
designed in edge-less transistors shape, by guard rings.

\paragraph[Characterization tests]{ Characterization tests}

\begin{figure}
\centering  \includegraphics[width=.7\textwidth]{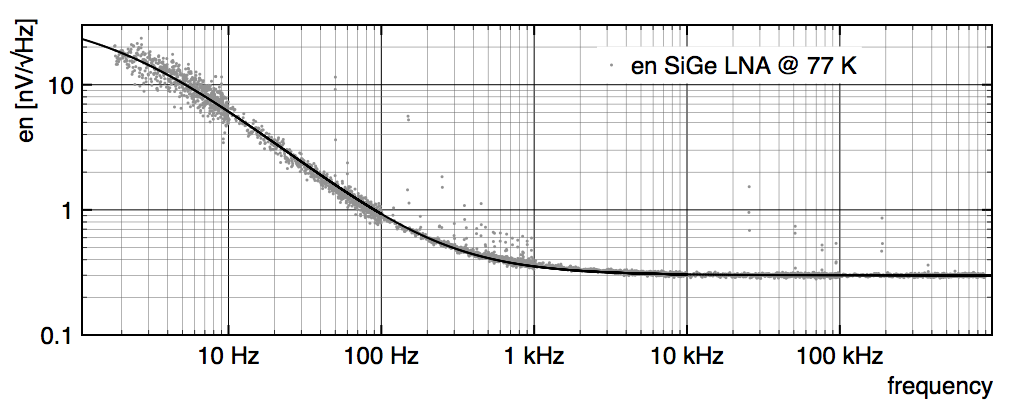}
\caption{Multiplexed LNA (low noise amplification) equivalent input
noise voltage measurement at 77K.\label{fig:asiclna}}
\end{figure}

Low noise multiplexed amplifier characterizations have been investigated using a vector analyzer. A white
noise level of 0.3 nV/$\sqrt{Hz}$ with a differential voltage gain of 200 and a bandwidth of 6 MHz were measured at 77K, as shown on Figure~\ref{fig:asiclna}. 

\begin{figure}
\centering
\includegraphics[width=13.cm]{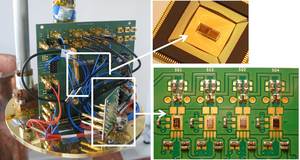} 
\caption{Experimental setup for the cryogenic ASIC
characterizations and the functional tests performed with 4 ``StarCryo'' SQUID chips bonded on a PCB and SMD addressing
capacitors.\label{fig:asicsetup}}
\end{figure}

The behavior of integrated DC biasing sources has been also investigated down to 4.2K, with the setup shown on Figure~\ref{fig:asicsetup}. Each of them is operational at
cryogenic temperature with expected values except for the source involved into the AC SQUID biasing operation. To
overcome this issue, the AC SQUID biasing circuit will be referenced to an external current source.

\begin{figure}
 \includegraphics[width=16cm]{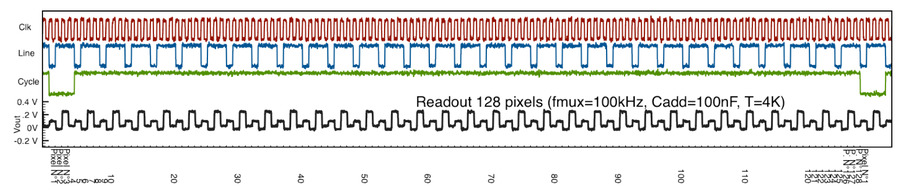} 
\caption[Preliminary clocking validation at 4.2K of the multiplexer]{Preliminary clocking validation at 4.2K
of the multiplexer: Clk (clock); Line (synchronize the SQUID switching current source to the multiplexed LNA); Cycle
(give the start - pixel 1 - of the full multiplexing cycle); Vout is the multiplexed signal of 128 pixels (SQUID stage
replaced by 128 resistors biased through capacitors in accordance to the bias reversal and alternatively
positive/negative amplified by the 4 multiplexed inputs LNA. 4 different offsets are noticeable).\label{fig:asicclock}
}\end{figure}

\begin{figure}\centering
 \includegraphics[width=14.cm]{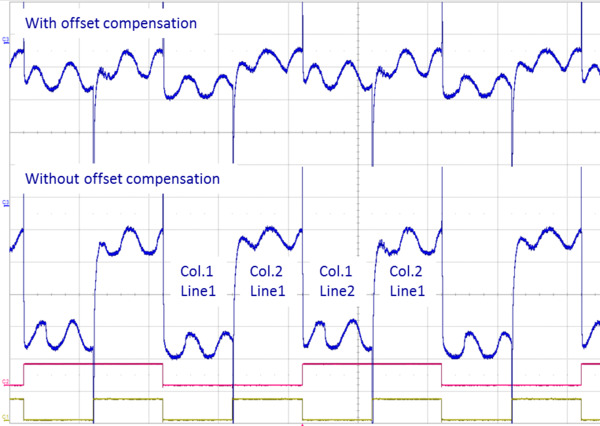} 
\caption[Validation at 4.2K of the AC SQUID
biasing operation through addressing capacitors (100nF) and dynamic offset compensation principle]{Validation at 4.2K of the AC SQUID
biasing operation through addressing capacitors (100nF) and dynamic offset compensation principle. The tests are
performed on an array of 2 columns of 2 SQUIDs in series associated to the cryogenic ASIC which includes the needed
switched current sources, multiplexed low noise amplifier and a digital sequencing circuit referenced to an external
clock signal. The clock frequency is here set to 2 kHz. Signals 1and 2 are synchronization signals of the SQUID
switching current source and the multiplexed LNA respectively. Signal 3 is the measured multiplexed output signal, with
and without dynamic offset compensation, corresponding to periodic sine like SQUID characteristics of each pixel
obtained by applying a large ramp signal into their feedback coil.\label{fig:asicbias}
}\end{figure}

\begin{figure}\centering
 \includegraphics[width=15cm]{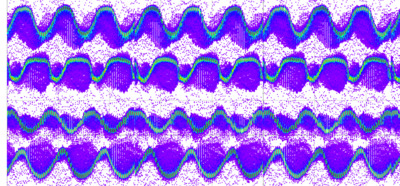} 
\caption[Output signal measured at 100
kHz of multiplexing clock frequency]{Output signal measured at 100
kHz of multiplexing clock frequency. Each periodic sine signal corresponds to sampled SQUID characteristics of each
pixel obtained by applying a large ramp signal with low frequency (20Hz) into the SQUID feedback coils.\label{fig:asicout}
}\end{figure}

Functional tests have been also performed on a small array of 2 columns of 2 SQUID in series which consists in 4
``StarCryo'' SQUID chips bonded on a Printed Circuit Board (PCB) with Surface Mount Device (SMD) addressing capacitors
associated to our cryogenic ASIC for the readout and the multiplexing. These tests have validated the AC SQUID biasing
operation, the dynamic offset compensation principle and the overall multiplexing topology (switching AC current
sources, multiplexed LNA and digital clocking) as shown on Figures~\ref{fig:asicclock}, \ref{fig:asicbias} and \ref{fig:asicout}.

\subsubsection{{Warm electronics}}

The room temperature (RT) readout electronics is designed to control and adjust the operating
biasing and feedback to TESs and their associated SQUIDs. Furthermore, it readouts the signal from the cold
multiplexing ASIC, computes the scientific signal and sends it compressed to the data acquisition system. Finally the
RT electronic readouts the thermometers needed to monitor the cold stages of the instrument. This electronic makes
ample use of the FPGA (programmable logic circuits) listed below.

Each board is associated to the cold electronics (ASIC) to manage 128 pixels, so that an
ensemble of 16 boards covers the full focal plane of 2048 pixels.

\begin{figure}
\centering \includegraphics[width=13.cm]{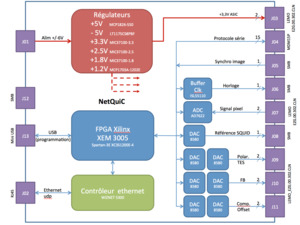} 
\caption{The NetQuiC board architecture \label{fig:NetQuiC}}
\end{figure}

This  board, called NetQuiC, and shown on Figure~\ref{fig:NetQuiC}, is built around a XEM3005 board from Opal Kelly that includes a Xilinx Spartan 3E FPGA. This FPGA programmed in VHDL    embeds: 

\begin{itemize}
\item {
Asics control}
\item {
Management of the TCP/IP connection with the PC. }
\item {
Acquisition of scientific signal with the ADC }
\item {
Bias generation }
\item {
Digital Flux Locked Loop (FLL) control. }
\end{itemize}

\begin{figure}
\centering
 \includegraphics[width=16.cm]{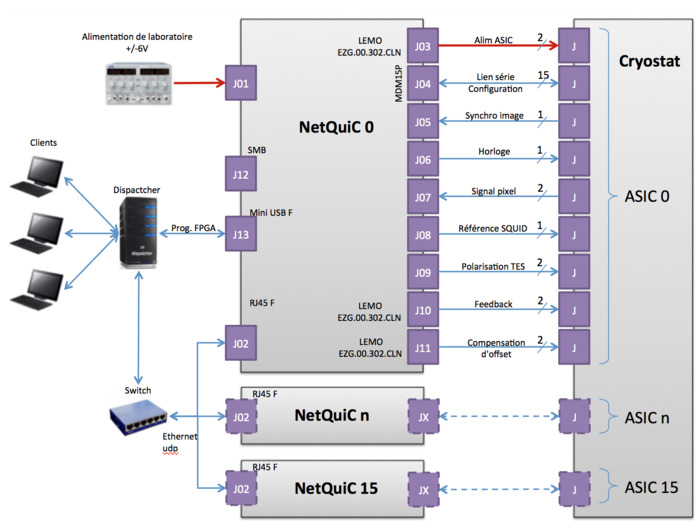} 
\caption{Warm electronic overall architecture\label{fig:warmelec}}\end{figure}

Figure~\ref{fig:warmelec} shows the architecture of the QUBIC experiment warm electronics. It includes 16 NetQuiC boards, one
for each ASIC. The boards are connected to a PC via a network switch. The PC is in charge of the data storage. 

\subsubsection{QUBIC Studio, readout and control software }
\label{qubicstudio}

We have designed a single interface to deal with the readout, the control command software and the data storage (cf. section~\ref{datastore}): the
QUBIC Studio. We made the choice to use the generic EGSE tool, called ``Dispatcher'', and developed at IRAP. This
real-time-oriented generic tool is widely used on various experiments such as Solar Orbiter, SVOM/ECLAIRS, PILOT. It
includes the tools described in the following sections.

This software includes a user-friendly interface to manage the connection with the readout electronics, the
management of the control command and Housekeeping data :
\begin{itemize}
\item TES matrix thermometers 
\item blind TESs
\item cryostat compressor, ``He tubes'' and ``cold heads'' thermometers
\item calibration source control parameters 
\item mount's motor control parameters 
\end{itemize}
 and the visualization of the scientific data. The QUBIC Studio
also includes an internal scripting capability allowing us to build simple transfer functions acting on scientific or
HK data, but also to develop calibration sequences associating control commands and acquisition functions. 
Users can connect to the QUBIC Studio as multiple clients to access HK/raw/scientific data or to set control-commands of
different subsystems of the QUBIC instrument. 
The QUBIC Studio software also provides an efficient session-history capability, so that it is possible to store the
global settings of the instrument at a given date, and re-launched directly from this setup. It is also possible to
rerun any data sequence already observed and stored.

\paragraph{Architecture visualisation tool}

\begin{figure}
\centering  \includegraphics[width=7.692cm,height=9.694cm]{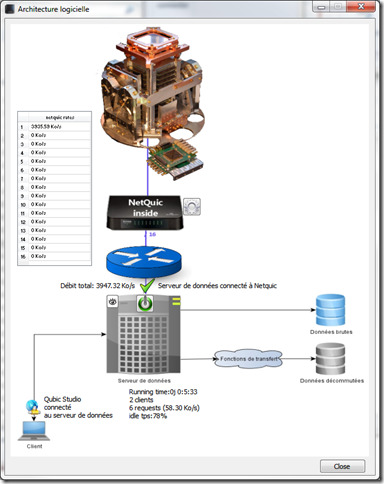}
\caption{View of architecture visualization tool \label{fig:archvisu}}\end{figure}

It allows to monitor data rates in the parts of the acquisition system, and check the connexion between the various
subsystems: the readout electronics, the switches between boards, the user-clients, and the data storage disk. A typical screen capture is shown on Figure~\ref{fig:archvisu}.

\paragraph{The focal plan visualization tool}

\begin{figure}
\centering  \includegraphics[width=12.cm]{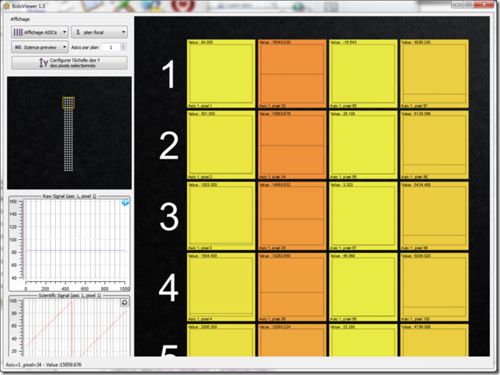}
\includegraphics[width=13cm]{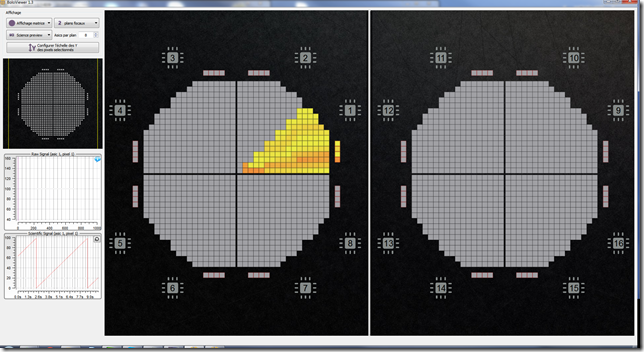}
\caption{View of focal plane visualization tool \label{fig:focplanvisu}}\end{figure}

This tool allows a global visualisation of the ASICs and associated scientific signals. It can display the
scientific signal, i.e. integrated over the raw samples for each detector, the noise level of the scientific signal, 
or the raw signal of the current detectors. This can be used to detect bad pixels or check the sanity of the full
readout chain very easily. The display can switch between an ASIC-centered view or a full focal plane overview, as
shown on Figure~\ref{fig:focplanvisu}.

\paragraph{Interface of the control command.}

\begin{figure}
\centering
 \includegraphics[width=15.148cm,height=7.458cm]{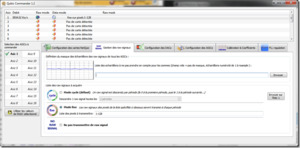} 
\caption{View of the interface of the control command\label{fig:controlinterf}}\end{figure}

The control command interface, shown on Figure~\ref{fig:controlinterf} allows us to initialize the following subsystems: 
ASICs,  FLL (Flux Locked Loop) regulation, NetQuiC boards, DACs,
raw Signal format, calibration and coefficients setting, horns switches,
calibration facilities

Concerning the ASICs, we can set and control the polarization biases of the TES and the SQUIDS, as well as the digital
FLL regulation parameters. These parameters may be automatically optimized by an internal script launched by the user. 
The scientific signal is processed in real time in the FPGA boards and sent to the QUBIC-Studio acquisition system.
Starting from the multiplexed signal coming from the ASICs, the scientific signal is de-multiplexed in the FPGA and
defined for each pixel as the sum of the raw signal over NSample, taking into account the rejection of data samples
defined by a mask, which can be tuned by the user. 

Because the raw signals represent a large amount of data (since they are defined at high sampling rate), and since they are not always needed,  they are not always  transmitted to  the QUBIC Studio acquisition system. However these raw
signals can be useful to investigate the sanity of given pixels or of the multiplexing chain, or also to test the
e.m. crosstalk between pixels. Hence we have defined three modes: 
\begin{itemize}
\item No Raw signal: nothing is returned except the scientific signal for the 128 pixels

\item Cycle  Mode: One raw signal is returned for each packet of 128 scientific pixel data, cycling over a list of
pixels provided by the user

\item Fixed Mode: The raw signals of a fixed list of pixels are returned for each packet of 128 scientific pixel data.
This mode can be used to return the full raw signals by selecting all 128 pixels.

\end{itemize}

The software includes much more features like:
raw data analyzer, data storage, logbook, scripting capabilities,
HK visualization (...).

\subsubsection{Data storage}
\label{datastore}

Concerning the data storage, the QUBIC Studio software presents two options. The first one consists in storing the
binary data as received from the FPGA boards. These data are compressed and put less constrains on the disk space
capacity. The second option consists in storing the data after interpretation \ by the software. These data are
directly usable for the users, and already formatted to be looked at and analyzed. However, this format will require
more disk space, which can be critical depending on the choice of the instrument setup. Indeed, the choice of the
NSample parameter leads the total volume of the data, as shown in the tables~\ref{table16} and ~\ref{table17}. The smaller this parameter NSample,
the larger the amount of data. Depending on the site of the instrument, this could be more or less critical. This is
the reason why the NSample parameter is not fixed yet in the current setup of the instrument. It can be set in a
reasonable range [10,1000] through the QUBIC Studio interface.

Data rate and storage of the scientific and raw data are respectively summarized in Table~\ref{table16} and ~\ref{table17}.

\begin{table}
\begin{tabular}{|m{1.726cm}|m{1.153cm}|m{1.2249999cm}|m{1.051cm}|m{1.111cm}|m{1.1409999cm}|m{1.134cm}|m{1.0949999cm}|m{1.0309999cm}|m{1.0819999cm}|m{1.416cm}|m{1.788cm}|}
\hline
\multicolumn{5}{|m{7.066cm}|}{~
} &
\multicolumn{2}{m{2.475cm}|}{{ Data Rate}} &
\multicolumn{5}{m{7.2120004cm}|}{{ Data Storage}}\\\hline
{ NSample} &
{ fpack} &
{ Tpack} &
{ facq} &
{ Tacq} &
{ 1 ASIC} &
{ FFP} &
{ 1min} &
{ 1h} &
{ 1day} &
{ 1month} &
{ 12months}\\\hline
~
 &
{ Hz} &
{ ms} &
{ kHz} &
{ us} &
{ ko/s} &
{ Mo/s} &
{ Mo} &
{ Go} &
{ To} &
{ To} &
{ To}\\\hline
{ 10} &
{ 1560} &
{ 0.64} &
{ 200} &
{ 5} &
{ 782} &
{ 12.2} &
{ 732} &
{ 42.9} &
{ 1000} &
{ 31} &
{ 374}\\\hline
{ 100} &
{ 156} &
{ 6.4} &
{ 20} &
{ 50} &
{ 78.2} &
{ 1.2} &
{ 73} &
{ 4.3} &
{ 100} &
{ 3.1} &
{ 37.4}\\\hline
{ 1000} &
{ 15.6} &
{ 64} &
{ 2} &
{ 500} &
{ 8} &
{ 0.122} &
{ 7.3} &
{ 0.43} &
{ 10} &
{ 0.31} &
{ 3.7}\\\hline
\end{tabular}
\caption{data rate and storage of scientific data.\label{table16}}

\end{table}

\begin{table}
\begin{tabular}{|m{2.0609999cm}|m{1.0489999cm}|m{1.77cm}|m{1.5869999cm}|m{1.64cm}|m{1.5799999cm}|m{1.5799999cm}|m{1.645cm}|m{1.686cm}|}
\hline
\multicolumn{2}{|m{3.3099997cm}|}{~
} &
\multicolumn{2}{m{3.557cm}|}{{ Data Rate}} &
\multicolumn{5}{m{8.931cm}|}{{ Data Storage}}\\\hline
{ Raw Mode} &
{ Nraw} &
{ 1 ASIC} &
{ FFP} &
{ 1min} &
{ 1h} &
{ 1 day} &
{ 1 month} &
{ 12 months}\\\hline
{ Fix / Cycle} &
{ 1} &
{ 32 ko/s} &
{ 492 ko/s} &
{ 30 Mo} &
{ 1.7 Go} &
{ 42 Go} &
{ 1.3 To} &
{ 15.1 To}\\\hline
{ Fix / Cycle} &
{ 5} &
{ 160 ko/s} &
{ 2.5 Mo/s} &
{ 150 Mo} &
{ 9 Go} &
{ 207 Go} &
{ 6.3 To} &
{ 75.4 To}\\\hline
{ Full / Fixed} &
{ 128} &
{ 4 Mo/s} &
{ 61 Mo/s} &
{ 3.7 Go} &
{ 215 Go} &
{ 5.2 To} &
{ 156 To} &
{ 1.9 Po}\\\hline
\end{tabular}
\caption{data rate and storage of raw data.\label{table17}}

\end{table}

\subsubsection{Detection chain: validation of a quarter of focal plane and readout system}
\label{subsec:quat_FP_test}

\paragraph{Description}

The tests that aim to validate QUBIC's focal plane are all performed on 256 TES sub-arrays at APC. 
The sub-array is placed on the coldest stage of a pulse-tube dilution cryostat that can cool the sample down to 100 mK. 
The next link in the readout chain is the SQUID. Each TES is inductively coupled to a SQUID, totalizing 256 SQUIDs cooled down to 
1 K. Then, a SiGe ASIC at 40 K is used to control and amplify the signal from 128 SQUIDs in a TDM scheme. Two of them are thus used to characterize a quarter of focal plane.

\paragraph{Principle on a TES}

\begin{figure}
\begin{tabular}{cc}
\includegraphics[width=.4\textwidth]{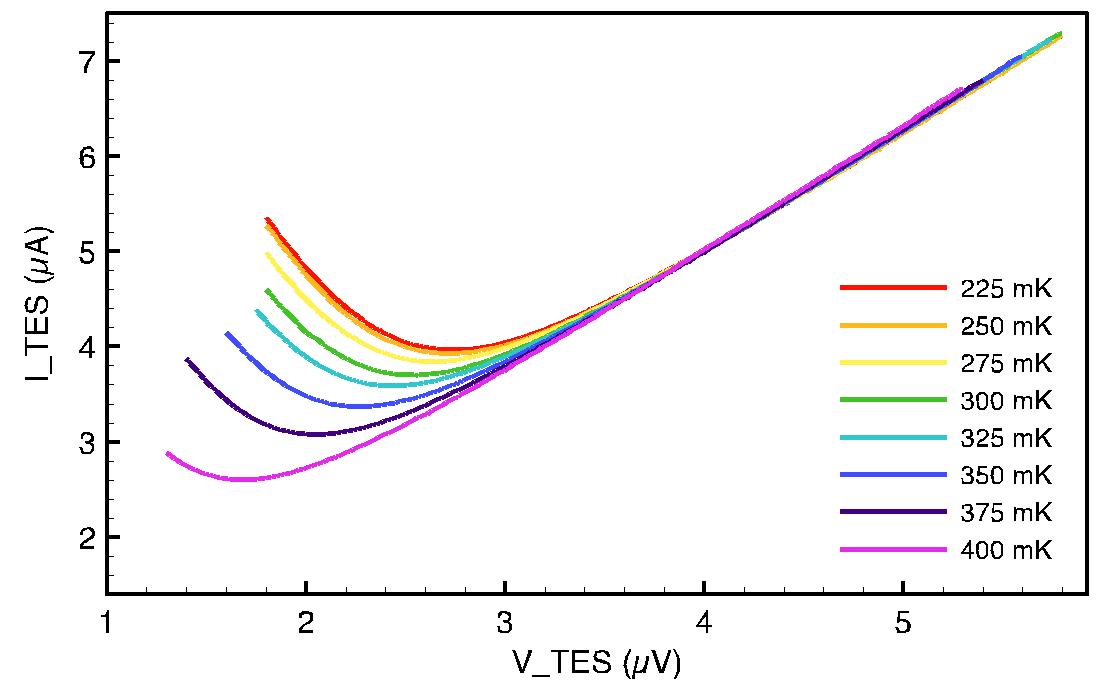}
&
\includegraphics[width=.4\textwidth]{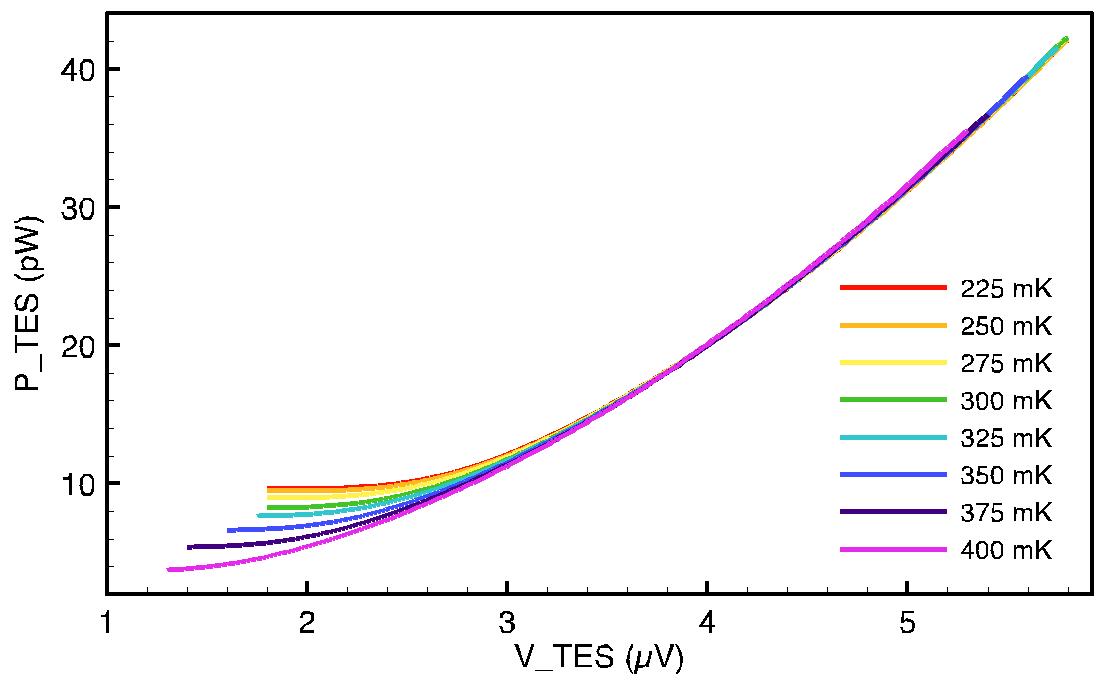}
\end{tabular}
\caption[I-V  and P-V  curves of a TES (n$^o$60, ASIC2, P41 array) at different Tbath < Tc]{I-V (left) and P-V (right) curves of a TES (n$^o$60, ASIC2, P41 array) at different Tbath < Tc. \label{fig:ivpvtest}}
\end{figure}

Being voltage-biased, the TESs work at a stable and controlled temperature thanks to a strong electrothermal feedback: any fluctuation of power of thermal origin is compensated by an opposite contribution of power by electrical dissipation. The total power across the TES is therefore quasi-constant. A way to ascertain it is to proceed to I-V and corresponding $P-V$ measurements. Figure \ref{fig:ivpvtest} displays these curves for a single pixel at different bath temperatures below its critical temperature Tc. 

\begin{figure}
\begin{tabular}{cc}
\includegraphics[width=.75\textwidth]{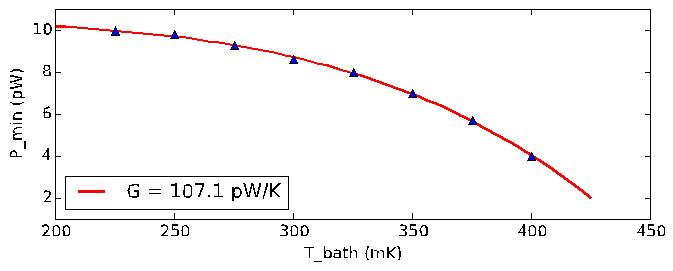}
\end{tabular}
\caption{Fitted measurements of plateau power at different regulation temperatures (TES n$^o$60) \label{fig:plateau_fit_tes_rd}}
\end{figure}

When regulated at a temperature lower than its Tc, a TES can be forced into its normal state by maintaining sufficiently high bias voltage (right part of the curves). While decreasing the voltage (from right to left), the TES is first in its normal state and shows a resistive metallic behaviour that follows Ohm's law, thus the I-V curve is firstly a straight line. Then the TES tends to transit to its superconducting state and the electrothermal feedback starts to take place. This is when the I-V curve reaches its minimum. Once the feedback is operational, the TES is auto-regulated and works at quasi-constant power (Pmin), which can be witnessed on the P-V plateaux. Meanwhile, the TES goes further on its transition and its resistance continues to drop, which leads to a portion of parabola on the left part of the I-V curve. One can also notice that the cooler the TES is regulated, the further it is from its transition and the higher the needed power has to be to bring the TES to its normal state (Pmin at 225 mK > Pmin at 400 mK).
The $I-V$ and $P-V$ curves of a bolometer provide a way to run through several of its stable states and therefore to recover some of its static parameters, such as its thermal conductance G and an evaluation of its theoretical NEP. To do so, the plateau power Pmin of the P-V curve is measured at different bath temperatures. These data are then fitted after $Pmin = K(T_0^{n+1} - T_{bath}^{n+1})$ as seen in Figure~\ref{fig:plateau_fit_tes_rd}.

The fitting returns the K, n and $T_0$ quantities, assuming that $T_0$ is the critical temperature $T_c$, and $G(T_0)$ is calculated from them. It is then simple to deduce as an approximation from the G term the theoretical NEP of the bolometer given that the phonon noise is prevailing, using 
$NEP_GÂ =Â (\gamma 4k_B T_0^2G)^{1/2}$.

\paragraph{Testing of an array}

\begin{figure}
\begin{center}
\includegraphics[width=.75\textwidth]{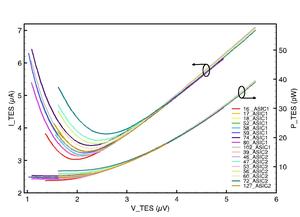}
\end{center}
\caption[I-V and P-V  curves of TESs on P41 array at $T_{bath} = 350 mK$]{I-V (upper curves, left axis) and P-V (lower curves, right axis) curves of TESs on P41 array at $T_{bath} = 350 mK$\label{fig:iv_and_pv_test}}
\end{figure}

The same analysis has been performed on about twenty randomly distributed TESs on a 256-pixel array. Figure \ref{fig:iv_and_pv_test} is a graph summing up the I-V (upper curves, left axis) and P-V (lower curves, right axis) measurements of these pixels. For clarity purpose, the result for only one regulation temperature ($T_{bath} = 350 mK$) is displayed, combining two experiments (one with ASIC1 addressing the first half of the 256 pixels and one with ASIC2 for the second half).

\begin{figure}
\begin{center}
\includegraphics[width=.75\textwidth]{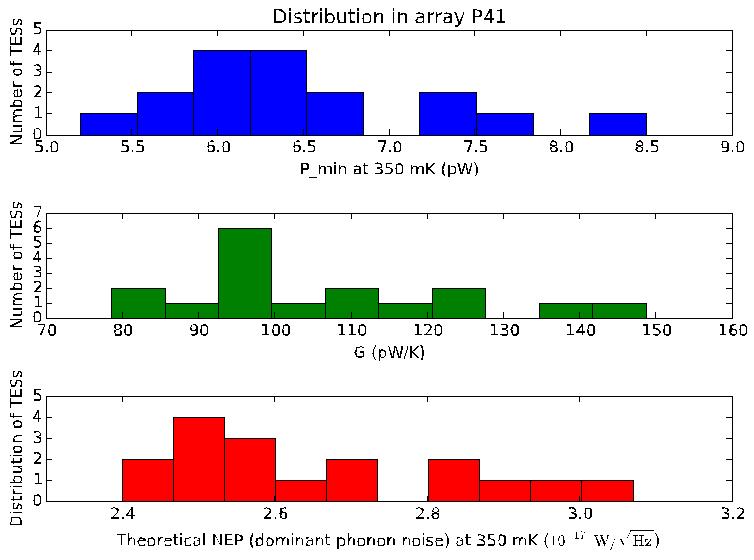}
\end{center}
\caption{Histograms of $P_{min}$, G and approximation of theoretical NEP of selected pixels of a 256 TES array obtained from I-V measurements at different bath temperatures. \label{iv_and_pv_tst_histo}}

\end{figure}

The picture shows first that the TESs are working excellently (expected shape of the I-V and P-V curves) and that the strong electrothermal feedback is efficient: all the pixels reach a constant minimum power when they are on their transition to their superconducting state (left part of the P-V curves). In terms of homogeneity, all the pixels exhibit the same behaviour on their normal state.
However, with the voltage decreasing and the TESs entering their transition, disparities can be noticed on the I-V curves showing that some pixels are getting superconducting while others are still at the beginning of their transition. But considering the P-V curves (that are a different manner to present the same data as the I-V curves), the disparities are less evident.
In order to have a comparison criterion, the Pmin of all the tested pixels has been collected at $T_{bath} = 350 mK$ and gathered in the histogram of Figure \ref{iv_and_pv_tst_histo}. The figure shows that most of the pixels (75\%) have a Pmin around $6.2 pW \pm 10\%$, which is a very good homogeneity. The small difference among the TESs may come from variations of the thickness of the suspending legs caused by anisotropies of etching during the manufacturing process. This would lead to disparities in the thermal conductivity G. As in Figure \ref{fig:plateau_fit_tes_rd}, G is calculated for each pixel from the fitting of Pmin at different bath temperatures and gathered in a histogram with an average value of 106 pW/K. A first approximation of the NEP distribution, deduced from the values of G is also given with an average of $2.6 \times 10^{-17} W/\sqrt{Hz}$ at 350 mK. The value largely meets the QUBIC requirements of an electrical 
$NEP < 5 \times 10^{-17} W/\sqrt{Hz}$, even for the few pixels whose NEP is a little scattered from the main distribution. 

In order to be consistent with the saturation power for the 220GHz channel that is multimoded, the critical temperature will be increased for these TESs. Since the photon noise is higher in this channel, there will be no performance degradation.
%\todocomment{ESB: Are we consistent, in terms of saturation power, with the multimoded solution on the 220GHz. In the chorrillo case I am afraid not}

%auto-ignore
\subsection{Mount System and Baffling}

\subsubsection{Mount system }

The QUBIC mount is a standard Alt-azimuthal astronomical mount. This will be able to support
1000kg and the total weight is expected to be 2000 kg with 2m x 2m x 2m approximate size. The mount will implement 3
rotational axis (Altitude, Azimuth and boresight) and is being designed under the responsibility of the Duch consortium
NIKHEF, Leiden University, and TNO with collaborators in LAL/IN2P3 Orsay who could also be interested by contributing.

\begin{figure}
\centering
\includegraphics[width=6.cm]{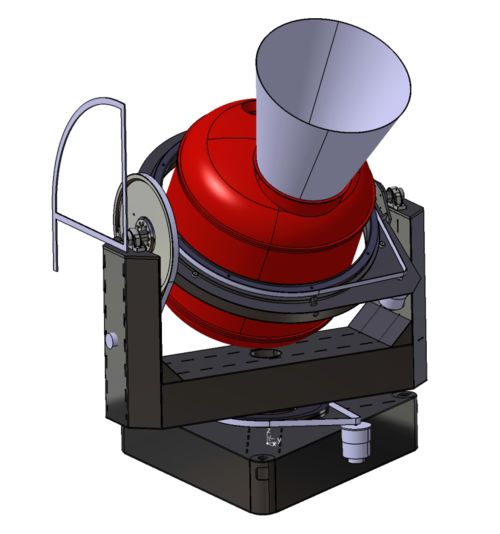} 
\caption[Mount]{Mount: preliminary design of the QUBIC experiment on its mount
(Acknowledgement NIKHEF/TNO).\label{fig72}}
\end{figure}

The mount, supporting the cryostat, will be installed on top of a  platform  (see Figure~\ref{fig72}).

\subsubsection{External Baffling}
\paragraph{Description}
The radiation shielding solution adopted for QUBIC instrument is composed by a Forebaffle (FB)
and a Ground Shield (GS), see section \ref{bkm:Ref314316307} about their analysis and configurations. Both the shields
will be manufactured by a selected external supplier. The final dimensions of FB are dictated by the cryostat window
while for GS by the mount design and the expected instrument platform.

FB will be realised in a single conical structure with a dismountable flare, both manufactured
with an aluminium alloy. The inner surface is covered with a 10 mm thick
Eccosorb\footnote{\ http://www.eccosorb.com} dielectric sheet, a material with high lossy
absorptive properties suitable for the microwaves. 

\begin{figure}
\centering
 \includegraphics[width=15.09cm,height=8.105cm]{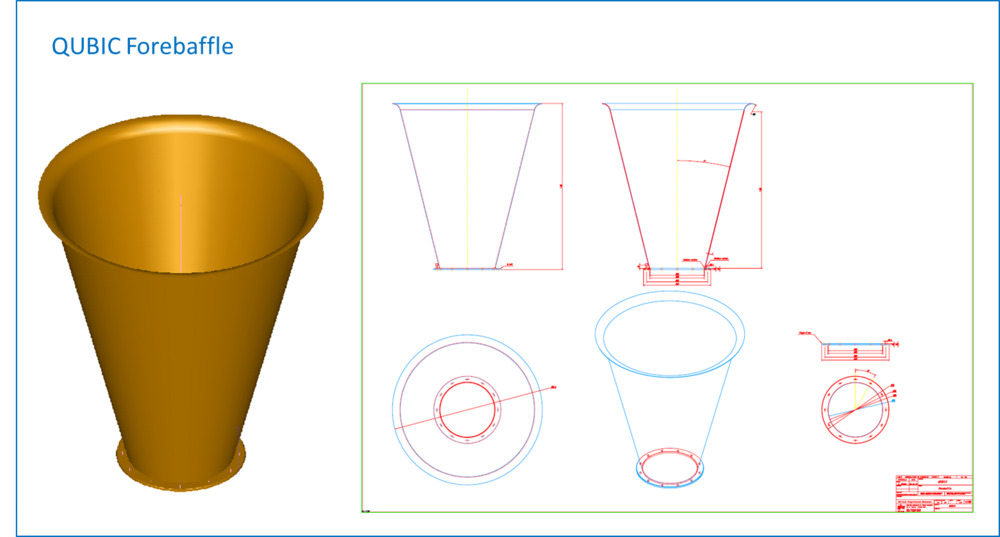} 
 \includegraphics[width=15.09cm,height=10.781cm]{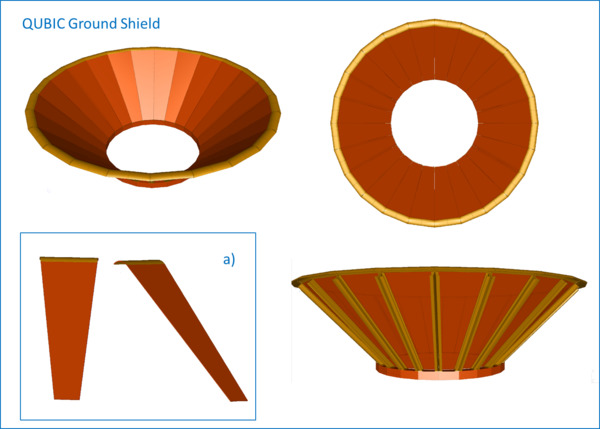} 
\caption[QUBIC Forebaffle and Ground Shield  mechanical drawings]{QUBIC Forebaffle
(top) and Ground Shield (bottom) mechanical drawings.\label{fig95}}
\end{figure}

GS will be manufactured with several flat panels, petals, to fit a conical shape (see panel a)
in Figure~\ref{fig95}. The inner surface of each petal is a sheet of aluminium ensuring low emissivity and high reflectivity to
point the beam spillovers to the cold sky instead of to the ground. The edge of the GS has a flare for the same reasons
of FB.

Preliminary drawings of both shields are shown in Figure~\ref{fig95}.
There are no strict requirements on the two shields from an optical point of view while a certain rigidity must be
ensured to meet the required geometry.

Similar solutions have been already adopted for FB and GS at BICEP intrument with %analogous
severe ambient conditions near the  South Pole Station\footnote{The cylindrical FB and the GS of BICEP3 are shown 
in  \url{http://bicep.caltech.edu/~yuki/shield/}}.
%{\notes would a reference only be enough ? strange
%to take picture from other experiments in a TDR}

%\begin{figure}
%\centering
% \includegraphics[width=6.897cm,height=4.897cm]{QUBICTDRcompilation-img167}
%\includegraphics[width=8.678cm,height=4.875cm]{QUBICTDRcompilation-img168.jpg} 
%\hypertarget{RefHeadingToc314323127}{}\caption{BICEP3 forebaffle
%and ground shield operating at South Pole
%Station{\ \url{http://bicep.caltech.edu/~yuki/shield/}\label{fig96}}}
%\end{figure}

\begin{figure}
\centering
\includegraphics[width=7.994cm,height=5.38cm]{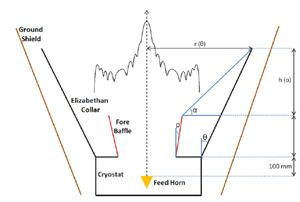}
\includegraphics[width=6.953cm,height=4.826cm]{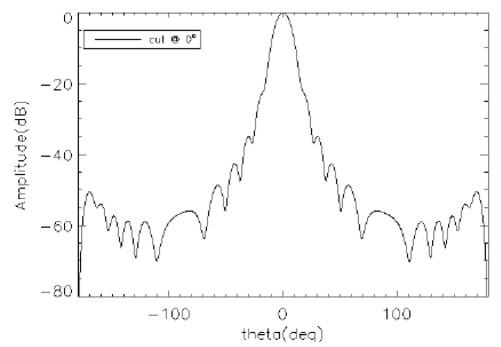} 
\hypertarget{RefHeadingToc314323059}{}
\caption[Sketch of the
QUBIC shielding configuration and cut of the hybrid conical horn beam pattern used in MultiGTD simulations]{Sketch of the
QUBIC shielding configuration (left) and cut of the hybrid conical horn beam pattern used in MultiGTD simulations
(right).\label{fig27}}
\end{figure}

\paragraph{{\incomplete Simulations}}
\label{bkm:Ref314316307}{
{In order to reduce the possible contamination derived by the presence of unwanted sources,
such as Sun, Moon and ground, a study of the shielding system for the first module of the QUBIC experiment has been
realized in terms of geometry and employed materials for the shields manufacturing~\cite{Buzi}. The study has
been performed with the commercial softwares GRASP and CHAMP combining MultiGTD (Geometric Theory of Diffraction) and
MoM (Method of Moments) approaches to infer the pattern of the instrument beam up to sidelobes at the lowest frequency,
150 GHz, where the impact is higher. In order to have a conservative approach in the estimation of the QUBIC instrument
spillovers, we have investigated the impact of the shielding configuration on the beam pattern central feed horn of the
feed-horn array. The shielding configuration is schematically shown on the left of Figure~\ref{fig27} while on the right a cut
of the beam pattern of a Hybrid Conical Horn employed in the MultiGTD analysis.}}

\begin{figure}
\centering
\includegraphics[width=7.795cm,height=5.595cm]{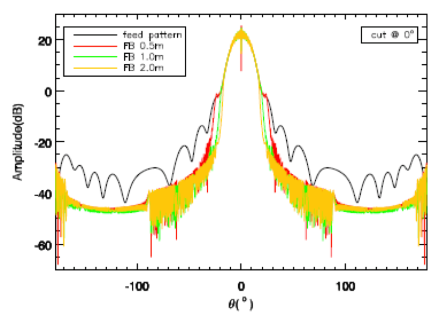} 
\includegraphics[width=7.754cm,height=5.581cm]{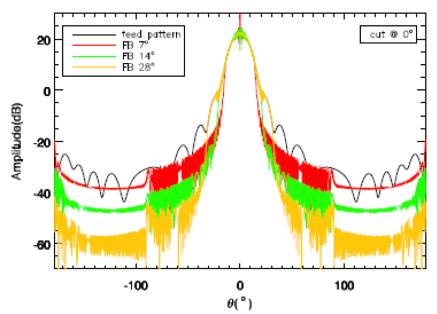} 
\hypertarget{RefHeadingToc314323060}{}\caption[Cuts of the
patterns (at 0 degrees) for the on-axis feedhorn varying FB height and aperture angle]{Cuts of the
patterns (at 0 degrees) for the on-axis feedhorn varying: FB height, for an aperture angle equals to 14 degrees (left
panel) and FB aperture angle, for the defined FB height (h=1m) (right panel). The colors code is specified in the
legend.\label{fig28}}
\end{figure}

{
{The Forebaffle geometry has been optimized to minimize sidelobes varying its height in the
range 0.5-2 meter and aperture angle from 7 up to 28 degrees. The different patterns are shown in Figure~\ref{fig28}.}}

{
FB heights larger than 1 meter and an aperture angle larger than 14 degrees seem to be no advantageous, even from a
mechanical manufacturing point of view.}

{
In addition to a reflective solution, we studied an absorbitive one to highlight the different impact on the pattern.
Also a cylindrical shape has been considered for comparison.}

{
The MultiGTD approach does not allow to analyze reflectors covered by dielectric materials with defined electrical
properties. To overcome this restriction, we have performed our simulations with the help of the commercial software
CHAMP, which allows to analyze rotationally symmetric scatterer using the Method of Moment (MoM) approach.}

{
{Regarding the absorptive solution, we have considered the possibility to cover the inner
surface of the forebaffle with a 10 mm thick Eccosorb}\footnote{\ http://www.eccosorb.com}{
dielectric sheet, a material with high lossy absorptive properties suitable for the microwaves. We assumed the
following electrical parameters: electric permittivity 3.54, magnetic permeability 1 and tangent loss 0.057. The impact
on the beam pattern of the central feedhorn with a conical or a cylindrical forebaffle, for the two different
investigated solutions, is reported in Figure~\ref{fig29}.

\begin{figure}
\centering  
\includegraphics[width=7.835cm,height=5.565cm]{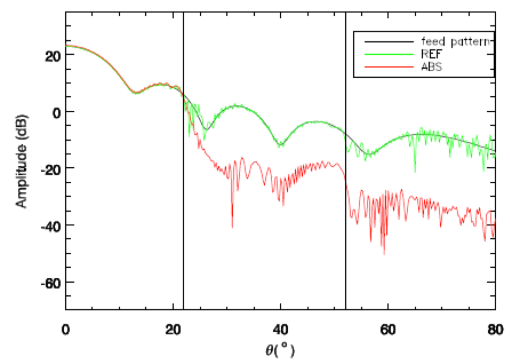} 
\includegraphics[width=7.775cm,height=5.544cm]{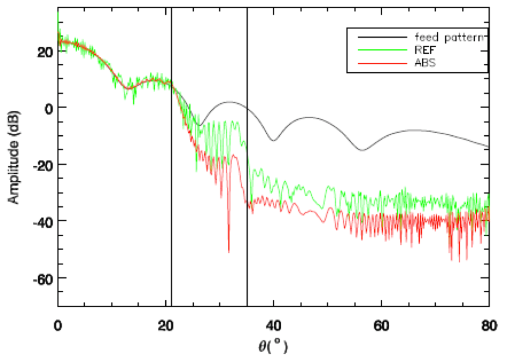} 
\hypertarget{RefHeadingToc314323061}{}\caption[Cuts of the beam pattern with and without a cylindrical shield and with and without a conical shield]{Cuts of
the beam pattern with and without a cylindrical shield (left panel) and with and without a conical shield (right
panel). The colors are referring to: Feed beam pattern (black line), reflective (green line) and absorptive (red line)
internal surface solutions.\label{fig29}}
\end{figure}

{
{For both configurations the presence of an absorptive inner surface leads to increase the
sidelobes rejection for angles larger than 20 degrees from boresight direction, respect to the nominal feed beam
pattern and the reflective solution. This effect results to be more evident for the conical-shaped shield, as shown in
the right panel of Figure~\ref{fig29}. }}

{
{Same analysis has been performed by adding a flared edge at the entrance aperture of the
forebaffle, hereafter Flare, for both configurations, with the aim of increasing sidelobes rejection. Three values for
the curvature radius, R = 25$\lambda $, 75$\lambda $ and 150$\lambda $ ($\lambda $=2 mm), have been analysed to study
the impact on the beam pattern.}}

{
{The insertion of a Flare in our configurations has allowed a further reduction of sidelobes at
angles larger than 30 degrees from boresight, as shown in Figure~\ref{fig30}
and Figure~\ref{fig31}.

\begin{figure}
\centering  \includegraphics[width=7.47cm,height=5.278cm]{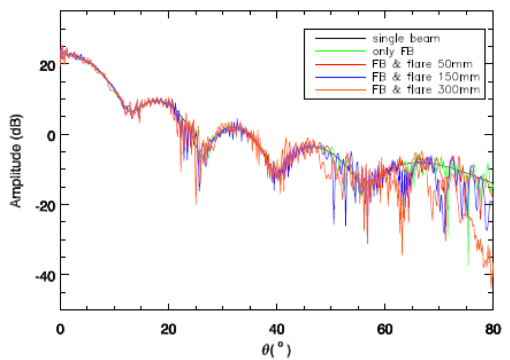} 
\includegraphics[width=7.343cm,height=5.302cm]{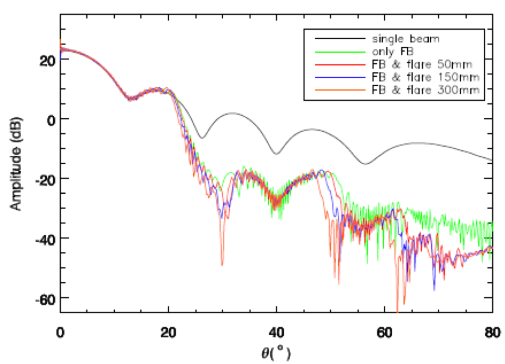} 
\hypertarget{RefHeadingToc314323062}{}\caption[Beam pattern cuts for the cylindrical forebaffle including the Flare]{Beam pattern cuts for the cylindrical forebaffle including the Flare: reflective inner surface (left panel) and
absorptive inner surface (right panel). All the colored curves are defined in the legend.
\label{fig30}}\end{figure}

\begin{figure}
\centering  \includegraphics[width=7.364cm,height=5.163cm]{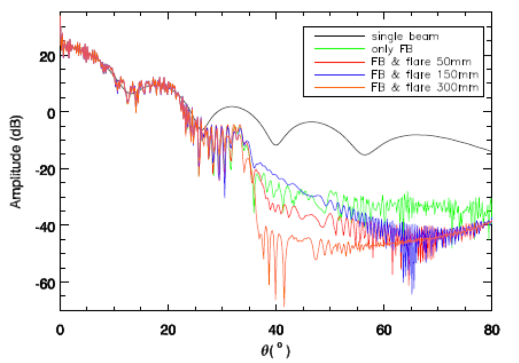} 
\includegraphics[width=7.119cm,height=5.249cm]{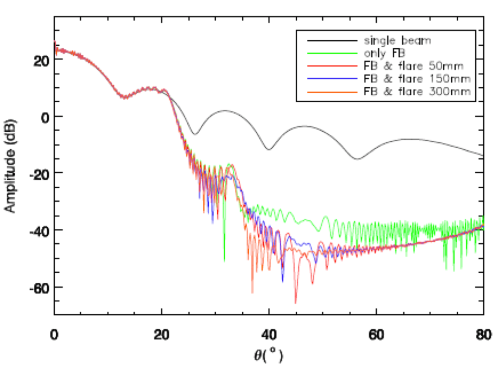} 
\hypertarget{RefHeadingToc314323063}{}\caption[Beam pattern cuts for the conical forebaffle including the Flare]{Beam pattern cuts for the conical forebaffle including the Flare: reflective inner surface (left panel) and
absorptive inner surface (right panel). All the colored curves are defined in the legend.
\label{fig31}}\end{figure}

Similarly to the configuration without flare, both absorptive solutions seem to show better performance in terms of
sidelobes amplitude. In the absorbing configuration, the flare's dimension seems to have a small impact on sidelobes
drop, this allows to choose a flare with the smallest curvature radius (R=50 mm), which implies a more simple
mechanical fabrication.

The additional radiative loading on a single detector, given by the emission of the Eccosorb
sheet in the inner surface of the forebaffle, has been evaluated, at 150 GHz, in the following way:

%{\centering  \includegraphics[width=12.561cm,height=1.191cm]{QUBICTDRcompilation-img057}
%\par}

\begin{equation}
P_{\mathrm{FB}}\ =\ A_h t_{tot} \epsilon_{opt} \frac{N_h}{N_{pixels}}\Delta \nu BB(\nu,T_{sheet})\int_{D\theta}\int_{D\varphi} AR(\theta,\phi)\sin{\theta}\cos{\theta}d\theta d\varphi
\end{equation}

%{\notes\bf check formula}

where $D\varphi $ and
$D\theta $ are the ranges of angles within
which the central feed horn subtends the FB.

The power collected by each pixel from the absorbing FB is $\approx$0.07 pW (blue line in Figure
~\ref{fig32}), lower than the expected saturation power of TES detectors, %(about 20 pW\todocomment{ESB: check consistency with previous section}), 
consequently it is not an issue in the
total power budgeted of the QUBIC experiment.

This value has been compared with the equivalent radiative loading coming from the FB in the reflective configuration.
For this case we have assumed that the loading reaching the focal plane is given by the emission of the atmosphere
reflected by the forebaffle inside the instrument.

{
{The numerical evaluation of the loading induced by the atmospheric emission
(}{\textbf{\textit{T}}}{\textbf{\textit{\textsubscript{atmo}}}}{=240
K) reflected on a single detector by the reflective FB, has been carried out in the following way:}}

\begin{equation}
P_{atmo}(z)\ = \ A_h t_{tot} \epsilon_{opt} \frac{N_h}{N_{pixels}}\Delta \nu BB_{atmo}(\nu,T_{atmo})\epsilon(z,pwv)\int_{D\theta}\int_{D\varphi} AR(\theta,\phi)\sin{\theta}\cos{\theta}d\theta d\varphi
\end{equation}

{
{where $BB_{atmo}(\nu,T_{atmo})\epsilon(z,pwv)$ is the assumed atmospheric brightness model where the zenith emissivity has been
modelled assuming the simple model of stratified layers:}}

%{\centering  \includegraphics[width=5.796cm,height=0.891cm]{QUBICTDRcompilation-img059}
\par}
\begin{equation}
\epsilon(z,pwv)\ =\ 1-e^{\left[-\tau_0(pwv,\Delta\nu)\sec{z}\right]}
\end{equation}

The zenith atmospheric opacity at 150 GHz, $\tau_0$ is derived from pwv, as described in~\cite{Gregori},
including also a finite spectral bandwidth. The pwv values exploited in our analysis are plotted in Figure~\ref{fig33}, as
expected at Dome C.

\begin{figure}
\centering  \includegraphics[width=7.811cm,height=5.697cm]{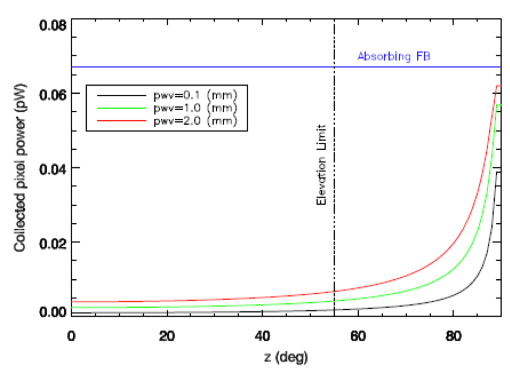}
\hypertarget{RefHeadingToc314323064}{}\caption[Power collected by a single pixel on the focal plane]{Power collected by a single pixel on the focal plane: the blue line is referring to the emission from an absorbitive
forebaffle while the other colored lines (see color code in the legend) are referring to the atmospheric emission (for
different pwv values) reflected by a reflective FB changing QUBIC zenithal angle. The expected maximum zenith angle is
55 degrees.
\label{fig32}}\end{figure}

\begin{figure}
\centering  \includegraphics[width=8.19cm,height=4.611cm]{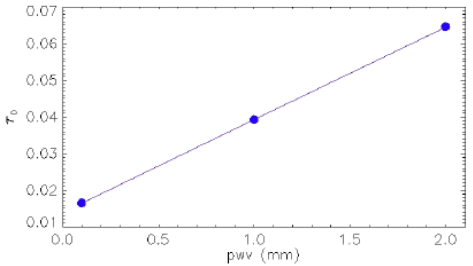} 
\hypertarget{RefHeadingToc314323065}{}\caption{Zenith atmospheric opacity at 150 GHz as a function of the pwv content, from 0.1 mm up to a maximum value of 2 mm. According to atmospheric models for Chajnantor, to get the opacities at 220 GHz, these values should be scaled by  $\sim 1.83$. \label{fig33}}\end{figure}

{
{The additional radiative loading collected by each pixel deriving by the atmospheric loading
reflected on the forebaffle, by varying QUBIC zenith angle and pwv values, is shown in Figure~\ref{fig32}.

The power collected by each pixel, by varying QUBIC elevation angle, defined as the sum
between the atmospheric emission and the FB emission in reflecting and absorbing configuration, is shown in Figure~\ref{fig34}%\todocomment{ESB: what about 220 GHZ?}.
This plot shows as the atmospheric emission provides the main contribution in terms of radiative loading on each
detector, about one order of magnitude, compared with the loading due to the emission of the FB in absorbing
configurations. 

For all these reasons, we selected the absorbing FB conical configuration with flare as the best solution in terms of
sidelobes rejection of the feed beam pattern.
\begin{figure}
\centering  \includegraphics[width=8.8cm,height=6.285cm]{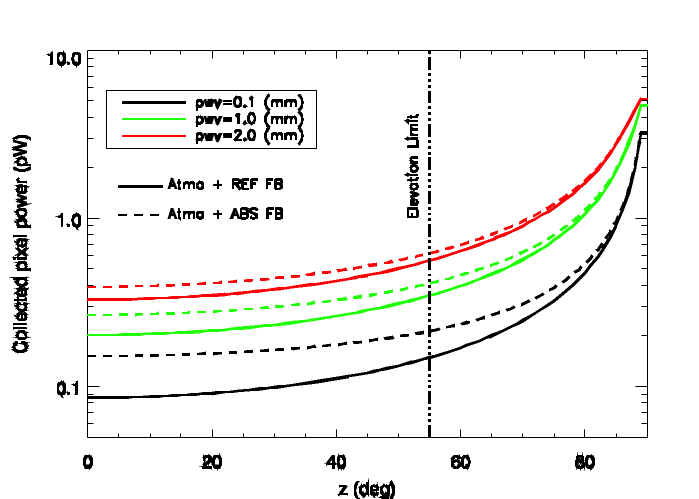}
\caption[Power collected by a single detector due to atmospheric emission plus FB emission in reflecting and absorbing
configurations]{Power collected by a single detector due to atmospheric emission plus FB emission in reflecting and absorbing
configurations, by varying QUBIC zenith angle. The colored lines are referring to 3 different pwv values as in Figure~\ref{fig33}.\label{fig34}}
\end{figure}

The whole instrument is oriented by an altazimuth mount and it is expected to be surrounded by a ground shield (GS) in
order to minimize the brightness contrast between the sky and the ground. 

\begin{figure}
\centering  \includegraphics[width=7.6cm]{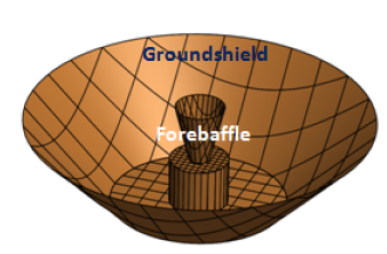} 
\includegraphics[width=8.2cm]{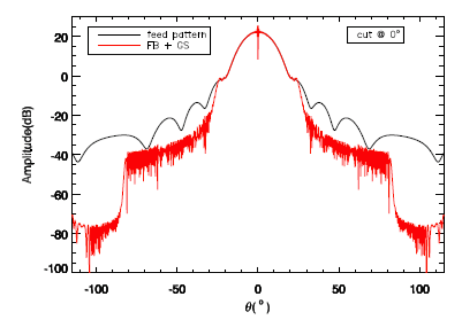} 
\caption[QUBIC shielding components: a forebaffle (FB) and a ground shield (GS) ;  Beam pattern cuts of the central feed with the forebaffle alone and including also the ground
shield]{Left : QUBIC shielding components: a forebaffle (FB), fixed on the cryostat window, and a ground shield (GS), fixed
on the ground. Right : Beam pattern cuts of the central feed with the forebaffle alone and including also the ground
shield are shown. The FB is pointing towards the zenith. The color code is described in the legend.
\label{fig35}}
\end{figure}

We considered a reflective shield with a conical shape and a full aperture angle of 90 degrees
(45 degrees from the vertical) with the base 3 m in diameter (see sketch on the left in Figure~\ref{fig35}) and an aperture of 6
m in diameter with a flared edge. The final dimensions are expected to be tuned with the mount design and the geometry
of the platform where the instrument will be installed.

The impact of to the reflective conical FB inside the GS on the central feed horn beam
pattern, assuming the instrument pointing toward the zenith, is shown in the plot on the right of Figure~\ref{fig35}. The
cut-off experienced by the beam pattern at {\textpm}80° are given by the presence of the GS edge.

\begin{table}[th]
\centering
\begin{tabular}{|m{2.7cm}|m{3.1cm}|m{2.9cm}|m{2.9cm}|}
\hline
{ @ z=0°} &
{ No Shields} &
{ With FB} &
{ FB+GS}\\
\hline
{\bfseries Ground} &
\centering{ 627 mK} &
\centering{ 117 mK} &
\centering\arraybslash{ 2mK}\\
{\bfseries Sun} &
\centering{ 3 mK} &
\centering{ 328 $\mu $K} &
\centering\arraybslash{ 7$\mu $K}\\
{\bfseries Moon} &
\centering{ 70 $\mu $K} &
\centering{ 7 $\mu $K} &
\centering\arraybslash{ 0.1$\mu $K}\\
\hline
\end{tabular}
\caption{spillover contributions.}
\label{table15}
\end{table}

In Table~\ref{table15} the spillover contributions in terms of brightness temperature due to the main
contaminants are listed with and without the presence of the shields. 

%auto-ignore
\subsection{Technological demonstrator}
\label{techno}

The development of the QUBIC instrument is a multi-step process (cf. section~\ref{devplan_instr}). Once the detection
chain has been validated, the next step will be the fabrication, integration and test
of the technological demonstrator, followed by the  
construction of the full first
module of the instrument. 
%The final step will be the fabrication and implementation of five additional
%modules.

To meet this last challenge, taking into account the financial situation, technological advancement
and schedule issues of each partner of the QUBIC collaboration, it was indeed decided after the June 2015 review to define the QUBIC technological
demonstrator as the final instrument is, with only five downsizings: a reduced number of pixels, a reduced number of
horns and switches, a reduced diameter of filters, a reduced number of pulse tubes, a reduced diameter of mirrors.

The decrease in the number of pixels is determined by the fact that we have no more funding to buy superconducting
cables to connect them to the cold electronics and to fabricate the focal plane mechanics. Due to this fact we will
re-use the 256 pixels array (and its mechanical support) already tested for the validation of the detection chain. For
this former validation we had some spare cables, we are currently assessing the possibility to connect an additional
256 pixels (possibly at 220 GHz), supported by a replica of the mechanical support, fabricated in LAL and APC
mechanical workshops.

The decrease in the number of horns and switches is due to the fact that the Milano
laboratories anyway manufacture an 8x8 back-to-back array in order to fully validate their process of fabrication and
test in view of the 20x20 final array. Using this 8x8 array for the demonstrator will save time and will reinforce the
validation process for this sub-system. The switches of the 8x8 array will not be equipped of chokes, since chokes are
usefull only for decreasing the return losses.

In the final instrument two pulse tubes cryo-coolers will be used: one for the focal planes, the other one for the 1K
box. Due to the reduced numbers of pixels and corresponding electronics and their related thermal loadings, the
demonstrator could be equipped with only one pulse tube. This will delay also the financial burden on the Roma
laboratory.

The decrease in the diameter of filters is due to the fact that as of November 2015, the Cardiff installations are not
able to produce filters with a diameter bigger than 300 mm. Improvements of these equipment will be done in a near
future but to save time and decrease risks the decision has been taken to install 300 mm filters in the technological
demonstrator.

The decrease in the diameters of mirrors is due to the fact that Milano University has
capabilities to manufacture mirrors with a diameter up to 400 mm. The mirrors of the QUBIC instrument have a diameter
of 600 mm. So it was decided in order to save time, to use for the demonstrator 400 mm diameter mirrors, made in-house
at Milano. 
%This is made in agreement with optical simulations, showing that such mirrors are compatible with the
%reduced number of horns.

All these downsizings are optically compatible with each other, according to simulations.

A comparison of the characteristics of the technical demonstrator and of the first module is summarized on Table~\ref{table33}.

\begin{table}
\begin{tabular}{|m{5.215cm}|m{5.215cm}|m{5.2330003cm}|}
\hline
\centering{\bfseries Sub-systems} &
\centering{\bfseries Technological demonstrator} &
\centering\arraybslash{\bfseries Final instrument}\\\hline
\centering{ Cryostat} &
\centering{ Nominal} &
\centering\arraybslash{ Nominal}\\\hline
\centering{ Down to 4K cryo-coolers} &
\centering{ One pulse tube} &
\centering\arraybslash{ Two pulse tubes}\\\hline
\centering{ Down to 320mK cryo-coolers} &
\centering{ Nominal} &
\centering\arraybslash{ Nominal}\\\hline
\centering{ Mirrors} &
\centering{ 400 mm diameter} &
\centering\arraybslash{ Nominal}\\\hline
\centering{ 1K box, cold stop} &
\centering{ Nominal} &
\centering\arraybslash{ Nominal}\\\hline
\centering{ Focal planes} &
{\centering 256 pixels at 150 GHz (option with 512 pixels).\par}

\centering{ Option with 512 pixels.} &
\centering\arraybslash{ 2 focal planes, 1024 pixels each (one at 150 GHz, one at 220
GHz)}\\\hline
\centering{ Horns and switches back-to-back array} &
{\centering 8x8\par}

\centering{ without chokes} &
{\centering 20x20\par}

\centering\arraybslash{ with dual band chokes}\\\hline
\centering{ Filters, HWP, Dichroic} &
\centering{180  to 280  mm diameter} &
\centering\arraybslash{ Nominal}\\\hline
\centering{ Cold electronics} &
\centering{ Nominal, but for just 256 or 512 pixels} &
\centering\arraybslash{ Nominal}\\\hline
\centering{ Warm electronics} &
\centering{ Nominal, but just for 256 or 512 pixels} &
\centering\arraybslash{ Nominal}\\\hline
\centering{ Data storage} &
\centering{ Nominal, but just for the volume of data produced by 256 or 512 pixels} &
\centering\arraybslash{ Nominal}\\\hline
\centering{ QUBIC-Studio (control and readout)} &
\centering{ Nominal} &
\centering\arraybslash{ Nominal}\\\hline
\centering{ Mount} &
\centering{ Possibly a fake mount} &
\centering\arraybslash{ Nominal }\\\hline
\end{tabular}
\caption{Respective definitions of the technological
demonstrator and the final instrument.
\label{table33}}
\end{table}

\section{Calibration, Operation Modes and data processing}

Before the installation of QUBIC on site (cf. section~\ref{sites}), the instrument will undergo a serie
of measurements for its characterisation that are described in section~\ref{calib}. 

Once on-site and besides the commissioning mode where the performances of the instrument will be assessed and tuned using calibration
sources and sky data, QUBIC will be operated in two distinct modes: self-calibration and normal observation mode 
(cf. section~\ref{operation}).

%auto-ignore
\subsection{Instrument testing and calibration operations}
\label{calib}
The objectives of the calibration is the determination of the main parameters of the instrument ``in lab'' (after its
integration and before its shipment to the exploitation site), and the verification that the extracted values are in
conformity with the requirements. The operations will be performed at APC and are divided in three steps:

\begin{enumerate}
\item {
First, the measurements will be achieved with the entrance window of the cryostat closed by a metal plate: this phase is
here-after called ``Blind cryostat'' configuration.}
\item {
Then, the metal plate will be removed. A density filter installed inside the cryostat will insure that there is no
saturation of the detectors (from the 300K). This configuration is called ``opened cryostat''}
\item {
Finally, while the first two steps will be done inside the APC Hall, the instrument will be transported outside pointing
at the calibration source placed on the roof of a building facing the Hall entrance. }
\end{enumerate}

The sub-system warm functional tests are supposed to have been performed successfully previously to the calibration
start-up stage. The test plan is built keeping in mind that the number of cooling downs should be kept as low as
possible. We estimate that we need fifteen days to cool down the instrument to 320mK.

\subsubsection{Cryogenic measurements and functional tests}
During the cooling downs (for both the blind and the opened configurations of the cryostat)
successive tests with respect to the thermal behaviour of the instrument will be pursued. Beyond the follow up of the
achieved temperatures on the different stages (40K, 4K, 1K and 320mK), we will measure the thermal conductance of each
stage:

\begin{itemize}
\item {
320mK: through regulation stage}
\item {
1K: through heater}
\item {
4K: through heater and switches}
\item {
40K: through ASIC}
\end{itemize}

We will also cross-check the thermal stability of the cryogenic chain. A comparison of the
results with respect to the model of the thermal transfer within the cryostat will be assessed. 

For different temperatures configurations, we will also measure the R(T) behaviour of the
detectors.

In parallel functional tests on the use of the half wave plate and of the switch array will be performed. We will then be
able to check the EMI/EMC their use induces in the instrument. 
%It has to be noted that at Dome C there is no fixed/solid ground
%plane: all electronics have to be properly grounded accordingly. 
Eventual repercussions on the cryostat temperature at
the different stages, and the corresponding amount of time needed to come back to the optimal values will be extracted
from the housekeeping data.

\subsubsection{Detector characteristics determination}

\begin{table}
\centering
\begin{tabular}{|m{6.729cm}|}
\hline
{ Measurement } \\\hline
{ R(T) and I(V) curves determination} \\\hline
{ Noise measurements (NEP, slope, fknee)} \\\hline
{ CrossTalk} \\\hline
{ EMI/EMC Compatibility} \\\hline
\end{tabular}
\caption{Characteristics of the instrument that will be
measured during the first cooling downs of the QUBIC instrument, for different loads on the detectors.}
\label{table26}
\end{table}

When the nominal values for the temperature of the different stages will be reached, the I(V) curves will be determined,
as well as the working points of the detectors. A comparison of the measurements of the integrated instrument with
respect to already existing ones on the response of the electronics, the SQUIDS and the TES will be assessed~\cite{Martino}.
This procedure is already mastered by IRAP/APC and has been applied during the TES tests that have been performed at
APC. The I(V) curves will be measured in parallel on all detectors.

We will then measure their corresponding noise characteristics: their NEP, and the slope and fknee values of the noise
spectrum. Those characteristics will be determined for different V to better constrain the Instrument Model. Specific 
characteristics of the noise, such as the cross-talk will also be assessed using the blind detectors. Finally 
we will redo an EMI/EMC test while monitoring the detector response.

All the measurements described above can be done under different load configurations, which can be mimicked either
 by playing (within some extend) with the different cryogenic stages, and/or  by the comparison between
the two cryostat configurations (blind or opened). This would permit to further refine the Instrument Model. 
%Option A
%is nice to have but not mandatory and will be removed from the calibration plan if we are running out of time.
%{\bf precise pros and cons ? A more detailed \& accurate but time consuming, B straightforward but conditions a bit uncertain and not mastered ?}
Depending on the availability of a temporary mounting system, the stability of the temperature stages and the noise on
the detectors will be measured for different inclinations of the cryostat to check the elevation domain in which the
instrument will work (in which the pulse tube is efficient enough).

\subsubsection{With the calibration setup}
Once we are sure that the previous measurements are within the requirements for the temperatures as well as for the
noise parameters, the next step is to use internal and external sources:
\begin{itemize}
\item{} Carbon fiber sources, within the cryostat (see below)
\item{} the Calibration source (cf. section~\ref{calibSource})
\item{} a Fourier Transform Spectrometer (FTS)
\end{itemize}
to further determine the detectors characteristics: the corresponding parameters are summarized in the
Table~\ref{table27}.

\begin{table}
\centering
\begin{tabular}{|m{6.677cm}|}
\hline
\centering\arraybslash{ Detectors Intercalibration \& Cross Talk}\\\hline
\centering\arraybslash{ Band Pass Spectral measurements}\\\hline
\centering\arraybslash{ Absolute Response}\\\hline
\centering\arraybslash{ Synthetic beam reconstruction}\\\hline
\centering\arraybslash{ Polarisation angle recovery}\\\hline
\centering\arraybslash{ Self-Calibration checks}\\\hline
\centering\arraybslash{ Time constants}\\\hline
\centering\arraybslash{ Detector Linearity}\\\hline
\end{tabular}
\caption{Characteristics of the QUBIC instrument that
will be measured with the Carbon fiber source, the Calibration source and/or the FTS.
}
\label{table27}
\end{table}

The intercalibration of all detectors will be first tackled with Carbon fibers sources emitting in the IR~\cite{Henrot}. A couple of such sources will be placed at the edge of the feedhorn array to monitor detectors responce and also to check the allignment of the combiner along pointing directions 
 (this can even be done in the blind cryostat configuration), and further determined with
the Calibration source with the instrument installed outside. 

At least an upper limit on the time constants will also be measured with the Carbon fiber sources (as it was done for the
Planck-HFI calibration~\cite{hficalib}). Since the time constant should not be an issue for QUBIC this test is rather a cross-check
that a real measurement. 

The absolute response will be measured with the Calibration source outside of the APC Hall. While proceeding with the
HWP we will also be able to recover the polarisation angle of the source. 

The percent level is expected on the spectral filter knowledge [ref]: band pass spectral measurements are therefore
needed to make sure QUBIC is within the requirements. Inside the APC Hall a first validation of the spectral response
will be tested with a vectorial analyser, and the measurements will be refined with the Calibration source by varying
its frequency and correcting for its emission spectrum.

With the use of the mounting system, a complementary test on the eventual impact of \ gravity effects on the instrument
will be checked by scanning the Calibration source and reconstructing the pointing of the focal plane. The synthetic
beam (cf. Section~\ref{beamsynth}), and the self calibration  (cf. Section~\ref{selfcalib}) procedures will finally be tested while pointing at the Calibration source
outside the Hall.

\begin{figure}
\centering
 \includegraphics[width=10.668cm,height=8.001cm]{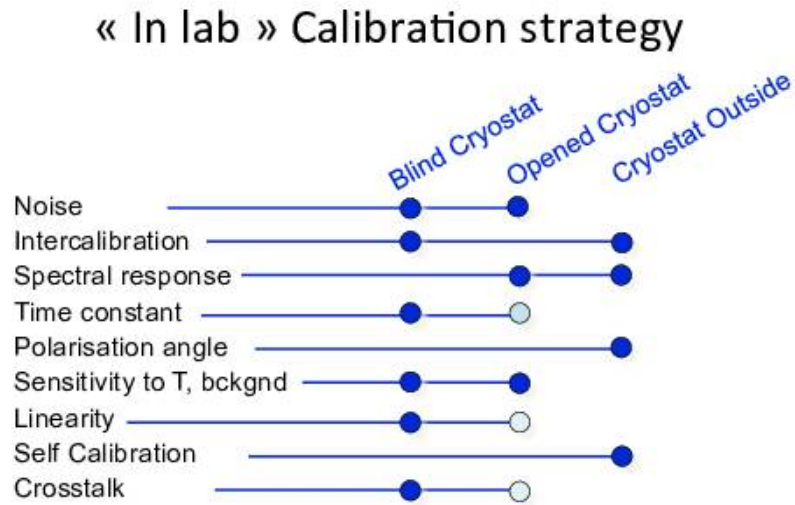} 
\hypertarget{RefHeadingToc314323128}{}
\caption{``In lab'' Calibration strategy}\label{fig97}
\end{figure}

The different parameters and the cryostat configuration in which they are supposed to be measured are illustrated in Figure~\ref{fig97}. The blue dots are mandatory while the ligh blue ones are nice to have measurements.

%\subsubsection{Organization during test phase}
%{\notes peut etre qu'on peut enlever ca ??}
%The data analysis for the calibration will be done either directly within QUBIC Studio (cf. section~\ref{qubicstudio}) if possible, or outside for the
%more complicated analysis which will need an offline treatment : for this last step the existing python interface to
%QUBIC Studio will be used. 

%Each measurement will be assigned to an in-charge person who will need to be at APC for the corresponding data taking
%period and will commit to perform the needed data analysis, up to the required parameters extraction. It includes the
%derivation before the data taking period of the main parameters requirements from Monte Carlo simulations. 

%During the duration of the calibration phase, a daily (max 1h) meeting is planned on the morning. It should be attended
%(at least by phone) by the calibration responsible, a cryogenic expert, the in-charge person of the measurements
%planned this day plus the one of the previous day. A short summary of the previous day data taking will be given and
%decisions will be taken for the coming day. 

%The output of the calibration of the QUBIC instrument will be written in a paper and submitted to a pear review
%journal.

%auto-ignore
\subsection{Modes of operations}
\label{operation}

%%%%%{\notes{Figure ~\ref{fig111} is never referenced in the text}}

\subsubsection{Self-calibration mode}
\label{selfnormal}
The self-calibration mode consists in observing and scanning an artificial polarized source
with QUBIC, opening and closing the horns switches in order to extract the fringe patterns for each of the 
interferometric baselines. We have shown in~\cite{Bigot}~\cite{BigotSazy:2012tr}
that such a
procedure allows constraining tightly the instrumental systematics. This is the operation mode that is specific to
QUBIC and cannot be performed with an usual imager. This mode is the one that we consider as the primary advantage of
QUBIC in controlling better the instrumental systematics. The level of control on the instrumental systematics depends
on the amount of time spent performing self-calibration. We therefore consider that up to 50\% of the observation time
could be invested in self-calibration if needed. The exact amount will be determined by analysing the self-calibration
data in order to balance between systematics and statistical uncertainties. In~\cite{BigotSazy:2012tr}, we have shown on rather simplistic simulations
that we hope to reduce the level of systematics by more than an order of magnitude by spending 2.5\% of the observation
time performing self-calibration (see Section~\ref{selfcalib} and  Figure~\ref{selfcalres} for more details).

%\begin{figure}
%\centering
% \includegraphics[width=15.242cm,height=9.34cm]{QUBICTDRcompilation-img/QUBICTDRcompilation-img185} 
%\hypertarget{RefHeadingToc314323140}{}\caption{Reduction of
%systematics from Self-Calibration}\label{fig109}
%\end{figure}

\subsubsection{Observation mode}
\label{bkm:Ref311965574}
QUBIC aims at observing regions of the sky that show the lowest amount of dust contamination possible. Two such regions
are considered up to now :

\begin{itemize}
\item {
The well known and well observed «~BICEP2 region~» with equatorial coordinates (RA = 0 deg, DEC = -57 deg). This region
is known to be contaminated by dust at a level $r\approx${}0.2 for which our simulations (based on power law assumption for the
dust component) have shown that we can achieve a $\sigma(r)\sim 0.02 $ from Argentina using our two bands 150
and 220 GHz and adding the Planck 353 GHz maps.}
\item {
{The region with equatorial coordinates (RA = 8.7 deg, DEC = -41.7 deg) where the analysis of
Planck data have shown an apparent lower lovel of dust contamination, although with significant uncertainties~\cite{Adam:2014bub}.}}
\end{itemize}

We prefer to delay the final decision of the observed field to the beginning of the operations for QUBIC as more precise
information might be available by then on the relative advantages of both regions in terms of dust contamination. They
can equivalently be observed by QUBIC in Argentina and Antartica with the Azimuth and Elevation
ranges (accounting for a lower limit of 30 degrees elevation for operations in order to avoid a too high atmospheric
emission) given in Table~\ref{table32}

\begin{table}
\begin{tabular}{|m{3.991cm}|m{2.55cm}|m{2.552cm}|m{3.143cm}|m{3.162cm}|}
\hline
~
 &
\multicolumn{2}{m{5.302cm}|}{{ Concordia (Antarctica)}} &
\multicolumn{2}{m{6.505cm}|}{{ San Antonio de los Cobres (Argentina)}}\\\hline
~
 &
{ Azimuth} &
{ Elevation} &
{ Azimuth} &
{ Elevation}\\\hline
{ BICEP2 Field} &
{ 0-360 deg.} &
{ 55 +/- 15 deg.} &
{ 140-220 deg.} &
{ 45 +/- 15 deg.}\\\hline
{ Planck Clean Field} &
{ 0-360 deg.} &
{ 45 +/- 15 deg.} &
{ 130-250 deg.} &
{ 50 +/- 20 deg.}\\\hline
\end{tabular}
\caption{Azimuth
and elevation ranges for the two considered fields for QUBIC, for Dome C (left) and Argentina (right).
\label{table32}}
\end{table}

The observations will be performed through the so-called ``constant elevation scanning'' that is done in the following
manner:
\begin{enumerate}
\item {
{Constant elevation sweeps in azimuth around the azimuth of the center of the observed field
with an amplitude of $\Delta $az (typically 30 degrees) at the angular speed vaz (typically 1 deg/sec) for a given
number of sweeps $N_{sweeps}$ (typically 300). During such a series of sweeps, the elevation of center of the field varies
through the field of view of the instrument and eventually the instrument starts sweeping the outer regions of the
field.}}
\item {
{\ After Nsweeps, one recenters the elevation of the instrument to that of the observed field
and starts the constant elevation azimuth sweeps again. In sky coordinates, these new sweeps now exhibit an angle with
respect to the previous ones, so that after 24 hours, all the regions in the sky have been observed with all angles
(see Figure~\ref{fig110}).}}
\item {
At the end of each azimuth sweep, the Half-Wave-Plate is rotated by one position, corresponding to 11.25 degrees.}
\item {
An additional continuous or stepped slow rotation around the optical axis is envisioned (so we require the instrument
mount to be able to perform it) in order to further modulate the signal, but further simulations are required to
optimize it, accounting for the atmospheric fluctuations.}
\end{enumerate}

We only take data when the elevation of the instrument is between $\approx${}30 and $\approx${}70 degrees,
the lower bound is set by the air-mass that becomes too large, and induces too high photon noise below 30 degrees,
while the upper bound is set by the inclination range allowed by the Pulse-Tube-Cooler (cf. section~\ref{cryo}) +/- $\approx${}20 degrees around the
central position with elevation 50 degrees (the exact range will be defined during commissioning). Such a scanning
strategy allows to cover $\approx${}1\% of the whole sky in 24 hours. This is exactly the sky coverage that is optimal for
setting an upper limit (meaning including only noise variance, the signal being considered to be 0 therefore without
sample variance) on the B-modes signal at multipoles around 100 (using the recombination bump).

\begin{figure}
\centering
\includegraphics[width=7.938cm,height=5.62cm]{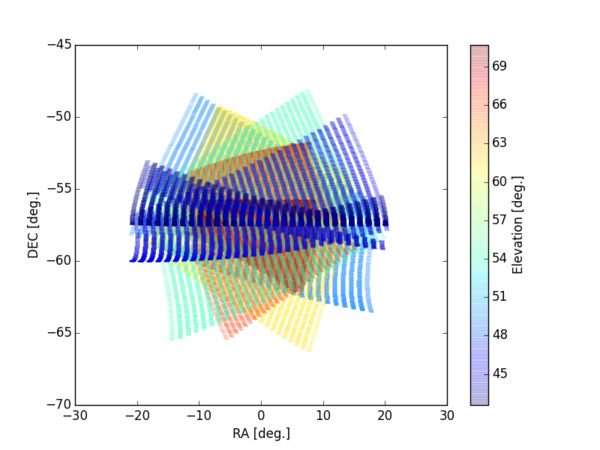}
\includegraphics[width=5.479cm,height=5.526cm]{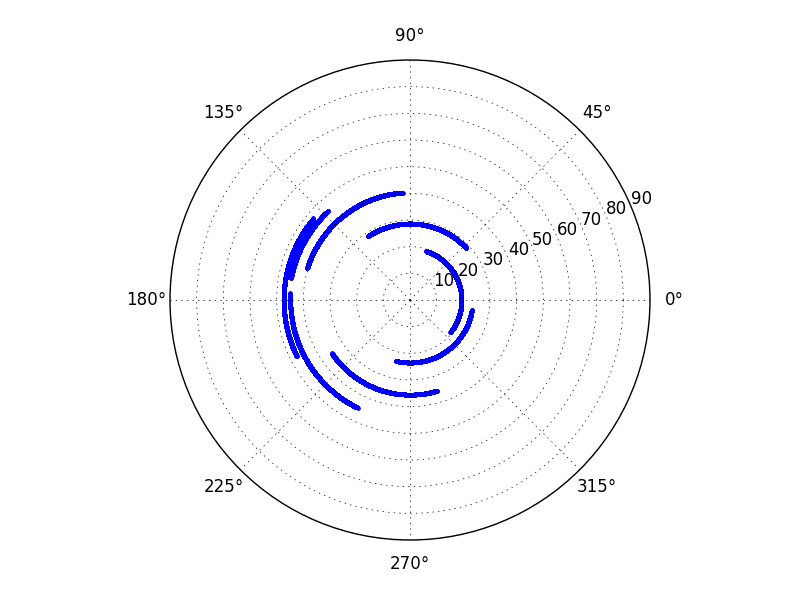}
\caption[Scanning Strategy in local coordinates  and
sweeps in RA, DEC (right) for a location in Dome C, Antarctica]{Scanning Strategy in local coordinates (left) and
sweeps in RA, DEC (right) for a location in Dome C, Antarctica.\label{fig110}}
\end{figure}

In the eventuality where a B-mode signal is detected by another experiment before us, or by
ourselves, one may want to increase the sky coverage in order to reduce the sample variance from the primordial B-modes.
This is easily achievable by performing the same scanning strategy with slight shift in RA, DEC for the center of the
field over a few days. This can be seen in Figure~\ref{fig112} where 2.4\% of the full sky is achieved with 10 days (whereas 
in 1 day, 0.9\% are covered as shown in Figure~\ref{fig111}).

The actual scanning strategy baseline is still evolving with the progress of the simulations and will be definitely
frozen only when observations will start.

\begin{figure}\centering
 \includegraphics[width=8.894cm,height=5.636cm]{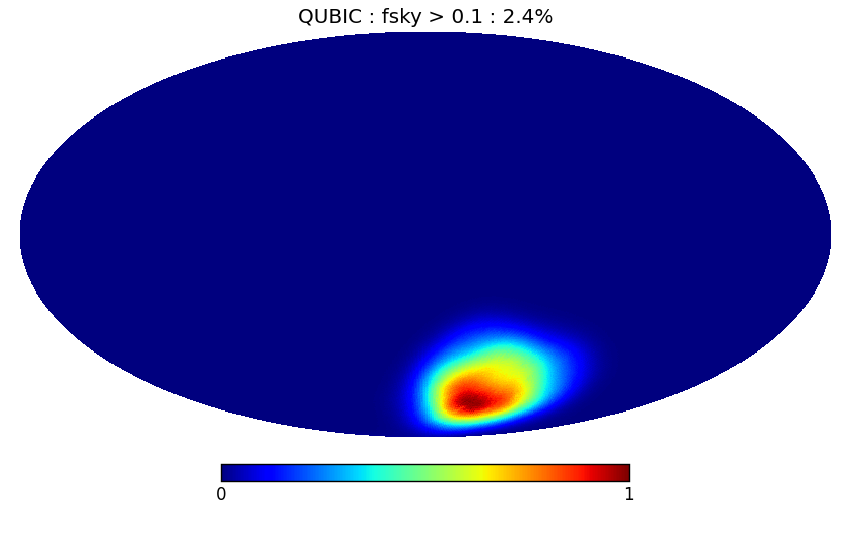} 
\includegraphics[width=6.359cm,height=5.913cm]{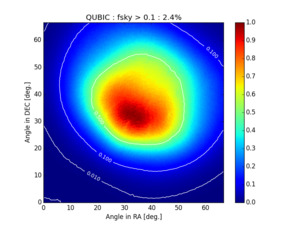} 
\hypertarget{RefHeadingToc314323142}{}\caption[Coverage achieved when adding successive
coverages over 10 days]{Coverage achieved when adding successive
coverages over 10 days, from dome C, each slightly shifted with respect to the original one (placed on a circle in RA, DEC with
radius 12 degrees).\label{fig112}}\end{figure}

\begin{figure}
\centering
 \includegraphics[width=8.894cm,height=5.65cm]{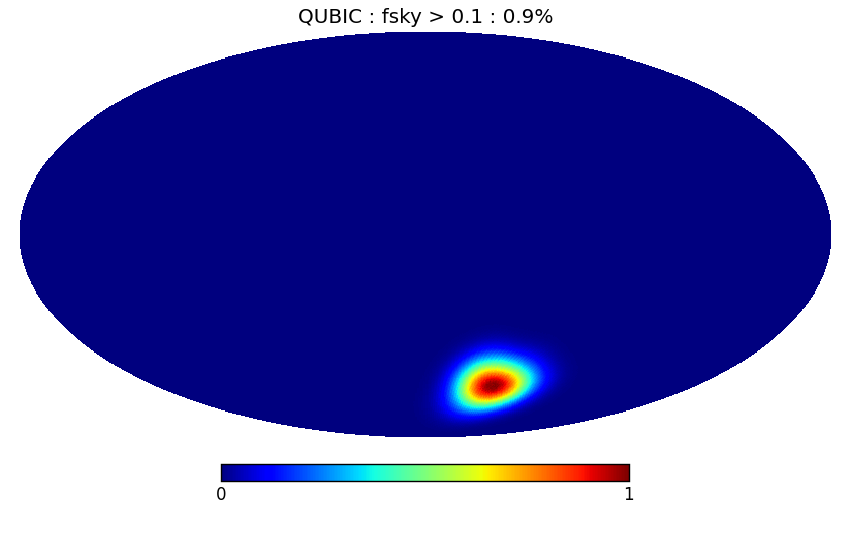} 
\includegraphics[width=6.359cm,height=5.943cm]{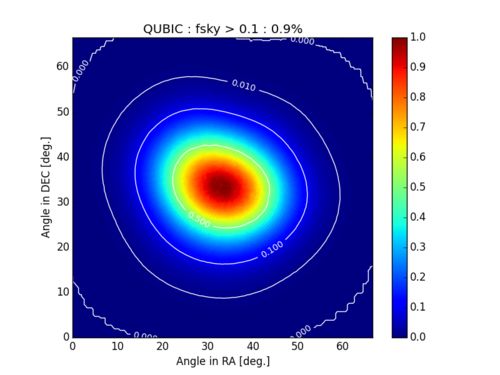} 
\hypertarget{RefHeadingToc314323141}{}
\caption[coverage in Sky Coordinates for a 24h scanning around a single spot on
the sky]{Coverage in Sky Coordinates (Galactic on the left, Equatorial on the right) for a 24h scanning around a single spot on
the sky. Colors represent the relative coverage. The scanning strategy is calculated for a location in Dome C,
Antarctica.\label{fig111}}
\end{figure}

%auto-ignore
\subsection{Self-calibration}
%{\notes{a lot of things are said twice here..need a reshaffling of text}}
%{\notes{we have quite change the text here...please make sure we have not make any mistake :-)}}

The self-calibration system, which is introduced in 
section ~\ref{selfcalsect}, is composed of millimetric wave sources (the self-calibration sources cf. section~\ref{calibSource2}),
placed on top of a calibration tower (cf. section~\ref{support}) at $\approx$ 45m from the experiment. The sources will be observed by the QUBIC
instrument 
with an elevation angle of
$\approx$ 30°-45° from the horizon.

\subsubsection{The self-calibration procedure}
\label{selfcalsect}

As discussed in section~\ref{selfcalib}, the interferometric nature of QUBIC will allow to track down systematic
effects such as a misalignment of horns, focal plane, mirrors, but also non-homogeneous temperature (...)

Theoretically, in case the
instrument doesn't have any systematics effect, all equivalent baselines (meaning any combination of a pair of
back-to-back horns separated by an identical distance) would behave and transmit the signal the same way (meaning
giving the same fringes on the focal planes). It was shown in~\cite{Bigot}, that 
a comparison between the real fringes obtained with equivalent 
baselines allows to estimate the systematic
effects of the instrument, and is called the self-calibration process. 

In order to practically implement such a procedure, all back-to-back horns are
equipped with RF switches located between input horn and output horn (cf. section~\ref{switch}).
By closing horns two by
two, the behavior of the instrument can be analyzed and systematics can be withdrawn from its transfer function. 

In order to implement the self-calibration procedure, a well characterized signal emitting at
150 and 220 GHz and polarized, has to be received by the instrument. 

\subsubsection{Self-calibration sources}
\label{calibSource2}

The main drivers that set the self-calibration sources
specifications are the high polarization degree of the sources (practically 100\% for
a corrugated horn excited through a rectangular waveguide), and the available power at the output. During the
self-calibration phase, we need to characterize the different realizations of the same Fourier mode (through different
horn pairs), which at the end consist in mapping the corresponding interference fringes on the focal plane: the higher
the S/N ratio of fringe detection, the higher the level of the accuracy of systematic effects retrieval. A study has
been done to show that a high S/N can be achieved as a result of a good balance between the FWHM of the sources, sources
power, and the fraction of saturated detectors in the focal plane.

More specifically, the self-calibration sources should emit in the two frequency bands observed by QUBIC with a bandwidth of 25$\%$.
For this purpose two different self-calibration sources are required, one for each frequency band.
The signal which should be linearly polarized should induce a cross-polarisation smaller than -30dB.
It is expected to be energetic enough to be detected and read-out
by the instrument, but sufficiently tiny not to saturate the detectors. To
assure the possibility of detecting an interference fringe with a S/N ratio
{\textgreater} 20000, this leads to an expected 
emitted power between 1 and 5~mW on the whole frequency band, with a flatness
of 3dB. In addition the signal should be very stable, with a power drift smaller than 1$\%$ per hour.
The main characteristics are given in table~\ref{table12}.

%{\notes{the numbers given in table~\ref{table12} do not match does given here -> what are the correct ones ?}} 

The typical solution matches those of off-the-shelf mm-wave sources, which 
is to use a microwave synthesizer in a baseband up to 20 GHz, which feeds
a cascade of amplifiers/multipliers up to the desired frequency band, with the desired power level. Gunn oscillators
endowed with a PLL are another possible option. The same technology will be used for both 150 and 220 GHz bands.
Standard corrugated horns can be manufactured with the desired beam (scalar horns with
10$^\circ$ beam are typically purchased
off-the-shelf).

The sources
will also be used for the Calibration operations (cf. section~\ref{calib}).

The self-calibration sources device must cope with the extreme weather conditions in the chosen exploitation site.
They will be maintained within a 30cm x
30cm x 30cm and 10kg weight thermally insulated box that will be designed within the collaboration.
%A sheltering
%warm-up system will avoid snow deposition on the source itself. 

They must have a high reliability and availability, since the self-calibration will represent an important
percentage of observation time (cf. section~\ref{selfnormal}).

%{\notes A call for tender should be issued in early 2016 for the 150 GHz source (funded by DIM-ACAV). -> not sure it should
%appear here}

\subsubsection{Self-calibration sources support}
\label{support}

The sources location needs be chosen so that the instrument would receive the signal as if the source was
in far field conditions. For this purpose it will be mounted on top of a mast able to maintain the
source at least 45 meters high and 45 meters away from QUBIC. 

The baseline (conservative) plan for the QUBIC
calibration tower is an exact copy of the American Tower installed on Dome C,  but we are also investigating the possibility to use a much
lighter solution with much closer struts. A preliminary design was produced by the ITAS company with
supporting struts limited to a distance of 10m from the tower (see Figure~\ref{fig51}). ITAS company is working also on
different solutions including one without struts. 

%The QUBIC collaboration is also undertaking feasibility studies for
%the use of tethered source on balloons and drones in place of the tower. In case QUBIC will be positioned in Dome C by
%the American Tower, this would be calibration tower with no need to build any additional 45m tower in Dome C.

%\begin{figure}
%\centering
% \includegraphics[width=4.658cm,height=8.035cm]{QUBICTDRcompilation-img/QUBICTDRcompilation-img086.jpg}
%\includegraphics[width=10.407cm,height=8.035cm]{QUBICTDRcompilation-img/QUBICTDRcompilation-img087.png} 
%\hypertarget{RefHeadingToc314323082}{}
%\caption[Image of the 45m tall American
%Tower,  preliminary design of the QUBIC calibration tower]{left: an image of the 45m tall American
%Tower. Right: preliminary design of the QUBIC calibration tower (acknowledgement ITAS).\label{fig50}}
%\end{figure}

\begin{figure}
\centering
\includegraphics[width=4.658cm,height=8.035cm]{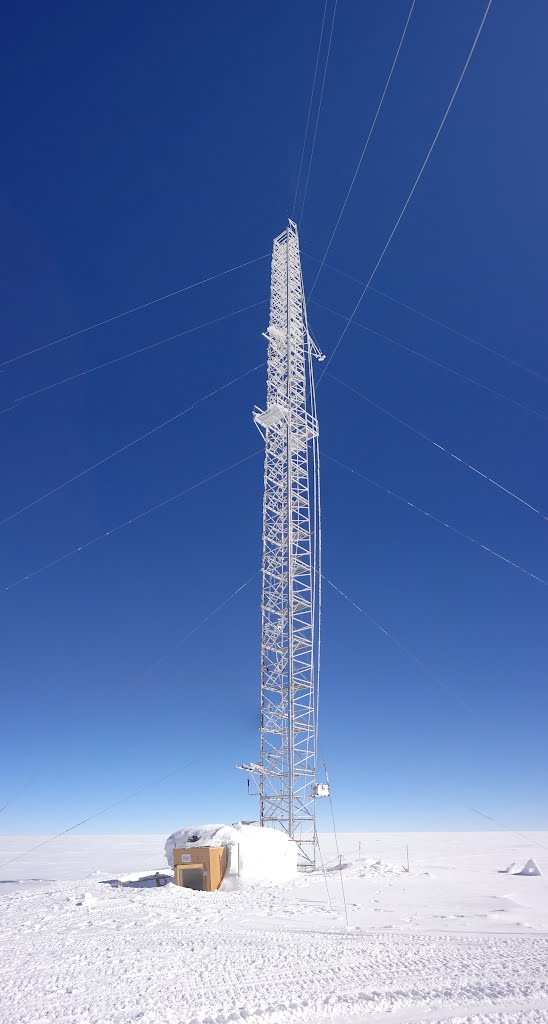}
\includegraphics[width=3.658cm,height=8.035cm]{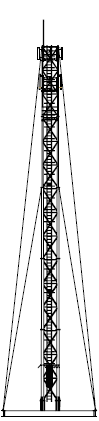}
\hypertarget{RefHeadingToc314323083}{}
\caption[The {\textquotedbl}American
tower{\textquotedbl} at Concordia Station and a schematic of a tower with less footprint]{the {\textquotedbl}American
tower{\textquotedbl} at Concordia Station, a guy-cable tower (left); a schematic of a tower with less footprint but
with a basement loaded with concrete (right, courtesy of ITAS company)} \label{fig51}
\end{figure}

%{\notes{ not sure it should appear here:
%A call for tender will be issued in 2016 to provide this tower, for an expected cost between 50 and 90 k{\texteuro}.}}

\subsubsection{Full beam calibration source}

The principal axis of the polarimeter must be known with high accuracy. To this purpose we will use a full-beam calibrator, consisting of a dielectric sheet stretched across the beam, in the fore-baffle outside the cryostat window, at 45\degree\ incidence, and a room-temperature blackbody. 
The sheet transmits a large part of the atmospheric and CMB signals, and reflects (partially polarizing) a small fraction of the emission of a room-temperature blackbody (see Figure\ref{fig:full_beamcal_sketch}).  

\begin{figure}
\centering
\includegraphics[width=.5\textwidth]{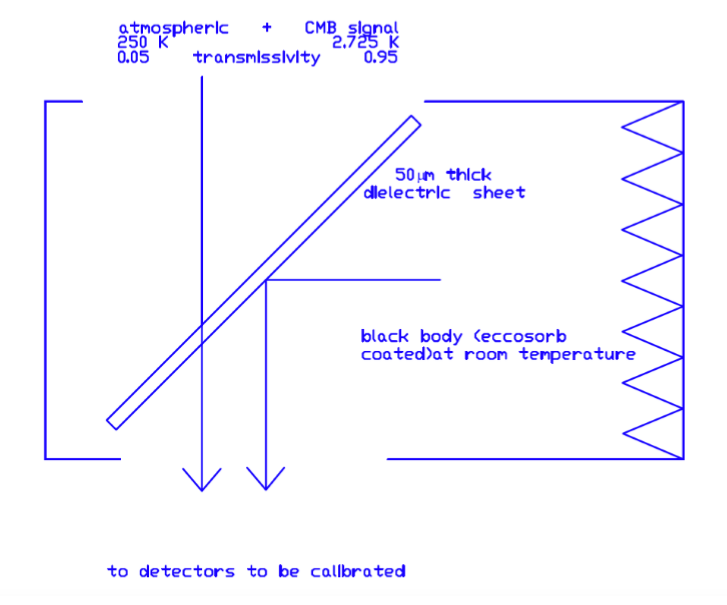}
\caption{Operating principle of the full-beam calibrator.\label{fig:full_beamcal_sketch}}
\end{figure}

The polarization properties of the calibration signal produced in this way can be computed with good accuracy. The assembly of the 45\degree\ dielectric sheet and room temperature blackbody can be rigidly rotated in  controlled matter around the optical axis of QUBIC. For each position of the calibration assembly, QUBIC observes with all the detectors and for all the positions of the HWP, so that a position angle of the polarimeter can be estimated fitting the data. 
In Figure~\ref{fig:full_beamcal_pol} we show that the polarized signal produced in the D-band by the calibrator does not depend significantly on atmospheric emission.

\begin{figure}
\centering
\includegraphics[width=.5\textwidth]{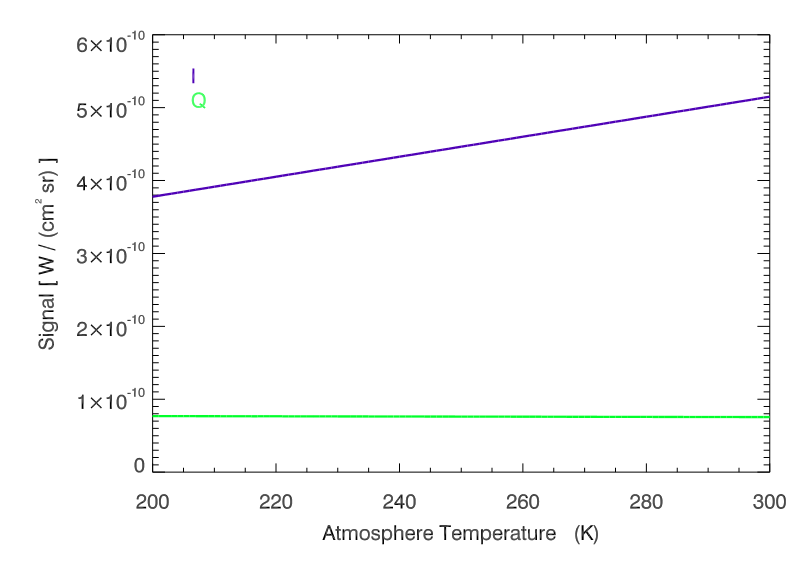}
\caption{Stokes parameters I and Q of the radiation produced by the full beam calibrator in the 150 GHz band of QUBIC.\label{fig:full_beamcal_pol}}
\end{figure}

The mechanical design of this calibration system will be finalized together with the mechanical design of the forebaffle, to allow for mounting on the same holes, on the top flange of the QUBIC cryostat.

\section{Operation site}
\label{sites}

%auto-ignore
Dome C was the initial foreseen site for the installation of the QUBIC first module. In view of the complexity of its logistics and cost, 
and of its requirements in term of the experiment reliability, we have however decided to deploy this first module from 
another site, Alto Chorillos in the Puna province of Argentina, which still offers excellent observation conditions while 
being much less demanding on logistics and reliability. 
To assess this choice, we have thoroughly compared the forecasted QUBIC sensivity in both sites. This comparison is detailed in Section~\ref{siteComparison}. 
More details on the Chorillos site are then given in Section~\ref{argentina}.

\subsection{Sites comparison}
\label{siteComparison}

This section presents a detailed comparison of Dome C and Chorillos, in terms of sensitivity, in order to help assessing
the advantages and drawbacks of each site. We compare the sensitivity that will be achieved by QUBIC using realistic
meteorological conditions for both sites (obtained from measurements) and the scanning strategies optimized for each
site.

Regarding the scanning strategy, the main difference between the two sites is that the target
fields for QUBIC are visible above 30 degrees and below 70 degrees elevation 100\% of the time in Concordia,
Antarctica, while only 40\% of the time in San Antonio de los Cobres, Argentina (see section \ref{bkm:Ref311965574} for
details about the scanning strategy).

In order to calculate the sensitivity for both sites, one needs to have a detailed knowledge of the atmospheric water
vapour content as emission from water is the main driver for the amount of photons QUBIC receives, hence the photon
noise we measure. We have used the following data for this study:
\begin{figure}
\centering  
\includegraphics[width=11.414cm,height=6.692cm]{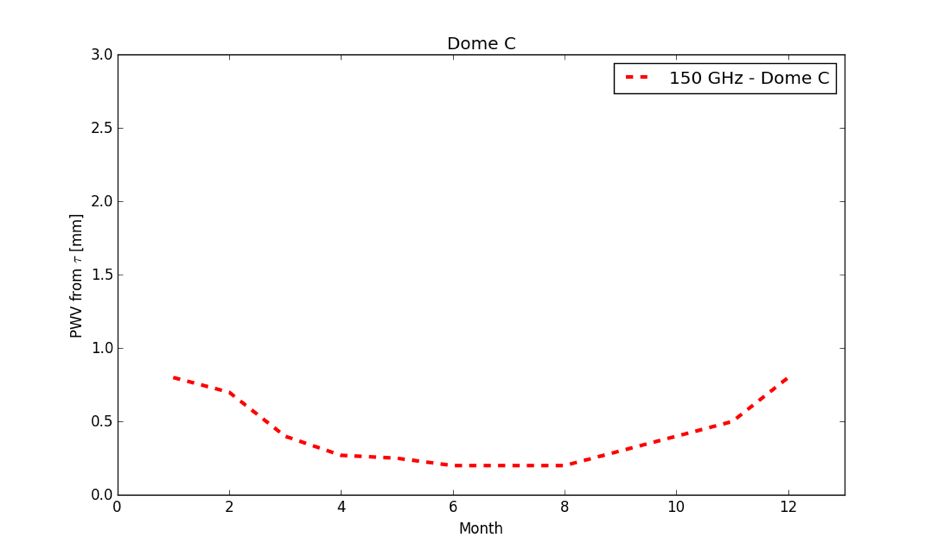}
\hypertarget{RefHeadingToc314323133}{}
\caption[PWV data for Concordia from balloon soundings at Concordia]{PWV data for Concordia from balloon soundings at Concordia~\cite{Puddu}.\label{fig102b}}
\end{figure}

\begin{figure}
\centering
\includegraphics[width=11.412cm,height=6.71cm]{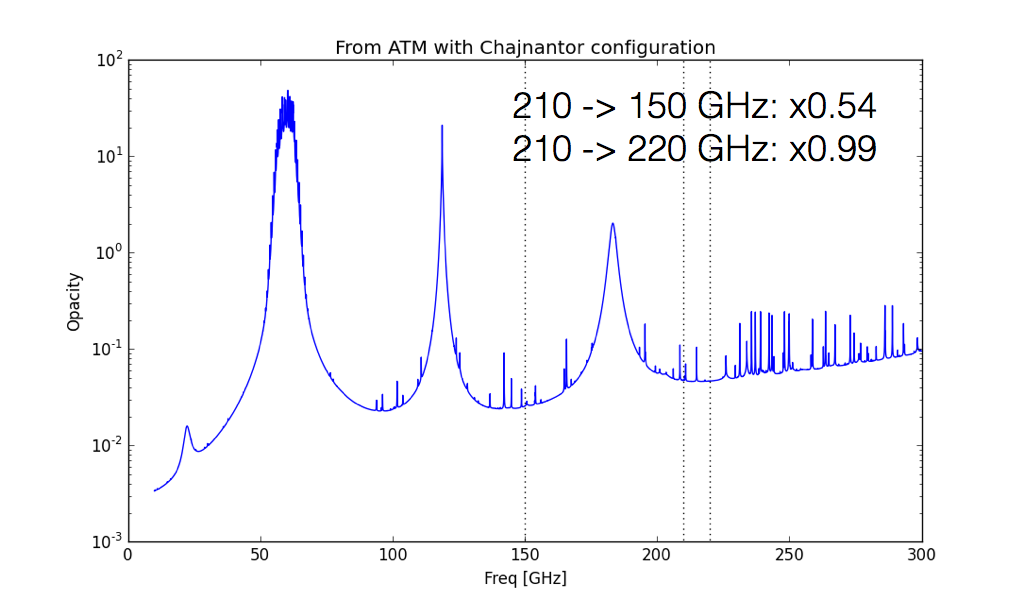}

\caption{Atmospheric spectrum from ATM with parameters
optimized for Chajnantor Atmosphere\label{fig103b}}
\end{figure}

\begin{figure}
\centering
\includegraphics[width=13.328cm,height=7.835cm]{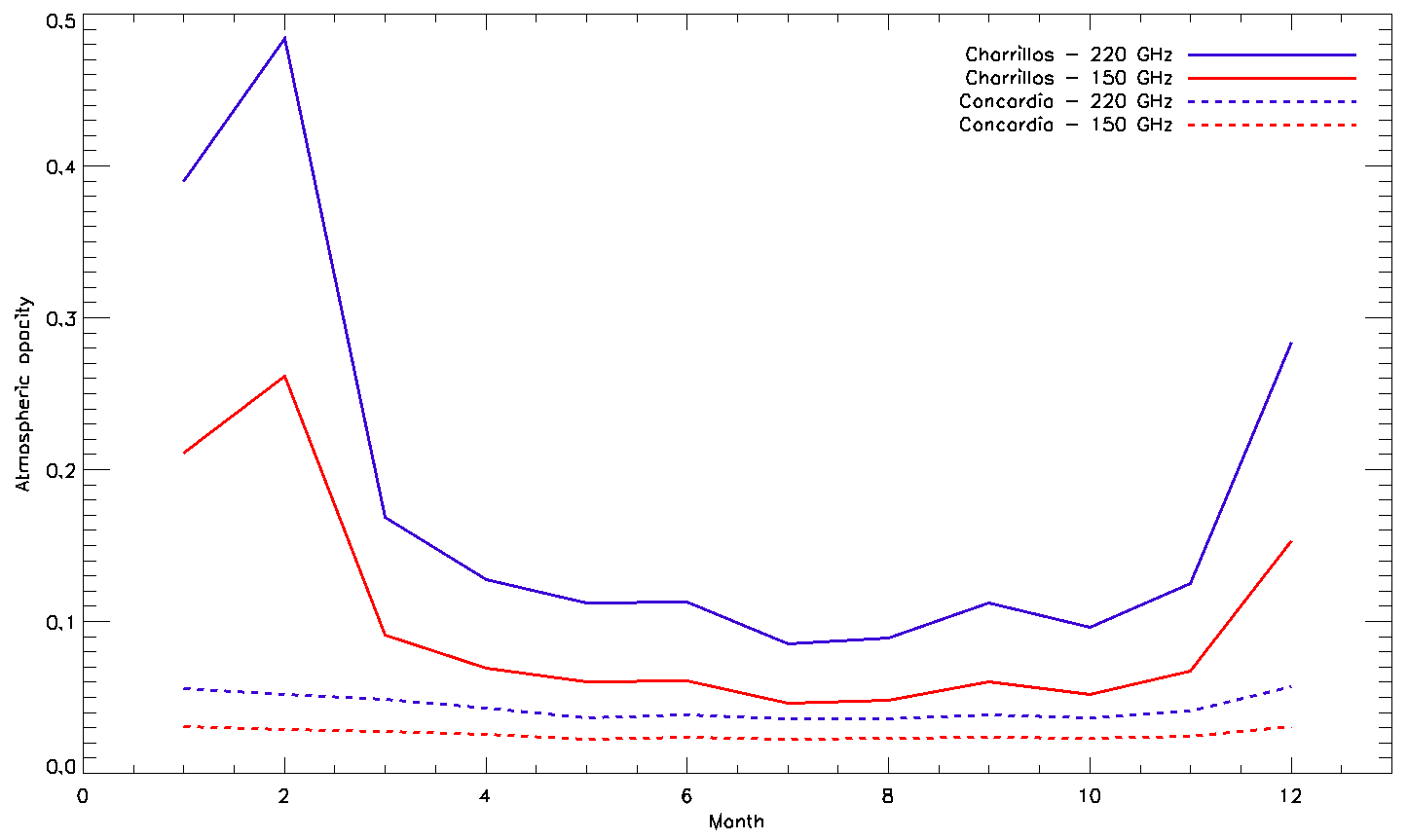}
\caption{Atmospheric Opacity at 150 and 220 GHz derived at the LLAMA site over more than three years.\label{fig104} }
\end{figure}

\begin{itemize}
\item For Concordia, Antarctica: 

\begin{itemize}
\item Precipitable Water Vapour (PWV) data in Concordia are derived in~\cite{Battistelli}, direct Radiosoundings data provided by the Routine Meteorological Observations Research Project
at Concordia station,
employed to feed Atmospheric Transmission at Microwaves (ATM) code to generate synthetic spectra as in~\cite{Gregori} and balloon soundings as in~\cite{Puddu}.

%Precipitable Water Vapour (PWV) data directly obtained from balloon soundings in Concordia~\cite{puddu}.
%\todocomment{Marco's estimate seem more optimistic. Shell we include this, maybe smoothing
%them?}

\item When needed, conversion from PWV to atmospheric opacity (usually labeled $\tau $) at 150 GHz using the
modeling performed for Concordia in~\cite{Battistelli} (eq. 5).
\item The atmospheric temperatures in Concordia were obtained along the year using the data
published in~\cite{Aristidi}.
\item When needed, conversion from $\tau $(150 GHz) to $\tau $(220 GHz) using an atmospheric emission spectrum
from the ATM code. 
Precipitable Water Vapour (PWV) data directly and from direct analysis of Radiosoundings data provided by the Routine Meteorological Observations Research Project
at Concordia station, corrected by temperature and humidity errors and dry biases and then
employed to feed Atmospheric Transmission at Microwaves (ATM) code to generate synthetic spectra in the wide spectral range from 100 GHz to 2 THz as in~\cite{Gregori}.

%Unfortunately, this spectrum was obtained under «Chajnantor» configuration (hence Chile, not
%Antarctica) due to the absence of an Antarctica configuration in the code. The result we will get will therefore
%slightly pessimistic at 220 GHz for Concordia. We expect the effect to be less than 10\%.

\end{itemize}
\item For San Antonio de los Cobres, Argentina:

\begin{itemize}
\item $\tau $(210 GHz) measurements performed for the LLAMA site testing over more than three years kindly transmitted by
Marcello Arnal (PI of LLAMA, from IAR, La Plata).
\item 
Extrapolation to $\tau $(150 GHz) from $\tau $(220 GHz)
using an atmospheric emission spectrum from the
ATM code
%using the same ATM atmospheric model as
%above, 
optimized for Chajnantor (see Figure~\ref{fig103b}). Final results are shown on Figure~\ref{fig104}.
%It is therefore expected to be accurate.

\end{itemize}
\end{itemize}

We combine all the results together to obtain a compared atmospheric emissivity along the year
for the two sites, at both of our observations frequencies. This is shown in Figure~\ref{fig105}. 

\begin{figure}
\centering
\includegraphics[width=14.049cm,height=7.601cm]{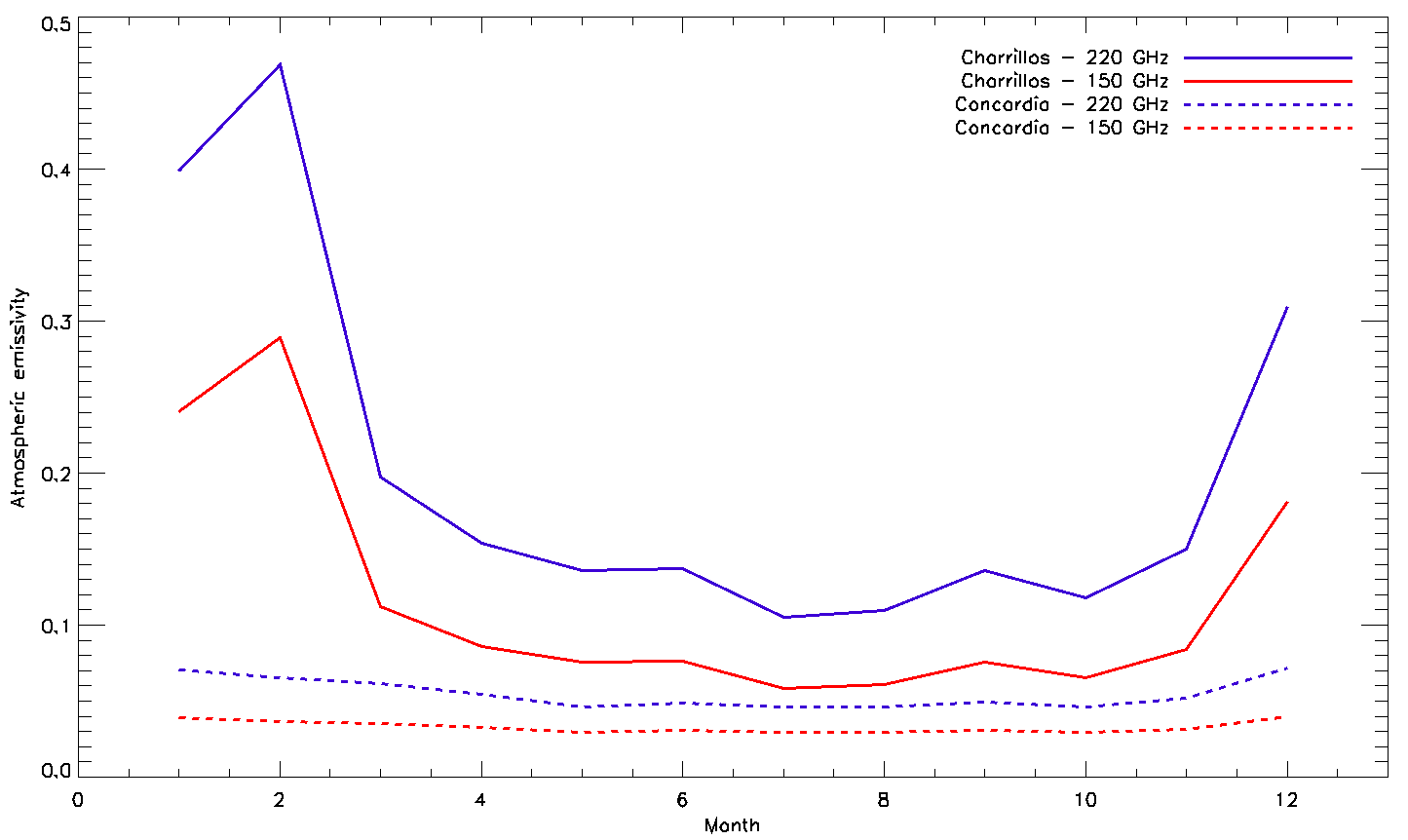}
\caption{ Atmospheric
emissivities compared for both Concordia and Argentinian sites.\label{fig105}}
\end{figure}

The atmospheric emissivity, along with the temperature of the atmosphere along the year allows
to forecast the background radiation arriving from the atmosphere on the QUBIC focal plane, it is straightforward to
calculate the photon Noise Equivalent Power (NEP) on the detectors from this quantity. The actual noise of the detector
is the quadratic sum of this photon noise and of the intrinsic noise of the detectors. Once the total NEP is
calculated, it is straightforward to convert it into Noise Equivalent Temperature (NET) as shown in Figure~\ref{fig106}.

%\todocomment{Was the atmosphere temperature included in this calculation? It is surprising that
%(more than) twice the emissivity only gives a slightly higher NET}

\begin{figure}
\centering
 \includegraphics[width=15.963cm,height=9.79cm]{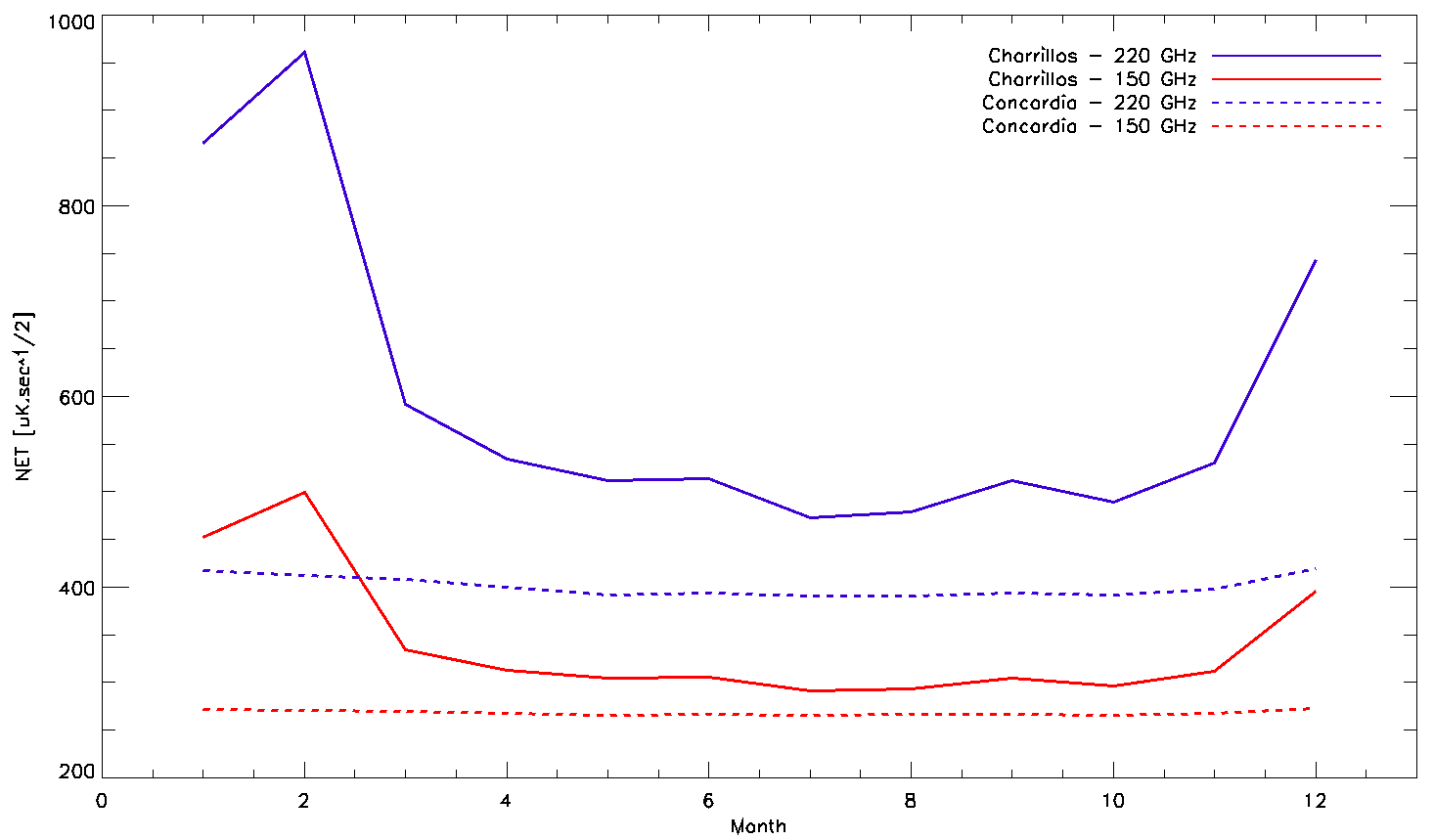} 
\caption{Noise Equivalent Temperatures forecasted from our
study at 150 and 220 GHz for QUBIC in both possible sites.\label{fig106}}
\end{figure}

Unsurprisingly, Concordia offers the highest sensitivity with an average polarized NET at 150
GHz of {\textbf{\ }} 278 $\mu
$K.sec{\textsuperscript{1/2}} and 461 $\mu
$K.sec{\textsuperscript{1/2 }} at 220 GHz while the numbers are 305
$\mu $K.sec \textsuperscript{1/2}} at 150 GHz and 552 $\mu
$K.sec \textsuperscript{1/2 } at 220 GHz in Argentina for the best
nine months in the year. During the so-called Bolivian Summer, the atmospheric conditions are strongly
degraded in Argentina.

In addition, atmospheric conditions in Dome C have been demonstrated to be more stable. This
is not accounted in the present analysis.   %\todocomment{citation needed}

In order to accurately assess the relative sensitivity of the two sites, one therefore needs to precisely account for
the observation efficiency that is going to be rather different in the two sites:

\begin{itemize}
\item 
The target fields are only visible 40\% of the time in Argentina, while 100\% of the time in Antarctica. Note that
self-calibration can be performed during the periods where the target field is not visible from Argentina.
\item 
The fridge cooling the detectors to 320mK needs to be recycled for 4 hours every 24 hours wherever the instrument is
located.
\item 
During the Bolivian Summer (three months), it appears difficult to make useful observations,
however these months will be chosen for doing upgrades and maintenance of the instrument. Similarly, but to a lesser
extent, the strong activity in Concordia during at least one month in summer makes it hard to pursue useful observations, this
time will be better used with upgrades and maintenance of the instrument.
\end{itemize}

\begin{table}
\centering
\begin{tabular}{|m{5.616cm}|m{1.0009999cm}m{1.0009999cm}m{1.0009999cm}|m{1.0009999cm}m{1.0009999cm}m{1.028cm}|}
\hline
~
 &
\multicolumn{3}{m{3.4029999cm}|}{{\bfseries Concordia}} &
\multicolumn{3}{m{3.43cm}|}{{\bfseries Chorillos}}\\\hline
{\bfseries Bad months} &
\multicolumn{3}{m{3.4029999cm}|}{{ 1}} &
\multicolumn{3}{m{3.43cm}|}{{ 3}}\\\hline
{\bfseries Monthly Observation efficiency} &
\multicolumn{3}{m{3.4029999cm}|}{{ 92~\%}} &
\multicolumn{3}{m{3.43cm}|}{{ 75~\%}}\\\hline
{\bfseries Hours below 30 deg.} &
\multicolumn{3}{m{3.4029999cm}|}{{ 0}} &
\multicolumn{3}{m{3.43cm}|}{{ 14.4}}\\\hline
{\bfseries Hours Fridge Cycling} &
\multicolumn{3}{m{3.4029999cm}|}{{ 4}} &
\multicolumn{3}{m{3.43cm}|}{{ 4}}\\\hline
{\bfseries Hours Self Calibration} &
\multicolumn{1}{m{1.0009999cm}|}{{ 0}} &
\multicolumn{1}{m{1.0009999cm}|}{{ 6}} &
{ 12} &
\multicolumn{1}{m{1.0009999cm}|}{{ 0}} &
\multicolumn{1}{m{1.0009999cm}|}{{ 6}} &
{ 12}\\\hline
{\bfseries Hours Field Observation} &
\multicolumn{1}{m{1.0009999cm}|}{{ 20}} &
\multicolumn{1}{m{1.0009999cm}|}{{ 14}} &
{ 8} &
\multicolumn{1}{m{1.0009999cm}|}{{ 9.6}} &
\multicolumn{1}{m{1.0009999cm}|}{{ 9.6}} &
{ 8}\\\hline
{\bfseries Daily Observation Efficiency} &
\multicolumn{1}{m{1.0009999cm}|}{{ 83~\%}} &
\multicolumn{1}{m{1.0009999cm}|}{{ 58~\%}} &
{ 33~\%} &
\multicolumn{1}{m{1.0009999cm}|}{{ 40~\%}} &
\multicolumn{1}{m{1.0009999cm}|}{{ 40~\%}} &
{ 33~\%}\\\hline
{\bfseries Total Observation Efficiency} &
\multicolumn{1}{m{1.0009999cm}|}{{ 76~\%}} &
\multicolumn{1}{m{1.0009999cm}|}{{ 53~\%}} &
{ 28~\%} &
\multicolumn{1}{m{1.0009999cm}|}{{ 31~\%}} &
\multicolumn{1}{m{1.0009999cm}|}{{ 30~\%}} &
{ 25~\%}\\\hline
\end{tabular}
\caption{Relative observation efficiency for the two sites\label{tab:obseffcmp}}

\end{table}

These considerations on the observation efficiency are summarized in Table \ref{tab:obseffcmp} with various
hypotheses regarding the amount of time one will spend performing self-calibration.
 Assuming a negligible time spent self-calibration, the observation efficiency in Concordia is more than
twice higher than in Argentina, while with a large amount of self-calibration, both sites achieve the same observation
efficiency. This occurs because we manage to perform self-calibration when the target field is below the observation
limit.

\begin{figure}
\centering
\includegraphics[width=11.846cm,height=6.883cm]{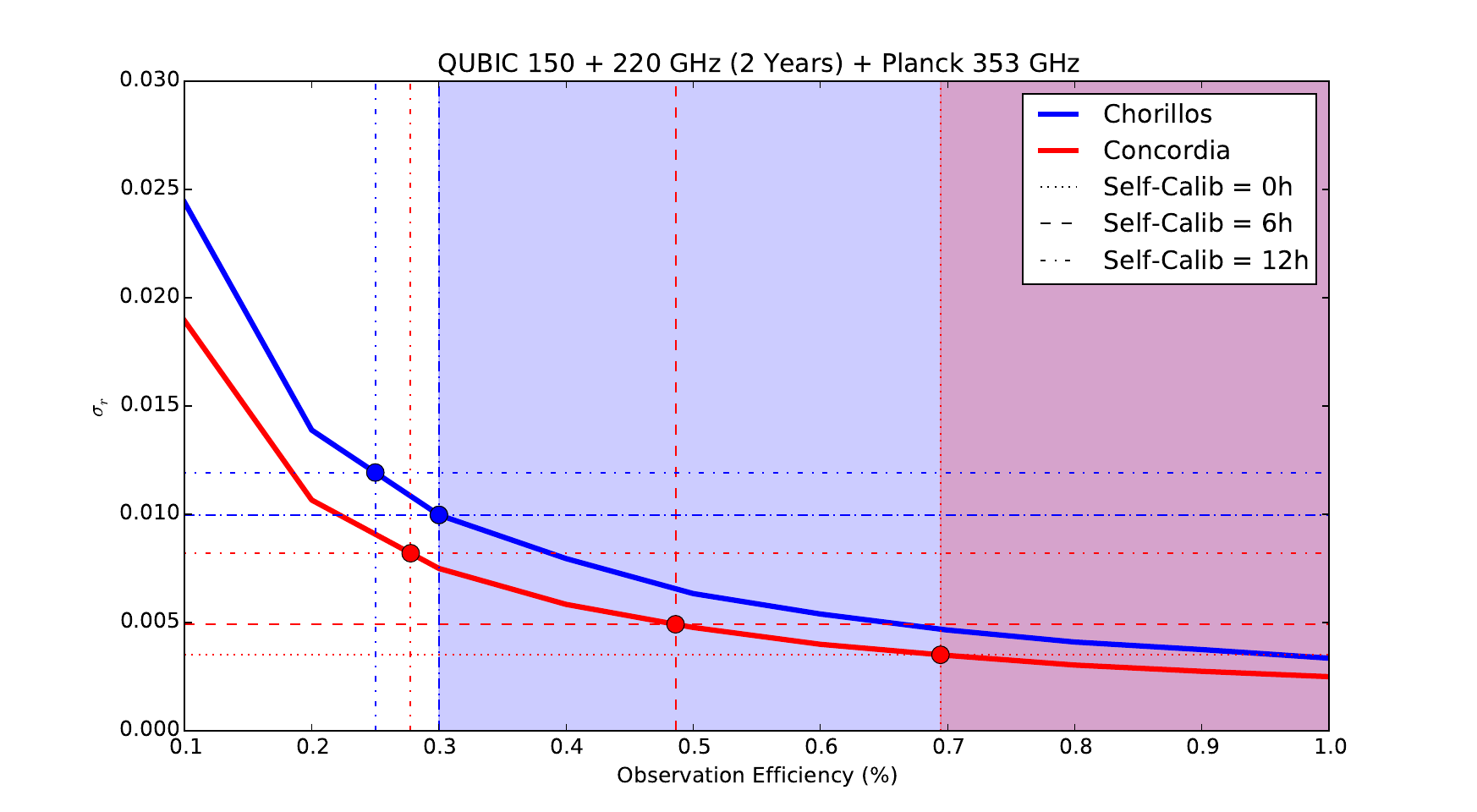}
\caption{Relative sensitivity of the two sites as a
function of the observation efficiency\label{fig107}. }
\end{figure}

These results and the Noise Equivalent Temperatures from Figure~\ref{fig106} are then used to feed the
sensitivity forecast program that was used to predict sensitivities for QUBIC (in section \ref{bkm:Ref311965747}). We
run this program for various values of the observation efficiency from 0 to 1 and obtain the corresponding minimum
tensor-to-scalar ratio achievable in two years for the two different sites. The results are shown in Figure~\ref{fig107}.%\todocomment{Elia:replace this with the latest MCMC simulation}. 
We see that installing QUBIC in Argentina will lead to a reduction of
sensitivity of a factor between 1.4 and 2.8 wrt Dome C, for 0h and 12h per day spent doing  self-calibration.% \todocomment{1.3 and 2.8  : these are the ratios with 12h and 0h of calibration, please check the numbers on the latest MCMC simulation}.

%{
%{\textit{I think the number to compare are those with 0.7 efficiency in Concordia and 0.3
%efficiency in Argentina. \ }}}

%auto-ignore

\subsection{San Antonio de los Cobres, Argentina}
\label{argentina}
As mentioned earlier, 
we have chosen to install the  QUBIC first module in Argentina, near the city of San Antonio de los Cobres, in the Salta Province. This site has coordinates (24\degree\ 11' 11.7'' S; 66\degree\  28' 40.8'' W ) and an altitude of 4869m a.s.l. It is located in the plateau known as ``La Puna'' in Argentina, and ``Atacama'' in Chile. The site is located  $\sim$180km from the Chajnantor site where other millimeter-wave experiments are located (ALMA, ACTPol, PolarBear) and offers similar atmospheric properties (see Fig.\ref{fig100}).

\begin{figure}[!h]
\centering
\includegraphics[width=7.cm,height=6.6cm]{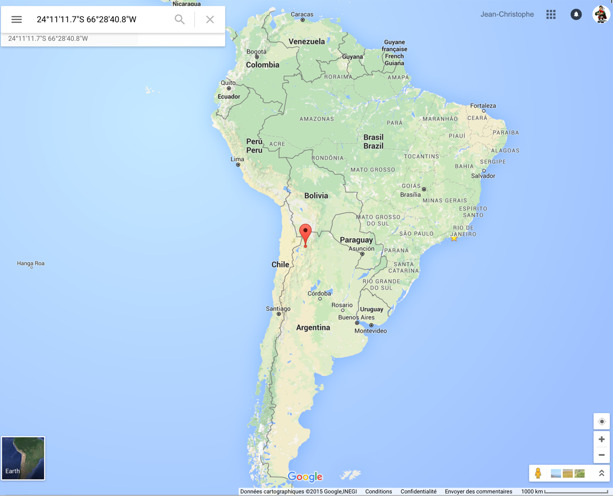}
\includegraphics[width=8.2cm,height=6.6cm]{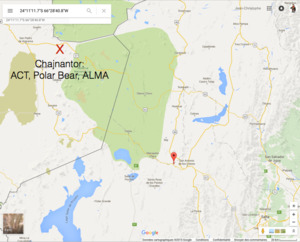} 

\caption{ Alto Chorillos site in Argentina, near San Antonio de los Cobres, province of Salta. The position for other astronomical instruments location in Chile is also shown (red cross). \label{fig100}}
\end{figure}

In Figure \ref{fig101} we can see the position of the chosen site for QUBIC along with  the LLAMA position. LLAMA is a project for the installation of an ALMA-like antenna  lead by an Argentina-Brazilian collaboration\footnote{http://www.iar.unlp.edu.ar/llama-web}. This project had conducted the site characterization studies, in particular related to atmospheric opacity shown in Section ~\ref{siteComparison}. In Figure ~\ref{fig101} is posible to see the relative position of QUBIC and LLAMA. The magenta polygon is the 400 hectares area allocated by the government of the Salta province to CONICET for scientific use only. The corresponding decree can be found in the official web site of the government: \url{http://boletinoficialsalta.gob.ar/NewDetalleDecreto.php?nro_decreto=824/14}. In this figure we can also see the gas pipeline (green line) that will feed the gas generators for LLAMA and the Vega lagoon, from which the water needed for both instruments can be extracted.  A high-speed internet connection to San Antonio de los Cobres will be available soon thanks to the installation of an optical fiber in the near future. 

\begin{figure}[!h]
\centering
\includegraphics[width=7.939cm,height=5.239cm]{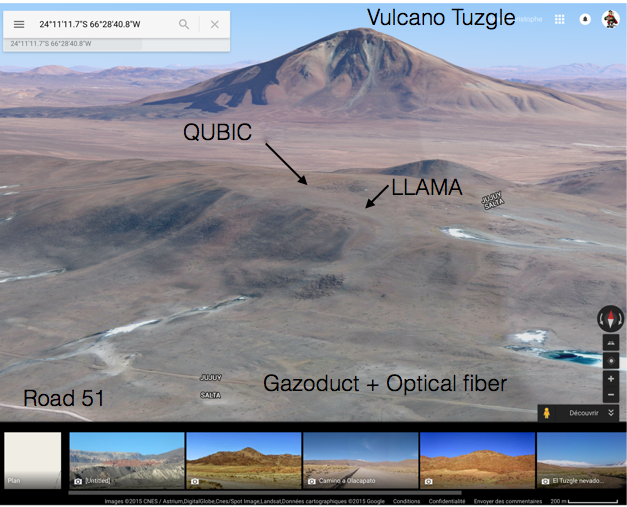} 
\includegraphics[width=7.606cm,height=5.209cm]{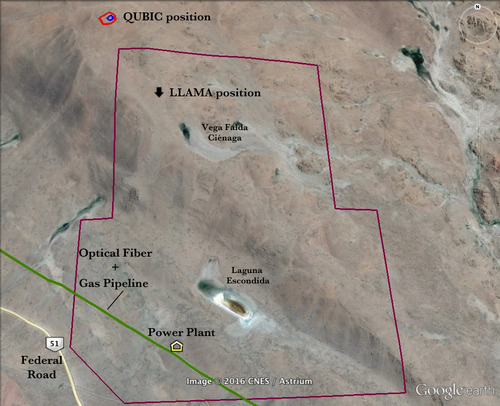} 

\caption{ Views of the chosen QUBIC site in Argentina along with the LLAMA~\cite{LLAMA} location and logistic installations planned to be installed during 2017 and funded by MinCyT and the Province of Salta. In magenta the limits of the allocated area for scientific use from the Salta government.\label{fig101}}
\end{figure}

Even if the chosen position for the instrument is outside the initially scientific allocated area, QUBIC could always benefit from the LLAMA installation associated logistics (i.~e. access road, electrical power, water and internet network) already funded and expected to be constructed during 2017. The schedule for this infrastructure to be ready is the following : 

\begin{itemize}
	\item Electricity, connectivity		$\longrightarrow$		  April 2017
	\item Access road 			  	$\longrightarrow$ 		  July 2017
	\item First Stage of Headquarters   	$\longrightarrow$    	December 2016. 
\end{itemize}

The LLAMA project has also offered the possibility of using the headquarters they have planned in the city for installing the data storage and control center for QUBIC (50 m$^2$ granted), as well as bedrooms, access to the workshop and a clean room. Dr. Marcelo Arnal, Argentinian PI of the LLAMA project, has expressed his agreement to the installation of QUBIC on the same grounds. 

The shipping of the instrument and the access to it will be clearly easier than in the original antarctic choice, as Chorillos site is only a 45 minutes drive away from San Antonio de los Cobres, and at a 3.5 hours drive from Salta, the capital of the province, where a large airport and university facilities are available. A smaller airport is also directly available in San Antonio de los Cobres. Access to the site is granted 365 days a year with less than 24 hours trip from Europe. In San Antonio de los Cobres are hotels and the basic services (i.~e. hospital, bank, restaurants, gas station, shops). 

Regarding  the general atmospheric conditions, in addition to atmospheric opacity, already studied in section \ref{siteComparison}, temperature, humidity and wind speed have been monitored on site for several years. Except during the Bolivian Summer (December to March period) the values of these parameters are within the specifications for smooth operation (see Figure~\ref{fig102}). In order to confirm the values taken in LLAMA site and to monitor the atmospheric conditions during at least one year, a weather station will be installed on QUBIC site in the next months. 

\begin{figure}[!h]
\centering
\includegraphics[width=16.3cm,height=12.2cm]{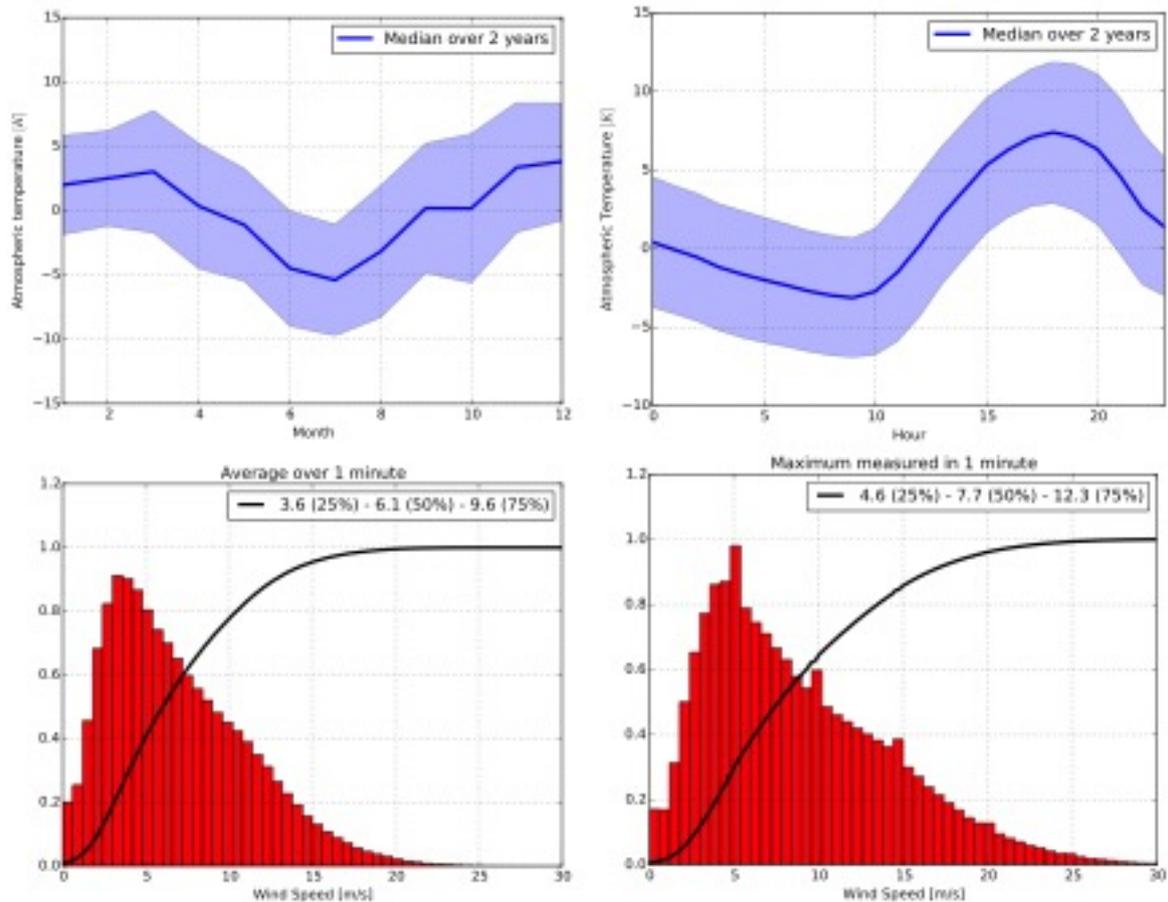} 

\caption{ Top, left: Annual Temperature variation at Alto Chorillos site. Top, right: Daily temperature distribution. Bottom, left: 1-minute average wind speed distribution. Bottom, right: Maximum wind speed in 1-minute distribution.\label{fig102}}
\end{figure}

The installation of QUBIC in Argentina is strongly supported by the National Ministry of Science and Technology (MINCyT), the government of Salta Province, the National Scientific and Technical Research Council (CONICET) and National Commission of Atomic Energy (CNEA). Researchers from different Argentinean institutions are involved in QUBIC now, participating not only on site development but also on instrumental aspects. The  Qubic-Argentina Collaboration is formed by researchers from the Argentinean Institute of Radioastronomy (IAR), the Institute of Detection Technologies and Astroparticles (ITeDA), the Bariloche Atomic Center (CAB) and the Department of Aeronautics and the Faculty of Astronomy and Geophysics  of the University of La Plata (FCAGLP, UNLP).

At this moment, 500,000 USD has been already allocated by MINCyT for the first stage of QUBIC installation and negotiations between the different Argentinean agencies supporting QUBIC in Argentina are carrying on in order to define the global structure of the collaboration and the contribution from each party to the project.

\section{Organisation}
%auto-ignore

\subsection{Management}
\subsubsection{Management of the collaboration}

The organization of the QUBIC collaboration is based on two main governing bodies: the QUBIC Steering Committee and the
QUBIC Project Office. The Steering Committee is in charge of deciding crucial orientations and decisions; the Project
Office is in charge of the implementation into the project of these decisions.
The Steering Committee is composed according to the track record of the QUBIC project with representatives from
participating countries (France, Italy, United Kingdom, Ireland, United States of America) and the managing executives
of the project (the Project Scientist, Project Manager, Instrument Scientist and Logistic Coordinator). 
The Steering Committee is the governing body choosing between the most crucial options during the life of the project:
funding, schedule, introduction or exit of new partners, Intellectual Property issues, release of general
specification, general strategy, choice of key personnel, publication policy.

The Project Office is in charge of concretely executing the decisions of the Steering Committee. It is the executing
governing body of the project. It is composed with the Project Scientist, the Project Manager, the Instrument
Scientist, the System Scientist or Engineer, the Logistics Coordinator and the CAD and Mechanical Architect.
 The Project Office takes the day-to-day technical and managerial decisions and coordinates  with the
involved laboratories the implementation of its own decisions and of those decided by the Steering Committee.

The institutions participating to the QUBIC Collaboration, and their respective responsabilities (if any, besides scientific exploitation) are described in table ~\ref{tablerespo}.

\begin{table}
\centering
{\scriptsize
\begin{tabular}{|l|l|l|}
\hline
France	& CNRS / IN2P3 / APC &	Leadership, technical coordination and management \\
&  & Detection chain system\\
&  & AIT activities\\
&  & Calibration system\\
&  & Instrument design\\
&  & Science simulations\\
&  & Mechanical design, assembly and test of RF switches\\
&  & Sub-K control electronics \\
 &  	CNRS / IN2P3 / CSNSM	& TES detectors \\
 & 	CNRS / IN2P3 / LAL & 	Calibration and test of the instrument\\
&  & Carbon fiber sources \\
 & 	CNRS / INSU / IRAP & 	Slow control\\
&  & Readout software\\
&  & Data storage \\
 & 	CNRS / INSU / IAS & 	- \\
\hline
Italy & 	Università di Roma La Sapienza	 & Optical group coordination\\
&  & Cryostat\\
&  & Cryostat control electronics\\
&  & Cryogenics \\
&  & Cryostat window\\
&  & HWP rotation system\\
&  & Full beam source\\
&  & Baffling\\
&  & Test of mirrors \\
 & 	Universita degli Studi di Milano & 	Horns fabrication \\
 & 	Universita degli Studi di Milano-Bicocca & 	Fabrication, slow control of RF switches\\
&  & Test of horns\\
&  & Fabrication of mirrors \\
\hline
U.K. & 	Cardiff University & 	RF design of switches\\
&  & Half Wave Plate\\
&  & Polarizer\\
&  & Dichroïc\\
&  & Filters \\
 &	Manchester University / JB Technology Group &	RF design of horns\\
 &	Manchester University / Advanced  &  	1K and Sub-K fridges \\
& \ \ \ \ \ Technology Team & Heat-switches \\
\hline
Ireland	& National University of Ireland, Maynooth &	 Optical simulations of the instrument \\
\hline
USA &	Wisconsin University & 	-  \\
 &	Brown University &	- \\
 &	Richmond University &	- \\
\hline
Argentina & IAR La Plata 				& Logistics - Site Development \\
 & ITEDA Buenos Aires 		& Logistics - Site Development \\
 & ITEDA Mendoza 		& Logistics - Site Development \\
 & CNEA CAB (Bariloche)		& Cryogenic maintenance\\
 & GEMA La Plata University 	& Mount design and fabrication \\
\hline
\end{tabular}
}
\caption{Main characteristics of Institutes in the QUBIC collaboration \label{tablerespo}}
\end{table}

\subsubsection{Collaboration Agreement (CA) / Memorandum of Understanding (MoU)}
In April 2015 a Collaboration Agreement has been signed between the members of the QUBIC Collaboration. 
This Collaboration Agreement is focused on the construction and implementation of the first module of the QUBIC
experiment. It deals with the description of the collaboration, the detailed commitments of each stakeholders, the
global view of the funding situation, the schedule, the organization of the consortium, the exploitation of the
instrument, the access to scientific data, the publication policy.

A Memorandum of Understanding (MoU) more focused on commitments on funding and manpower will be written and signed as a
second step in the formalization of the QUBIC collaboration.

\subsubsection{Publication policy}

\textit{Research Projects }}{undertaken in the collaboration must be
announced to the collaboration through the mailing list specifying the topic, project leader, and known collaborators.
The list of \textit{Research Projects }will be maintained by the
\textit{Project Scientist }and made available on the
Collaboration internal website. Any member of the QUBIC collaboration is entitled to work on any of the
\textit{Research Projects } undertaken. 

Publications are expected to be the result of these
\textit{Research Projects }and need to be reviewed within the
collaboration at least a month before being submitted to a journal.

Authorship of the publications: 

\begin{itemize}
\item {
{Any member of the QUBIC collaboration can request that his/her name be added to the list of
author of an article. It is the responsibility of the }{\textit{Research Project
}}{leader to accept or not this request depending on his contribution to this specific
research project. The ordering of the list of authors is left to the choice of the specific
}{\textit{Research Project }}{members. }}
\item {
{QUBIC members having the }{\textit{Architect
}}{status have the right to choose to be co-author of any of the publications. The
}{\textit{Architect }}{status is granted by the
}{\textit{Steering Committee}}{ based on significant contributions to
any stage of the experiment (design, construction, integration, operation or analysis). }}
\item {
The default policy for signature of papers is: first the main authors, then other involved people in alphabetic order.
}
\end{itemize}

Any disagreement should be resolved through the }{\textit{Steering
Committee }}{whose decision would be final.

In addition to the \textit{Research
Projects}, The QUBIC \textit{Project
Scientist}, in agreement with the \textit{Steering
Committee}, may establish \textit{Key Projects
}and \textit{Working Groups }whose
work could also result in a publication. The authorship for such projects will be discussed in the
\textit{Steering Committee }and would by default be the whole
QUBIC collaboration in alphabetical order.

\subsection{Organization }

\subsubsection{Product
Breakdown Structure (PBS)}

The QUBIC system is structured as in Table~\ref{table34}.
\begin{table}
\centering
{\scriptsize
\begin{tabular}{|m{5.242cm}|m{4.801cm}|m{5.3190002cm}|}
\hline
{\bfseries Sub-system} &
{\bfseries Component} &
{\bfseries Items}\\\hline
{\bfseries Detection chain} &
{ Focal plane} &
{ TES}\\\hline
~
 &
~
 &
{ Mechanical support}\\\hline
~
 &
{ Cold electronics} &
{ ASIC, SQUID, Wiring}\\\hline
~
% &
%~
% &
%{ SQUID}\\\hline
%~
% &
%~
% &
%{ Wiring}\\\hline
~
 &
{ Warm electronics} &
{ Amplifier,  FPGA board}\\\hline
%~
% &
%~
% &
%{ FPGA board}\\\hline
{\bfseries Optical system} &
{ Window} &
~
\\\hline
~
 &
{ Horns / switches array} &
{ Horns, Switches, Mechanics,}\\%\hline
%~
% &
%~
% &
%{ Switches}\\\hline
%~
% &
%~
% &
%{ Mechanics}\\\hline
~
 &
~
 &
{ Slow control and readout }\\\hline
~
 &
{ Filters } &
{ Thermal filters,  Band defining filters, Half wave plate, Dichroic }\\\hline
%~
% &
%~
% &
%{ Band defining filters}\\\hline
%~
% &
%~
% &
%{ Half wave plate, Dichroic}\\\hline
%~
% &
%~
% &
%{ Dichroic}\\\hline
~
 &
{ Mirrors} &
{ Primary, Secondary, Hexapods}\\\hline
%~
% &
%~
% &
%{ Secondary mirror}\\\hline
%~
% &
%~
% &
%{ Hexapods}\\\hline
~
 &
{ Cold stop box} &
{ 1K Box, Optical covers}\\\hline
%~
% &
%~
% &
%{ Optical covers}\\\hline
~
 &
{ External baffling} &
{ Ground shield, Fore baffle}\\\hline
%~
% &
%~
% &
%{ Fore baffle}\\\hline
{\bfseries Cryostat system} &
{ Vacuum vessel} &
~
\\\hline
~
 &
{ Thermal screens} &
~
\\\hline
~
 &
{ Cryogenics} &
{ 4K fridge, 1K fridge, 320mK fridges, Heat switches}\\\hline
%~
% &
%~
% &
%{ 1K fridge}\\\hline
%~
% &
%~
% &
%{ 300mK fridges}\\\hline
%~
% &
%~
% &
%{ Heat switches}\\\hline
~
 &
{ HWP rotator} &
~
\\\hline
~
 &
{ Slow control} &
~
\\\hline
{\bfseries Mount system} &
{ Mechanics } &
~
\\\hline
~
 &
{ Motors } &
~
\\\hline
~
 &
{ Slow control} &
~
\\\hline
~
 &
{ Cables guide} &
~
\\\hline
{\bfseries Infrastructure} &
{ Laboratory} &
~
\\\hline
~
 &
{ Service shelter} &
~
\\\hline
~
 &
{ Instrument shelter} &
~
\\\hline
{\bfseries Control, readout, data storage} &
{ QUBIC-Studio software} &
~
\\\hline
~
 &
{ Control computer} &
~
\\\hline
~
 &
{ Data storage} &
{ RAID system, Back-up system}\\\hline
%~
% &
%~
% &
%{ Back-up system}\\\hline
~
 &
{ External wiring} &
~
\\\hline
~
 &
{ Science simulation software} &
~
\\\hline
~
 &
{ Data treatment software} &
~
\\\hline
~
 &
{ Self calibration software} &
~
\\\hline
~
 &
{ Computing hardware system} &
~
\\\hline
{\bfseries Calibration system} &
{ Calibration sources} &
~
\\\hline
~
 &
{ Calibration source support} &
~
\\\hline
~
 &
{ Full beam source} &
~
\\\hline
~
 &
{Carbon fiber sources} &
~
\\\hline
 &
{ Slow control} &
~
\\\hline
{\bfseries AIT system} &
{ Electronics test system} &
~
\\\hline
~
 &
{ Optical test system} &
~
\\\hline
~
 &
{ Thermal test system} &
~
\\\hline
~
 &
{ Assembly system} &
~
\\\hline
{\bfseries Logistics systems} &
{ Handling system} &
~
\\\hline
~
 &
{ Containers} &
~
\\\hline
\end{tabular}
}
\caption{PBS of the QUBIC project\label{table34}. when present, comma separated lists in the {\bf Items} column indicate separate products.}

\end{table}

\subsubsection{Works Breakdown Structure / Works packages}

The different tasks for design, fabrication, tests for each component of QUBIC are described in the chart shown on Figure~\ref{fig113}.

Otherwise explicitly stated, a component is designed, fabricated and tested by the same laboratory.

Besides activities for providing the components parts of QUBIC, integration activities, under the responsibility of APC,
have been integrated in the chart. 
The global management and coordination of the experiment is also under the responsibility of APC.

Each laboratory is represented by its own color code in the graph.

\begin{figure}
\centering
\includegraphics[width=.95\textwidth]{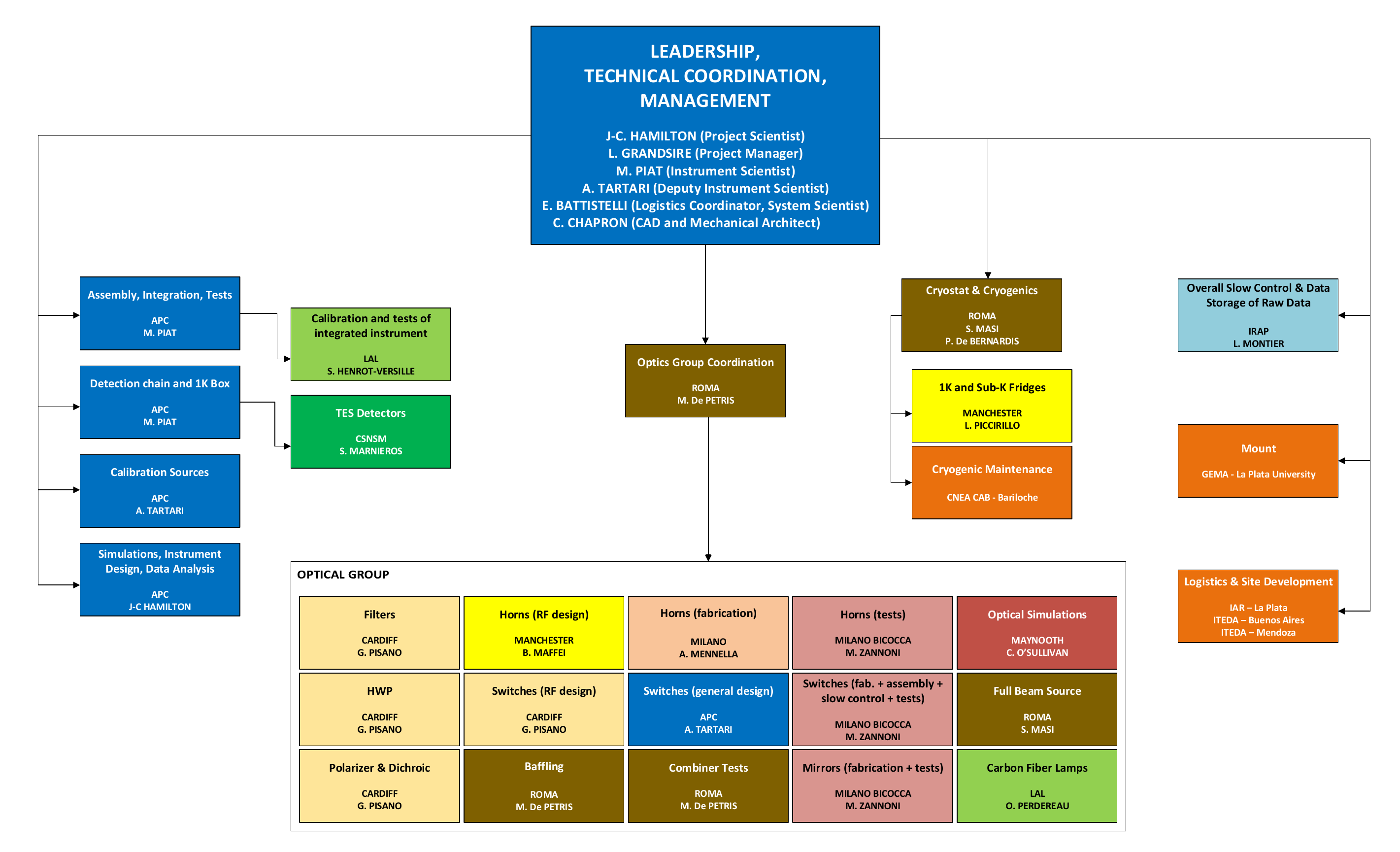} 

\caption{Work Breakdwon Structure of QUBIC \label{fig113}}
\end{figure}

%auto-ignore
\subsection{Development plan}
\label{devplan_instr}
\hypertarget{RefHeadingToc314323003}{}{
{The development of the QUBIC instrument is a multi-step process. After i) the first R\&D
works, it was decided in a review held in 2013 to ii) validate the detection chain. Now this validation is almost done.
The next steps are iii) the validation of a technological demonstrator, and then iv) the construction of the full first
module of the instrument. The final step will be v) the fabrication and implementation of the five additional
modules.}}

\subsubsection{Heritage and first R\&D works }
\hypertarget{RefHeadingToc314323004}{}{
{Since 2008, many Research and Development works on the following topics have path the way to
the construction of QUBIC.}}

\begin{itemize}
\item { Proof of principle concept of bolometric interferometry:}
\end{itemize}
{
{MBI-4 and BRAIN have been successful demonstrator of the concept of bolometric interferometry.
}}

{
The merge of this two projects and their respective teams gave birth to the QUBIC project.}

\begin{itemize}
\item {
Exploitation site:}
\end{itemize}
{
{BRAIN has been successfully installed and exploited in the Concordia Station in Antarctica in
2006. This implantation has been of great value with respect to the exploitation of such an instrument and its main
sub-systems (such as the pulse tube cryo-coolers) in the harsh conditions of Antarctica.}}

{
{Besides this installation measurements have been made on the quality of the sky at Concordia
Station by Dr Alessandro Schillaci in 2010 and 2011.}}

\begin{itemize}
\item {
Detectors:}
\end{itemize}
{
{The BSD collaboration, granted by an ANR, is a collaboration between the laboratories involved
in QUBIC detectors development (APC, CSNSM). By establishing this collaboration, these laboratories have been able to
procure equipment and materials for production or test of TES arrays.}}

{
{One of the first result of this collaboration and of first R\&D was the production of a 23
pixels TES array in 2011.}}

\begin{itemize}
\item {
Switches:}
\end{itemize}
{
{First tests were made in 2012 on a single switch. Integration of the electro-magnet with the
blade that close the wave guide, good functionality in liquid nitrogen and in 4K cryostat of the unit, development of
the electronic system able to perform the switches and detect their position, first measurements of the cut-off when the blade
close the waveguide were the main results of this R\&D.}}

\begin{itemize}
\item {
Horns:}
\end{itemize}
{
{First developments were done on the RF design of the corrugated horns at Manchester in 2010.
\ Parallel developments were undertaken in APC on the mechanical design and fabrication of such horns. Use of platelets
(instead of electroforming) was investigated in order to decrease costs of fabrication. In 2012 Milano team of Marco
Bersanelli, who provides the platelets horns for the Planck \ satellite joins the QUBIC collaboration. Their experience
and industrial networking is of great value for that matter.}}

\begin{itemize}
\item {
Read-out electronics:}
\end{itemize}
{
The heritage from the control and read-out system of the PILOT experiment, designed by IRAP, has been fully integrated
in QUBIC-Studio.}

\subsubsection{Validation of the detection chain}
\hypertarget{RefHeadingToc314323005}{}{
The QUBIC project was reviewed in summer 2013 by IN2P3, INSU and IPEV. The main conclusion of the review group was that
the detection has to be validated before giving the project a formal go-ahead.}

{
The validation of the detection chain was focused on the following issues:}

\begin{itemize}
\item {
Definition of the numerous steps of fabrication for the production of an array of 256 TES,}
\item {
Production and test of this array (checking short-circuit or cut in the routing, control of superconducting transition
temperature, decrease of the percentage of damaged pixels),}
\item {
Finalization of the cold and warm electronic read-out (definition, purchase, fabrication and tests of SQUIDs, ASICs,
PCBs, superconducting and cryogenics cables and connectors, warm amplifier, warm electronic PCB with embedded FPGA,
software system QUBIC-Studio for control and read-out),}
\item {
Mechanical, thermal and electrical integration of all these elements into a dilution-cooled test cryostat at APC,}
\item {
Functional tests of this detection chain (measurement of Tc, ability of reading transition on the system),}
\item {
Characterization of the TES (measurement of V(I) curves)}
\item {
Measurement of the noise level of the system.}
\end{itemize}

%{
%{\bf All of these points are now validated, with the exception of the very last one. To validate
%this last point, the sampling rate needs to be increased. In order to do that, we need to implement numerical FLL in
%the FPGA and develop a new read-out PCB. These modifications should be implemented in early 2016 and be concluded just
%afterwards by a new campaign of test. }}
%\todocomment{update this -> FLL etc are improved !}

%{
%In the spring of 2016, the detection chain of the QUBIC has been  technically validated. }

All of these steps have been successfully performed. The noise level measurement has been done in spring 2016 by increasing the sampling rate, integrating a numerical FLL in the FPGA using a updated version of the read-out PCB. However, the noise level measured $10^{-16} W/\sqrt{Hz}$.  is a factor two above the $5.10^{-17} W/\sqrt{Hz}$ requirement. This is now understood and is due to noise aliasing. This explanation has been demonstrated by increasing the read-out rate yielding to a reduced noise level. The natural solution is to add superconducting inductances in front of the SQUIDs which is currently being implemented on a new test TES array. The expected noise reduction is at least a factor of two (therefore reaching the requirement) or even better.
The detection chain of QUBIC is therefore technically validated although the ultimate noise will only be reached after the test of the new TES array incorporating the inductances to limit the noise aliasing.

We can already say that this validation process allows us to have a far better understanding of the QUBIC detection
chain. Procurement of parts, efficiency of potential suppliers, cost issues, technical difficulties and
incompatibilities into the industrial process of fabrication, thermal issues are matters that have been greatly
clarified.

\subsubsection{Validation of a technological demonstrator}

In June 2015, a new review assessed the progress in the QUBIC project and especially the status of the detection chain
validation.

A major conclusion of this review was to encourage the fabrication of a demonstrator of the final QUBIC instrument with
the already available funding, before end of 2016. Details on the configuration of this demonstrator are given in Section~\ref{techno}. 
The QUBIC technological demonstrator, when assembled and tested will validate the full instrument design and test it electrically, thermally and
optically. 

The test plam of the technological demonstrator is similar to that of the first module, which is described in Section~\ref{calib}.

\subsubsection{Construction and
implementation of the first module}

The final results of the detection chain will be available in 2016 and the technological
demonstrator of the instrument will be assembled in early 2017 and will undergo first tests during this same year. This way, it is expected
that funding agencies and authorities of each partners will give the go-ahead and funding and manpower for the
fabrication of the final first module (i.e instrument, mount and logistics).

With the additional funding for superconducting and cryogenics electrical cables, for mechanics of focal planes, final
filters and additional pulse-tube, the assembly and tests of the first module are expected to take
place during the second half of 2017 in the assembly hall of APC.

When assembled, the instrument will be tested following the plan described in Section~\ref{calib}.

After all these steps an important milestone will be the acceptance review of the QUBIC first module in order to get the
approval from authorities before shipping it to the exploitation site.

Shipping the first module, the mount and all ancillary materials to the exploitation site 
will take a few months.

Once arrived on site the instrument will be visually inspected in order to check that its integrity has been kept during
the shipping. Then the mount will be installed on its platform. Afterwards the instrument is put on the mount. Once done, the commissioning phase begins.

When the instrument is fully checked in its good health and performances, the scientific operations begin. They are
expected to take two years.

\subsection{Schedule }

\subsubsection{Construction and test of the technological demonstrator}

The fabrication, procurement and test of each sub-systems of the technological demonstrator 
 is expected to be done between october 2016 and the beginning of 2017. 
Tests are expected to start soon afterwards.

\subsubsection{Construction of and test the first module instrument}

The construction of the first module will start as soon as the tests on the technological demonstrator will be finished
and that requested funding to achieve the instrument (cabling, mechanics for focal plane, additional pulse tube) are
delivered in due time by the authorities. The goal is to finish assembly and tests by end 2017.

%auto-ignore

\subsection{Costs and funding}

 \subsubsection{Costs an funding for R\&D and validation of the detection chain}

Main R\&D and cost for construction and test of a prototype detection chain have been funded by the main 
contributions listed in Table~\ref{table36}.

\begin{table}
\begin{tabular}{|m{6.242cm}|m{1.55cm}|m{7.8719997cm}|}
\hline
{\bfseries Agency} &
{\bfseries Grant} &
{\bfseries Funded sub-system }

{\bfseries (hardware, software or temporary manpower)}\\\hline
{ ANR} &
{ 640 k{\texteuro}} &
{ SQUIDs, ASICs, cold and warm electronics, focal plane mechanics, horns and switches,
QUBIC-Studio software, equipment for TES production, temporary manpower, missions.}\\\hline
{ PNRA} &
{ 293 k{\texteuro}} &
{ Cryostat and cryogenics, horns, switches, optics.}\\\hline
{ Labex UnivEarths} &
{ 231 k{\texteuro}} &
{ Manpower, ASIC, misc. material}\\\hline
{ CNRS (INSU, IN2P3, PNCG, CSAA)} &
{ 98 k{\texteuro}} &
{ Missions, superconducting wiring, mechanics, electronics.}\\\hline
\end{tabular}
\caption{Fundings for R\&D and validation of
detection chain\label{table36}
}
\end{table}

\subsubsection{Costs and funding for construction and test of the technological
demonstrator}
The construction of the technological demonstrator is expected to be constructed without any additional funding (at
least for the French contribution).

\subsubsection{Cost and funding for
the construction of the first module}
\label{cost1stmod}

The cost for construction of the first module is likely to be shared between the agencies listed in Table~\ref{table37}.

\begin{table}
\begin{tabular}{|m{3.8cm}|m{4.55cm}|m{6.9cm}|}
\hline
{\bfseries Agency} &
{\bfseries Grant} &
{\bfseries Funded sub-system }

{\bfseries (hardware, software or temporary manpower)}\\\hline
{ DIM-ACAV (France)} &
{ 130 k{\texteuro}} &
{ Calibration tower, 150 GHz calibration source.}\\\hline
{ CNRS (IN2P3, INSU, France)} &
{ 270 k{\texteuro} (under examination)} &
{ Focal plane mechanics, wiring, 220 GHz calibration source.}\\\hline
{ PNRA (Italy)} &
{ 400 k{\texteuro} (under examination, maybe  INFN)} &
{ Pulse tube, cryostat, optics, horns, switches.}\\\hline
{ CNRS (France)} &
{ 60 k{\texteuro}} &
{ {Temporary manpower for AIT at APC}}\\\hline
{MINCYT (Argentina)} &{ 500 k{\$}}

&{ Mount \& logistics}\\\hline
\end{tabular}
\caption{Funding for the construction of the first
instrument\label{table37}}

\end{table}

\subsubsection{Cost and funding for the implementation on site of the first module}
Regarding implementation on site and logistics (buildings and shelters, manpower for
installation of infrastructure, shipping of goods, consumed power, etc.) the agencies in charge of the sites are
expected to take these costs as their participation to the project. 
A rough estimate of these logistics costs can be
read in section \ref{argentina}.%\todocomment{Make sure costs are given in this section !!}

Regarding scientific exploitation, an application has been sent to the ANR 2016 call for that purpose.

\begin{table}
\begin{tabular}{|m{5.215cm}|m{5.215cm}|m{5.2330003cm}|}
\hline
{\bfseries Agency(ies)} &
{\bfseries Grant} &
{\bfseries Funded sub-system }

{\bfseries (hardware, software or temporary manpower)}\\\hline
{ TBD } &
{ 806 k{\texteuro}} &
{ Science exploitation: 8 year of post-doc and manpower, calibration, RAID system, missions for
exploitation.}\\\hline
\end{tabular}
\caption{Estimation of costs for scientific
exploitation\label{table38}.}

\end{table}

\subsection{QUBIC Evolution to a CMB Stage IV experiment}
The technology proposed by QUBIC is innovative and therefore includes a balance of risk (novel technology) and virtues (high sensitivity, higher immunity to systematic effects, one of the main concerns in deep microwave background polarimetry). Therefore, definitive statements on what we envision beyond the first module can only occur after a precise assessment of its results after at least few months of observations. 
We can however depict a roadmap under the assumption that the first modules gives results as predicted and demonstrates the interest of bolometric interferometry.

For such a roadmap, various options can be considered:
\begin{enumerate}
\item Evolving the technology of QUBIC with modules comprising more horns and detectors in order to significantly increase the sensitivity and angular resolution (set by the size of the primary horn array). The final sensitivity would deserve a detailed study that can only be done when the results of the first module will be known.
\item Keeping the size of the first module for future modules operating at a total of three different frequencies (90, 150 and 220 GHz). In such a scenario, the achieved sensitivity is easier to estimate. One could imagine than the installation of the second module could start two years after the installation of the first one and then, one additional module could be added each year until we reach six modules and operate until 2025 following the example scheme in Table~\ref{tabdep}. The resulting number of (focal planes).(year) is 9 at 90 GHz, 21 at 150 GHz and 20 at 220 GHz. Keeping the same number of focal planes at 150 and 220 GHz allows us to extrapolate the achieved sensitivity from the one that was calculated for a single module while the 9 modules at 90 GHz are supposed to handle the contamination from synchrotron B-polarization. In such a scenario, we could achieve a sensitivity increase of a factor 10 with respect to the 2 (focal planes).(year) we have with the first module, and therefore reach r=0.002
\end{enumerate}
Of course the numbers given above are just for the sake of illustration and definitve decisions will be based on the results of the first module. We will certainly learn a lot from it and may want to upgrade or modify some of its hardware. For instance, the TES bolometers could be replaced by MKIDs if these turn out to become more competitive in the next few years. Various teams in QUBIC are already involved in R\&D on MKIDs.

\begin{table}[h]
\begin{center}
\begin{tabular}{l | l | l}
Module & Start & End \\
\hline
Module 1: 150/220 GHz 	&2018 	&2025 \\
Module 2: 150/220 GHz 	&2019 	&2025 \\
Module 3: 90 GHz 		&2020 	&2025 \\
Module 4: 150 GHz 		&2021 	&2025 \\
Module 5: 220 GHz 		&2022 	&2025 \\
Module 6: 90 GHz 		&2023 	&2025 \\
\end{tabular}
\caption{Possible deployment schedule for future modules (just an example). This would result in 2025 into a number of (focal plane).(year) distributed as follows: 9 at 90 GHz, 21 at 150 GHz and 20 at 220 GHz. \label{tabdep}}
\end{center}
\end{table}

\begin{figure}[h]
  \centering
 \includegraphics[width=8.cm]{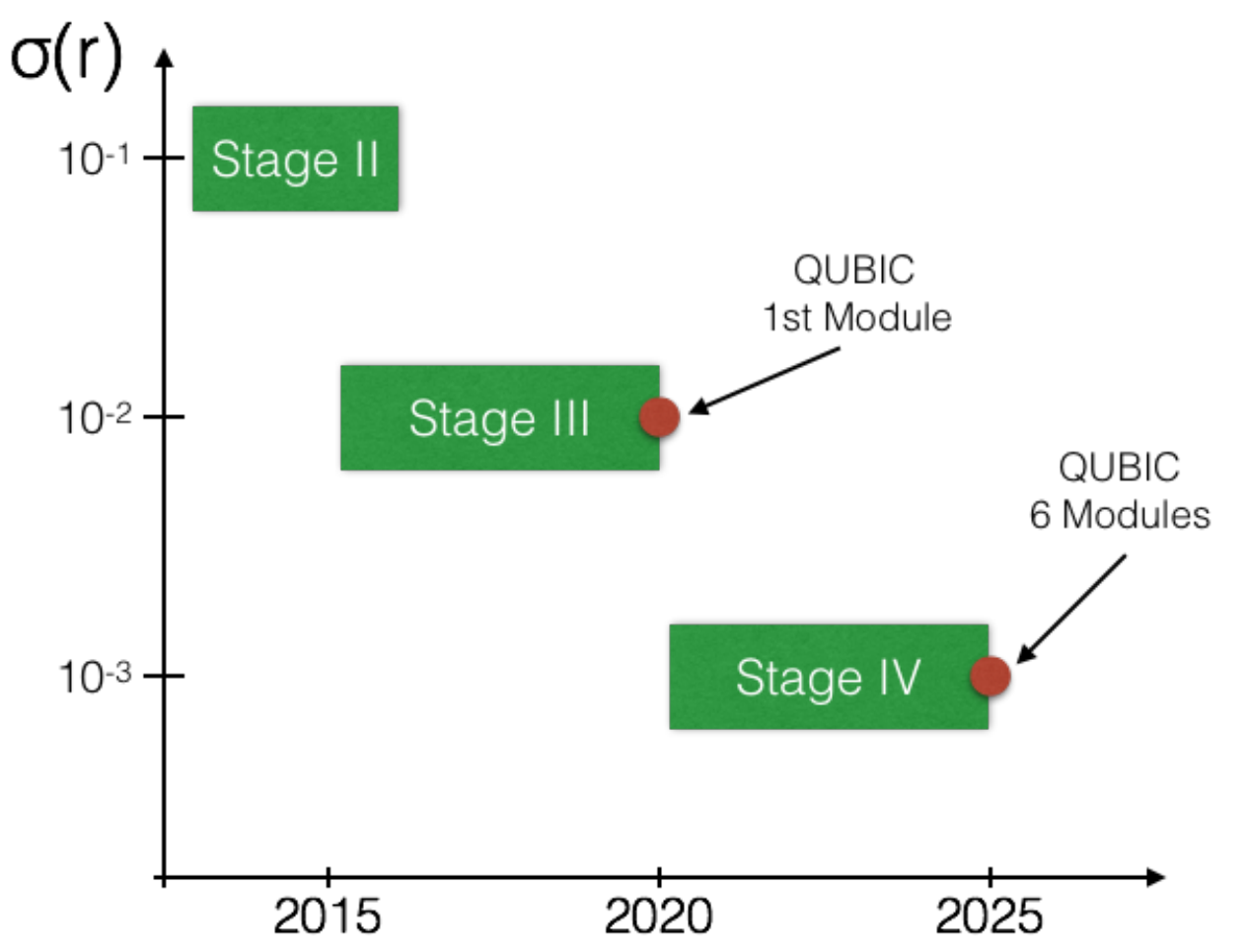} 
\caption{One sigma sensitivities for QUBIC (red dots) compared with the anticipated sensitivities of Stage III and Stage IV experiments. \label{QUBICstages}}
\end{figure}
\newpage

\newpage
\clearpage
\listoffigures 
\clearpage
\listoftables 
\newpage
%auto-ignore

\newpage
\renewcommand\refname{References}

\newpage
%COMMENTED BY OP 
%\section{List of Figures}
%\addcontentsline{toc}{section}{\protect\numberline{}List of Figures}%
%\renewcommand\listfigurename{}
%\listoffigures

%\newpage
%\section{{List of authors [F0E0] J-C or Andrea}}
%\addcontentsline{toc}{section}{\protect\numberline{}List of Authors}%
%To be added when we have converged on the list of Authors/Architects}
%\include{QubicAuthorList}

\newpage
\appendix
%auto-ignore
\section{{Annexe: Experiment location in Dome C }}

\subsection{Studies of the deployment of a QUBIC module at Dome C, Antarctica}
\label{domec}

This annex details the implementation studies that have been performed to install a QUBIC module
at Dome C, which offers exquisite atmospheric
characteristics. 

A QUBIC module may be hosted in the Concordia base, at Dome C, which is ran jointly by French IPEV and
Italian PNRA. The site is located at an altitude of 3233m a.s.l. with coordinates (75° 06[2032] 00[2033] S 123°
19[2032] 58[2033] E). The base was created in 1995 and started winterovering in 2005. It is now one of the three inland
Antarctica stations operating all year through. In summer, Concordia can host around 60 people, while around 15 persons
are present during winter. The Meteorological conditions in Concordia can be very hard, with temperatures as low as
-80C in winter. This limits the access to the site to only \~{}three months per year during Antarctic summer and thus
the implementation of an experiment in Dome C has to be carefully planned in advance (see section \ref{implement_domec}).
Views from the station are shown in Figure~\ref{fig98}.

The QUBIC collaboration has started logistic and site testing operations in Dome C as early as in
2006 with the BRAIN-pathfinder experiment \cite{Polenta}. Since then we have demonstrated
the possibility to use a dry (cryocooler based) cryogenic system 
in such
environmental conditions. We also carried out detailed site atmospheric testing demonstrating the stability of the
atmospheric emission, the low level of opacity and water vapour content, and the lack of polarized emission from the
atmosphere itself~\cite{Battistelli}. During summer 2014-2015 further site testing has been
performed with the CASPER experiment~\cite{depetris}. Thanks to this experiment, we now have information about the short timescale atmospheric fluctuations (sampling time 10s) that are
useful to complement the daily radiosounding data.

%{{Marco,
%please add something here}}}

\begin{figure}
\centering
\includegraphics[width=5.2cm]{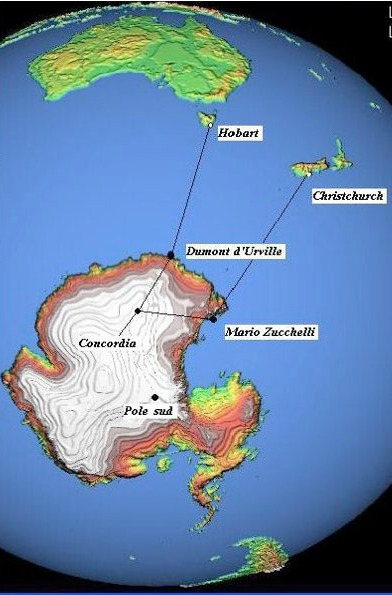} //
\includegraphics[width=10.5cm]{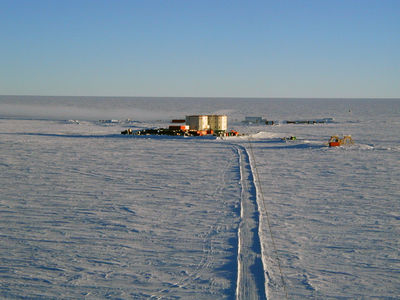} 
\caption{ Location and view of the Concordia Station in Dome C, Antarctica\label{fig98}}
\end{figure}

%{{%The actual location of QUBIC within the Concordia station is still under discussion with IPEV
%and PNRA, based on easiness of access for winter-overers during winter, interferences with other experiments and
%logistics considerations for IPEV and PNRA.}}
The study of the location of a QUBIC module within the Concordia station has been the subject of the 2015-2016 summer campaign. During this campaign a detailed analysis based on the needs of the QUBIC experiment, on the easiness of access for winter-overers during winter, on the interference with other experiments and logistic considerations for IPEV and PNRA has been performed. The discussion is still ongoing but a preferential location has been selected
%A QUBIC module is formed of three main facilities: the experiment module, the calibration tower, and the data acquisition/storage
%unit. The QUBIC experiment will be constituted of two modules: a scientific module mainly dedicated to data acquisition
%and clean lab services and an experiment module where the main instrument will be located and will take data from.%}
(cf. Section~\ref{domec_details}). The
most favored solution appears to be the one labeled as ``QUBIC 3bis'' where a QUBIC module would be installed at 50-70m
North-East from the so-called American Tower which would then be usable for hosting our calibration source and would
prevent us from reinstalling such a tower in Concordia. The distance from the scientific module to the Concordia base would be 950m, a
rather large distance for winter-overers, but this issue would be mitigated by the use of the Astronomy laboratory,
located at a reasonable distance between the module and the station, for data acquisition, storage and control of the
experiment. This location also has the advantage of leaving plenty of freedom for the possible installation of more
QUBIC modules in the future. Flags were installed on this location during the summer campaign 2015-2016 (in Dec. 2015)
by the QUBIC Logistic Coordinator. The ``QUBIC 3bis'' site is shown in Figure~\ref{fig99}.

\begin{figure}
\centering
 \includegraphics[width=7.615cm,height=5.278cm]{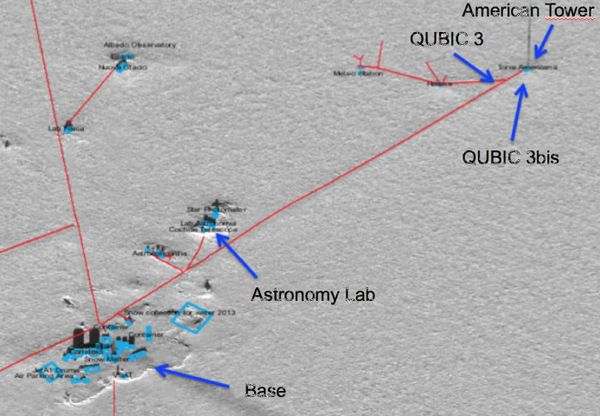}
\includegraphics[width=8.163cm,height=5.196cm]{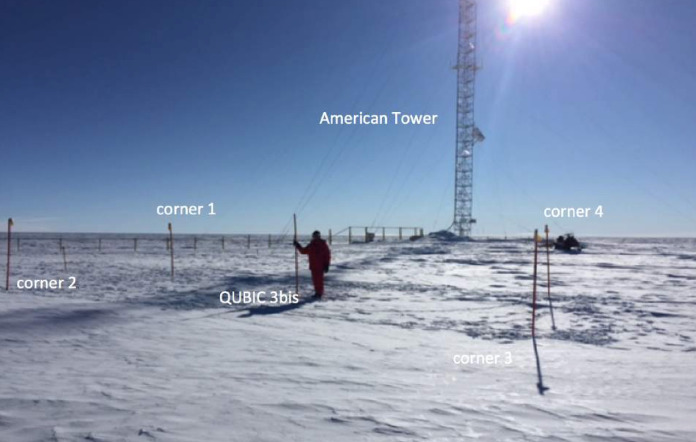} 
\hypertarget{RefHeadingToc314323130}{}
\caption[location of a QUBIC module in the case of QUBIC 3 and 3bis option]{(left) location of a QUBIC module in the case of QUBIC 3 and 3bis options.  (right)  picture of the flags positioned in the QUBIC 3bis option indicating the shelter corners.\label{fig99}}
\end{figure}

\subsubsection{Implementation issues at Dome C}
\label{implement_domec}
In this section we describe our default implementation plan for a QUBIC module in Dome C. Any modification is
possible  in coordination between PNRA, IPEV and the QUBIC collaboration.
% however, for scientific reasons
%(i.e. competition with other groups focusing on the same scientific target) we cannot postpone the experiment
%commissioning after summer 2017-2018, as otherwise it would seriously reduce the scientific impact of QUBIC results,
%although the experiment would still remain significant. For this reason, and considering the financial and technical
%effort already put in place, it is of paramount importance that the commissioning of the experiment and the beginning
%of observations takes place on summer 2017-2018. 

{
The envisaged location for the first module presented in this document has to be suitable to sustain from one
side the full development of QUBIC and from the other side to ensure the minimum impact of the full implemented
project with respect to the other ongoing projects.}

\subsubsection{Milestones}
\hypertarget{RefHeadingToc314322994}{}

\begin{table}
\begin{center}
\begin{tabular}{|m{8.027cm}|m{7.5930004cm}|}
\hline
\multicolumn{2}{|m{15.820001cm}|}{{\bfseries\itshape
MILESTONES/TIMETABLE}
}\\\hline
%\multicolumn{2}{|m{15.820001cm}|}{[]{\bfseries\itshape }
%}
\hline
{\bfseries\itshape ITEM}
 &
{\bfseries\itshape TIME}
\\\hline
%\hline
{ Definition of the experiment site}
 &
{ 09-12/2015}
\\\hline
{ Prep. of the hill for the
instrument and cal. tower}
 &
{ 11/2015-01/2016}
\\\hline
{ Construction of the experiment module
pillars }
 &
{ 12/YY-02/(YY+1)}
\\\hline
{ Routing for the cables (exp. + cal.
tower)}
 &
{ 12/YY-02/(YY+1)}
\\\hline
{ Insertion of the mount concrete pillars}
 &
{ 12/YY-02/(YY+1)}
\\\hline
{ Transportation of the experiment module}
 &
{ 08/YY-02/(YY+1) (R0 or R1 + T2 or T3)}
\\\hline
{ Transportation of the calibration tower}
 &
{ 08/YY-02/(YY+1) (R0 or R1 + T3 or
T1 (YY+1)-(YY+2))}
\\\hline
{ Building the experiment module }
 &
{ 11-12/(YY+1)}
\\\hline
{ Building the calibration tower}
 &
{ 11/(YY+1)-01/(YY+2)}
\\\hline
{ Transportation of the Mount}
 &
{ 08/(YY+1)-12/(YY+1) (Basler)}
\\\hline
{ Transportation of the cryostat}
 &
{ 08/(YY+1)-12/(YY+1) (Basler)}
\\\hline
{ Mounting of the mount+cryostat}
 &
{ 01/(YY+2)}
\\\hline
{ Commissioning and beginning of
observations}
 &
{ 01-02/(YY+2)}
\\\hline
\end{tabular}
\end{center}
\caption{Milestones for the implementation of QUBIC
in Concordia Station, YY stands for the year for which the experiment module would be available for shipping. \label{table29}}
\end{table}

A preliminary installation timetable is shown in Table~\ref{table29}.
The logistic should evaluate if there are additional costs to ensure as much as possible that these milestones can be
respected.

\subsubsection{Materials transportation}

Our plan for materials transportation has been designed assuming the typical time schedule for
delivery of Cargo at Concordia as in  Figure~\ref{fig108}. It is outlined in Table~\ref{table30},
%\todocomment{OP: pas realiste, depart en aout 2016 ?!} 
and its cost is estimated in Table~\ref{table31}. We have assumed a shipping cost of 6 {\texteuro}/kg for ship+traverse shipment and 25 {\texteuro}/kg for ship+Basler
plane shipment.

\begin{figure}
\centering
 \includegraphics[width=16.565cm,height=10.347cm]{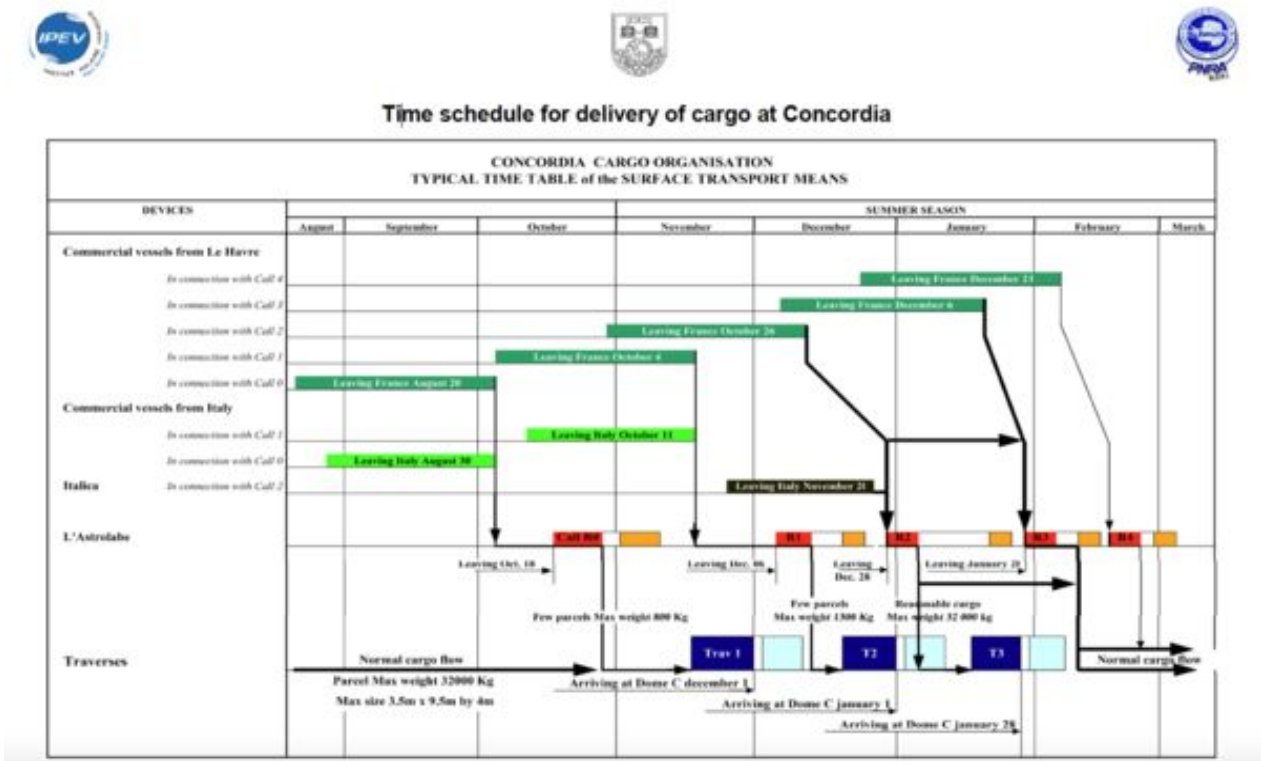} 
\caption{Materials
delivery: typical time schedule for cargo delivery at Concordia.\label{fig108}}
\end{figure}

%{
%We have assumed a shipping cost of 6 {\texteuro}/kg for ship+traverse shipment and 25 {\texteuro}/kg for ship+Basler
%plane shipment.}

\begin{table}
{\scriptsize
\begin{tabular}{|m{4.083cm}|m{3.7619998cm}|m{5.556cm}|m{3.065cm}|}
\hline
\multicolumn{4}{|m{17.065998cm}|}{{\bfseries\itshape
TRANSPORTATION OF MATERIALS}
}\\\hline
\\\hline
{\bfseries\itshape ITEM}
 &
{\bfseries\itshape CRATES AND WEIGHT}
 &
{\bfseries\itshape WAY OF TRANSPORTATION*}
 &
{\bfseries\itshape COSTS**}
\\\hline
\\\hline
{ CALIBRATION TOWER}
 &
{ 5 crates 2,7m x 1,4m x 1,7m each, 600kg each}
{{IMPORTANT: note needed in case the QUBIC 3 location is selected }}
 &
\begin{itemize}
\item { {commercial vessel from Le Havre on August
20}{\textsuperscript{th}}{ 2016}}
\item { Astrolabe R0, R1 2016-2017}
\item { Traverse T3 2016-2017 (in case of use helicopters to traverse the see) or T1 2017-2018;}
\end{itemize}
 &
{{18.000{\texteuro}}}
\\\hline
{ EXPERIMENT MODULE}
 &
 { {total weight
}{$\sim$ 7500kg }}
 &
\begin{itemize}
\item { {commercial vessel from Le Havre on August
20}{\textsuperscript{th}}{ 2016}}
\item { Astrolabe R0, R1 2016-2017}
\item { Traverse \ T2 or T3 2016-2017 (crucial to have the material ready at the beginning of
2017-2018);}
\end{itemize}
 &
{ 45.000{\texteuro}}
\\\hline
{ 1/2 EQUIPMENT}
 &
{ $\sim$10 crates of 0,6m
x 0,7m x 0,7m and 50kg each}
 &
\begin{itemize}
\item { {commercial vessel from Le Havre on August
20}{\textsuperscript{th}}{ 2016}w}{\textsuperscript{th}}
\item { Astrolabe R0, R1 2016-2017}
\item { Traverse T3 2016-2017 (in case of use helicopters to traverse the see) or T1 2017-2018;
}
\end{itemize}
 &
{ 3.000{\texteuro}}
\\\hline
{ 2/2 EQUIPMENT}
 &
{ {$\sim$10 crates of 0,6m
x 0,7m x 0,7m and 50kg each }}
 &
\begin{itemize}
\item { Commercial Vessel on August 2017}
\item { C130 (New Zealand-MZS) + Basler plane MZS-Dome C on November 2017 }
\end{itemize}
 &
{ 12.500{\texteuro}}
\\\hline
{ CRYOSTAT}
 &
 { 1 crate 1,7m x 1,7m x 1,42m; 600kg }
 &
\begin{itemize}
\item { Commercial Vessel on August 2017}
\item { {C130 (New Zealand-MZS) + Basler plane MZS-Dome C on November
2017}}
\end{itemize}
 &
{ 15.000{\texteuro}}
\\\hline
{ MOUNT \ }
 &
 { 2 crates 1,7m x 1,7m x 1,42m; 1000kg each (2 separate flights)}
 &
\begin{itemize}
\item { Commercial Vessel on August 2017}
\item { C130 (New Zealand-MZS) + Basler plane MZS-Dome C on November 2017 }
\end{itemize}
 &
{ 50.000{\texteuro}}
\\\hline
{ TOTAL}
 &
 &
 &
{ 143.500 {\texteuro}}
\\\hline
\end{tabular}
}
\caption{Tentative schedule with a departure date on  August, 20{\textsuperscript{th}} 2016 and estimation of costs for transportation of
materials. \label{table30}}
\end{table}

%{\textbf{\textit{*}}}{\textit
%Our transportation plan has been defined
%assuming the typical time schedule for delivery of Cargo at Concordia as in Figure~\ref{fig108}. 

%{
%{\textbf{\textit{**}}}{\textit{We have assumed a shipping cost of 6
%{\texteuro}/kg for ship+traverse transportation and 25 {\texteuro}/kg for ship+Basler plane transportation.}}}

\subsubsection{Resume of logistics costs}

%\todocomment{some of what follows seems ill-placed ... does not concern site ?}
%{
%The first module of QUBIC experiment is mainly funded. What is missing is:}

%\begin{itemize}
%\item {
%{the telescope mount (300.000{\texteuro}), for which we have an agreement with a Dutch
%consortium;}}
%\item {
%{the finalization of part of the focal plane, requested and pending upon the IN2P3, INSU and
%IPEV review to be held in 2016. This is estimated to be around 270.000{\texteuro} (including the detectors, cables,
%cryogenic mechanics, calibration source at 220GHz);}}
%\item {
%{the logistic and infrastructure costs, including the experiment module, with the exception of
%the calibration system which is funded in France by }{\textit{\textcolor{black}{DIM-ACAV
%2015;}}}}
%\item {
%{\textcolor{black}{
In case of the installation of a QUBIC module in location option 3, we estimate the logistic costs to 
$\sim$ 154.000{\texteuro}.

In the following we have performed an attempt to resume the total logistic costs of a QUBIC module shipped and
installed
at Dome C. This is intended to be a
preliminary estimation to be detailed, validated and recalculated by PNRA and IPEV.

{
Costs assumptions:}

\begin{itemize}
\item {
Shipping cost of 6 {\texteuro}/kg for ship+traverse and 25 {\texteuro}/kg for ship+Basler plane; }
\item {
kWh cost assumed to be 1,52 {\texteuro} per kWh in Concordia;}
\item {
Summer logistic and scientific personnel salary in Concordia 9000 {\texteuro} gross/month (i.e. 300 {\texteuro} per
Man-Day/MD);}
\item {
Winter Over personnel salary in Concordia 12.500 {\texteuro} gross/month;}
\item {
{Part of the costs and manpower assumptions have been derived from~\cite{ipevdoc}.}}
\end{itemize}

{
{For the mount and construction of a QUBIC module we have done the following
assumptions:}}

\begin{itemize}
\item {
Preliminary study: 45 MD on 09/2015-11/2015; }
\item {
Preparation of the surface (hill) for the instrument: 10 MD on summer 2015-2016;}
\item {
Construction of the experiment module piles and concrete pillars: 20 MD on summer (YY-YY+1); }
\item {
Routing for the cables and service (exp. + cal. tower): 40 MD on summer (YY-YY+1);}
\item {
Building the experiment module: 40 MD on summer (YY+1-YY+2); }
\item {
Building the calibration tower: 20 MD on summer (YY+1-YY+2);}
\item {
Mounting the mount and cryostat: 10 MD on summer (YY+1-YY+2);}
\end{itemize}

{
For the scientific personnel we have assumed the presence in Dome C of the following people:}

\begin{itemize}
\item {
1 scientific person on summer 2015-2016 (1 x 30 MD);}
\item {
2 scientific people on summer (YY/YY+1) (2 x 30 MD);}
\item {
6 scientific people on summer (YY+1/YY+2) (6 x 60 MD);}
\item {
1 winter-over during year (YY+2);}
\item {
1 winter-over during year (YY+3).}
\end{itemize}

{
{In Table~\ref{table31} we have listed approximate logistic costs deriving from:}}

\begin{itemize}
\item {
isolated experiment module design and procurement;}
\item {
transportation as detailed above in section 4;}
\item {
mounting and construction of the 1st module as detailed above;}
\item {
calibration tower construction;}
\item {
scientific personnel;}
\item {
power and service connections;}
\item {
running costs for 2 years including winterovers.}
\end{itemize}

\begin{table}
{\scriptsize
\begin{tabular}{|m{4.492cm}|m{1.55cm}|m{1.55cm}|m{1.5519999cm}|m{1.55cm}|m{1.55cm}|m{1.6099999cm}|m{1.8179998cm}|}
\hline
{\bfseries ITEM}
 &
{\bfseries Timing}
 &
{\bfseries 2015-16}
 &
{\bfseries YY-YY+2}
 &
{\bfseries YY+1-YY+2}
 &
{\bfseries YY+2-YY+3}
 &
{\bfseries YY+3-YY+4}
 &
{\bfseries Total cost}
\\\hline
%\\\hline
{\bfseries\itshape Experiment module:}
 &
{ }
 &
{ }
 &
{ }
 &
{ }
 &
{ }
 &
{ }
 &
{\bfseries 130.000{\texteuro}}
\\\hline
{\itshape Structure on pillars}
 &
{ 1-7/2016}
 &
{ 65.000{\texteuro}}
 &
{ }
 &
{ }
 &
{ }
 &
{ }
 &
{ }
\\\hline
{\itshape Isolated module}
 &
{ 1-7/2016}
 &
{ 45.000{\texteuro}}
 &
{ }
 &
{ }
 &
{ }
 &
{ }
 &
{ }
\\\hline
{\itshape Service (electricity,...)}
 &
{ 1-7/2016}
 &
{ 10.000{\texteuro}}
 &
{ }
 &
{ }
 &
{ }
 &
{ }
 &
{ }
\\\hline
{\itshape Flexible cover (Boot)}
 &
{ 1-7/2016}
 &
{ 10.000{\texteuro}}
 &
{ }
 &
{ }
 &
{ }
 &
{ }
 &
{ }
\\\hline\hline
%{ }
% &
%{ }
% &
%{ }
% &
%{ }
% &
%{ }
% &
%{ }
% &
%{ }
% &
%{ }
%\\\hline
{\bfseries\itshape Transportation:}
 &
{ }
 &
{ }
 &
{ }
 &
{ }
 &
{ }
 &
{ }
 &
{\bfseries 143.500{\texteuro}}
\\\hline
{\itshape Calibration tower (traverse)}
 &
{ sum (YY/YY+1)}
 &
{ }
 &
{ 18.000{\texteuro}}
 &
{ }
 &
{ }
 &
{ }
 &
{ }
\\\hline
{\itshape Exp. module (traverse)}
 &
{ sum (YY/YY+1)}
 &
{ }
 &
{ 45.000{\texteuro}}
 &
{ }
 &
{ }
 &
{ }
 &
{ }
\\\hline
{\itshape 1/2 Equipment (traverse) }
 &
{ sum (YY/YY+1)}
 &
{ }
 &
{ 3.000{\texteuro}}
 &
{ }
 &
{ }
 &
{ }
 &
{ }
\\\hline
{\itshape 2/2 Equipment (Basler) }
 &
{ sum (YY+1/YY+2)}
 &
{ }
 &
{ }
 &
{ 12.500{\texteuro}}
 &
{ }
 &
{ }
 &
{ }
\\\hline
{\itshape Cryostat (Basler)}
 &
{ sum (YY+1/YY+2)}
 &
{ }
 &
{ }
 &
{ 15.000{\texteuro}}
 &
{ }
 &
{ }
 &
{ }
\\\hline
{\itshape Mount (Basler)}
 &
{ sum (YY+1/YY+2)}
 &
{ }
 &
{ }
 &
{ 50.000{\texteuro}}
 &
{ }
 &
{ }
 &
{ }
\\\hline
%{ }
% &
%{ }
% &
%{ }
% &
%{ }
% &
%{ }
% &
%{ }
% &
%{ }
% &
%{ }
%\\
\hline
{\bfseries\itshape 1st module construction:}
 &
{ }
 &
{ }
 &
{ }
 &
{ }
 &
{ }
 &
{ }
 &
{\bfseries 39.900{\texteuro}}
\\\hline
{\itshape Preliminary study (45 MD)}
 &
{ 9-12/2015}
 &
{ 13.500{\texteuro}}
 &
{ }
 &
{ }
 &
{ }
 &
{ }
 &
{ }
\\\hline
{\itshape Surface preparation (10 MD)}
 &
{ sum 15-16}
 &
{ 3.000{\texteuro}}
 &
{ }
 &
{ }
 &
{ }
 &
{ }
 &
{ }
\\\hline
{\itshape Platform n pillars (20 MD)}
 &
{ sum (YY/YY+1)}
 &
{ }
 &
{ 6.000{\texteuro}}
 &
{ }
 &
{ }
 &
{ }
 &
{ }
\\\hline
{\itshape Exp. mod. construction (40 MD)}
 &
{ sum (YY+1/YY+2)}
 &
{ }
 &
{ }
 &
{ 10.800{\texteuro}}
 &
{ }
 &
{ }
 &
{ }
\\\hline
{{\textit{Connections (12 MD) }}}
 &
{ sum (YY+1/YY+2)}
 &
{ }
 &
{ }
 &
{ 3.600{\texteuro}}
 &
{ }
 &
{ }
 &
{ }
\\\hline
{\itshape Experiment mounting (10 MD)}
 &
{ sum (YY+1/YY+2)}
 &
{ }
 &
{ }
 &
{ 3.000{\texteuro}}
 &
{ }
 &
{ }
 &
{ }
\\\hline
%{ }
% &
%{ }
% &
%{ }
% &
%{ }
% &
%{ }
% &
%{ }
% &
%{ }
% &
%{ }
%\\
\hline
{\bfseries\itshape Cal. tower construction (20MD)}
 &
{ }
 &
{ }
 &
{ }
 &
{ }
 &
{ }
 &
{ }
 &
{\bfseries 6.000{\texteuro}}
\\\hline
{ }
 &
{ sum (YY+1/YY+2)}
 &
{ }
 &
{ }
 &
{ 6.000{\texteuro}}
 &
{ }
 &
{ }
 &
{ }
\\\hline
%{ }
% &
%{ }
% &
%{ }
% &
%{ }
% &
%{ }
% &
%{ }
% &
%{ }
% &
%{ }
%\\
\hline
{\bfseries\itshape Scientific personnel (except wo):}
 &
{ }
 &
{ }
 &
{ }
 &
{ }
 &
{ }
 &
{ }
 &
{\bfseries 135.000{\texteuro}}
\\\hline
{\itshape 1 scientific (1 x 30MD)}
 &
{{sum (YY/YY+1)}}
 &
{ 9.000{\texteuro}}
 &
{ }
 &
{ }
 &
{ }
 &
{ }
 &
{ }
\\\hline
{\itshape 2 scientific (2 x 30MD)}
 &
{ sum (YY+1/YY+2)}
 &
{ }
 &
{ 18.000{\texteuro}}
 &
{ }
 &
{ }
 &
{ }
 &
{ }
\\\hline
{\itshape 6 scientific (6 x 60MD)}
 &
{ sum (YY+2/YY+3)}
 &
{ }
 &
{ }
 &
{ 108.000{\texteuro}}
 &
{ }
 &
{ }
 &
{ }
\\\hline
\hline
{\bfseries\itshape Service connections:}
 &
{ }
 &
{ }
 &
{ }
 &
{ }
 &
{ }
 &
{ }
 &
{\bfseries 13.800{\texteuro}}
\\\hline
{\itshape power (20 MD)}
 &
{ sum (YY/YY+1)}
 &
{ }
 &
{ 6.000{\texteuro}}
 &
{ }
 &
{ }
 &
{ }
 &
{ }
\\\hline
{\itshape other connections (20 MD)}
 &
{ sum (YY/YY+1)}
 &
{ }
 &
{ 6.000{\texteuro}}
 &
{ }
 &
{ }
 &
{ }
 &
{ }
\\\hline
{{\textit{organization (6 MD)}}}
 &
{ sum (YY/YY+1)}
 &
{ }
 &
{ 1.800{\texteuro}}
 &
{ }
 &
{ }
 &
{ }
 &
{ }
\\\hline
\hline
{\bfseries\itshape Running costs (2 years):}
 &
{ }
 &
{ }
 &
{ }
 &
{ }
 &
{ }
 &
{ }
 &
{\bfseries 699.456{\texteuro}}
\\\hline
{{\textit{winter-over
(1}}{\textit{\textsuperscript{st}}}{\textit{ season)}}}
 &
{ win YY+2}
 &
{ }
 &
{ }
 &
{ }
 &
{ 150.000{\texteuro}}
 &
{ }
 &
{ }
\\\hline
{\itshape power cons. (15kW x 365 days)}
 &
{ win YY+2}
 &
{ }
 &
{ }
 &
{ }
 &
{ 199.728{\texteuro}}
 &
{ }
 &
{ }
\\\hline
{{\textit{\textcolor{black}{winter-over
(2}}}{\textit{\textcolor{black}{\textsuperscript{nd}}}}{\textit{\textcolor{black}{
season)}}}}
 &
{ win YY+3}
 &
{ }
 &
{ }
 &
{ }
 &
{ }
 &
{ 150.000{\texteuro}}
 &
{ }
\\\hline
{{\textit{\textcolor{black}{power cons. (15kW x 365
days)}}}}
 &
{ win YY+3}
 &
{ }
 &
{ }
 &
{ }
 &
{ }
 &
{ 199.728{\texteuro}}
 &
{ }
\\\hline
\hline
{\bfseries\itshape TOTAL COSTS:}
 &
{\bfseries }
 &
{\bfseries 155.500{\texteuro}}
 &
{\bfseries 103.800{\texteuro}}
 &
{\bfseries 208.900{\texteuro}}
 &
{\bfseries 349.728{\texteuro}}
 &
{\bfseries 349.728{\texteuro}}
 &
{\bfseries 1.167.656{\texteuro}}
\\\hline
\end{tabular}
}
\caption{Logistic
costs: Logistic costs foreseen for the QUBIC experiment. We have divided the costs within the years. \label{table31}}

\end{table}
%\subsubsection{{Experiment location in Dome C [F0E0] Elia}}

%\subsubsection{{Buildings and facilities [F0E0] Elia}}
%\hypertarget{RefHeadingToc314322962}{}{

%\label{domeclocation}
%\hypertarget{RefHeadingToc314322961}{}\label{bkm:Ref314312141}{

\subsection{Possible implementation at Dome C}
\label{domec_details}
For a module location in Dome C three options are envisaged, the most advisable of which is
by the American Tower, 57m toward east (QUBIC 3bis option). Exact position has to be defined by IPEV, PNRA and QUBIC
collaboration considering needs of other experiments in the surrounding. See Figure \ref{antarticamap} for a detailed map of the area. 
\begin{figure}[h]
{\centering  \includegraphics[width=.5\textwidth]{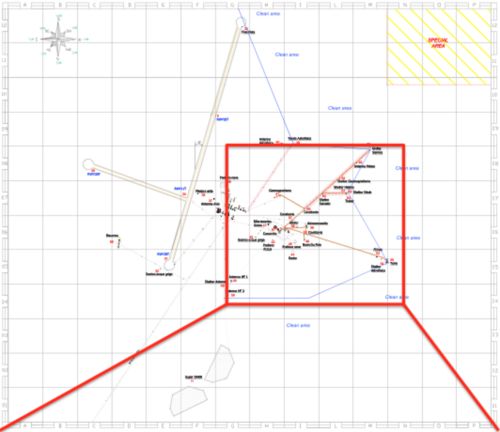}
\par}
{\centering  \includegraphics[width=.5\textwidth]{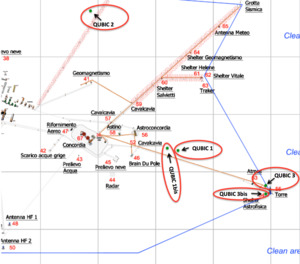}
\par}

\caption{\label{antarticamap} Map of
the Antarctic base Concordia-Dome C with an inset where the possible locations of a QUBIC module are
highlighted.}

\end{figure}

During 2012-2013 summer campaign, the QUBIC collaboration 
has started a detailed and fruitful scouting activity in Concordia for the definition of the QUBIC experiment localization. During
the 2015-2016 summer campaign, the QUBIC collaboration was represented, in Dome C by its logistic
coordinator, and this campaign aimed at finalizing this activity in the most coordinated and
agreed way between the QUBIC collaboration, IPEV, PNRA as well as the other experiments and collaborations present or 
planned in Concordia station.

The conclusion of this study is that the QUBIC 3bis location seems to us the best trade-off of all possible solutions from the point of view of the
interference problems, quality of the QUBIC observations, and for timing and economic reasons.

\begin{itemize}
\item {
{\textbf{QUBIC 1 option}}{ is located approximately 450m from the Dome
C base toward the west direction. The 45m calibration tower needs to be located \~{}50m from the experiment, at its
west. The idea behind this solution, is to position a QUBIC module at 100m from the Astronomy lab toward west,
and use the Astronomy lab itself as data acquisition/storage with minimal continuous occupation. The possible
interference between QUBIC, IRAIT, BSRN and the experiments located in the shelter Physics, shelter Atmo, and shelter
Astronomy have been investigated and have driven the site localization in that area. Despite most of the experiments
have minimal or null impact for QUBIC, there is some visual impact, especially for IRAIT, which could however be
minimized with a distance of the calibration tower \~{}150m and with BSRN which may need to be repositioned. The QUBIC
1bis solution is very similar and relies on the fact to reuse an existing facility, the Star-Photomer/German Dome,
sited approximately 9m west of IRAIT, whose shelter does not fit the QUBIC requirements but the platform sits on a
structure of three lightened concrete pillars identical to the one designed for the QUBIC platform. The main draw-back
of the QUBIC 1bis solution is the optical pollution produced by QUBIC to IRAIT and vice-versa. This has been simulated
and found to be minimized (although not cancelled) if ground shields are placed and the relative position between QUBIC
and IRAIT allows each experiment to keep the other below 15° elevation. Both these solutions (QUBIC1 and 1bis) are
technically feasible and the QUBIC collaboration foresees them as positive. We should however bear in mind that they
result in a compromise which would reduce the scientific capabilities both of QUBIC and of IRAIT and may require the
repositioning of BSRN.}}
\end{itemize}

\begin{table}
\begin{center}
\begin{tabular}{|m{7.1540003cm}|m{7.715cm}|}
\hline
{\bfseries Advantages / Gain} &
{\bfseries Disadvantages / Mitigation}\\\hline
~
 &
~
\\\hline
{ Short distance from the base / Easier operations for winter-over} &
{ Less immune from Dome C base optical pollution due to its vicinity to the Base}\\\hline
{ Only one shelter to build (i.e. the data acquisition module would be the Astronomy lab.) } &
{ Interaction with BSRN: possible need to reposition the experiment}\\\hline
{ Large bandwidth connection between experiment module and Astronomy lab} &
{ Interaction with IRAIT: reduce the sky area directly accessible to IRAIT }\\\hline
{ For QUBIC 1bis: reuse the existing concrete pillars structure} &
{ For QUBIC 1bis: direct air pollution on and from IRAIT}\\\hline
~
 &
{ Need to build the calibration tower}\\\hline
~
 &
{ Not straight-forward to have a plan for the hypothetical QUBIC next generation formed by 6
identical modules}\\\hline
\end{tabular}
\end{center}
\caption{Pro and cons of QUBIC 1 option}
\end{table}

\begin{itemize}
\item {
{\textbf{QUBIC 2 option}}{ was already selected during 2012-2013
summer campaign. In this option the experiment would be located 650m south form the Dome C base and the calibration
tower either toward east or toward west, 50m from the experiment itself. The interference between geo-magnetic and
seismology activities have been studied and an agreement has been found implying the construction of a new road going
around the magnetic quite area and the necessity to build the QUBIC calibration tower in aluminium in order to reduce
magnetic interference. This region is unfortunately currently not well served by infrastructure and, in addition, QUBIC
implementation would require the construction of two shelters, one for the experiment, and one as data
acquisition/storage. This service would be expensive and time consuming especially in the light of the constraints set
by other impacted activities. Unless there are no other choices, at the present status we consider this option not
viable because it would be too expensive.}}
% and does not fit the current schedule of QUBIC experiment (i.e. beginning of
%observations is foreseen during summer campaign 2017-2018).}}
\end{itemize}

\begin{table}
\centering
\begin{minipage}{15.21cm}
\begin{center}
\tablefirsthead{}
\tablehead{}
\tabletail{}
\tablelasttail{}
\begin{tabular}{|m{6.743cm}|m{8.067cm}|}
\hline
{\bfseries Advantages / Gain} &
{\bfseries Disadvantages / Mitigation}\\\hline
~
 &
~
\\\hline
{ quite and free-from-interference observational site} &
{ need to build a new road going around the Amagnetic area}\\\hline
{ straight-forward to implement the hypothetical next generation 6-modules experiment} &
{ expensive and difficult operation due to the constraints created by the interference with
geomagnetic operations }\\\hline
~
 &
{ need to build a calibration tower}\\\hline
~
 &
{ need to build 2 brand new shelters}\\\hline
\end{tabular}
\end{center}
\end{minipage}
\caption{ Pro and cons of QUBIC 2 option}
\end{table}
\begin{itemize}
\item {
{\textbf{QUBIC 3 option}}{ (see Figure QUBIC 3) foresees a QUBIC module installation
at 950m west from the Dome C base, by the existing American Tower. The idea behind this solution, is to
position the QUBIC experiment shelter as close as possible to the American tower (\~{}60m from it) at its east, clearly
outside the clean area, in order to use the American Tower as Calibration Tower and being able to observe the
calibration source with as high as possible elevation angle. The use of the American Tower as calibration tower would
result in no need to build an additional tower with clear economic and time advantages. The impact of our calibration
source on the tower would be minimal. Also, we foresee the possibility to use the Astronomy lab as data
acquisition/storage with minimal continuous occupation. Within this }{solution we selected two
possible locations: QUBIC 3 location would result at 60/70 m from the American Tower, south-east of it, while QUBIC
3bis location would be at 50/60m from the Tower, north-east from it. From a recognition performed in order to establish
the possible interference between QUBIC and other experiments mounted on the American Tower, it seems clear that,
despite most of the experiments have no impact to and from QUBIC, the positioning of QUBIC toward south (main wind
direction in Dome C, QUBIC 3 option) would have an impact due to the change of the snow accumulation in the surrounding
area (although QUBIC would anyway be outside the clean area). On the other hand, positioning a QUBIC module on the north(-east)
side of the tower (QUBIC 3bis option), designing the experiment module in such a way that the snow accumulation would
be kept under control, and keeping the experiment module as low as possible, would result in negligible effect on the
activities on-going from the American Tower. There are several other advantages of this solution including the immunity
from air pollution due to the Dome C base. The main draw back of this solution is the distance from the base. For this
reason, the instrument operations will be totally remotized and the winterover operations will be limited to
three-times-a-week visits to the experiment besides the hypothesis of instrumental emergencies. }}
\end{itemize}

This solution (QUBIC 3bis) seems to us the best trade-off of all possible solutions from the point of view of the
interference problems, quality of the QUBIC observations, and for timing and economic reasons.

\begin{table}
%\centering
%\begin{minipage}{15.055cm}
\begin{center}
%\tablefirsthead{}
%\tablehead{}
%\tabletail{}
%\tablelasttail{}
\begin{tabular}{|m{6.665cm}|m{7.99cm}|}
\hline
{\bfseries Advantages / Gain} &
{\bfseries Disadvantages / Mitigation}\\\hline
~
 &
~
\\
\hline
{ Use the American Tower as Calibration tower / no need to build another 45m tower} &
{ Distance from the base $\sim$ 950m with more complicated winterover procedures / fully remote
experiment + dedicated winterover}\\\hline
{ Free from optical pollution due to the Dome C base} &
{ Dedicated/redundant network connection needed}\\\hline
{ Use of the Astronomy lab and data acquisition-storage unit / no need to build another data
acquisition shelter} &
{ Interference with experiment on the American Tower / place the experiment on the north-east
side (QUBIC 3bis) at $\sim$ 60m form the tower + build the experiment shelter as low as possible taking care of the snow
accumulation}\\\hline
{ Freedom in the planning of the hypothetical next generation 6-modules QUBIC } &
~
\\
\hline
\end{tabular}
\caption{ Pro and cons of QUBIC 3 option}
\end{center}
%\end{minipage}
\end{table}

\begin{table}
\begin{tabular}{|l|l|l|l|}
\hline
{ LOCATION} &
{ Lat.} &
{ Lon.} &
{ Notes}\\\hline

{ BASE} &
{ S 75° 06' 00.1'{}'} &
{ E 123° 19' 57.4'{}'} &
~
\\\hline
{ ASTRONOMY LAB} &
{ S 75° 05' 59.3'{}'} &
{ E 123° 19' 13.4'{}'} &
{ \~{}350m from the base}\\\hline
{ AMERICAN TOWER} &
{ S 75° 05' 50.7'{}'} &
{ E 123° 17' 58.4'{}'} &
{ \~{}970m from the base}
~
\\\hline
{ \color{orange}QUBIC 1} &
{ \color{orange}S 75° 05' 58.2'{}'} &
{ \color{orange}E 123° 18' 59.7'{}'} &
{ \color{orange}at $\sim$100m from Astronomy Lab}\\\hline
{ \color{orange}CALIBRATION TOWER 1} &
{ \color{orange}S 75° 05' 57.4'{}'} &
{ \color{orange}E 123° 18' 53.0'{}'} &
{ \color{orange}$\sim$50m from the experiment}\\\hline
{\color{orange} QUBIC 1bis} &
{ \color{orange}S 75° 06' 01.5'{}'} &
{ \color{orange}E 123° 19' 09.5'{}'} &
{ \color{orange}reuse German-Dome platform}\\\hline
{ \color{orange}CALIBRATION TOWER 1bis} &
{ \color{orange}S 75° 06' 01.5'{}'} &
{ \color{orange}E 123° 19' 03.3'{}'} &
{ \color{orange}$\sim$50m from the experiment}\\\hline
{ \color{red}QUBIC 2} &
{ \color{red}S 75° 06' 22.6'{}'} &
{ \color{red}E 123° 19' 48.7'{}'} &
{ \color{red}$\sim$650m from the base}\\\hline
{ \color{red}CALIBRATION TOWER 2} &
{ \color{red}S 75° 06' 22.6'{}'} &
{ \color{red}E 123° 19' 42.5'{}'} &
{ \color{red} $\sim$ 50m from the experiment}\\\hline
{ \color{orange}QUBIC 3} &
{ \color{orange}S 75° 05' 51.9'{}'} &
{ \color{orange}E 123° 18' 07.1'{}'} &
{ \color{orange}$\sim$950m form the base}\\\hline
{ \color{green}QUBIC 3bis} &
{ \color{green}S 75° 05' 50.3'{}'} &
{ \color{green}E 123° 18' 05.5'{}'} &
{ \color{green}reduce interference}\\\hline
{ \color{green}QUBIC 3bis corner 1} &
{ \color{green}S 75° 05' 50.5'{}'} &
{\color{green} E 123° 18' 05.1'{}'} &
{ \color{green}SW corner of the shelter}\\\hline
{ \color{green}QUBIC 3bis corner 2} &
{ \color{green}S 75° 05' 50.5'{}'} &
{ \color{green}E 123° 18' 06.2'{}'} &
{ \color{green}SE corner of the shelter}\\\hline
{ \color{green}QUBIC 3bis corner 3} &
{ \color{green}S 75° 05' 50.0'{}'} &
{ \color{green}E 123° 18' 06.1'{}'} &
{ \color{green}NE corner of the shelter}\\\hline
{ \color{green}QUBIC 3bis corner 4} &
{ \color{green}S 75° 05' 50.0'{}'} &
{ \color{green}E 123° 18' 05.2'{}'} &
{ \color{green}NW corner of the shelter}\\\hline
{ \color{green}CALIBRATION TOWER 3/ } &
{ \color{green}S 75° 05' 50.7'{}'} &
{ \color{green}E 123° 17' 58.4'{}'} &
{ \color{green}no need to build another 45m tower}\\
{ \color{green}AMERICAN TOWER } & & & \\ \hline
\end{tabular}

\caption{
Possible locations of a QUBIC module
with coordinates. In red we highlight not viable solutions (unless no other options are possible), in orange 
we highlight possible solutions with some limitations and loss on efficiency, in green we highlight our preferred
solution which, from our point of view, is the best trade-off solution with minimal interference. 
}
\end{table}

\newpage
\section{ATRIUM-77335}


\section*{Supplementary material}

# B   ATRIUM-77335

Living document

| | SPECIFICATIONS / REQUIREMENTS | Réf : | I-042261 |
|---|---|---|---|
| 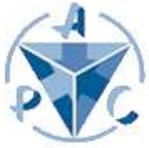 | | Version : | 1 |
| | | Date : | 03/11/2015 |
| | | Page : | 1/5 |

# Requirements for implantation of first module of QUBIC in San Antonio de los Cobres

| General Specifications | | |
|---|---|---|
| **Item** | **Requirement** | **Comment** |
| Site Altitude | 4969m a.s.l. | |
| Site location | S : 24 deg 11' 11.71 ''<br>O : 66 deg 28' 40.79'' | Plateau visited on Sept 22nd above LLAMA site |
| Electrical Power | 25 kW | Pulse Tube Cooler : 10 kW<br>Electronics/Computer : 5 kW<br>Mount (estimation) : 10 kW |
| Internet Bandwidth | 5 MB/sec | To San Antonio de los Cobres and to the rest of the world |

| Infrastructure on the mountain | | |
|---|---|---|
| **Item** | **Requirement** | **Comment** |
| Road | Road to access the QUBIC site from the LLAMA one : a few hundred meters to be added (To be specified by experts) | |
| Planar area | 100m in the EW direction<br>20m in the NS direction | Concrete ground or not depends on soil quality. To be assessed by experts accounting for Instrument + mount weight |
| Shelter for Instrument | • 8m in the EW direction<br>• 5.5m in the NS direction<br>• Basement supports the cryostat + mount ~ 3 tons<br>• Window in the ceiling to point the cryostat window outwards<br>• Circular ground shield installed on the roof to prevent straylight from the ground on the cryostat window. | • Large width is chosen to have the ability to add more modules later<br>• Hole in the window needs an « insulation coat » making it thermally/water/sand insulated (à la BICEP2 – see images below)<br>• detailed specifications of the ground shield being finalized by Marco de Petris, Roma- see pictures for an example<br>• May also be used as a temporary shelter for people working on the site. |
| Compressor Shelter | 3x3m ~ 15-20m from the Instrument Shelter | Will host the compressors and noisy equipment.<br>May also be used as a temporary shelter for people working on the site. |
| Calibration Tower | 45 m height<br>45 meters North from Module | Hardware Funded by France DIM-ACAV. Manpower for installation needed. |

| | SPECIFICATIONS / REQUIREMENTS | Réf : | I-042261 |
|---|---|---|---|
| 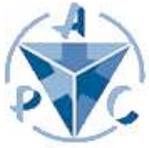 | | Version : | 1 |
| | | Date : | 03/11/2015 |
| | | Page : | 2/5 |

| Infrastructure in San Antonio de los Cobres (inside LLAMA building) | | |
|---|---|---|
| **Item** | **Requirement** | **Comment** |
| Offices and labs | 50 sqm permanently allocated to QUBIC | |
| Clean Room | Access when needed for instrument upgrade | |
| Bedrooms | Two bedrooms permanently | |
| Kitchen / Restrooms / common facilities | Shared access with LLAMA | |

| Integration / Operations | | |
|---|---|---|
| **Item** | **Requirement** | **Comment** |
| Integration | <ul><li>Personal from Argentina and their QUBIC partners to perform integration</li><li>Cranes, trucks, transportation in Argentina to be funded by Argentina</li><li>Integration hangar (around 50 sqm) in Salta (or San Antonio de los Cobres) to integrate and test Mount+Instrument prior installation on site.</li></ul> | |
| Operation | Two trained technicians permanently in San Antonio de los Cobres with regular visits to the instrument (frequency and duties to be detailed later) | |
| Cryogenic Maintenance | Under the responsibility of San Carlos de Bariloche Team | |

Important questions to be answered :
- Assess cost of the above infrastructure
- Assess date at which it could be available for QUBIC installation
- Precise various aspects of the above table by interaction with experts (especially J. Viramonte)

## Illustrations:

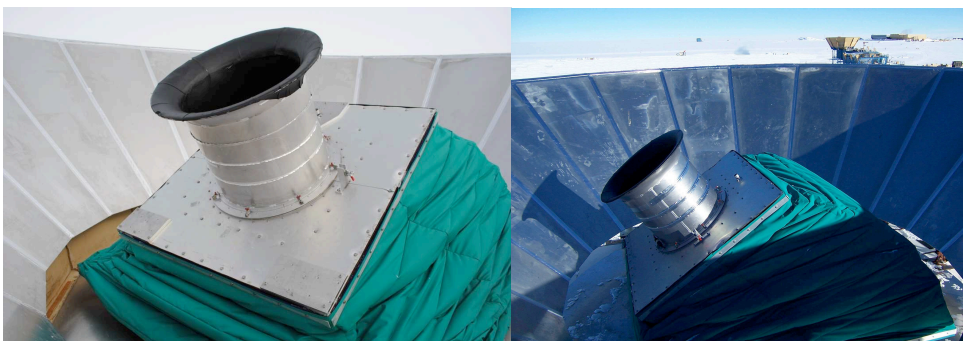

**Figure 1: BICEP2 Insulating coat, forebaffle (black) and ground shield (metal)**

| | | Réf : | I-042261 |
|---|---|---|---|
| | SPECIFICATIONS / REQUIREMENTS | Version : | 1 |
| | | Date : | 03/11/2015 |
| | | Page : | 3/5 |

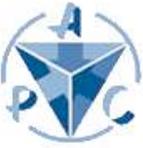

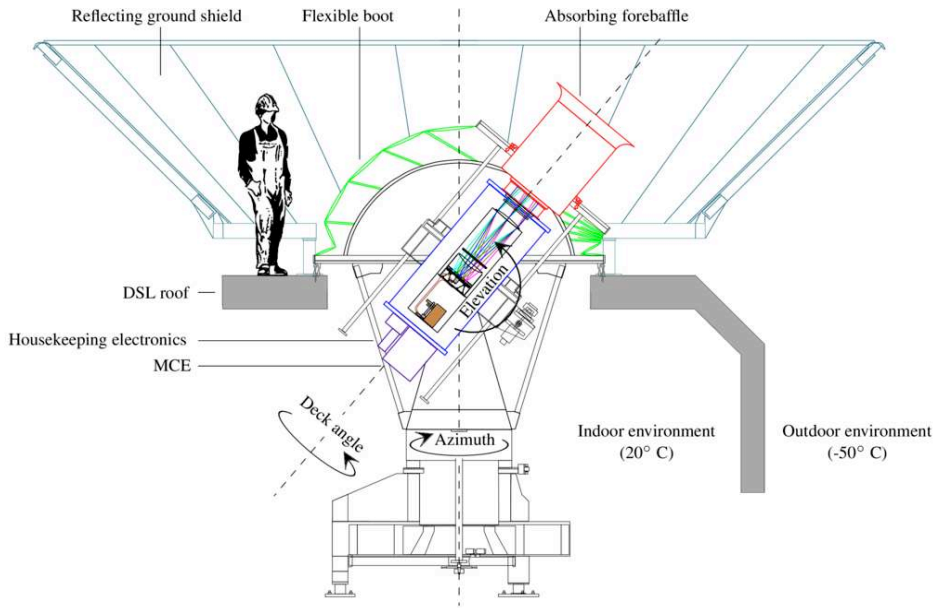

**Figure 2: BICEP2 Shelter + mount sketch**

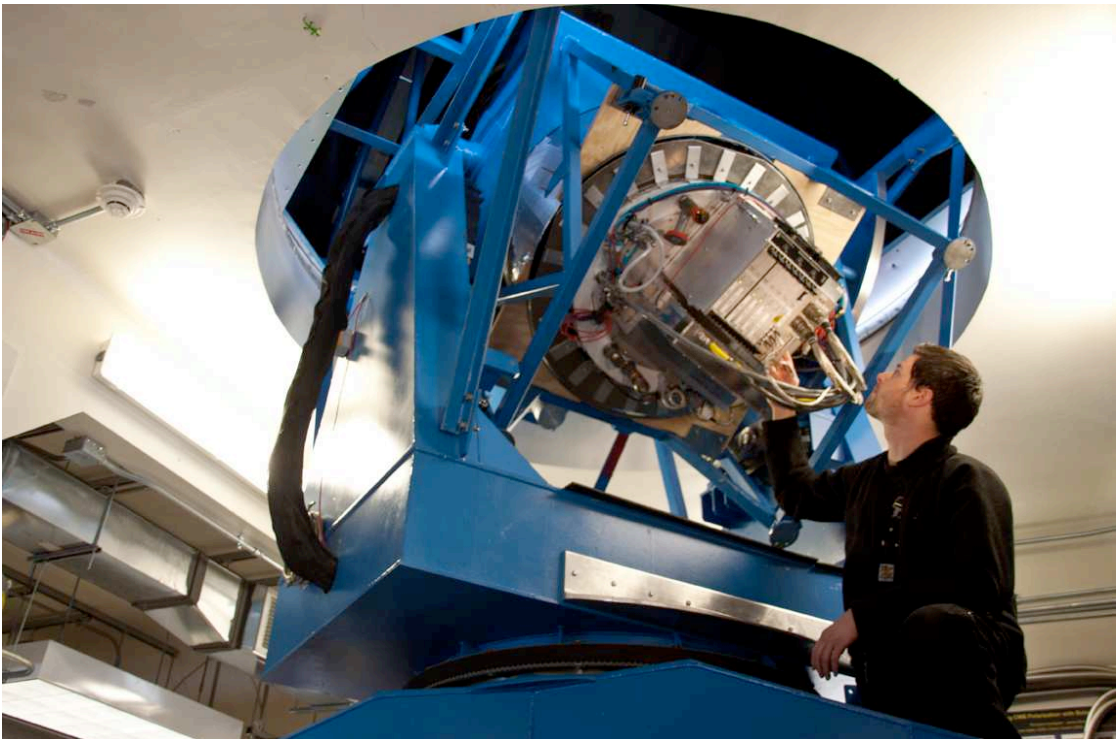

**Figure 3: BICEP2 inside of the shelter with view of the ceiling hole, mount and instrument**

| | SPECIFICATIONS / REQUIREMENTS | Réf : | I-042261 |
|---|---|---|---|
| | | Version : | 1 |
| | | Date : | 03/11/2015 |
| | | Page : | 4/5 |

Figure 4: QUBIC implantation map at Alto Chorillo



Figure 5 : schematic front view of the instrument module

Living document



\newpage
\todototoc
\listoftodos

\end{document}